%% file: arik_thesis.tex
\documentclass[defaultstyle, 11pt]{thesis}

\usepackage{hyperref}		% PDF hyperreferences??
\usepackage{amsmath,amsfonts,amssymb,amsthm}
\usepackage{mathtools}
\usepackage{stmaryrd}       % knuth
\usepackage{dsfont}         % font for \one
\usepackage{expdlist}
\usepackage{physics}
\usepackage{verbatim}
\usepackage{xspace}
\usepackage{subfigure}
\usepackage{scalerel,stackengine}
\usepackage{ulem} %to strike the words
\usepackage{caption}

\usepackage{graphicx,xcolor}
\hypersetup{pdfstartview=}
\usepackage{url}

\usepackage{verbatim}
\usepackage{alltt}
\usepackage{moreverb}

\usepackage{xr} 
\usepackage{float}

\title{Quantum State Characterization Using Measurement Configurations Inspired by Homodyne Detection}

\author{Arik}{Avagyan}

\otherdegrees{B.A., University of Illinois Urbana-Champaign, 2014}

\degree{Doctor of Philosophy}		%  #1 {long descr.}
	{Ph.D., Physics}		%  #2 {short descr.}

\dept{Department of}			%  #1 {designation}
	{Physics}		%  #2 {name}

\advisor{}{Emanuel Knill and Scott Glancy}			%  #2 {name}
\reader{Joshua Combes}		%  2nd person to sign thesis
\readerThree{Murray Holland}		%  3rd person to sign thesis
\readerFour{L. Krister Shalm}
\readerFive{James K. Thompson}

\abstract{  \OnePageChapter	% because it is very short

  In the standard homodyne configuration, an unknown optical state is
  combined with a local oscillator (LO) on a beam splitter (BS).  Good
  quadrature measurements require a high-amplitude LO and two
  high-efficiency photodiodes whose signals are subtracted and
  normalized. By changing the LO phase, it is then possible to infer
  the optical state in the mode matching the LO.  For quantum
  information processing, the states of interest are in well-separated
  modes, corresponding to a pulsed configuration with one relevant LO
  mode per measurement.
  
   We theoretically investigate what can be learned about the
  unknown optical state by counting photons in one or both outgoing
  paths after the BS, keeping the LO mode
  fixed but choosing its phase and magnitude. We consider measurement configurations where the BS acts differently on different sets of matching modes. When the BS acts identically on all matching modes it is possible to determine the content of
  the unknown optical state in the mode matching the LO conditional on
  each number of photons in the orthogonal modes on the same path. In particular, if both the phase and the intensity of the LO can be varied, then the statistics of just one of the counters is enough to infer these parameters, while in the case of an LO with fixed intensity both detectors are needed to accomplish this. Our results are derived by demonstrating a bijection, or lack thereof, between the probability distributions over the space of outcomes of the counter(s) and certain parameters of the unknown state for different measurement configuration.
  
   We report an experiment that was conducted to demonstrate the theory in the case where the BS acts differently depending on the polarization. Due to certain irregularities observed in the data, which could not be traced back to the experimental conditions, we decided to halt the data analysis midway. Nevertheless, the experiment provides lessons for designing a better experiment in the future with greater control and access to calibration of the various parameters. Further, the associated numerical modeling, which we report in significant detail, might also be found to be useful in the future.
	}

\dedication[Dedication]{	% NEVER use \OnePageChapter here.

To my family and friends.

	}

\acknowledgements{	\OnePageChapter	

First and foremost I am deeply grateful to Manny Knill and Scott Glancy for their support and mentorship all these years. It took me a long time to gain some measure of maturity as a researcher compared to my peers. I am certain that without their kindness and patience with my growth I would not have made it through those first few difficult years. 

I am indebted to my collaborators on this project: Hilma Vasconcelos, Thomas Gerrits, Mike Mazurek and David Phillips, from whom I learned a great deal about quantum optics and about the myriad challenges of conducting good experiments.

My graduate experience would have been very poor indeed if not for all the friends and colleagues I got acquainted with over the years. I would like to thank, among many others, my past and current groupmates Karl Mayer, Jim Van Meter, Peter Bierhorst, Charlie Baldwin, Mohammad Alhejji, Shawn Geller, Alex Kwiatkowski, Ezad Shojaee, Akshay Seshadri and May An van de Poll for so many educational conversations and interactions, as well as for the good times. 

Last but not least, I would like to thank the (metaphorical) lucky stars for being blessed with such a wonderful family and friends. My parents and grandparents, my aunt, my siblings and my friends acquired during childhood and during later years have been a continuous source of support, wisdom and joy throughout this journey.

	}

\ToCisShort	% use this only for 1-page Table of Contents

\LoFisShort	% use this only for 1-page Table of Figures
\LoTisShort	% use this only for 1-page Table of Tables
\input Thesis_defs.tex

\begin{document}

\input chapter_background.tex

\input chapter_quantum_optics.tex

\input chapter_symmetries.tex

\input chapter_invertibility.tex

\input chapter_simulations.tex

\input chapter_experiment.tex

%%%%%%%%%   then the Bibliography, if any   %%%%%%%%%
%\normalem
\bibliographystyle{siam}	% or "siam", or "alpha", etc.
%\nocite{*}		% list all refs in database, cited or not
\bibliography{arik_thesis}		% Bib database in "refs.bib"

%%%%%%%%%   then the Appendices, if any   %%%%%%%%%
\appendix
\input appendix_opordering.tex

\end{document}

%% file: Thesis_defs.tex
\providecommand{\ignore}[1]{}

\newif\ifcmnt
\cmnttrue
\ifdefined\cmntsoff\cmntfalse\fi

\providecommand{\vcentcolon}{\mathrel{\mathop{:}}}
\newcommand{\normord}[1]{\vcentcolon\mathrel{#1}\vcentcolon}
\newcommand{\anormord}[1]{\mathrel{\vdots}\mathrel{#1}\mathrel{\vdots}}

\ifcmnt

\else

\fi

\newcommand{\cmplx}{\mathbb{C}}

\newcommand{\nats}{\mathbb{N}}

\newcommand{\cA}{\mathcal{A}}
\newcommand{\cB}{\mathcal{B}}
\newcommand{\cC}{\mathcal{C}}

\newcommand{\cF}{\mathcal{F}}

\newcommand{\cH}{\mathcal{H}}

\newcommand{\cL}{\mathcal{L}}
\newcommand{\cM}{\mathcal{M}}
\newcommand{\cN}{\mathcal{N}}

\newcommand{\cP}{\mathcal{P}}

\newcommand{\cR}{\mathcal{R}}

\newcommand{\cT}{\mathcal{T}}

\newcommand{\one}{\mathds{1}}

\numberwithin{equation}{section}
\newtheorem{theorem}{Theorem}[section]
\newtheorem*{theorem*}{Theorem}

\newtheorem*{problem*}{Problem}
\newtheorem{lemma}[theorem]{Lemma}
\newtheorem*{lemma*}{Lemma}
\newtheorem{corollary}[theorem]{Corollary}
\newtheorem{definition}[theorem]{Definition}

\newcommand\at[2]{\left.#1\right|_{#2}}
\newtheorem{prop}{Proposition}

\stackMath
\newcommand\reallywidehat[1]{%
\savestack{\tmpbox}{\stretchto{%
  \scaleto{%
    \scalerel*[\widthof{\ensuremath{#1}}]{\kern.1pt\mathchar"0362\kern.1pt}%
    {\rule{0ex}{\textheight}}%WIDTH-LIMITED CIRCUMFLEX
  }{\textheight}% 
}{2.4ex}}%
\stackon[-6.9pt]{#1}{\tmpbox}%
}

%% file: chapter_background.tex
\chapter{Overview}\label{chap background}

The modern field of measurement science in quantum optics is a result of the interplay between several broad strands of experimental and theoretical developments during the past several decades. We are in the midst of fast-moving technological progress in the production of devices, procedures and platforms allowing for ever-better generation, control and storage of states of light, and for more accurate and precise measurement and recording. Taking a very incomplete (and very biased) historic tour, one can trace the origin of the field to the discovery of the photoelectric effect by Hertz in 1887 \cite{hertz1887ueber} and its analysis by Einstein in 1905 \cite{einstein1905uber}, which provided one of the first pieces of evidence for the non-classical nature of light. This laid the groundwork for the creation of photomultiplier tubes in the 1930s which may be considered the first photon detection devices. The creation of the laser in the 1950s and the experiments to measure its photon-number properties led to the discovery that even at high intensities the statistics of photon counts is Poissonian - something that cannot be explained by a classical model \cite{bachor2019guide}. The work of Glauber \cite{glauber1963quantum} and others in the 1960s led to the concept of the coherent state, which describes the light produced by a laser. The theoretical work made predictions about the existence of other states of light, including states exhibiting sub-Poissonian statistics of photon counts (or photon anti-bunching). The 1970s and the 1980s saw the theoretical and experimental discoveries of exploiting the energy level structure of the atoms in atomic beams or clouds or in non-linear crystals to transform laser light into different states, including squeezed states, which exhibit a degree of freedom with lower noise than the shot noise of the laser \cite{bachor2019guide}. These developments have heralded the field of quantum metrology, where the experiments are designed such that the degree of freedom with suppressed noise carries the information about the parameters of interest and thus allows for more accurate and precise measurement. They have also created new possibilities for generating quantum states, for testing the unique features of quantum physics such as non-locality and entanglement, and, more recently, in quantum computing and quantum communication \cite{bachor2019guide, slussarenko2019photonic}.

On the theoretical side, the works of Glauber, Sudarshan \cite{glauber1953coherent, sudarshan1963equivalence} and others advanced the phase-space formulation of quantum mechanics and showed its great usefulness as an alternative description of states of light. The concept of a ``mode" as an independent degree of freedom of the electromagnetic field was developed and elucidated, and was shown to be indispensable for modeling quantum optical phenomena. In particular, any practical model of an experiment assumes that a finite set of (orthogonal) modes carry all the relevant degrees of freedom that describe the experiment. The last decades of the past century also saw important advances in our understanding of the notion of the observable in quantum theory, and more rigorous ways of describing the relationship between the measurement apparatus and the physical system in general, which contributed to better theoretical descriptions of measurements in quantum optics as well \cite{busch1995oper}. Last but not least, the advent of the field of quantum information theory has brought about new mathematical tools and concepts, which have led to new designs for experiments that can infer parameters of interest with better statistical precision \cite{weedbrook2012gaussian, lvovsky2009continuous}.

One experimental measurement technique that has proven incredibly powerful is the standard homodyne detection \cite{leonhardt:qc1997a}. In this scheme the unknown state is interfered on a balanced beam splitter (BS) with a controlled high intensity coherent state, usually called the ``local oscillator'' (LO) or simply the ``phase reference'', at different values of the relative phase covering the whole angle range. The outgoing beams are absorbed by photodiodes, and the resulting electrical signals are temporally integrated and subtracted from each other. The output signal, in the limit of infinite LO intensity, corresponds to a measurement of the quadrature of the unknown state in the mode matching the LO at the relative phase~\cite{LEONHARDT2}. Knowing the quadrature distribution for all relative phase values is sufficient to reconstruct the reduced density matrix of the unknown state in the mode matching the LO \cite{LEONHARDT2, MAURODARIANO2003205}.

There are some obvious limitations with this technique, however. For instance, it is not practical to implement in certain settings where a high-intensity LO cannot be used. For example, there are a number of experiments exploring photon-photon interactions within atomic clouds where the LO must be passed through the cloud and therefore cannot be so strong as to cause unwanted disturbances \cite{maxwell:qc2013a,Beck:qc2016, peyronel:qc2013a,Lukin}. In particular, \cite{peyronel:qc2013a} and \cite{Lukin} utilized a weak LO together with photon counters to measure the phase information of the photons leaving the clouds. Another shortcoming of the standard homodyne detection is that it allows one to learn only the reduced state in the mode matching the LO and does not give any information about the parts of the state occupying the orthogonal mode space.

In a different line of developments, the technology of detecting (and counting) photons has come a long way since the advent of photomultiplier tubes. While photomultiplier tubes are still used today, many different types of devices based on semi-conductors have emerged and are in development, and devices based on superconducting materials have become very popular as well \cite{migdall2013experimental, slussarenko2019photonic}. Among the latter are the transition-edge-sensors (TESs) that are built around maintaining a superconducting material at its critical temperature of transition, such that the heat generated during the absorption of a single photon increases the resistance of the material enough to be measurable \cite{gerrits2016super}. The experiment we describe in this thesis utilizes such TESs as photon counters. State-of-the-art TESs can distinguish, with decreasing accuracy for higher photon numbers, up to twenty photons in a single pulse.

Advancements in photon counting technologies create the possibility of combining photon counting with an interferometer that uses a LO, as in the standard homodyne detection, to build new types of measurement devices. Given the low maximum photon number resolution of modern photon counters, the LO needs to have low intensity. These kinds of measurement devices are sometimes grouped under the label of ``hybrid" or ``weak-field homodyne" detectors \cite{migdall2013experimental}. A number of promising studies have been conducted with this idea in mind. One of the first such
studies showed that, in theory, an unknown state in a single mode can be
  determined with a single photon counter and an unbalanced BS
\cite{wallentowitz1996unbalanced}. 
On the experimental side, such schemes have proven useful to measure the two-photon phase coherence of
parametrically down-converted photons \cite{Koashi}, to measure
Bell inequality violations \cite{Walmsley1}, to demonstrate the
creation of superpositions of a single photon state with vacuum with well-defined phases \cite{Resch}, to carry out state discrimination
\cite{Olivares2,Olivares3,Muller}, and to provide lower bounds on the
amount of entanglement in bipartite states \cite{Walmsley3}. In
addition, there have been several experimental studies demonstrating
the applicability of such schemes for the reconstruction of unknown
states occupying a single mode \cite{Olivares1, Olivares4, Walmsley2, Walmsley4, Thekkadath}, as well as two-mode squeezed states \cite{Walmsley5}. Several studies \cite{zhang2012recursive, xu2020exp} focused on the experimental characterization of the observables associated with weak-field homodyne detectors, again assuming a single mode is being measured. Another study \cite{Shadow} showed that an arbitrary multimode state can be determined, in principle, using a single detector provided one can prepare the LO in any mode.

In the measurement schemes described above, the experimental data comes from the photon detectors in the setups. These detectors, in general, respond to photons in a large set of modes, and therefore, to match the experiments to the theoretical models these studies assume (sometimes implicitly) all modes aside from the modes assumed in the model are in vacuum. However, the existence of experimental imperfections during the preparation and the evolution of the physical system normally results in photons being present in other modes, and taking them into account can provide experimentally relevant information. In addition, if the number of photons in the other modes is too large, the methods above will give poor estimates. Also, the measurement techniques requiring the preparation of the LO in many different modes are hard to realize in practice as the number of measurements required can be potentially very large, and it may be difficult to prepare the LO in all of the required modes. These factors motivate us to consider a family of measurement configurations based on photon counting and using a LO, and where no assumptions are made about which modes are occupied by the unknown state.

Broadly speaking, in designing and building any measurement device one faces tradeoffs between the simplicity of the design, the time needed for the desired measurement, and the information content of the measurement. For example, the standard homodyne detector does not have a complex design and can only characterize the reduced state in the mode matching the LO given enough measurement shots, and if one wants to learn more about the unknown state one must prepare the LO in many different modes and take enough shots for each mode, which significantly increases the duration of the measurement. Conversely, one can perhaps imagine a complex design using multiple interferometers and photon counters that would allow one to learn a significant amount about an unknown multimode state given a reasonably small number of shots. But such a measurement scheme would be very challenging to characterize, let alone to build in practice. Therefore, when coming up with measurement schemes that use photon counters and LOs to characterize an unknown multimode state, it is reasonable to start by considering a fairly simple family of measurement configurations (described below), many members of which should not be too challenging to realize in the laboratory. The characterization of this family of configurations is the main goal of this thesis.

The main contributions of this thesis are the characterization of the observables associated with a certain family of idealized measurement configurations, and the analysis of an experiment that was performed by David S. Phillips, Thomas Gerrits and Michael Mazurek at NIST. In each configuration an LO in a coherent state of a well-defined mode is mixed on a BS with the unknown multimode state, and either both outputs of the BS are measured by photon counters or only one of the counters is used. We do not constrain the BS to act identically on all matching modes. When the BS acts identically on all matching modes, it is surprisingly possible to obtain more information than is available with the standard homodyne detection. In particular, it is possible to determine the contents of the unknown state in the mode matching the LO conditional on each number of photons present in the orthogonal mode space. We find that the measurement statistics of only one counter for enough different LO amplitudes can determine this information. It is required that both the magnitude and the phase of the LO can be controlled. A finite number of probe amplitudes are sufficient when the unknown state has a bound on its maximum photon number. The sufficiency of the single counter configuration is a nice surprise since it can help in the experimental conditions where it is 
difficult to ensure that the same spatial mode or modes are measured
by the two photon counters, which increases the ambiguity of the interpretation of the measurement. Using only one counter removes this ambiguity as, regardless of which modes it measures, such a configuration can be, in principle, taken into account by our multimode model.

When both counters are used with a BS that acts identically on all matching modes, a fixed-magnitude LO is sufficient to determine the contents of the unknown state in the mode matching the LO conditional on each number of photons present in the orthogonal mode space, provided the measurement statistics are available for every phase. 

When the BS has several different actions on the total set of matching modes, the information about the unknown state that can be obtained, in principle, from the measurement statistics is significantly larger. When both counters are used, and the BS enacts two different kinds of transformations on different matching pairs of modes, the most that can be learned about the unknown state is the content of the mode matching the LO conditional on each pair of numbers of photons present in the two orthogonal mode subspaces differentiated by the two different actions of the BS. A finite number of LO amplitudes are sufficient if the unknown state has bounded maximum photon number. The measurement configuration modeling the experiment is the practical adaptation of this configuration. The BS has different splitting ratios for different polarizations, which can be modeled as having two different actions on the set of matching spatiotemporal modes. We find the accuracy of the reconstruction is not bad, at least for one of the prepared unknown states that we focused our analysis on (the fidelities with respect to the best guess of the real state are around $98.5\%$ using non-intersecting datasets). However, we suspect that a significantly higher accuracy is achievable, since we found systemic discrepancies in the data that we could not explain with the experimental conditions and which had the biggest negative contribution to the quality of the data. We were forced to abandon the analysis midway, but as of this writing plans are being made for another demonstration experiment, and an effort is put on the design to allow for better characterization and control of the various parts of the measurement configuration.

The results of our theoretical investigations are important in that they show that using photon counters with a weak LO, as well as a BS with up to two or three kinds of different actions on the set of matching modes, can allow one to learn, in principle, significantly more about the state than using standard homodyne detection. Here is conclusive evidence that substituting photon counters for the photo-current detectors in the standard homodyne scheme provides an advantage in the information content of the measurement, without increasing the difficulty of the characterization of the device significantly, or maybe at all. 

The first half of the thesis is devoted to the theoretical analysis of the family of idealized measurement configurations. The broad question we ask is what can be in principle determined about the unknown state from the measurement statistics of photon counts for a given measurement configuration. Chap.~\ref{chap quantum optics} is devoted to introducing the physical concepts and terminology that are used to set up the problem, as well as to introduce certain mathematical results and techniques that we use in our proofs. Among the former belong the descriptions of the Fock space associated with a finite set of orthogonal modes and of the set of operators on this space. For example, the unitary operator associated with the BS, which is a passive linear transformation (PLT), as well as the observables associated with the photon counters are described here. We also introduce several facts about the representations of PLTs on Fock space. 

We spend significant time in this chapter discussing normally ordered polynomials of the creation and annihilation operators of the modes and how these transform under PLTs. By taking the expectation of such an operator with an arbitrary multimode coherent state, one obtains a complex-valued multivariate polynomial. There is a one-to-one correspondence between these two polynomial spaces, which we exploit by representing operators as complex-valued functions. At the end of the chapter we describe the concept of an operator-valued generating function, which is heavily utilized in deriving our results in Chap.~\ref{chap invertibility}. We build such generating functions by encoding operators as the coefficients of the power series of one or several (symbolic) variables. The isomorphism between the polynomials of mode operators and complex valued functions induces an isomorphism between operator-valued generating functions and generating functions with complex coefficients. The latter, under appropriate conditions, have a simple functional representation in a neighborhood of the origin of the variable(s). We exploit this closed form to simplify our calculations, which are aimed at understanding the connections between the operators in different paths within the measurement configuration.

Chap.~\ref{chap sym and twirl} starts with the mathematical description of the family of idealized measurement configurations and introduces a standardization of the action of the BS due to inherent symmetries of the corresponding measurement configuration. The consequences of these symmetries are further explored and, as a result, the set of unknown states is divided into equivalence classes, where each state in a given equivalence class produces the same measurement statistics as the other states in that class. For a given measurement configuration, we find a particularly simple member in each equivalence class, which we call the twirled state (Thms.~\ref{theorem 1.2} and~\ref{theorem 1.2p}). The twirled state is a convex combination of density matrices in the mode matching the LO, each multiplied by a tensor product of normalized projectors onto orthogonal Fock spaces. These tensor products of projectors are determined by the action of the BS. For the important class of BSs that enact the same transformation on all matching modes, each density matrix in the mode matching the LO in the convex combination is in a tensor product with the normalized identity on a subspace of the rest of the modes characterized by a particular total photon number. The main theoretical aim is then reformulated as whether a given measurement configuration can be used to determine the corresponding twirled state under certain restrictions, or lack thereof, on the preparation of the LO.

Chap.~\ref{chap invertibility} is where the question of the relationship between the unknown parameters of the twirled state and the measurement statistics is investigated for different measurement configurations. It is assumed that we have access to infinite data for each measurement configuration and the given LO amplitude, so that the probability distribution over the outcomes of the photon counter(s) is known. Glossing over some mathematical technicalities, the expectations of the powers of the total photon number observable(s) at the output path(s) of the BS capture all the information in the measurement statistics for the given LO and unknown states. We build an operator-valued generating function from the powers of the total photon number operator(s). The BS relates these observables with operators on the input paths of the BS. There is a corresponding transformation of the operator-valued generating function. By taking the partial trace over the modes of the LO path one obtains a generating function on the unknown state. The expectations of the operators generated by this function contain all of the information in the measurement statistics. Thus, if the expectations of these sets of operators for a given set of LO amplitudes do not determine the twirled state, then the twirled state cannot be determined, and vice versa.

When the BS transforms all matching modes in an identical manner, we find that, if one can control both the magnitude and the phase of the LO, then even the measurement statistics of the configuration with one counter can determine the corresponding twirled state. More specifically, even if the twirled state doesn't have a bound on its maximum photon number, the measurement statistics for all LO amplitudes in a neighborhood of the origin of the complex plane determine the twirled state (Thm.~\ref{thm:1CntrNbhd} and Cor.~\ref{cor:1CntrNbhd}). If the twirled state has a maximum photon number $N$, then the measurement statistics of an order of $N^2$ different LO amplitudes are sufficient to determine it (Thm.~\ref{thm:1CntrNbnd}, Prop.~\ref{cor:1CntrNbnd} and Cor.~\ref{cor:twirldet}). When the magnitude of the LO is fixed, however, we find that one counter is not sufficient to determine the twirled state unless $N=1$ (Thm.~\ref{thm: magfixed_onedet_trivpart}). With a fixed-magnitude LO the measurement configuration with two counters produces measurement statistics that do determine the twirled state, assuming that the measurement statistics for all phases of the LO are available (Thm.~\ref{thm: twodet_R_fixed}). 
 
\begin{figure}[!htb]
    \centering
\includegraphics[width=160mm]{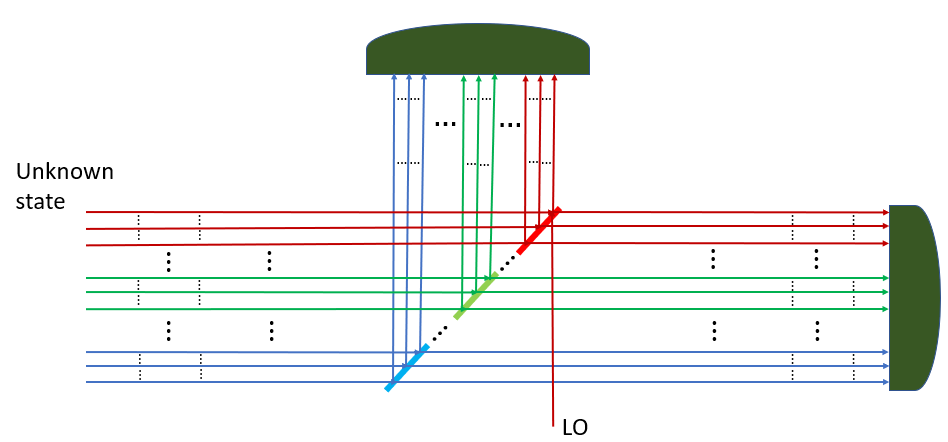}
    \caption{Diagram of a measurement configuration with a BS that acts differently on different sets of modes. The multiple BSs in the middle of the diagram should be thought of as a single physical BS. The diagram represents each mode occupied by the unknown state by a single arrow, and only the mode corresponding to the uppermost arrow matches the LO. The "red BS" denotes the transformation that is enacted on the LO and its matching mode, as well as on a set of other modes that might or might not be occupied by the unknown state. The outputs of the BS are measured by photon counters that cannot distinguish the modes. We also consider configurations where one of the counters is absent. }
    \label{fig: meas_config_mult_BS}
\end{figure}

The situation is more complex when the (standardized version of the) BS does not act identically on all matching modes (see Fig.~\ref{fig: meas_config_mult_BS} as a guide for visualizing the statements that follow). In this case, whether the modes of the unknown state that transform in the same way as the mode matching the LO are assumed to be in vacuum plays a significant role. Also, the number of different BS actions on different sets of matching modes is found to be crucial. More specifically, if the BS is such that the mode matching the LO and the LO mode are the only matching pair undergoing a particular transformation, then the twirled state can be determined only if the rest of the matching modes undergo up to two different kinds of transformations (Lem.~\ref{lem: M_K<=>rho_S1=1} and Thm.~\ref{thm 2det_nontrivpart, S_1=1}). If there is at least one more matching pair of modes that undergo the same transformation as the mode matching the LO, then the twirled state can be determined only if the BS enacts up to two different transformations on the total set of matching modes (Lem.~\ref{lem: M_K<=>rho_S1>1} and Thm.~\ref{thm 2det_nontrivpart}). These two statements assume that both counters are used, and, when the twirled state can be determined, the measurement statistics for all values of the LO amplitude in a neighborhood of the origin of the complex plane are sufficient for determinacy. When a single photon counter is used, and the measurement statistics for all values of LO amplitude are assumed to be available, the twirled state cannot be determined by the measurement statistics of a measurement configuration with a BS that has more than one kind of action on the set of matching modes and there is at least one other pair of matching modes besides the pair associated with the LO that undergoes the same transformation as the latter pair (Thm.~\ref{thm 1det_nontrivpart}). Under the same conditions but with a BS which acts uniquely on the mode matching the LO and the LO, the measurement statistics determine the twirled state if the rest of the matching modes are transformed identically by the BS (Thm.~\ref{thm 1det_nontrivpart}).

The second half of the thesis is devoted to describing the work we did to create a numerical model for certain members of the family of measurement configurations described above (Chap.~\ref{chap simulations}), and to the analysis of the experiment which was designed to be a practical validation for some of the theoretical results (Chap.~\ref{chap experiment}). We numerically modeled the measurement configurations where the BS acts identically on all matching modes and the measurement configurations where the BS has two different actions on the set of matching modes - namely, a subset of the matching pairs of modes experience one kind of transformation, while the rest experience another kind of transformation. To each measurement configuration and to a given LO amplitude, we associate a positive-operator-valued-measure (POVM). Each element of the POVM corresponds to a particular measurement outcome of the counter(s). Given a set of samples for different LO amplitudes, we first construct the associated set of POVMs, make sure that their expectations determine the twirled state, and then construct a likelihood function and perform maximum likelihood estimation to obtain an estimate of the twirled state. We perform a set of simulations for the measurement configuration with a BS that is balanced on all matching modes for different LO and twirled states, in order to examine the accuracy of the numerical model and of the estimation procedure for some of the simplest kinds of statistical models possible with this family of measurement configurations. We find that for a reasonably large number of simulated samples ($\geq 10^6$) and for twirled states with few photons, the estimates agree closely with the corresponding true states (generally with a quantum fidelity of $>99.8\%$).

The analysis of the experiment is broadly divided into two parts. In part one, we describe how the data for performing the estimation were generated from the values extracted from the TESs during the experiments, while the second part describes how we choose the POVMs and perform reconstruction with the corresponding samples that we produce from the generated data. The idealized measurement configuration corresponding to the experiment has both photon counters, allows for adjusting both the magnitude and the phase of the LO, and has a BS with two different actions on the set of matching modes. We assume the part of the state in an orthogonal mode to the LO but experiencing the same action by the BS is not in vacuum. Experimentally, we used a polarizing BS fiber with different splitting ratios for the two orthogonal polarizations. TESs are used as photon counters, and the set of unknown states are all prepared as coherent states. We describe the challenges and issues that we encountered during the analysis. Most importantly, we report a certain set of discrepancies that we found in the data that could not be explained by the experimental conditions, which eventually forced us to abandon the analysis. Even with these discrepancies, we achieved $\approx 0.985$ values of the fidelity between the reconstructed estimates and our best guess of the true state for different sets of LO amplitudes with the state that we looked at before abandoning the analysis for the rest of the prepared states. We hope that the lessons learned during the analysis of this experiment will help in setting up and performing a more controlled demonstration in the future, as well as in finding ways to use such a setup as a measurement device in various kinds of experiments.

%% file: chapter_quantum_optics.tex
\chapter{Quantum Optics}\label{chap quantum optics}

In this chapter, we describe the mathematical machinery of quantum optics that we use to characterize the measurement configurations and to derive our results. We suppose that the physical system can be described by assuming that it occupies only a finite number of orthogonal modes. We usually denote the number of orthogonal modes by $S$. Our calculations assume that the
optical states have density operators for which all polynomials of
mode operators (defined in Sec.~\ref{qo intro}) have well-defined expectations; specifically, the
family of states with rapidly decaying Wigner functions. This
family of states is defined and characterized in
Ref.~\cite{hernandez2021rapidly} and contains all states with bounded
photon number, Gaussian states, and all finite superpositions and mixtures
of these states. This family of states is dense in the space of states.

We start by giving a brief overview of the concept of a mode as it is used in linear quantum optics and introduce our main notation in Sec.~\ref{qo intro}. We then describe the observable associated with a photon counter that does not distinguish the modes in Sec.~\ref{qo pc}. We introduce some basic results in representation theory of finite-dimensional unitary groups in Sec.~\ref{qo rep_theory}. In Sec.~\ref{qo PLT} we introduce an important class of unitary operators, called passive linear transformations (PLTs), and describe some of their properties that are relevant for deriving our results in later chapters. Sec.~\ref{qo order} is devoted to understanding the concept of normal ordering of polynomials of mode operators and their properties, and describes an isomorphism between the space of normally ordered polynomials of mode operators and the space of complex-valued multivariate polynomials. The next section describes the Husimi representation of an optical state, and explores the conditions under which the moments of the Husimi function, which are related to the expectations of anti-normally ordered monomials of mode operators, determine the Husimi function. In Sec.~\ref{qo genfun} we describe the concept of a generating function that has coefficients that are polynomials of mode operators. This framework is heavily utilized in our analysis to investigate the relationship between the distribution of measurement outcomes and the parameters describing the unknown state. Finally, in the last section, we sketch the mathematical formulation of ideal standard homodyne detection due to its similarities with the measurement configurations we study in this work.

\section{Introduction}\label{qo intro}

In the classical theory of the electromagnetic (EM) field in infinite free space, the field configurations in spacetime can be decomposed as superpositions of ``simple" solutions to the Maxwell's equations in the absence of charges and currents. The simple solutions we have in mind are plane waves that are defined by a single frequency, an axis of propagation of the EM field, and a polarization that is orthogonal to the direction of propagation. Given an axis of propagation, one can construct an orthogonal basis of solutions, such that any beam on that axis can be written as a superposition of the bases. This is essentially a Fourier decomposition of each polarization component of the EM wave along that propagation axis in terms of components corresponding to particular frequencies and momenta. These are classical modes of the EM wave. The standard (canonical) quantization procedure substitutes the amplitudes of the components in the Fourier expansion with separate creation and annihilation operators. Very roughly, to each creation and annihilation operator one associates a harmonic oscillator (HO) with a frequency equal to that of the corresponding Fourier component. These are modes in the quantum theory of the free EM field, and the time evolution of the state-space of a given mode is produced by the corresponding HO Hamiltonian. A more mathematically rigorous explanation (and treatment) in terms of operator-valued distributions can be found in Ref.~\cite{haag1992local}.

The concept of the mode can be extended to include systems with creation operators that are arbitrary linear superpositions of the creation operators of the HOs associated with the frequency-momentum components of the free EM field. Provided the weights are normalized, the resulting creation and annihilation operators obey the canonical commutation relations, and one can treat the associated system as a HO, even though it does not have a fixed frequency of oscillation. Any state in such a system can be written as a linear superposition of states of the HOs that compose it, and the time evolution of the state under the free field Hamiltonian is described by the contributions of the Hamiltonians of these HOs to the relative phases in the superposition.

The above discussion can be extended to free EM fields confined in a cavity or in a transmission line such as an optical fiber, which are the usual settings of quantum optics. This is in principle done by canonically quantizing the configurations of the EM field allowed by the boundary conditions. ``In principle" because the effects of some of the optical elements taking part in defining the boundary conditions - a beam splitter (BS) for instance, on a spatially localized wavepacket are usually treated as instantaneous unitary operations on the state-space (or on the space of operators in the Heisenberg picture) at a particular point of time during the propagation of the wavepacket. In contrast, the walls of an optical fiber are treated as a confining boundary that effectively restricts the directions of the propagation of the EM wave to a single axis. Since the number of modes is infinite or continuous, one cannot explicitly model the whole system, and the assumption that the state-space is not in vacuum for only a finite number of modes is usually made. More information can be found in textbooks such as Ref.~\cite{steck2007quantum}.

If the state lives in $S$ orthogonal modes, let us order these modes in some arbitrary way, and denote the creation and annihilation operators of the $i$'th mode by $a_i^\dagger$ and $a_i$, respectively. Notice, we choose not to use the ``hat" symbol for these operators to reduce notational clutter. We refer to these operators as ``mode operators". For the $i$'th mode the corresponding quadrature operators are defined as $q_i = \frac{1}{\sqrt{2}} (a_i+a_i^\dagger)$ and $p_i = \frac{1}{\sqrt{2} i} (a_i- a_i^\dagger)$. The mode operators satisfy the commutation relations $[a_i,a_j^\dagger] = \delta_{ij}$ and $[a_i,a_j] = 0$ for $1 \leq \forall i,j \leq S$. For a system occupying $S$ orthogonal modes, we denote the vector of annihilation operators $(a_1,\hdots,a_S)^T$ by $\vec{a}$, and the vector of creation operators $(a_1^\dagger,\hdots,a_S^\dagger)^T$ by $\vec{a^\dagger}$. We imagine these as columns vectors so that we can multiply them from the left by a matrix. Thus, $(\vec{a})^\dagger$ is not the same as $\vec{a^\dagger}$; namely, the former is the row vector composed of the $a_i^\dagger$. We use the term ``mode space" to refer to the vector space composed of all linear combinations of annihilation operators, where each $a_i$ is regarded as a separate basis element.  

A Fock state where $n_i$ photons occupy mode $i$ is denoted in the ket-bra notation as $\ket{n_1}\otimes \cdots \otimes \ket{n_S} \equiv \ket{n_1,\hdots,n_S}$. $n = \sum_{i=1}^S n_i$ is called the total photon number of the corresponding Fock state. Greek characters inside the kets or bras are used to denote coherent states. The Fock space associated with the modes $a_1,\hdots,a_S$ is composed of all linear combinations of Fock states. We denote these spaces by $\cF(a_1,\hdots,a_S)$. The Fock space associated with the combination of two mode spaces is composed of the vectors in the tensor product of the Fock spaces associated with these mode spaces. Namely, $\cF(a_1,\hdots,a_{S}) =  \cF(a_1,\hdots,a_d) \otimes \cF(a_{d+1},\hdots,a_S) $ for any $d <S$. Let us further denote the subspace of $\cF(a_1,\hdots,a_S)$ spanned by the Fock states with total photon number $n$ by $\cF_n(a_1,\hdots,a_S)$.

We can compose the annihilation and creation operators to form polynomials of mode operators. By the latter we mean operators of the form $\sum_{i} c_i M_i(\vec{a},\vec{a^\dagger}) $, where the $c_i$ are real or complex coefficients and the $M_i$ are monomials of the mode operators, that is, each $M_i$ is a particular product of a finite number of mode operators. For example, the ``total photon number" observable is such a polynomial, given by $\hat{n} = \sum_{i=1}^S a_i^\dagger a_i$.
 
\section{Photon Counting}\label{qo pc}

Photon counters are physical devices that are built to measure the number of photons in a beam of light. Our results are appropriate for spatially localized beams of light that are absorbed by the photon counter in a finite time. In that case, one can talk about the photon-number distribution of the modes comprising the wavepacket. The photon counters we consider are only sensitive to the total photon number of the wavepacket. We define the ideal photon counter as a hypothetical device that, upon absorption of a wavepacket, outputs a non-negative integer, and the output $k$ is associated with the projector onto the subspace of Fock space with total photon number $k$. We denote these projectors by $\hat{D}_k$, so that when the number of modes is one $\hat{D}_k = \ket{k}\bra{k}$, while for $S$ modes $\hat{D}_k = \sum_{k_1+\hdots +k_s=k}\ket{\vec{k}}\bra{\vec{k}}$, where $\vec{k} = (k_1,\hdots,k_S)$. The probability of the output $k$ is given by Born's rule - the trace of the product of the density operator of the state with  $\hat{D}_k$. We assume that, in the infinite data limit, the relative frequency of each outcome converges to the corresponding probability. In our theoretical investigations we often make the assumption that the number of measurements (shots) is infinite, and, hence, the probabilities of the outcomes are exactly available.

We define a non-ideal counter as a device associated with a set of outcomes and a corresponding set of operators, where the latter are linearly dependent on the $\hat{D}_k$. The probability of each outcome is similarly given by Born's rule using the corresponding operator. It can be that some, or all, of the operators associated with a particular non-ideal counter are expressed as infinite linear combinations of the $\hat{D}_k$. For our theoretical results to hold for non-ideal counters, it is essential that the probability distribution over its outcomes determines the probability distribution over the outcomes of the ideal counter for the same state. This implies that the non-ideal counter must have the property that the map relating the operators associated with it to the $\hat{D}_k$ is invertible. Therefore, the expectation of any operator that can be constructed from the $\hat{D}_k$ is available in the infinite data limit. When referring to the measurement statistics or to the observables associated with a counter in Chaps.~\ref{chap sym and twirl} and~\ref{chap invertibility} we make the implicit assumption that the counter satisfies the aforementioned property.

In Chap.~\ref{chap invertibility} we also consider a click detector that does not distinguish the modes. Such a detector produces two different outcomes - it outputs $0$ when no photons are registered, and $1$ when one or more photons are registered. The operator associated with the outcome $0$ is $\hat{D}_0 $, and the operator associated with the outcome $1$ is \(\sum_{k=1}^{\infty}\hat{D}_k \). It can be seen that click detectors can be modeled as photon counters, where the only information about the outcome of the measurement that is kept is whether no photons were measured or otherwise. In Chap.~\ref{chap simulations} we consider ideal counters that cannot distinguish photon numbers greater than some number $N_c$. The operators associated with the outcomes of such counters are then given by the $\hat{D}_k$ when $k \leq N_c$ photons are registered, and by $\sum_{k=N_c+1}^\infty \hat{D}_k$ when a photon number greater than $N_c$ is registered. Since the $\hat{D}_k$ have the property that  $\sum_{k=0}^\infty \hat{D}_k = I$, where $I$ is the identity, we can write the projector $\sum_{k=N_c+1}^\infty \hat{D}_k$ as $I - \sum_{k=0}^{N_c} \hat{D}_k$. We denote this projector by $\hat{D}_{>}$, where the dependence on $N_c$ is implicit. In Chap.~\ref{chap experiment} we attempt to model real-world counters that have losses. The operators associated with these counters are linear combinations of the operators associated with the ideal counter that can distinguish up to $N_c$ photons.

As mentioned earlier, the family of states we are interested in have well-defined expectations w.r.t. polynomials of mode operators. In particular, this means that the expectations of the non-negative integer powers of $\hat{n}$ (referred to as the moments of $\hat{n}$) are well-defined for these states. This property is preserved when any two states satisfying the property are interfered by a BS. In particular, we consider measurement configurations where the state is interfered on a BS with a coherent state and the outputs are measured by photon counters. Thus, the products of the moments of $\hat{n}$ in the joint outputs of the BS are well-defined.

\section{Representation Theory Basics}\label{qo rep_theory}

In this thesis, we make use of the representations of the family of finite dimensional unitary groups. We denote these by $U(S)$, where $S$ is the size of the dimension. Here we introduce Schur's lemma, or, rather, a specific formulation of it applied to $U(S)$, and a corollary. More information can be found in standard textbooks such as \cite{fulton:qc1991a}. A representation of $U(S)$ is a tuple composed of a vector space $V$, a subgroup of $\textrm{Aut}(V)$ ($\textrm{Aut}(V)$ is the group of automorphisms of $V$), and a homomorphism $\phi: U(S) \mapsto \textrm{Aut}(V)$ which associates to each element $U \in U(S)$ an element $A_U \in \textrm{Aut}(V)$. An intertwiner $f$ between two representations $(V,\phi)$ and $(V',\phi')$ is a linear map from $V$ to $V'$ that commutes with the action of $U$. Namely, for $\forall U \in U(S)$, $f(A_U(\cdot)) = A'_U(f(\cdot))$, where $A_U$ and $A'_U$ are the elements in $\textrm{Aut}(V)$ and $\textrm{Aut}(V')$ associated with $U$, respectively. If for a representation $(V,\phi)$ there exists a linear subspace $W \subseteq V$ that is invariant under $\phi(U(S))$, we say that $(W,\phi)$ is a sub-representation of $(V,\phi)$. Every representation has itself and the zero vector space as trivial sub-representations. An irreducible representation (irrep) is a representation that has no non-trivial sub-representations. The version of Schur's lemma that we state and use is a result about the classification of finite-dimensional irreps of $U(S)$. 

\begin{lemma}\label{lem Schur}
Let $(V,\phi)$ and $(V',\phi')$ be finite-dimensional irreps of $U(S)$, where $V$ and $V'$ are over the field of complex numbers $\mathbb{C}$. Assume $f : V \mapsto V'$ is an intertwiner between the representations.
\begin{enumerate}
\item Either $f$ is an isomorphism, in which case $V$ and $V'$ must have the same dimension, or $f=0$.
\item If $V=V'$ and $\phi = \phi'$, then $f=\lambda I$ for some $\lambda \in \mathbb{C}$, where $I$ is the identity.
\end{enumerate}
\end{lemma}
\begin{proof}
See, for example, Lem.~1.7 in \cite{fulton:qc1991a}.
\end{proof}

Schur's lemma can be extended straightforwardly to intertwiners between more general representations. In particular, let $\left( (V_i,\phi_i) \right)_i$ be a (possible infinite) sequence of mutually non-isomorphic finite-dimensional irreps of $U(S)$. They can be combined to form a representation $(V,\phi) = ( \bigoplus_i V_i, \bigoplus_i \phi_i  )$. 
\begin{corollary}\label{cor shur_lem}
Let $f :V \mapsto V$ be an intertwiner from $(V,\phi)$ to itself. Then $f$ has the form $f = \bigoplus \lambda_i I_i$, where $\lambda_i\in \mathbb{C}$ and $I_i$ is the identity on $V_i$. 
\end{corollary}
\begin{proof}
This follows from the fact that the restriction of $f$ to any $V_i$ must be of the form $\lambda_i I_i$ by Schur's lemma. Here, by the restriction of $f$ to $V_i$ we mean the map that results by restricting the domain of $f$ to $V_i$, while the co-domain remains $V$.
\end{proof}
We show in the next section that the irreps of $U(S)$ in Fock space correspond to the subspaces with fixed total photon number. These subspaces are mutually orthogonal and together span the whole Fock space. We denote these subspaces by $\cF_n$, where $n$ is the total photon number of each vector in the subspace. Then, according to Cor.~\ref{cor shur_lem} the intertwiners of the representations of $U(S)$ in Fock space are spanned by the projectors onto the $\cF_n$. This fact is used in the proof of Thm.~\ref{theorem 1.2}.

Finally, it should be mentioned that we use the same symbols for the unitary groups and for their representations on state-space. It should be clear from context what the referent is.

\section{Passive Linear Transformations}\label{qo PLT}
In the next chapter where we study the symmetries of the measurement configurations, we repeatedly consider passive linear transformations (PLTs) on multimode states. These are the unitary operations that can be performed by using only BSs and phase shifters. Given $S$ orthogonal annihilation mode operators $\vec{a} = (a_1,\hdots,a_S)^T$, the action of a PLT $U$ in the Heisenberg picture transforms the $a_i$ linearly into each other according to $U^\dagger a_i U = (U_M \vec{a})_i$. Here $U_M$ is the $S \times S$ unitary matrix associated with $U$, and the subscript $i$ denotes the $i$'th row of $U_M \vec{a}$. Namely, $(U_M \vec{a})_i = \sum_{j=1}^S U_{M,ij} a_j $. We use the subscript $M$ in $U_M$ to denote the action of the unitary $U$ in mode space. This convention is used throughout the thesis. $U$ transforms a multimode coherent state $\ket{\vec{\alpha}} = \ket{\alpha_1}\otimes \hdots \otimes \ket{\alpha_S}$ into another coherent state according to $U \ket{\vec{\alpha}} = \ket{U_M \vec{\alpha}}$. When referring to a particular PLT, we either use the symbol denoting its action in state-space (we already did this in this paragraph) or the symbol denoting is action in mode-space - that is, we either say ``PLT $U$" or ``PLT $U_M$", but it is understood that we mean the same underlying object. 

One has to be careful when working with vectors of operators. In particular, it is important to notice that
\begin{align}\label{eq: identities_PLT_creat}
U^\dagger a^\dagger_i U &= (U^\dagger a_i U)^\dagger \nonumber\\
& = ((U_M \vec{a})_i)^\dagger \nonumber \\
&=\sum_{j=1}^S U^*_{M,ij} a^\dagger_j \nonumber \\
&  = (U_M^* \vec{a^\dagger})_i  = (  (\vec{a})^\dagger U^\dagger_M )_i.
\end{align}
 $U$ commutes with the total number operator $\hat{n} = \sum_{i=1}^S a_i^\dagger a_i$:
\begin{align}\label{eq U_commute_n}
U^\dagger \hat{n} U &=  \sum_{i=1}^S U^\dagger a_i^\dagger U U^\dagger a_i U =  \sum_{i=1}^S (U^\dagger a_i U)^\dagger U^\dagger a_i U = \sum_{i=1}^S (U_M \vec{a})_i^\dagger (U_M \vec{a})_i \nonumber \\
& =  \sum_{i=1}^S \sum_{j=1}^S U^*_{M,ij} a^\dagger_j \sum_{k=1}^S U_{M,ik} a_k =  \sum_{j=1}^S  \sum_{k=1}^S a^\dagger_j  a_k \left[ \sum_{i=1}^S U^*_{M,ij} U_{M,ik} \right] \nonumber \\
& =  \sum_{j=1}^S  \sum_{k=1}^S a^\dagger_j  a_k \delta_{jk} = \sum_{j=1}^S a_j^\dagger a_j= \hat{n}.
\end{align}
An immediate consequence of this is that acting with a PLT on a state before measurement by a photon counter that does not distinguish the modes will not affect the measurement outcome probabilities. This follows from the fact that the probabilities of the outcomes are associated with the eigenspaces of $\hat{n}$, and the latter are invariant under PLTs. Another consequence is that a PLT commutes with arbitrary powers of $\hat{n}$, as well as with any operator that can be expressed as a linear combination of powers of $\hat{n}$ (that is, a polynomial of $\hat{n}$).

An important result we need is that the subspace of states with a given total number of photons forms an irreducible representation of the group of PLTs. 
\begin{lemma}\label{lem Uirreps}
Let $\cF_n$ denote the space of vectors spanned by the Fock states $\ket{n_1,\hdots,n_S} $ where $\sum_{i=1}^S n_i =n$. Then $\cF_n$ is an irreducible representation of $U(S)$.
\end{lemma}
\begin{proof}
We write $\ket{n_1,\hdots,n_S}$ as $\frac{1}{\sqrt{n_1! \cdots n_S!}} (a_1^\dagger)^{n_1}\cdots (a_S^\dagger)^{n_S} \ket{\vec{0}}$. This makes evident that there is a one-to-one correspondence between the basis vectors of $\cF_n$ and the set of products of the $a_i^\dagger$ where the number of creation operators in the product is $n$. This correspondence extends to a linear bijection between $\cF_n$ and the space spanned by the set of monomials $\{ (a_1^\dagger)^{n_1}\cdots (a_S^\dagger)^{n_S} \}_{n_1+\hdots+n_S=n}$. This is because for any $\ket{\vec{n}} = \ket{n_1,\hdots,n_S}$ and $\ket{\vec{m}} = \ket{m_1,\hdots,m_S}$ one can write their arbitrary linear combination $c_1 \ket{\vec{n}} + c_2 \ket{\vec{m}}$ as
\begin{align}
c_1 \ket{\vec{n}} + c_2 \ket{\vec{m}} & =  \frac{c_1}{\sqrt{n_1! \cdots n_S!}} (a_1^\dagger)^{n_1}\cdots (a_S^\dagger)^{n_S} \ket{\vec{0}} + \frac{c_2}{\sqrt{m_1! \cdots m_S!}} (a_1^\dagger)^{m_1}\cdots (a_S^\dagger)^{m_S} \ket{\vec{0}} \nonumber \\
& = \left ( \frac{c_1}{\sqrt{n_1! \cdots n_S!}} (a_1^\dagger)^{n_1}\cdots (a_S^\dagger)^{n_S} + \frac{c_2}{\sqrt{m_1! \cdots m_S!}} (a_1^\dagger)^{m_1}\cdots (a_S^\dagger)^{m_S} \right) \ket{\vec{0}}.
\end{align}
The space spanned by $\{ (a_1^\dagger)^{n_1}\cdots (a_S^\dagger)^{n_S} \}_{n_1+\hdots+n_S=n}$ can be identified with the complex vector space of homogeneous polynomials of degree $n$ and in $S$ variables - namely, with the set of polynomials in $S$ variables where each term in the linear combination has the same total power $n$. We denote this vector space by $\cP_h(S,n)$. 

Now, the action on $\ket{n_1,\hdots,n_S}$ of an arbitrary PLT $U$ on $S$ modes results in
\begin{align}\label{lem Uirreps eq}
U \ket{n_1,\hdots,n_S} &= \frac{1}{\sqrt{n_1! \cdots n_S!}} (U a_1^\dagger U^\dagger)^{n_1}\cdots (U a_S^\dagger U^\dagger)^{n_S} \ket{\vec{0}}  \nonumber \\
& = \frac{1}{\sqrt{n_1! \cdots n_S!}} (\sum_{j=1}^S U_{M,1j} a^\dagger_j  )^{n_1}\cdots (\sum_{j=1}^S U_{M,Sj} a^\dagger_j  )^{n_S} \ket{\vec{0}}.
\end{align}
 Eq.~\ref{lem Uirreps eq} shows that the action of a PLT on a basis vector of $\cF_n$ can be identified with its action on the corresponding monomial in $\cP_h(S,n)$. To clarify, by the ``action" of a PLT in these two representations we mean the transformation performed by the automorphism associated with that PLT in the corresponding representation. Thus, by linear extension we can identify the action of a PLT on $\cF_n$ with its action on $\cP_h(S,n)$. That $\cP_h(S,n)$ forms an irreducible representation of $U(S)$ is a standard result that can be found in textbooks such as \cite[Chap.~6,11]{fulton:qc1991a}. By the demonstrated isomorphism between the action of $U(S)$ on $\cP_h(S,n)$ and its action on $\cF_n$, this implies that $\cF_n$ is also an irreducible representation of $U(S)$.
\end{proof}

For our analysis, we use operator-valued generating functions whose coefficients are
polynomials of mode operators. 
A consequence of Lem. \ref{lem Uirreps} is that if a polynomial $P(\vec{a},\vec{a^\dagger})$ commutes with PLTs, then it is a linear combination of the powers of $\hat{n} = \sum_{i=1}^S a_i^\dagger a_i$.
\begin{corollary}\label{cor PLTinvpolynum}
  A polynomial of mode operators that is invariant under the action of any PLTs is a polynomial of the
  total number operator.
\end{corollary}
\begin{proof}
If $A$ is a polynomial of the total number operator, it is expressed as $A = p(\hat{n})$, where we treat $p(\cdot)$ as a polynomial that can take both operators and numbers as its argument. Then the spectral decomposition of $A$ is given as $A = \sum_{k=0}^\infty p(k) \hat{D}_k $, where the sum is over non-negative integers and $\hat{D}_k$ is the projector onto $\cF_{k}$, which is the space of states having a total of $k$ photons. We show that any operator that is invariant under PLTs must have a spectral decomposition of this form. The characterization of the irreducible subspaces $\cF_{k}$ for PLTs in Lem. \ref{lem Uirreps} implies
  that the commutant of the action of the group of PLTs are operators that act as
  multiplication by a scalar $\lambda_{k}$ on $\cF_{k}$. This is due to Schur's lemma (Lem.~\ref{lem Schur} in Sec.~\ref{qo rep_theory}). 
  
  If $A$ is a polynomial of mode operators and is in the commutant, the coefficient
  $\lambda_{k}$ can be computed as $\bra{k_1,\hdots,k_S}  A \ket{k_1,\hdots,k_S}$ with any basis state where $k_1+\hdots+k_S = k$. $A$ can be written as a linear combination of monomials, where the creation operators are to the right of the annihilation operators in each monomial. This is accomplished by repeated use of the commutation relations. Also, since the action of $A$ preserves $\cF_{k}$, each monomial in this expansion must have the same number of creation and annihilation operators. Thus, for any given $\ket{k_1,\hdots,k_S}$, $\bra{k_1,\hdots,k_S}  A \ket{k_1,\hdots,k_S}$ is a polynomial in the $k_i$. This is because the monomials in the expansion of $A$ where the number of creation and annihilation are not equal for any mode have vanishing expectations with the basis states $\ket{k_1,\hdots,k_S}$, and thus terms of the form $\sqrt{n+k_i}$ do not appear in the evaluation of $\bra{k_1,\hdots,k_S}  A \ket{k_1,\hdots,k_S}$. It is left to notice that since $\bra{k_1,\hdots,k_S}  A \ket{k_1,\hdots,k_S}$ evaluates to the same polynomial for each basis vector $(a_i^\dagger)^k/\sqrt{k!} \ket{\vec{0}}$, it must have a contracted form that depends only on $k$. 
\end{proof}

In our studies we consider PLTs that have separate actions on orthogonal subspaces of mode space. Writing $\vec{a} = \vec{a'} \oplus \vec{a''}$, where $\vec{a'} = (a_1,\hdots,a_d)^T$ and $\vec{a''} = (a_{d+1},\hdots,a_S)^T$ for some $d <S$, consider a PLT $U_M $ that has the form $U_M= U'_M \oplus U''_M$, where $U_M'$ acts on $\vec{a'}$ and $U_M''$ acts on $\vec{a''} $. Then, we claim that the action of $U_M$ in state space decomposes as a tensor product $U' \otimes U''$.
\begin{lemma}\label{lem: U=U'oU''}
If $U_M= U'_M \oplus U''_M$ then $U$ satisfies $U = U' \otimes U''$ in state space.
\end{lemma}
\begin{proof}
It suffices to consider the action of $U$ on an arbitrary Fock state $\ket{n_1,\hdots,n_S}$. Let us observe that for $i \leq d $, $U a_i^\dagger U^\dagger =  (U'\otimes I_2) a_i^{\dagger} (U^{'\dagger}\otimes I_2)$, where $I_2$ is the identity on modes $a_{d+1},\hdots,a_S$. Similarly, for $ d+1 \leq i \leq S$, $U a_i^\dagger U^\dagger =  (I_1 \otimes U'') a_i^{\dagger} (I_1 \otimes U^{''\dagger})$, where $I_1$ is the identity on the modes $a_1,\hdots,a_d$. Then, starting with Eq.~\ref{lem Uirreps eq}, we use the above observation to obtain
\begin{align}
& U \ket{n_1,\hdots,n_S} = \frac{1}{\sqrt{n_1! \cdots n_S!}} (U a_1^\dagger U^\dagger)^{n_1}\cdots (U a_S^\dagger U^\dagger)^{n_S} \ket{\vec{0}}  \nonumber \\ 
& = \frac{1}{\sqrt{n_1! \cdots n_S!}} \left(  \left( (U'\otimes I_2) a_1^{\dagger} (U^{'\dagger}\otimes I_2) \right)^{n_1}  \cdots  \left( (U'\otimes I_2) a_d^{\dagger} (U^{'\dagger}\otimes I_2) \right)^{n_d} \right) \nonumber \\
& \hdots \left( \left(  (I_1 \otimes U'') a_{d+1}^{\dagger} (I_1 \otimes U^{''\dagger}) \right)^{n_{d+1}}  \cdots \left(  (I_1 \otimes U'') a_{S}^{\dagger} (I_1 \otimes U^{''\dagger}) \right)^{n_{S}} \right) \ket{\vec{0}} \nonumber  \\
& = \frac{1}{\sqrt{n_1! \cdots n_S!}} (U'\otimes I_2) ( (a_1^{\dagger})^{n_1}\cdots (a_d^{\dagger})^{n_d}) (U^{'\dagger}\otimes I_2) (I_1 \otimes U^{''}) ( (a_{d+1}^{\dagger})^{n_{d+1}}\cdots (a_S^{\dagger})^{n_S})(I_1 \otimes U^{''\dagger})\ket{\vec{0}}.
\end{align}
In the last line we can take the operator $I_1 \otimes U''$ in the middle of the expression all the way to the left since it commutes with the operators $a_1^\dagger,\hdots,a_d^\dagger$. We can similarly take the operator $U^{'\dagger} \otimes I_2$ in the middle all the way to the right of the expression. Then,
\begin{align}
U \ket{n_1,\hdots,n_S} & =  \frac{1}{\sqrt{n_1! \cdots n_S!}} ( (I_1 \otimes U^{''}) (U'\otimes I_2) ( (a_1^{\dagger})^{n_1} \cdots (a_S^{\dagger})^{n_S}) ) (U^{'\dagger}\otimes I_2) (I_1 \otimes U^{''\dagger})\ket{\vec{0}} \nonumber \\
& = (U' \otimes U'')    \frac{1}{\sqrt{n_1! \cdots n_S!}}  ( (a_1^{\dagger})^{n_1} \cdots (a_S^{\dagger})^{n_S}) )  \ket{\vec{0}} \nonumber \\
& =  (U' \otimes U'')  \ket{n_1,\hdots,n_S}.
\end{align}
\end{proof}

\section{Normal and Anti-Normal Orderings}\label{qo order}  
For a given polynomial of mode operators, one can express it as a linear combination of normally ordered monomials of mode operators by using the commutation relations repeatedly. A normally ordered monomial is a product of the mode operators, where all creation operators appear to the left of the annihilation operators. This way one can associate to each polynomial of mode operators a polynomial in complex variables and their conjugates. In particular, we associate to the annihilation operator $a_i$ the complex
variable $\alpha_{i}$ and to the $a_i^\dagger$ the complex conjugate of $ \alpha_{i}$. We show below how this association can be accomplished by evaluation of expectations of the polynomials of mode operators over coherent states. 

It is necessary to distinguish between formal expressions in variables
representing mode operators and the operators obtained by evaluating
these expressions. 
The ordering of operator variables in formal
expressions matters, which is to say that we treat the variables as
being fully non-commutative in the absence of evaluation as operators
or other specific contexts. The normal ordering is a manipulation of an expression in variables as follows: if $P$ is a polynomial
expression, then $\normord{P}$ is the expression obtained by rearranging the
terms in each monomial by moving the variables representing creation
operators to the beginning of the monomial. We also distinguish
between $P$ and its value as an operator after evaluation by
substituting actual mode operators for the variables. If it is necessary to be clear, we use $\hat{P}$ to denote
the operator value of the expression $P$ when we substitute operators for the
variables. In this section, we freely use $\reallywidehat{\hdots}$ to denote operators for better clarity of the presentation, and use ``bold face" print for variables to distinguish them from operators. So, $\boldsymbol{a_i}$ and $\boldsymbol{a_i^{\dagger}}$ 
are mode operator variables, not actual mode operators.

For a system occupying a single mode, if $P(\boldsymbol{a},\boldsymbol{a}^{\dagger})$ is a polynomial
expression in the mode variables and their adjoints with normally
ordered terms, then
$\bra{\alpha}\reallywidehat{P}(\boldsymbol{a},\boldsymbol{a}^{\dagger})\ket{\alpha}
= P(\alpha,\alpha^*)$ where $\ket{\alpha}$ is a coherent state. This comes from the evaluation of expectations over a sum of operators being linear in the operators, and that each term in the sum of $\reallywidehat{P}(\boldsymbol{a},\boldsymbol{a}^{\dagger})$ is of the form $c_i (\hat{a}^\dagger)^n\hat{a}^m$ so that $\bra{\alpha}c_i (\hat{a}^\dagger)^n \hat{a}^m \ket{\alpha} = c_i (\alpha^*)^n \alpha^m$. If $P(\boldsymbol{a},\boldsymbol{a}^{\dagger})$ is an arbitrary polynomial, it holds that
$\bra{\alpha}\reallywidehat{\normord{P(\boldsymbol{a},\boldsymbol{a}^{\dagger})}}\ket{\alpha}
= P(\alpha,\alpha^*)$, where $\reallywidehat{\normord{P(\boldsymbol{a},\boldsymbol{a}^{\dagger})}}$ is the operator obtained by first applying the normal ordering manipulation to the expression $P(\boldsymbol{a},\boldsymbol{a}^{\dagger})$ and then substituting the operators $\hat{a}$ and $\hat{a}^\dagger$ in place of the variables $\boldsymbol{a}$ and $\boldsymbol{a}^\dagger$, respectively, as described in the paragraph above. On the other hand,
$\bra{\alpha}\reallywidehat{P}(\boldsymbol{a},\boldsymbol{a}^{\dagger})\ket{\alpha}=
Q(\alpha,\alpha^*)$ for some polynomial
$Q$, which can be constructed by repeatedly applying the commutation relations to
the terms of $P$ to express $P$ as a linear combination of normally ordered monomials of mode operators. 

This discussion straightforwardly extends to the multimode case. In particular, let us introduce the vector of complex variables $\vec{\alpha} = (\alpha_1,\hdots,\alpha_S)^T$ and the vectors of mode variables $\vec{\boldsymbol{a}} = (\boldsymbol{a_1},\hdots,\boldsymbol{a_S})^T$ and $\vec{\boldsymbol{a}^\dagger} = (\boldsymbol{a_1}^\dagger,\hdots,\boldsymbol{a_S}^\dagger)$. Then, for any polynomial of mode variables $ P(\vec{\boldsymbol{a}},\vec{\boldsymbol{a}^{\dagger}})$ it holds that
$\bra{\vec{\alpha}}\reallywidehat{\normord{P(\vec{\boldsymbol{a}},\vec{\boldsymbol{a}^{\dagger}})}}\ket{\vec{\alpha}}
= P(\vec{\alpha},\vec{\alpha}^*)$, where $\ket{\vec{\alpha}} = \ket{\alpha_1}\otimes \hdots \otimes \ket{\alpha_S}$, while $\bra{\vec{\alpha}} \reallywidehat{P}(\vec{\boldsymbol{a}},\vec{\boldsymbol{a}^{\dagger}}) \ket{\vec{\alpha}} = Q(\vec{\alpha},\vec{\alpha}^*)$ for a unique polynomial $Q(\vec{\alpha},\vec{\alpha}^*)$. As a consequence, there is a one-to-one correspondence between polynomials of
$\vec{\alpha},\vec{\alpha}^*$ and operators obtained by evaluating polynomials in $\vec{\boldsymbol{a}},\vec{\boldsymbol{a}^{\dagger}}$. 
  
Normal ordering commutes with PLTs. In particular, for a PLT $U$:
\begin{align}
   U^{\dagger}\reallywidehat{\normord{P(\vec{\boldsymbol{a}},\vec{\boldsymbol{a}^{\dagger}})}}U
   &=\reallywidehat{\normord{P(U_M\vec{\boldsymbol{a}},U^*_M\vec{\boldsymbol{a}^{\dagger})}}},
  \label{eq:polyexprcov}\\
 \bra{\vec{\alpha}}U^{\dagger}\reallywidehat{\normord{P(\vec{\boldsymbol{a}},\vec{\boldsymbol{a}^{\dagger}})}}U\ket{\vec{\alpha}}
  &= P(U_M\vec{\alpha}, (U_M\vec{\alpha})^*).
  \label{eq:polyampcov}
\end{align}
To see this, it suffices to consider an arbitrary monomial of mode variables $M(\vec{\boldsymbol{a}},\vec{\boldsymbol{a}^\dagger})$. The operator $\reallywidehat{\normord{M(\vec{\boldsymbol{a}},\vec{\boldsymbol{a}^\dagger})}}$ has the form $\prod_{i=1}^S (a_i^\dagger)^{k_i} \prod_{j=1}^S a_j^{l_j}$ for some integers $k_i,l_j$. Hence,
\begin{align}\label{eq:monexprcov}
U^\dagger\reallywidehat{\normord{M(\vec{\boldsymbol{a}},\vec{\boldsymbol{a}^\dagger})}}U &= U^{\dagger}\prod_{i=1}^S (a_i^\dagger)^{k_i} \prod_{j=1}^S a_j^{l_j}U \nonumber \\
&= \prod_{i=1}^S (U^{\dagger} a_i^\dagger U)^{k_i} \prod_{j=1}^S (U^{\dagger}a_j U)^{l_j}  \nonumber \\
&= \prod_{i=1}^S (U_M^* \vec{a^\dagger})_i^{k_i} \prod_{j=1}^S (U_M \vec{a})_j^{l_j} \nonumber \\
&= \reallywidehat{\normord{M(U_M\vec{\boldsymbol{a}},U_M^*\vec{\boldsymbol{a}^\dagger})}},
\end{align} 
where in the second line we inserted the identity $U^{\dagger} U$ between the consecutive mode operators in the product. 

Let us introduce the expression of mode variables that corresponds to the total number operator, $\boldsymbol{n} = \sum_{i=1}^S \boldsymbol{a_i}^\dagger \boldsymbol{a_i}$. Below, the operators $(\hat n)_{k}$ for $k\in\nats$ denote the falling factorials of the total number operator - that is, $(\hat
n)_{k}=\hat n(\hat n-1)\hdots(\hat n-k+1)$.
\begin{lemma}\label{lem:ntotnum}
Normally ordered powers of $\hat{n} = \sum_{i=1}^S a_i^\dagger a_i$ are equivalent to the corresponding falling factorials:
\begin{align}
\normord{\hat{n}^k}&=(\hat n)_{k}.
\end{align}
\end{lemma}

\begin{proof}
Notice that $ \normord{\hat{n}^k}= \reallywidehat{\normord{\boldsymbol{n}^{k}}} $. Due to the bijection between the polynomials of mode operators and the polynomials of complex variables, it suffices to verify that for all $\vec{\alpha}$:
\begin{align}
  \bra{\vec{\alpha}}
  \reallywidehat{\normord{\boldsymbol{n}^{k}}}\ket{\vec{\alpha}}
  &= |\vec{\alpha}|^{2k},
\end{align}
and
\begin{align}
  \bra{\vec{\alpha}}
  (\hat n)_{k}\ket{\vec{\alpha}}
  &= |\vec{\alpha}|^{2k}.
\end{align}
Let $\vec{e_1}=(1,0,\hdots,0)$ be the vector with $1$ in the first entry and $0$'s elsewhere.
Given $\vec{\alpha}$, define $\alpha=|\vec{\alpha}|$, and let $U$ be a PLT for which $U^\dagger \ket{\vec{\alpha}} = \ket{\alpha \vec{e_1}}$. We use the fact that the number operator commutes with a PLT in conjunction with equation \ref{eq:polyexprcov} in the following:
\begin{align}
  \bra{\vec{\alpha}}
  \reallywidehat{\normord{\boldsymbol{n}^{k}}}\ket{\vec{\alpha}}
  &=\bra{\alpha\vec{e_1}}U^{\dagger}
       \reallywidehat{\normord{\boldsymbol{n}^{k}}}U \ket{\alpha\vec{e_1}}
       \nonumber \\
  &=\bra{\alpha\vec{e_1}}
       \reallywidehat{\normord{\boldsymbol{n}^{k}}}\ket{\alpha\vec{e_1}}
       \nonumber \\
  &=\bra{\alpha \vec{e_1}}\reallywidehat{\normord{(\boldsymbol{a}_{1}^{\dagger}\boldsymbol{a}_{1})^{k}}}\ket{\alpha \vec{e_1}} \nonumber\\
  &=\bra{\alpha \vec{e_1}} (a_{1}^{\dagger})^{k}a_{1}^{k}\ket{\alpha \vec{e_1}}\nonumber \\
  &=\alpha^{2k} = |\vec{\alpha}|^{2k}.
\end{align}
When transitioning from $S$ modes to mode $1$ in the steps above, we implicitly expanded the expression $ \reallywidehat{\normord{\boldsymbol{n}^{k}}}$ in terms of the commuting mode operators of the different modes and used the fact that the modes of the state besides the first are in vacuum. Thus, terms not involving the first mode vanish due to the existence of an annihilation operator of at least one of the other modes acting on vacuum.
Similarly:
\begin{align}\label{eqtn 4 lem:ntotnum}
  \bra{\vec{\alpha}}
  (\hat n)_{k}\ket{\vec{\alpha}}
  &= \bra{\alpha\vec{e_1}}U^{\dagger}(\hat n)_{k}
  U \ket{\alpha\vec{e_1}}\nonumber \\
  &= \bra{\alpha\vec{e_1}}(\hat n)_{k}
  \ket{\alpha\vec{e_1}}\nonumber \\
  &= \bra{\alpha \vec{e_1}}(a_{1}^{\dagger}a_{1})_{k}\ket{\alpha \vec{e_1}}.
\end{align}
Direct computation shows that $(a_{1}^{\dagger}a_{1})_{k}=
{a_{1}^{\dagger}}^{k}a_{1}^{k}$.
\begin{align}
(a_{1}^{\dagger})^{k}a_{1}^{k} &=   \sum_{m=k}^{\infty} (m)\cdots(m-k+1)\ket{m}\bra{m} \nonumber \\
 & = \left[ \sum_{m_1 =1}^{\infty} m_1 \ket{m_1}\bra{m_1}\right]\left[  \sum_{m_2 =1}^{\infty} (m_2-1) \ket{m_2}\bra{m_2}  \right]\cdots \left[  \sum_{m_k =1}^{\infty} (m_k-k+1) \ket{m_k}\bra{m_k}  \right] \nonumber \\
 &=  (a_{1}^{\dagger}a_{1})(a_{1}^{\dagger}a_{1}-1)\cdots (a_{1}^{\dagger}a_{1}-k+1) =  (a_{1}^{\dagger}a_{1})_{k},
\end{align}
where the kets and bras correspond to Fock states in mode $1$. From this it follows that the expression \ref{eqtn 4 lem:ntotnum} evaluates to $\alpha^{2k}$.
\end{proof}
A similar result can be obtained for anti-normally ordered powers of the number operator using similar proof methodology. Anti-normally ordering a monomial of mode operators means shifting all creation operators to the right of the annihilation operators. Since we do not use that result in this thesis, we decided to put it in the appendix (App.~\ref{app antinormord}) for interested readers.

\section{The Husimi Function and the Moment Problem}\label{qo husimi}

The Husimi function is an alternate and equivalent representation of the state. We refer the reader to, for example, \cite[Chap.~3]{leonhardt:qc1997a} for a discussion of the Husimi function and its properties. Here we mention that for an arbitrary $S$ mode state $\rho$ its Husimi function is defined as $Q(\vec{\alpha},\vec{\alpha}^*) = 1/\pi^S \bra{\vec{\alpha}}\rho \ket{\vec{\alpha}}$, where $\ket{\vec{\alpha}} = \otimes_{i=1}^S \ket{\alpha_i}$, and can be treated as a probability distribution since it is non-negative and integrates to one. One of the most important properties of the Husimi function is that its complex moments correspond to the expectations of anti-normally ordered mode operators. A complex moment is defined as the expectation of any monomial in $\vec{\alpha}$ and $\vec{\alpha}^*$ w.r.t. the Husimi function. More specifically, for any anti-normally ordered monomial $ \reallywidehat{\anormord{M(\vec{\boldsymbol{a}},\vec{\boldsymbol{a}^\dagger})}} $, where the triple dots imply anti-normal ordering of the expression between them, one has $\tr( \rho \reallywidehat{\anormord{M(\vec{\boldsymbol{a}},\vec{\boldsymbol{a}^\dagger})}}) = \int d\vec{\alpha} d\vec{\alpha}^* Q(\vec{\alpha},\vec{\alpha}^*)  M(\vec{\alpha},\vec{\alpha}^*)$. 

In our investigations on the invertibility of the relationship between the measurement statistics and the unknown state, we often find ourselves facing the problem of whether, for a given unknown state, the expectations of a certain set of polynomials of mode operators determine the state. Since any polynomial of mode operators can be written as a finite linear combination of anti-normally ordered monomials of mode operators, the problem can be recast as whether the expectations of a given set of anti-normally ordered operators determine the state. Since the state can be identified with its Husimi function, a directly related problem is whether a given set of complex moments of the Husimi function determine it. For a general probability distribution, this is one of the main sub-problems grouped under the name ``the moment problem" \cite{schmudgen2017moment}. For a multivariate probability distribution over complex variables, Ref.~\cite[Chap.~15]{schmudgen2017moment} defines its ``complex moment sequence" to be the set of expectations of all monomials in the variables and their complex conjugates, ordered as a sequence. A complex moment sequence is called ``determinate" if it uniquely determines the distribution.

What kinds of states have Husimi functions with a determinate complex moment sequence? The relevant result is given by Thm.~15.11 in Ref.~\cite{schmudgen2017moment}. There the multivariate Carleman condition is stated, which is a sufficient condition for the determinacy of a complex moment sequence. Adapted to Husimi functions, the condition implies that the Husimi function of an $S$-mode state is determined by its moments if
\begin{align}\label{eq: carleman_cond}
\sum_{n=1}^\infty \left( \frac{1}{  \int d\vec{\alpha} d\vec{\alpha}^* Q(\vec{\alpha},\vec{\alpha}^*)  \abs{\alpha_j}^{2n} }
\right)^{\frac{1}{2n}}  = \infty 
\end{align}
for $j=1,\hdots,S$. An important class of states that satisfy this condition are the Gaussian states. This is because Gaussian states have a Gaussian Husimi function \cite[Chap.~4]{serafini2017quantum}, and a Gaussian distribution over complex variables and their conjugates has a determinate complex moment sequence \cite[Chaps.~14 and~15]{schmudgen2017moment}.

Another important class of states that satisfy the Carleman condition are the states with bounded photon number. To see this, write an arbitrary $S$-mode state $\rho$ with at most $N$ photons as $\rho = \sum_{n_1+\hdots+n_S=0}^N \sum_{m_1+\hdots+m_S=0}^N h_{\vec{n},\vec{m}} \ket{\vec{n}}\bra{\vec{m}}$. The corresponding Husimi function is
\begin{align}
Q_{\rho}(\vec{\alpha},\vec{\alpha}^*) &= 1/\pi^S \bra{\vec{\alpha}}\rho \ket{\vec{\alpha}} \nonumber \\
& = 1/\pi^S \sum_{n_1+\hdots+n_S=0}^N \sum_{m_1+\hdots+m_S=0}^N h_{\vec{n},\vec{m}}  \bra{\vec{\alpha}}\ket{\vec{n}}\bra{\vec{m}} \ket{\vec{\alpha}}  \nonumber \\
& = 1/\pi^S \sum_{n_1+\hdots+n_S=0}^N \sum_{m_1+\hdots+m_S=0}^N h_{\vec{n},\vec{m}} \left[ \prod_{i=1}^S \frac{(\alpha_i^*)^{n_i}}{\sqrt{n_i!}}  e^{-\abs{\alpha_i}^2/2}\right] \left[ \prod_{j=1}^S \frac{\alpha_j^{m_j}}{\sqrt{m_j!}}  e^{-\abs{\alpha_j}^2/2} \right] \nonumber \\
& = 1/\pi^S e^{-\abs{\vec{\alpha}}^2} \textrm{poly}(\vec{\alpha},\vec{\alpha}^*),
\end{align}
where $\textrm{poly}(\vec{\alpha},\vec{\alpha}^*)$ is a (real-valued) polynomial of the variables in $\vec{\alpha}$ and $\vec{\alpha}^*$ determined by the $h_{\vec{n},\vec{m}}$. Thus, the Husimi function of $\rho$ is a product of a polynomial with the Gaussian $1/\pi^S e^{-\abs{\vec{\alpha}}^2}$. Such distributions (product of a polynomial with a Gaussian) also have determinate complex moment sequences \cite[Chap.~15]{serafini2017quantum}.

Our results about invertibility of the relationship between the unknown state and the measurement statistics in Chap.~\ref{chap invertibility} assume the Husimi function of the state has a determinate moment sequence. This is because our results are derived by relating the moments of the Husimi function to the moments of the total number operator(s) of the photon counter(s) in the given measurement configuration. But the probability distribution over the measurement outcomes of the photon counter(s), which we assume as given, contains more information than its sequence of moments if the latter do not determine the distribution. Therefore, by using the information available in the moments of the total number operator(s) we are effectively constraining ourselves to a smaller set of distributions. Thus, the condition that the Husimi function has a determinate moment sequence should be viewed as a technical condition that is required with our proof methodology. It is likely that, if the states with Husimi functions with determinate moment sequences are determined by the corresponding measurement statistics of a given measurement configuration, then any state is determined as well by its corresponding measurement statistics. This intuition is based on the observation that the set of density matrices with finite maximum photon number in finite modes (which are associated with Husimi functions that are determined by their complex moments) is dense in the space of density operators in the same number of modes. We do not investigate this question in this thesis.

\section{Generating Functionology}\label{qo genfun}

We utilize operator-valued generating functions and their expectations for our investigations of what can be determined about the parameters of the unknown state from parts of or from the entire probability distribution of measurement outcomes. We need to define the concept of an ``operator-valued generating function". Usually, a generating function is a way of capturing a sequence of numbers. In particular, given a sequence of numbers, one constructs a power series in some variable $z$ where the coefficient of the monomial $z^{n-1}$ is the $n$'th member of the sequence. An operator-valued generating function is a way of encoding a sequence of operators as coefficients of the monomials of $z$. One is encouraged to think of the variable of the power series as being symbolic (or formal) since the power series might not converge anywhere outside the point $z=0$. The word ``generating" comes from the fact that the generating function ``generates" the family of operators (or numbers) in the corresponding sequence by evaluating the derivatives of different orders of the generating function at $z=0$ when the (analytic version of the) generating function has a non-zero radius of convergence. The evaluation of the derivative of any order at zero can also be thought of as a linear functional on the sequence of coefficients that returns the corresponding element of the sequence multiplied by a factorial. We refer to the latter notion of the derivative as the ``formal derivative" of a given order. The standard algebraic operations such as addition and multiplication of generating functions are well-defined. Compositions of two formal series are not always defined and require certain conditions, but we do not use compositions in this work. An introduction to generating functions can be found in ``Generatingfunctionlogy" by Wilf \cite{wilf2005generating}.

Operator-valued generating functions are useful, in part, because they can capture all the information about the probabilities of all outcomes of an observable. For example, the ideal photon counter can be represented by $F(z) = \sum_{n}\hat{D}_{n}z^{n}$ so that the coefficients of the monomials of $z$ correspond to the projectors onto the subspaces of Fock space with fixed total photon number. The corresponding generating function for a state $\rho$ is given by $f_{\rho}(z) = \tr(F(z)\rho)$,
with the trace being evaluated independently on each coefficient of
$F(z)$. For a given state, operator-valued generating functions can also be used to simultaneously contain the information about the expectations of multiple observables. For example, we make frequent use of the following generating function:
\begin{align}\label{eq: example F(z)}
  F(z) &= \sum_{k=0}^{\infty}(\hat n)_{k}\frac{1}{k!}z^{k}\\
  F(z,\vec{\alpha},\vec{\alpha}^*) &= \sum_{k=0}^\infty |\vec{\alpha}|^{2k} \frac{1}{k!}z^k = e^{|\vec{\alpha}|^2 z}.
\end{align}
Here 
$F(z,\vec{\alpha},\vec{\alpha}^*) =
\bra{\vec{\alpha}}F(z)\ket{\vec{\alpha}}$ by Lem. \ref{lem:ntotnum}, where
operator and state operations are performed term-by-term. The
expectations of the coefficients of $F(z)$ can be determined from the
photon number distribution, which is true also for any linear combination of the
coefficients. It is in this sense that $F(z)$ generates a family of
operators whose expectations are known if the photon number distribution is known.  
Because of the one-to-one correspondence between polynomials of
$\vec{\alpha}$ and $\vec{\alpha}^*$ and polynomials of mode operators, this
family of operators can also be derived from
$F(z,\vec{\alpha},\vec{\alpha}^*)$, which has a nice
exponential form. 

This correspondence between polynomials of $\vec{\alpha}$ and $\vec{\alpha}^*$ and normally ordered operators, when applied to generating functions, has a deeper significance. In particular, as mentioned earlier, one can obtain the $n$'th member of the sequence of a generating function by using the formal derivative of the $(n+1)$'th order, but if the power series of the generating function has a finite radius of convergence then this formal derivative is equal to the usual derivative of the $(n+1)$'th order. We are not aware of results about the convergence of $F(z)$ in Eq.~\ref{eq: example F(z)} to a well-defined operator in some neighborhood of $0$, but it is clear that its expectation with any coherent state does converge to a well-defined function. More generally, consider an arbitrary operator-valued generating function $F'(z)$ whose coefficients are polynomials in mode operators.  Assume we are given the expectation of $F'(z)$ for all coherent state amplitudes in a neighborhood of zero. Then, we could obtain the coefficients of $F(z)$ as follows. To obtain the $n$'th coefficient, we would evaluate the $(n-1)$'th order derivative of $\bra{\vec{\alpha}}F'(z)\ket{\vec{\alpha}} $ at $0$ and record this value for every $\alpha$. Thus, we would know the expectations of the $n$'th operator in the sequence for every coherent state. Since we know the operator is a polynomial of mode operators, it can be converted to a normally-ordered polynomial. The degree of this polynomial is not known, but since we know its expectation for all coherent states, the associated complex-valued polynomial is identified by this set of expectations, and, hence, the operator is identified as well. 

Given a measurement device and an associated operator-valued generating function, we call the latter ``observable'' if the coefficients are operators
whose expectations can be determined from the measurement outcome probability distributions. Similarly, we call the operator-valued coefficients ``observable" (notice that this is an adjective, so as not to confuse with the standard observables) if their expectations can be determined. We also use this terminology when referring to a complex valued generating function associated with an operator-valued generating function. For example, if $F(z)$ in Eq.~\ref{eq: example F(z)} is observable, we say that $F(z,\vec{\alpha},\vec{\alpha}^*)$ is observable as well, and similarly for their coefficients. We also work with generating functions of several variables. Such operator-valued generating functions have operators as the coefficients of the monomials of the variables in the power series notation. Partial traces for operator-valued generating functions associated with two or more modes are well-defined as well. Namely, each operator in the sequence is partially traced, and the resulting sequence denotes the partially traced generating function. We sometimes use the language of partially tracing a complex-valued generating function, by which we naturally mean partially tracing the corresponding operator-valued generating function and then substituting the operators with expressions of variables.

Given an arbitrary observable generating function $G(\vec{z}) = \sum_{i_1=0}^\infty \cdots \sum_{i_n=0}^\infty A_{i_1,\hdots,i_n} z_1^{i_1}\cdots z_n^{i_n}$ in the variables $\vec{z} = (z_1,\hdots,z_n)$, where the $ A_{i_1,\hdots,i_n} $ are operator-valued coefficients, we often perform certain transformations of the variables, or multiply $G(\vec{z})$ by a scalar valued generating function in the same variables to obtain a generating function with a different sequence of operators. We want the resulting sequence of operators to be observable as well. In order for this to happen, each coefficient of the new generating function must be in the span of the $ A_{i_1,\hdots,i_n} $. Let us investigate the effect of such transformations on the coefficients in some detail.

We start with examining the coefficients of $H(\vec{z}) G(\vec{z})$, where $H(\vec{z})$ is a (known) scalar-valued generating function: $H(\vec{z}) = \sum_{i_1=0}^\infty \cdots \sum_{i_n=0}^\infty h(i_1,\hdots,i_n) z_1^{i_1}\cdots z_n^{i_n}$. The coefficient of an arbitrary monomial $z_1^{k_1}\cdots z_n^{k_n}$ of $H(\vec{z}) G(\vec{z})$ can be obtained by multiplying the formal power series of $G(\vec{z})$ and $H(\vec{z})$ and collecting the terms with degree $k_i$ in $z_i$ for all $i=1,\hdots,n$. More specifically,
\begin{align}
\textrm{coeff}_{z_1^{k_1}\cdots z_n^{k_n}}H(\vec{z}) G(\vec{z}) & = \sum_{i_1 =0}^{k_1}\cdots \sum_{i_n =0}^{k_n} A_{i_1,\hdots,i_n} h(k_1-i_1,\hdots,k_n-i_n).
\end{align}
Thus, each coefficient is a linear combination of some of the $ A_{i_1,\hdots,i_n} $, and is therefore observable.

Next, consider a variable transformation $z_i = f_i(\vec{x})$ for some set of polynomial functions $f_i$ in the variables $\vec{x} = (x_1,\hdots,x_m)$ that vanish at $\vec{x} = \vec{0}$. Then, the generating function $H(\vec{x}) = G(f_1(\vec{x}),\hdots, f_n(\vec{x}) )$ is observable. More specifically, the coefficients of the monomials in $\vec{x}$ are observable. To see this, expand $H(\vec{x})$ in the power series:
\begin{align}\label{arbitrary eq 1}
H(\vec{x}) &= G(f_1(\vec{x}),\hdots, f_n(\vec{x}) ) \nonumber \\
& = \sum_{i_1=0}^\infty \cdots \sum_{i_n=0}^\infty A_{i_1,\hdots,i_n} f_1(\vec{x})^{i_1}\cdots f_n(\vec{x})^{i_n},
\end{align}
and collect the terms for every monomial $x_1^{k_1}\cdots x_m^{k_m}$.
Since the $f_i$ do not have a zeroth order term, the lowest order term in $f_1(\vec{x})^{i_1}\cdots f_n(\vec{x})^{i_n}$ has a total degree of at least $i_1+\hdots+i_n$. Therefore, when collecting the terms associated with $x_1^{k_1}\cdots x_m^{k_m}$ in the power series expansion of Eq.~\ref{arbitrary eq 1}, the terms $A_{i_1,\hdots,i_n} f_1(\vec{x})^{i_1}\cdots f_n(\vec{x})^{i_n}$ with $i_1+\hdots+i_n >k_1+\hdots +k_m$ can be ignored. As for the terms $A_{i_1,\hdots,i_n} f_1(\vec{x})^{i_1}\cdots f_n(\vec{x})^{i_n}$ with $i_1+\hdots+ i_n \leq k_1+\hdots +k_m$, one can first ascertain the coefficient of $x_1^{k_1}\cdots x_m^{k_m}$ in each $f_1(\vec{x})^{i_1}\cdots f_n(\vec{x})^{i_n}$. The coefficient of $x_1^{k_1}\cdots x_m^{k_m}$ of $H(\vec{x})$ is then the linear combination of the $A_{i_1,\hdots,i_n}$ with $i_1+\hdots + i_n \leq k_1+\hdots +k_m$, with the weight of $A_{i_1,\hdots,i_n}$ in the combination given by the coefficient of $x_1^{k_1}\cdots x_m^{k_m}$ in the corresponding $f_1(\vec{x})^{i_1}\cdots f_n(\vec{x})^{i_n}$. This implies the desired observability of $H(\vec{x})$.

If we take the $ A_{i_1,\hdots,i_n} $ to be normally ordered polynomials of mode operators in $S$ modes, then we can take the partial trace of $G(\vec{z})$ with a coherent state in some of the modes. If we do not fix the amplitude of the coherent state, then the resulting generating function can be considered as a formal power series in the original variables $\vec{z}$ as well as in the variables corresponding to the amplitude of the coherent state. More specifically, we can consider taking the partial trace with an arbitrary coherent state $\ket{\vec{\alpha}} = \ket{\alpha_1,\hdots,\alpha_m}$ for $m<S$, without loss of generality. Let us denote the monomial $\alpha_1^{k_1}\cdots \alpha_m^{k_m} (\alpha_1^*)^{l_1} \cdots (\alpha_m^*)^{l_m} $ by $\vec{\alpha}^{\vec{k}} (\vec{\alpha}^*)^{\vec{l}}$ for ease of notation. Then, the transformation of each $ A_{i_1,\hdots,i_n} $ is of the following form:
\begin{align}\label{arbitrary eq 2}
A_{i_1,\hdots,i_n} \mapsto  \sum_{\vec{k}}\sum_{\vec{l}} A'_{\vec{k},\vec{l},i_1,\hdots,i_n} \vec{\alpha}^{\vec{k}} (\vec{\alpha}^*)^{\vec{l}},
\end{align}
 where the $A'_{\vec{k},\vec{l},i_1,\hdots,i_n}$ are operators in modes $m+1,\hdots,S$, and the sum is over a finite index set. The corresponding transformation of $G(\vec{z})$ is then 
\begin{align}
G(\vec{z}) \mapsto  \sum_{i_1=0}^\infty \cdots \sum_{i_n=0}^\infty \sum_{\vec{k}}\sum_{\vec{l}} A'_{\vec{k},\vec{l},i_1,\hdots,i_n} \vec{\alpha}^{\vec{k}} (\vec{\alpha}^*)^{\vec{l}},
\end{align}
which can be treated as a generating function in the variables $\vec{z}$, $\vec{\alpha}$ and $\vec{\alpha}^*$ with a sequence of coefficients comprised of the $A'_{\vec{k},\vec{l},i_1,\hdots,i_n}$.

At the risk of being redundant, we again note here that the operator-valued generating functions we consider in this work all have expectations with coherent states that converge to analytic functions in some neighborhood of $0$. Thus, the derivatives of any order evaluated at $0$ are well-defined. Crucially, this also applies to differentiation with respect to those coherent state amplitudes which we choose to treat as variables, such as in the preceding paragraph.

\section{Homodyne Detection}\label{sec homo_tomo}

\begin{figure}[!htb]
    \centering
\includegraphics[width=80mm]{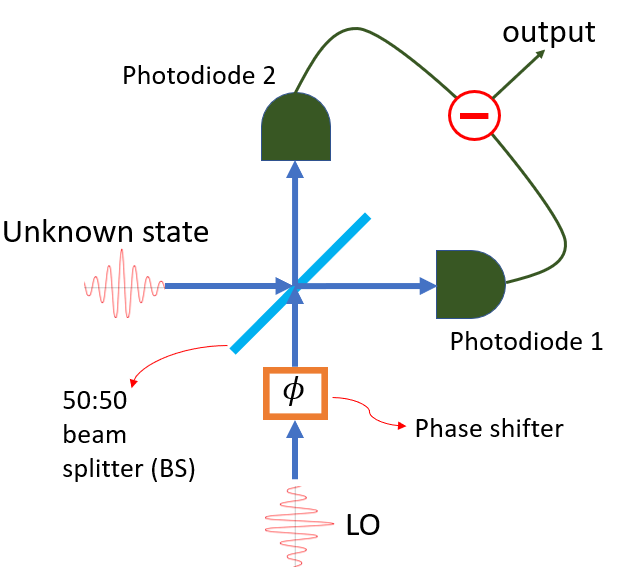}
    \caption{A diagram depicting pulsed homodyne detection. The accompanying description is in the main text.}
    \label{fig: homodyne_config}
\end{figure}

Here we give a brief overview of the pulsed version of standard homodyne detection due to its similarity to the measurement configurations we are interested in, as well as for serving as an inspiration to consider such measurement configurations. In the pulsed standard homodyne detection, the unknown state and the LO are wavepackets with finite temporal length (Fig.~\ref{fig: homodyne_config}). The LO is a high-amplitude laser in a coherent state $\ket{\alpha}$, and is interfered with the unknown state on a BS which has equal reflection and transmission coefficients (we also refer to such a BS as a ``balanced BS"). The outputs of the BS are measured by photodiodes which record the intensity over time, and the resulting signals are temporally integrated and subtracted. In the limit of infinitely high amplitude of the laser, the subtracted signal corresponds to a quadrature measurement of the reduced state in the mode matching the mode of the laser. Below we sketch a standard derivation of this result, which can be found in more detail in textbooks such as Ref.~\cite[Chap.~4]{leonhardt:qc1997a} or in Ref.~\cite[Chap.~18]{steck2007quantum}. 

Let us take mode $1$ as the mode occupied by the LO. Further, denote the mode operators corresponding to the unknown state by $a_i$, and the mode operators corresponding to the LO by $b_i$. The mode operators at the corresponding outputs of the BS are denoted by $a_i'$ and $b_i'$, respectively. These operators are related by the BS transformation according to $a_i' = 2^{-1/2} (a_i-b_i)$ and $a_i' = 2^{-1/2} (a_i+b_i)$. The difference of the output signals of the photodiodes is proportional to the observable $\hat{n}_{\Delta} = \sum_{i=1}^S (b_i')^\dagger b_i' -  \sum_{i=1}^S (a_i')^\dagger a_i' )$ in the narrowband limit (that is, in the limit where the central frequencies of the different modes are very close to each other). Substituting for the mode operators at the inputs of the BS one obtains
\begin{align}\label{eq: homotomo_ndiff}
 \hat{n}_{\Delta} & = \frac{1}{2} \sum_{i=1}^S \left( (a^\dagger_i+b_i^\dagger) (a_i+b_i) -   (a_i^\dagger-b_i^\dagger) (a_i-b_i)  \right) \nonumber \\
 &=  \sum_{i=1}^S \left( a^\dagger_i b_i +   a_i  b_i^\dagger \right).
\end{align}
At this point the ``classical" approximation of substituting the $b_i$ with the classical amplitude of the laser is performed, which results in
\begin{align}\label{eq: homotomo_approx}
 \hat{n}_{\Delta}  \approx a^\dagger_1 \alpha +   a_1 \alpha^*. 
\end{align}
One way that one can check that the approximation is valid for large values of $\abs{\alpha}$ is by comparing the expectations of the moments of Eq.~\ref{eq: homotomo_ndiff} with those of Eq.~\ref{eq: homotomo_approx} in the limit $\abs{\alpha} \rightarrow \infty$. Then, writing $\alpha = \abs{\alpha}e^{i \theta}$ and dividing  $\hat{n}_{\Delta}$ by $\sqrt{2} \abs{\alpha}$ one obtains the quadrature observable $\hat{q}_\theta = 2^{-1/2} (a^\dagger_1 e^{i\theta} +   a_1 e^{-i\theta})$ on the unknown state. Thus, by performing many measurements for a given value of $\theta$, one obtains an estimate of the probability distribution over the spectrum of $\hat{q}_\theta$. The reduced state in the mode matching the LO is determined by the set of probability distributions for all values of $\theta$, and thus one can obtain an estimate of this reduced state by performing many measurements for many different $\theta$ covering the angle range.

%% file: chapter_symmetries.tex
\chapter{Measurement Configurations and Their Symmetries }\label{chap sym and twirl}
In this chapter we first describe the family of measurement configurations we want to study in Sec.~\ref{sec meascon}. In Sec.~\ref{sec symset} we investigate the symmetries of the measurement configurations - in particular, for each measurement configuration we determine the set of unitary operators on the unknown state which leave the measurement statistics invariant. For each measurement configuration, the corresponding set of unitaries generate an equivalence class of states which produce the same measurement statistics. For each equivalence class there exists a particular simple representative which we derive in Sec.~\ref{sec twirl}. This representative is obtained by uniformly mixing the unknown state over all unitary operators that preserve the measurement statistics.

\section{Description of the Measurement Configurations}\label{sec meascon}

\begin{figure}
    \centering
\includegraphics[page =3, width=1\textwidth, trim={3cm 3cm 3cm 3cm}, clip]{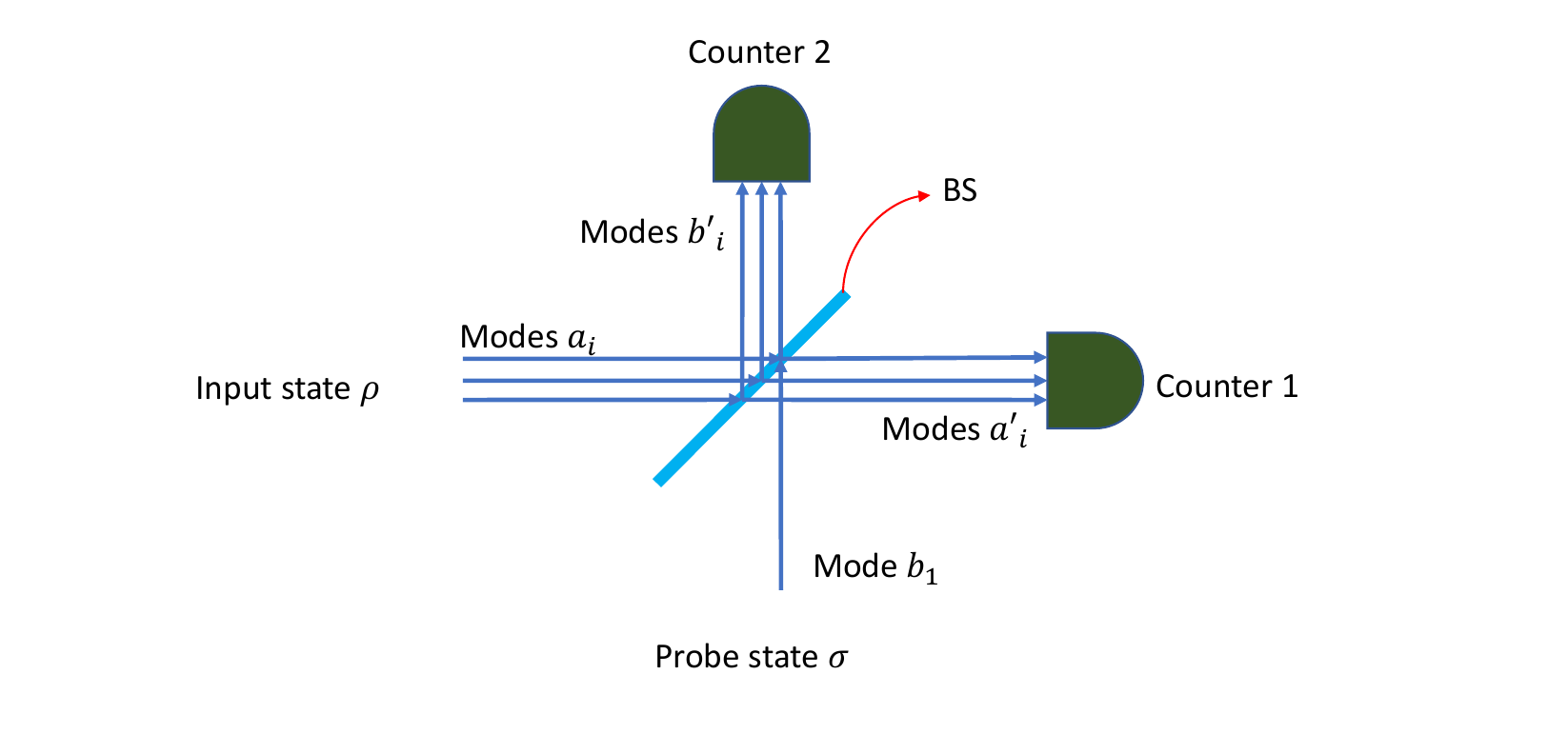}
    \caption{The family of measurement configurations with two counters: The probe \(\sigma\) in mode 1 is interfered with the unknown multimode state \(\rho\) on a BS and the outgoing modes are measured by two photon counters that do not distinguish the modes. The different configurations in the family differ by BS action.}
    \label{fig:bsconfig}
\end{figure}

The general form of the family of measurement configurations with two photon counters is
depicted in Fig.~\ref{fig:bsconfig}. The unknown state \(\rho\) and the probe state \(\sigma\) are directed
towards the inputs of a BS, and the reflected and transmitted modes
are measured by ideal photon counters that can distinguish perfectly
the total number of photons. The configurations in the family depend on the action of the BS, and on whether both detectors are utilized or only one of them is present (or, equivalently, the outcome of one of the detectors is forgotten during each measurement shot). We refer to a particular configuration in the family as a WFH (standing for weak-field homodyne) configuration. The intent of a measurement configuration is to learn the properties of $\rho$ by collecting data of the measurement outcomes of the counters for different probe states and keeping everything else fixed. We refer to the unknown state and its
modes as the ``input'' state and modes. The input state is assumed to
occupy \(S\) orthogonal modes with the vector of mode annihilation operators denoted by \(\vec{a} = (a_{1},...,a_{S})^T\). The corresponding vector of probe mode
operators is \(\vec{b} = (b_{1},...,b_{S})^T\). We treat these vectors
as row vectors and use transposition as necessary. We refer to the spaces consisting of the linear combinations of the $a_i$ and of the linear combinations of the $b_i$ as $\cA$ and $\cB$, respectively. The BS interferes
input mode \(a_{i}\) with probe mode \(b_{i}\).  The mode operators
after the BS are denoted by $\vec{a'}$ and $\vec{b'}$. The action of
the BS is associated with a unitary matrix $B$ so that, in the
Heisenberg picture, after the BS acts $(a_1, b_1,\hdots, a_S,b_S)^T \mapsto (a'_1, b'_1,\hdots, a'_S,b'_S)^T =
B(a_1, b_1,\hdots, a_S,b_S)^T$. The action of the BS in
the state space of all modes is denoted by \(U\) and transforms the
operators according to \( (a'_1, b'_1,\hdots, a'_S,b'_S)^T =
(U^{\dagger}a_{1} U,U^{\dagger}b_{1}U,\hdots,
U^{\dagger}a_{S}U, U^{\dagger}b_{S}U)^T =
B(a_1, b_1,\hdots, a_S,b_S)^T\).

The input state can be expressed in the Fock basis as \(\rho = \sum_{\vec{n},\vec{m}}h_{\vec{n},\vec{m}}\ket{\vec{n}}\bra{\vec{m}}\), where \(\ket{\vec{n}} = \ket{n_1,...,n_S}\) and \(\ket{\vec{m}} = \ket{m_1,...,m_S}\) are the Fock states with \(n_i\) and \(m_i\) photons in mode \(a_i\), respectively. The probe is assumed to occupy a single mode and its state is therefore of the form \(\sigma = \sigma_{1}\otimes \ket{\vec{0}}\bra{\vec {0}}\), where $\sigma_{1}$ is a state in mode $b_1$. On the output side of the BS, the photon counters are assumed to be lossless and count all incoming photons without distinguishing the modes. Hence, when \(k\) and \(l\) photons are registered by counters 1 and 2, respectively, the corresponding projector is given by:
\begin{equation} \label{eq D_kl}
    \hat{D}_{kl} =\hat{D}_k \otimes \hat{D}_l = \sum_{k_1+...+k_S=k}\sum_{l_1+...l_S=l}\ket{\vec{k}}\bra{\vec{k}}\otimes\ket{\vec{l}}\bra{\vec{l}},
\end{equation}
where \(\vec{k} = (k_1,...,k_S)\) and \(\vec{l} = (l_1,...,l_S)\). The probability of detecting \(k\) and \(l\) counts is given by
\begin{equation} \label{eq Born_D_kl}
    p_{kl} = \tr(U(\rho\otimes\sigma)U^\dagger \hat{D}_{kl}).
\end{equation}
This can be expressed in terms of partial traces over probe and input modes.
For this purpose, we label the subsystem consisting of the input modes by the subscript $\textsl{s}$ and
the subsystem consisting of the probe modes by the subscript $\textsl{p}$. We can apply the cyclicity property of the trace
and rewrite 
\begin{align}\label{eq Born_D_kl_2}
    p_{kl} &= \tr(U(\rho\otimes\sigma)U^\dagger \hat{D}_{kl})\nonumber\\
              &= \tr((\rho\otimes\sigma)( U^{\dagger} \hat{D}_{kl} U))\nonumber\\
              &= \tr_{\textsl{s}}(\rho\tr_{\textsl{p}}((\one\otimes\sigma)U^{\dagger}\hat{D}_{kl}U)).
\end{align}
Thus, for fixed $\sigma$, the measurement outcomes are associated with
the operators
\begin{align}\label{eq: Pi_kl_def}
\Pi_{kl}(\sigma) = \tr_{\textsl{p}}((\one\otimes\sigma)U^{\dagger}\hat{D}_{kl}U).  
\end{align}

For a WFH configuration with a single photon counter, we assume the counter 1 in Fig.~\ref{fig:bsconfig} is used, without loss of generality. Then, the probability of detecting $k$ photons is given by
\begin{align}\label{eq Born_D_k_2}
    p_{k} &= \tr(U(\rho\otimes\sigma)U^\dagger ( \hat{D}_{k} \otimes I))\nonumber\\
              &= \tr((\rho\otimes\sigma)( U^{\dagger} ( \hat{D}_{k} \otimes I) U))\nonumber\\
              &= \tr_{\textsl{s}}(\rho\tr_{\textsl{p}}((\one\otimes\sigma)U^{\dagger} ( \hat{D}_{k} \otimes I)  U)).
\end{align}
We can thus associate the operators
\begin{align}\label{eq: Pi_k_def}
\Pi_{k}(\sigma) = \tr_{\textsl{p}}((\one\otimes\sigma)U^{\dagger} ( \hat{D}_{k} \otimes I) U)  
\end{align}
with the measurement outcomes of the counter for fixed $\sigma$.

Before moving on to the next section, let us investigate the properties of the BS first. We find that we can significantly simplify the form of $B$. First, notice that because the BS does not interfere differently labeled modes, $B$ has the form 
\begin{align}
B &= \bigoplus_{i=1}^S \left[2 \times 2 \textrm{ unitary matrix}\right]_i
\end{align}
where the subscript $i$ indicates that the corresponding $2 \times 2$ unitary matrix acts on the vector of mode operators $(a_i,b_i)$. Let us, for now, parametrize the $i$'th unitary matrix as $B_i = \begin{pmatrix} \eta_i & \zeta_i \\ \zeta'_i & \eta'_i \\ \end{pmatrix}$. To exclude trivial BSs, we assume that $0<\abs{\eta_i}<1$ for all $i$. Now, the measurement outcome of a photon counter is not affected when a phase shifter is inserted into the path right before the counter (see Secs.~\ref{qo pc} and~\ref{qo PLT}). This means that applying an arbitrary phase shift $\phi_i$ ($\phi'_i$) on mode $a'_i$ ($b'_i$) at the outputs of the BS does not affect the outcome probabilities of the counters. This is equivalent to multiplying $B_i$ from the left by the matrix $\begin{pmatrix} e^{i \phi_i} & 0 \\ 0 & e^{i \phi'_i} \\ \end{pmatrix}$. Further, since for modes other than mode $b_1$ the probe is in vacuum, the measurement statistics cannot change when a phase shift is applied to those modes before they enter the BS. This is equivalent to multiplying $B_i$, for $i \neq 1$, from the right by the matrix $\begin{pmatrix} 1 & 0 \\ 0 &  e^{i \tilde \phi_i} \\ \end{pmatrix}$. As for $i=1$, we can do the same but keep track of the phase $\tilde \phi_1$ that we are using. This is possible since we can absorb $\tilde \phi_1$ into $\sigma_1$ and consider this modified probe state as the original probe state that appears in Eqs.~\ref{eq: Pi_kl_def} and~\ref{eq: Pi_k_def}. Let us compute the resulting transformation of the $B_i$ explicitly:
\begin{align}\label{eq: B_phase_altered}
B_i & \mapsto \begin{pmatrix} e^{i \phi_i} & 0 \\ 0 & e^{i \phi'_i} \\ \end{pmatrix}   \begin{pmatrix} \eta_i & \zeta_i \\ \zeta'_i & \eta'_i \\ \end{pmatrix} \begin{pmatrix} 1 & 0 \\ 0 &  e^{i \tilde \phi_i} \\ \end{pmatrix} \nonumber \\
& =  \begin{pmatrix} e^{i \phi_i} \eta_i & e^{i (\phi_i+\tilde \phi_i)} \zeta_i \\ e^{i \phi'_i} \zeta'_i & e^{i (\phi'_i+\tilde \phi_i)} \eta'_i \\ \end{pmatrix}.
\end{align}
Now, since $B_i$ is unitary, by looking at the entries of both sides of the identities $B_i B_i^\dagger = I $ and $ B_i^\dagger B_i = I$, where $I$ is the $2 \times 2$ identity, one can find that its parameters satisfy the relations $\abs{\eta_i} = \abs{\eta_i'}$, $\abs{\zeta_i} = \abs{\zeta_i'}$ and $\abs{\eta_i}^2 + \abs{\zeta_i}^2=1$, as well as the relations $ \eta_i^* \zeta_i'+\eta_i' \zeta_i^*=0$ and $ \eta_i \zeta_i^*+ (\eta_i')^* \zeta_i'=0$. By substituting the first two relations into the latter two relations, one finds that these produce the same constraint on the phases of the four parameters. More specifically, the four phases are related by the equation $\textrm{phase}(\eta_i)+\textrm{phase}(\eta'_i)-\textrm{phase}(\zeta_i)-\textrm{phase}(\zeta'_i) = \pm 2\pi$. Now, since Eq.~\ref{eq: B_phase_altered} introduces three independent phases into the parameters of $B$, this implies that we can always choose $\phi_i$, $\phi'_i$ and $\tilde \phi_i$ to be such that the right-hand side of Eq.~\ref{eq: B_phase_altered} is real. With our parametrization, this means that replacing the $B_i$ with the corresponding matrices $\begin{pmatrix} \abs{\eta_i} & \abs{\zeta_i} \\ \abs{\zeta_i} & -\abs{\eta_i} \\ \end{pmatrix}$, and modifying $\sigma_1$ accordingly, produces the same measurement statistics. If $\sigma_1$ is a coherent state, this modification results in the amplitude acquiring an additional phase equal to $\tilde \phi_1$. 

The above standardization of the $B_i$ means that if for $j \neq k$ the corresponding $\eta_j$ and $\eta_k$ have the same magnitude, then $B_j$ and $B_k$ can be replaced by the same matrix $\begin{pmatrix} \abs{\eta_j} & \abs{\zeta_j} \\ \abs{\zeta_j} & -\abs{\eta_j} \\ \end{pmatrix}$. Note that the parametrization of this matrix has a redundancy, as $\abs{\zeta_j}^2 =1- \abs{\eta_j}^2$. Let us perform this standardization for each $B_i$ in $B$, and rename $\abs{\eta_i}$ and $\abs{\zeta_i}$ into $\eta_i$ and $\zeta_i$, respectively, which are now both assumed to be real and positive. $B$ is then given by
\begin{align}
B &= \bigoplus_{i=1}^S \begin{pmatrix} \eta_i & \zeta_i \\ \zeta_i & -\eta_i \\ \end{pmatrix}.
\end{align}
If the $\eta_i$ are not all mutually distinct, we can further simplify the form of $B$ using the observation above. In particular, let us group the submatrices $B_i$ (after their standardization) that are identical together and count their numbers. We can introduce an ordered integer partition $S = S_1 + ... + S_K$ with $S_i \neq 0$. Here $K$ is the number of distinct submatrices and $S_i$ is the number of submatrices in the group labeled by $i$. Since our labeling of the modes besides mode $1$ was arbitrary, we can, without loss of generality, reshuffle and relabel the mode operators $a_i$ and $b_i$ of matching modes $2,\hdots,S$ such that first $S_1$ of the vectors $(a_i,b_i)^T$ transform by the same $2 \times 2$ submatrix (which is $B_1$ as mode $1$ is fixed), the next $S_2$ of them each transform by the same submatrix that is different from $B_1$, the next $S_3$ of them each transform by the same submatrix that is different from these two, and so on. With this ordering of the mode operators, $B$ has the form:
\begin{align} \label{eq: BS_general}
B & = \bigoplus_{i=1}^K I_{S_i}  \otimes  \begin{pmatrix}
\eta_i & \zeta_i \\ 
\zeta_i & -\eta_i \\
\end{pmatrix} ,
\end{align}
where each $I_{S_i}$ denotes the $S_i\times S_i$
identity matrix. We refer to $K$ as the ``size of the partition", and a particular ordered partition is sometimes denoted by $\lambda(S)$ for reference. We refer to a BS with a $B$ of this form (Eq.~\ref{eq: BS_general}) as a ``BS characterized by the partition $\sum_{i=1}^K S_i$", or, later in the thesis, as a ``BS with partition size $K$", when the values of the $S_i$ are irrelevant. This partition of $S$ produces a natural decomposition of $\cA$ and of $\cB$ - namely, $\cA = \bigoplus_{i=1}^K \cA_i$, where $\cA_i$ is given by all linear combinations of the $a_j$ with $S_{i-1}+1 \leq j \leq S_i$, and similarly for $\cB = \bigoplus_{i=1}^K \cB_i$. A physical example of such a $B$ is a BS with different reflection/transmission ratios depending on the polarization of the input modes.

\section{Symmetries of the Measurement Configurations}\label{sec symset}

In this section we determine the set of PLTs, which, when inserted
into the path of the input state before the BS, do not affect the measurement statistics for all probe states. For a given WFH configuration, the corresponding statistics-preserving set of PLTs depends only on the BS action - in particular, it depends on the ordered partition of $S$ characterizing the BS in Eq.~\ref{eq: BS_general}. More specifically, for a BS characterized by the partition $\sum_{i=1}^K S_i = S$ this set is composed of all PLTs that act trivially on mode $a_1$ and that act as automorphisms on $\cA_i$ for each $i$. This set forms a group as (1) the identity is in it, (2) the composition of any two PLTs of this form has the same properties, and (3) the same is true about the inverse of any PLT of this form since the inverse corresponds to its adjoint. We denote this group of PLTs by $U(\lambda(S))$ where $\lambda(S)$ stands for the ordered partition of the BS. $U(\lambda(S))$ gives rise to a natural equivalence relation on the set of input states - any two states that are related to each other by the action of a PLT in $U(\lambda(S))$ produce the same measurement statistics for all probe states. Further, since the set of states is convex, for a given state and for any subset of $U(\lambda(S))$, the convex combination of the set of transforms of this state under the action of the PLTs in the subset produces the same measurement statistics as that state. We use this fact in the next section to replace, or rather, represent the unknown state by a state with a simpler form.

We start by first showing that for a WFH configuration with both counters, a PLT in $U(\lambda(S))$ acting on the input modes preserves the measurement statistics for all input states. This statement trivially extends to WFH configurations with one counter.

\begin{prop}\label{thm: PLTpresmeasstat}
Consider a WFH configuration with two counters and with a BS unitary $B$ that is characterized by the ordered partition $\lambda(S)$. Then any PLT $W \in U(\lambda(S))$ on the input modes preserves the measurement statistics for all input and probe states.
\end{prop}

\begin{figure}[!htb]
    \centering
    \subfigure[]
    {
\includegraphics[page =1, width=0.75\columnwidth, trim={3.5cm 10cm 5.5cm 1.5cm}, clip]{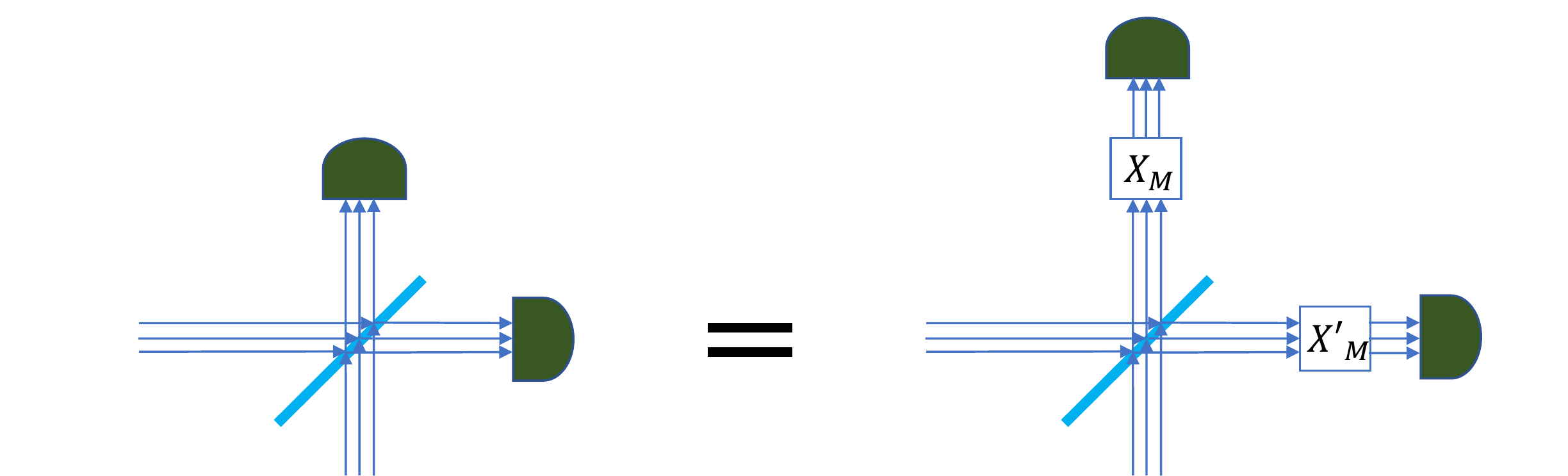}
    }
    \\
    \subfigure[]
    {
\includegraphics[page =1, width=0.75\columnwidth, trim={3.5cm 1.0cm 5.5cm 11cm}, clip,scale=0.8]{figures_1.pdf}
    }
    \\
    \subfigure[]
    {
\includegraphics[page =2, width=0.75\columnwidth, trim={3.5cm 8cm 5.2cm 3cm}, clip]{figures_1.pdf}
    }
    \\
  \subfigure[]
    {
\includegraphics[page =2, width=0.75\columnwidth, trim={3.5cm 1.5cm 5.2cm 12cm}, clip]{figures_1.pdf}
    }
    \caption{Symmetries represented as diagrams. a) Arbitrary PLTs $X_M$ and $X_M'$ in the output paths of the BS preserve the measurement statistics for all input states. b) The action of $X_M$ and $X_M'$ on the output paths of the BS is equivalent to the action of the PLT $Y_M$ (Eq.~\ref{equation 6}) on the joint input paths of the BS. c) If $X_M = X_M'$ and the $X_M$ in the signal (probe) path maps $\cA_i$ (the $\cB_i$) to itself for all $i$, then $Y_M$ simplifies to $X_M \oplus X_M$. d) If $X_M$ in addition to (c) has the form $X_M = 1 \oplus Z_M$ then its action on the probe state is trivial. Therefore, any PLT on the input modes of this form preserves the measurement statistics.}
  \label{fig:bsequiv}
\end{figure}

\begin{proof}
As explained in Sec.~\ref{qo PLT} of the last chapter, acting with a PLT on a state right before it is measured by a photon counter that does not distinguish the modes leaves the measurement statistics invariant. Thus, we are allowed to insert an arbitrary PLT in each output of the BS. This is pictorially represented by the equivalence of the two diagrams in Fig.~\ref{fig:bsequiv}a, where the PLTs in mode space on the output paths of the BS are denoted by \(X_M'\) and \(X_M\). To simplify our calculations we arrange the mode operators into a vector as $(a_1,\hdots,a_S,b_1,\hdots,b_S)^T$ so that the joint action of \(X_M'\) and \(X_M\) is expressed by their direct sum. With this ordering the action of $B$ is expressed as:
\begin{equation} 
B = \left[
\begin{array}{c|c}
   \Lambda_{\vec{\eta}} & \Lambda_{\vec{\zeta}}  \\
  \hline
    \Lambda_{\vec{\zeta}}  &   -\Lambda_{\vec{\eta}} 
\end{array}    \right],
\end{equation}\label{eqtn BSaltform}
where $\Lambda_{\vec{\eta}} = \bigoplus_{i=1}^K \eta_i I_{S_i}$ and similarly for the other quadrants. Since unitary matrices are invertible, there exists a PLT \(Y_M\) on the joint input paths of the BS such that:
\begin{equation} \label{equation 5}
    BY_M = (X_M'\oplus X_M)B.
\end{equation}
We can determine \(Y_M\) by multiplying both sides of Eq.~\ref{equation 5} with \(B^\dagger\):
\begin{align} \label{equation 6}
    Y_M &= B^\dagger(X_M'\oplus X_M)B=  \left[
\begin{array}{c|c}
   \Lambda_{\vec{\eta}} & \Lambda_{\vec{\zeta}} \\
  \hline
    \Lambda_{\vec{\zeta}}&  - \Lambda_{\vec{\eta}}
\end{array}    \right] \left[
\begin{array}{c|c}
   X'_M&0  \\
  \hline
   0  &  X_M 
\end{array}    \right]   \left[
\begin{array}{c|c}
   \Lambda_{\vec{\eta}} & \Lambda_{\vec{\zeta}}  \\
  \hline
    \Lambda_{\vec{\zeta}}  &  - \Lambda_{\vec{\eta}} 
\end{array}    \right]  \nonumber \\
 &= \left[ \begin{array}{c|c}
\Lambda_{\vec{\eta}} X'_M \Lambda_{\vec{\eta}}  + \Lambda_{\vec{\zeta}} X_M \Lambda_{\vec{\zeta}} & \Lambda_{\vec{\eta}} X'_M \Lambda_{\vec{\zeta}} - \Lambda_{\vec{\zeta}} X_M\Lambda_{\vec{\eta}} \\ 
  \hline
 \Lambda_{\vec{\zeta}} X'_M \Lambda_{\vec{\eta}} - \Lambda_{\vec{\eta}} X_M\Lambda_{\vec{\zeta}} & \Lambda_{\vec{\zeta}} X'_M \Lambda_{\vec{\zeta}}  + \Lambda_{\vec{\eta}} X_M \Lambda_{\vec{\eta}} \\
\end{array} \right].
\end{align}
So, any PLT of the form \(Y_M\) acting on the joint input and probe modes does not affect the measurement statistics (Fig.~\ref{fig:bsequiv}b). 

Let us set \(X'_M=X_M\). Further, let us consider an $X_M$ that has the form $X_M = \bigoplus_{i=1}^K X_{M,i}$, where each $X_{M,i}$ acts on the corresponding $\cA_i$ (or $\cB_i$ in the probe path). Then Eq.~\ref{equation 6} simplifies to \(Y_M = X_M \oplus X_M\) (Fig.~\ref{fig:bsequiv}c). To see this, notice that the form of the $X_M$ that we consider insures that $X_M$ commutes with $\Lambda_{\vec{\eta}}$ and $\Lambda_{\vec{\zeta}}$. Then, for example, the upper left quadrant of the last line of Eq.~\ref{equation 6} becomes $\Lambda_{\vec{\eta}}^* X_M \Lambda_{\vec{\eta}}  + \Lambda_{\vec{\zeta'}}^* X_M \Lambda_{\vec{\zeta'}} = \bigoplus_{i=1}^K \abs{\eta_i}^2 X_{M,i} + \abs{\zeta'}^2 X_{M,i} = \bigoplus_{i=1}^K  X_{M,i} $, where we used $\abs{\eta_i}^2+\abs{\zeta_i}^2=1$ for all $i$. The calculations for the other quadrants are done similarly. 

Then, if we further restrict \(X_M\) to act trivially on mode 1 the probe state is preserved since it only occupies the first mode. With our notation this implies requiring $X_{M,1}$ to have the form $X_{M,1} = 1 \oplus Z_{M,1}$, where $Z_{M,1}$ is empty if $S_1=1$ and otherwise acts on the mode subspace given by the linear combinations of $a_2,\hdots,a_{S_1}$. Then, $X_M$ can be written as $X_M = 1 \oplus Z_{M}$, where $Z_{M} = Z_{M,1} \oplus \left(\bigoplus_{i=2}^K X_{M,i} \right)$ acts on modes $a_2,\hdots, a_S$. Thus, we have shown that the measurement statistics are preserved when the input state is transformed by a PLT of the form $X_M = 1 \oplus Z_{M,1} \oplus \left( \bigoplus_{i=2}^K X_{M,i} \right)$ (Fig.~\ref{fig:bsequiv}d). It is left to notice that any $W \in U(\lambda(S))$ has the form of $X_M$ in mode space representation.

\end{proof}

\begin{corollary}\label{cor: PLTpresmeasstat}
Consider a WFH configuration with one counter and with a BS unitary $B$ that is characterized by the ordered partition $\lambda(S)$. Then any PLT $W \in U(\lambda(S))$ on the input modes preserves the measurement statistics for all input and probe states.
\end{corollary}
\begin{proof}
Prop.~\ref{thm: PLTpresmeasstat} shows that the second part of the corollary statement is true for a WFH configuration with a BS that is characterized by the same partition, but utilizes both counters. Then, the corollary follows from the fact that any WFH configuration with one counter can be represented by a WFH configuration with an identical BS and two counters, but where the outcome of counter 2 is always forgotten.
\end{proof}

Next, we want to show that a PLT that is not in $U(\lambda(S))$, when applied to the input modes, does not preserve the measurement statistics of counter 1 for all input states. We find the following observation useful.
\begin{lemma}\label{lem UcomDiagForm}
Let $\{a_{ij} \}_{ij}$, where $i =1,\hdots,K$, $j = 1,\hdots,S_i$, be a set of orthogonal modes. Further, define $\hat{n}_i = \sum_{j =1}^{S_i} a_{ij}^\dagger a_{ij} $, for all $1 \leq i \leq K$. If $\hat{\Lambda} = \sum_{i=1}^K \lambda_i \hat{n_i}$ for some non-zero $\lambda_i \in \mathbb{C}$, $\lambda_i \neq \lambda_j$ for $i \neq j$, then a PLT $W$ commutes with $\hat{\Lambda}$ iff its action in mode space decomposes as $W_M = \bigoplus_{i=1}^K W^i_{M}$, where $W^i_{M}$ acts on the vector of mode operators $\vec{a}_i = (a_{i1},\hdots,a_{iS_i} )^T$.
\end{lemma}

\begin{proof}
We first argue that a $W$ having the decomposition in the lemma statement commutes with $\hat{\Lambda}$. First, notice that this decomposition implies that $W^\dagger a_{ij} W = (W^i_M \vec{a}_i)_j =  \sum_{k =1}^{S_i} (W^i_M)_{jk} a_{ik}$ - that is, $W$ acts as a PLT on the mode space composed of the linear combinations of the $\{a_{ij} \}_{j=1}^{S_i}$. Then, according to Eq.~\ref{eq U_commute_n} of Sec.~\ref{qo PLT} this implies that $ \hat{n_i} W = W \hat{n_i}$. Therefore,
\begin{align}
W \sum_{j=1}^K \lambda_j \hat{n_j}  & =  \sum_{j=1}^K \lambda_j W \hat{n_j} \nonumber \\
& = \sum_{j=1}^K \lambda_j  \hat{n_j} W.
\end{align}

We prove the reverse now. Let us define $\vec{a} = (a_{11},\hdots,a_{1S_1},a_{21}\hdots,a_{2S_2},\hdots,a_{K1},\hdots,a_{KS_K})^T$ and label its components by double indices for ease of reference - that is, $(\vec{a})_{\underline{ij}} = a_{ij}$, where we underlined the indices to prevent confusion with matrix entries. Then $\hat{\Lambda}$ can be written as $\hat{\Lambda} = (\vec{a})^\dagger \Lambda_f \vec{a}$, where $\Lambda_f = \bigoplus_{i=1}^K \lambda_i I_{S_i}$ where $I_{S_i}$ is the $S_i \times S_i$ identity. Now, using the insights in Eq.~\ref{eq: identities_PLT_creat}, for an arbitrary $W$ we obtain the following set of identities:
\begin{align}
W^\dagger \hat{\Lambda} W &= W^\dagger \sum_{i=1}^K \lambda_i  \sum_{j =1}^{S_i} a_{ij}^\dagger a_{ij}  W \nonumber \\
& = \sum_{i=1}^K \lambda_i  \sum_{j =1}^{S_i}   \left( W^\dagger a_{ij}^\dagger  W \right) \left( W^\dagger  a_{ij}   W \right) \nonumber \\
& =    \sum_{i=1}^K \lambda_i      \sum_{j =1}^{S_i}   ( (\vec{a})^\dagger W^\dagger_M )_{\underline{ij}}      (W_M \vec{a})_{\underline{ij}} \nonumber \\
& = (\vec{a})^\dagger W^\dagger_M \Lambda_f W_M \vec{a}.
\end{align}
Thus, the requirement that $W^\dagger \hat{\Lambda} W = \hat{\Lambda}$ is equivalent to the requirement $W^\dagger_M \Lambda_f W_M  = \Lambda_f$. Now, let us compare the elements of $\Lambda_f W_M$ and $W_M \Lambda_f$ at the matrix entry location $(k,l)$. The indices of the rows and columns can be partitioned into consecutive blocks, with the $i$'th block having size $S_i$. Then, $(\Lambda_f W_M)_{kl} = \lambda_{i} W_{M,kl} $ where $i $ is the index of the block in which $k$ falls, and $( W_M \Lambda_f)_{kl} = \lambda_{j} W_{M,kl} $ where $j $ is the index of the block in which $l$ falls. Thus, we require $W_{M,kl} = 0$ whenever $k $ and $l$ do not fall in the same block to satisfy the commutativity of $W_M$ and $\Lambda_f$. This constrains $W_M$ to be block diagonal according to the decomposition in the lemma statement.
\end{proof}

In the following, \(\hat{n'}_1 = \sum_{i=1}^S (a'_i)^{\dagger} a'_i \) is the number operator associated with counter 1 in the Heisenberg picture. Similarly, we denote \(\hat{n}_1 = \sum_{i=1}^S a_i^{\dagger} a_i \).

\begin{prop}\label{thm PLTpresnumone}
Consider a WFH configuration with one or two counter(s) and with a BS unitary $B$ that is characterized by the ordered partition $\lambda(S)$. A PLT $W \notin U(\lambda(S))$ on the input modes does not preserve the expectation of $\hat{n'}_1$ for all input and probe states.
\end{prop}

\begin{proof}
Any PLT in $X \in U(\lambda(S))$ has the form $X_M = 1 \oplus Z_{M,1} \oplus \left( \bigoplus_{i=2}^K X_{M,i} \right)$, where $Z_{M,1}$ is empty if $S_1=1$ and otherwise acts on the mode subspace given by the linear combinations of $a_2,\hdots,a_{S_1}$, and each $X_{M,i}$ acts on the corresponding subspace $\cA_i$. We aim to show that the PLTs of the form of $X_M$ are the only PLTs acting on the input modes that preserve the expectation of the photon-number operator $\hat{n}_{1}$ at the output path going to counter 1. We prove by contradiction. Let us assume $W$ does preserve the expectation of $\hat{n'}_1$ for all input and probe states. Defining
\(\Pi_{\sigma} = \tr_{\textsl{p}}(\sigma U^\dagger \hat{n}_{1} U )\), in view of
Eqs.~\ref{eq Born_D_kl}~and~\ref{eq Born_D_kl_2},
the assumption is equivalent to the property that for all input states $\rho$:
\begin{align}\label{newequation 5}
\tr(\rho \Pi_{\sigma}) &= \tr(W\rho W^\dagger \Pi_{\sigma})  \nonumber \\
& = \tr(\rho W^{\dagger}\Pi_{\sigma} W).
\end{align}
By arbitrariness of $\rho$, this is equivalent to requiring that $ W^{\dagger}\Pi_{\sigma} W = \Pi_{\sigma}$.

Since $W \notin U(\lambda(S))$, $W_M$ does not have the direct sum decomposition as $X_M$ above. Then, either $W^{\dagger}a_{1}W$ contains terms involving mode operators other than those of mode 1, or there exists a mode $a_j$ in some $\cA_i$, such that $W^{\dagger}a_{j}W$ contains terms involving mode operators not in $\cA_i$, or both. To show
that $W$ does not commute with $\Pi_{\sigma}$ for some $\sigma$ it suffices to consider $\Pi_{\sigma}$ for coherent states $\sigma=\ket{\alpha}\bra{\alpha}\otimes\ket{\vec{0}}\bra{\vec{0}}$. With the BS coefficients defined in Eq.~\ref{eq: BS_general} we find that:
\begin{align} \label{newequation 7}
 U^\dagger (\sum_{i=1}^S a_i^{\dagger} a_i  )U & = \sum_{i=1}^S (U^\dagger a_i^{\dagger} U)( U^\dagger a_i U)   \nonumber \\
 &=  \sum_{i=1}^K \sum_{j = S_0+\hdots+ S_{i-1}+1}^{S_0+\hdots+S_i}  (\eta_i a_j^\dagger + \zeta_i  b_j^\dagger)(\eta_i a_j+ \zeta_i b_j)  ,
\end{align}
where in the second summation symbol in the second line we have defined $S_0 =0$. Putting Eq.~\ref{newequation 7} into the definition of $\Pi_\sigma$ and performing some algebraic manipulations, we obtain:
\begin{align} \label{newequation 8}
\Pi_{\ket{\alpha}\bra{\alpha}\otimes\ket{\vec{0}}\bra{\vec{0}}} &= \bra{\alpha}\bra{\vec{0}}  \sum_{i=1}^K \sum_{j = S_0+\hdots+S_{i-1}+1}^{S_0+\hdots+S_i}  (\eta_i a_j^\dagger + \zeta_i  b_j^\dagger)(\eta_i a_j+ \zeta_i b_j)  \ket{\alpha}\ket{\vec{0}} \nonumber \\
&=  \sum_{i=1}^K \sum_{j = S_0+\hdots+S_{i-1}+1}^{S_0+\hdots+S_i}  \bra{\alpha}\bra{\vec{0}} (\eta_i a_j^\dagger + \zeta_i  b_j^\dagger)(\eta_i a_j+ \zeta_i b_j)  \ket{\alpha}\ket{\vec{0}} \nonumber \\
&=   \zeta_1^2 \abs{\alpha}^2 + (\eta_1 \zeta_1) \alpha a_1^\dagger + (\eta_1 \zeta_1) \alpha^* a_1  +  \sum_{i=1}^K \sum_{j = S_0+\hdots+S_{i-1}+1}^{S_0+\hdots+S_i} \eta_i^2 a_j^\dagger a_j
\end{align}
where it is understood that the sandwiching by $\ket{\alpha}\ket{\vec{0}}$ acts as identity on $\cA$. The second and third summands are linear in $a_{1}$ and $a_{1}^{\dagger}$ and non-zero for $\alpha\ne 0$ and thus do not commute with a $W$ that does not act trivially on mode 1. The last sum does not commute with a $W$ the mode space representation of which cannot be decomposed as a direct sum of unitary actions on the mode subspaces $\cA_i$ according to Lem. \ref{lem UcomDiagForm}. 
\end{proof}

Now we combine the two propositions to state the main result of this section.
\begin{corollary}\label{cor PLTpresstat}
Given a WFH configuration (with either one or two counters) with a BS unitary $B$ that is characterized by the ordered partition $\lambda(S)$, a PLT $W$ on the input modes preserves the measurement statistics for all input and probe states iff $W\in U(\lambda(S))$.
\end{corollary}
\begin{proof}
According to Prop.~\ref{thm: PLTpresmeasstat} a PLT $W\in U(\lambda(S))$ on the input state $\rho$ preserves the measurement statistics for all $\rho$ and for all probe states. Prop.~\ref{thm PLTpresnumone} shows that a PLT $W \notin U(\lambda(S))$ on the input modes does not preserve the expectation of $\hat{n'}_1$, and, therefore, does not preserve the measurement statistics, for all probe and input states. Hence, the corollary statement.
\end{proof}

\section{Twirling the Input State}\label{sec twirl}

For a WFH configuration with a BS characterized by the partition $\lambda(S)$ the set of input states has a natural decomposition into equivalence classes. More specifically, the set of input states can first be partitioned into orbits under the action of $U({\lambda(S)})$. Then, for each orbit, consider the set of all convex combinations of the states in the orbit. These states cannot be distinguished from each other by the measurement statistics. Thus, each equivalence class is composed of a distinct set of orbits under the action of $U({\lambda(S)})$ that are related to each other through convex combinations. Each equivalence class has a particular simple member, which we call the ``twirled state". For an unknown state $\rho$, consider the state resulting from the following integration:
\begin{equation} \label{newequation 1}
\rho_t= \cT_\lambda (\rho) = \int_{X \in U(\lambda(S))} d\mu(X) X \rho X^\dagger.
\end{equation}
The integration is performed with respect to the normalized Haar measure \(\mu(X)\) on $U(\lambda(S))$. The interested reader can see, for example, Ref.~\cite{diestel2014joys} for a discussion of the Haar measure. For our purposes it is sufficient to know that \(\mu(X)\) is invariant under left and right multiplication by any element of $U(\lambda(S))$. We refer to $\cT$ as the ``twirling map", and $\rho_t$ is the twirled state associated with $\rho$ and belongs to the same equivalence class as $\rho_t$. 
\begin{lemma}\label{lem: rho_tIndistingrho}
\(\rho\) is indistinguishable from \(\rho_t\) by the WFH configuration. 
\end{lemma}
\begin{proof}
The channel $\cT$ maps $\rho$ into the uniformly weighed convex combination of all states in the same orbit as $\rho$ under the action of $U({\lambda(S)})$. According to Cor.~\ref{cor PLTpresstat} the states in the convex combination are indistinguishable from $\rho$. A convex combination of indistinguishable states is also indistinguishable from any of the states in the combination by linearity of Born's rule. 
\end{proof}

As described in Lem.~\ref{lem Uirreps} of Chap.~\ref{chap quantum optics} the action of $U(d)$ on the state-space of $d$-modes with total photon number $k$ forms an irreducible representation of $U(d)$. Thus, the action of $X \in U(d)$ on the total state-space decomposes as a direct sum of irreducible representations of $X$ on each subspace spanned by the states with the same total number of photons. Recall from Sec.~\ref{qo intro} that we denote the Fock space associated with a set of modes $a_1,\hdots, a_d$ by $\cF(a_1,\hdots,a_d)$, and by $\cF_n(a_1,\hdots,a_d)$ its subspace associated with total photon number $n$. In what follows, we take a liberal approach with the argument - more specifically, for each $2 \leq i \leq K$, $\cF(\cA_i) = \cF(a_{j+1},\hdots,a_{j+S_{i}})$, where $j = S_1+\hdots+S_{i-1}$, and similarly for the $\cF_n$. Further, we denote the uniformly mixed density matrix (that is, the identity operator normalized to be trace one) on $\cF_n(\cA_i)$, $2 \leq i \leq K$, by \(\mathbb{N}^i_{n}\), and on $\cF_n(a_2,\hdots, a_{S_1})$ by \(\mathbb{N}^1_{n}\). 

\begin{theorem} \label{theorem 1.2}
Consider a WFH configuration with a BS characterized by the ordered partition $\lambda(S) = \sum_{i=1}^K S_i$, where $S_1>1$. Then, the set of twirled states is the set of states of the form \(\rho_t =\sum_{i_1} \hdots \sum_{i_K} \chi_{\vec{i}} \otimes \left( \bigotimes_{k=1}^K \mathbb{N}_{i_k}^k \right) \), where the $\chi_{\vec{i}}$ are positive semi-definite operators
on mode 1 for each $\vec{i} = (i_1,\hdots,i_K)$.
\end{theorem}

\begin{proof}
We first show that $\cT_\lambda$ takes any operator to the commutant of $U(\lambda(S))$. For any $Y \in U(\lambda(S))$:
\begin{align}\label{eq: comutofT  }
Y \cT_\lambda (\cdot ) &= Y \int d\mu(X) X (\cdot ) X^\dagger =  \int d\mu(X) (YX) (\cdot ) X^\dagger =  \int d\mu(Y^\dagger Z ) Z (\cdot) (Y^\dagger Z)^\dagger \nonumber  \\ & = 
\int d\mu( Z ) Z (\cdot) Z^\dagger Y = \cT_\lambda (\cdot) Y,
\end{align}
where we have named $Z = YX$ and used the invariance of the Haar measure under multiplication by a group element. The statement of the theorem then follows from the application of Schur's lemma. In particular, any $\rho_t$ is an intertwiner from the representation of $U(\lambda(S))$ on state space to itself. Let us consider what the representation of $U(\lambda(S))$ on state space looks like. As shown in the last section, any $X \in U(\lambda(S))$ has the form $X_M = 1 \oplus Z_{M,1} \oplus \left( \bigoplus_{i=2}^K X_{M,i} \right)$ in mode space, and any $X_M$ of this form is in $U(\lambda(S))$ (see the proof of Thm~\ref{theorem 1.2} for details). Then, according to Lem.~\ref{lem: U=U'oU''} and the surrounding discussion, $X$ has a tensor product decomposition on state-space given by $X = I \otimes Z_{1} \otimes \left( \bigotimes_{i=2}^K X_{i} \right)$, and any $X$ having this form is in $U(\lambda(S))$. Here, $I$ is the identity on $\cF(a_1)$, $Z_1$ acts on $\cF(a_2,\hdots, a_{S_1})$, and, for $i \geq 2$, $X_i$ acts on $\cF(\cA_i)$. Therefore, the representation of $U(\lambda(S))$ on state-space is a tensor product of the trivial representation of $U(1)$ and of the representations of $U(S_1-1)$, $U(S_2),\hdots, U(S_K)$. 

Now, according to Cor.~\ref{cor shur_lem} and the ensuing discussion in Sec.~\ref{qo rep_theory}, the space of the intertwiners of $U(S_1-1)$ is spanned by the $\mathbb{N}^1_{i_1}$ for $i_1 = 0,1,2,\hdots$, while in case of $U(S_k)$, for $k >1$, it is spanned by the $\mathbb{N}^k_{i_k}$. Also, any operator on mode 1 alone is an intertwiner of $U(\lambda(S))$. The space of the intertwiners of $U(\lambda(S))$ must then be spanned by the tensor products of the $\mathbb{N}^k_{i_k}$ together with arbitrary action on mode 1. Any $\rho_t$ is precisely of this form, but satisfying additional constraints, such as the positivity and semi-definiteness of the $\chi_{\vec{i}}$ and having a unit trace due to being a density matrix.
\end{proof}

\begin{theorem} \label{theorem 1.2p}
Consider a WFH configuration with a BS characterized by the ordered partition $\lambda(S) = \sum_{i=1}^K S_i$, where $S_1=1$. Then, the set of twirled states is the set of states of the form \(\rho_t =\sum_{i_2} \hdots \sum_{i_K} \chi_{\vec{i}} \otimes \left( \bigotimes_{k=2}^K \mathbb{N}_{i_k}^k \right) \), where the $\chi_{\vec{i}}$ are positive semi-definite operators
on mode 1 for each $\vec{i} = (i_2,\hdots,i_K)$.
\end{theorem}
\begin{proof}
The proof is identical to that of Thm.~\ref{theorem 1.2}, but with the set of PLTs in $ U(\lambda(S))$ having the form $X = I \otimes \left( \bigotimes_{i=2}^K X_{i} \right)$, where $I$ is the identity on $\cF(a_1)$, and $X_i$ acts on $\cF(\cA_i)$. Thus, the representation of $U(\lambda(S))$ on state-space is a tensor product of the trivial representation of $U(1)$ and of the representations of $U(S_2),\hdots, U(S_K)$. Then, the form of $\rho_t$ follows by considering the set of intertwiners of this representation.
\end{proof}

The above results imply that a WFH
  configuration with a BS that is characterized by the partition $\sum_{i=1}^{K} S_i$ can at best determine the state which results from twirling the input state using the corresponding $U(\lambda(S))$. When performing state characterization with a WFH
  configuration, we could report the results using the parameters of the twirled
  state, but the twirled state depends on the total number of modes
  $S$, which is usually not known. According to the
  next two corollaries, in each equivalence class there exists an even simpler representative that occupies $K$ ($K+1$) modes when $S_1=1$ ($S_1>1$).
  
In what follows, for the given input state $\rho$ and the given WFH configuration, the $\chi_{\vec{i}}$ are the matrices appearing in the corresponding twirled state $\rho_t$, as in Thms.~\ref{theorem 1.2} and~\ref{theorem 1.2p}.
\begin{corollary}\label{cor K+1_twirlstate}
Let the BS of the WFH configuration be characterized by the ordered partition $\lambda(S) = \sum_{i=1}^K S_i$, where $S_1>1$. The state \(\tilde\rho_{K+1} = \sum_{i_1} \hdots \sum_{i_K} \chi_{\vec{i}} \otimes \left( \bigotimes_{k=1}^K \ket{i_k}\bra{i_k} \right) \), where the $\ket{i_1}$ are Fock states in mode $a_2$ while the $\ket{i_k}$ for $k>1$ are Fock states of mode $a_{S_1+\hdots+S_{k-1}+1}$, respectively, is indistinguishable from \(\rho\) by the WFH configuration.
\end{corollary}
\begin{proof}
This comes from the fact that twirling \(\tilde\rho_{K+1}\) results in the same twirled state \(\rho_t= \mathcal{T}_\lambda (\rho)\) as twirling $\rho$. In the description of $\tilde \rho_{K+1}$ we have suppressed the part of the state that is in vacuum in the $S$-mode space. In the following computation, for each $k>1$ where $S_k>1$, we ask the reader to imagine $\ket{i_k} \leftarrow \ket{i_k} \otimes \ket{0}_{k \perp}$, where $\ket{0}_{k \perp}$ is the vacuum state in the subspace of $\cF(\cA_k)$ that is orthogonal to all $\ket{i_k}$. As for $k=1$, the corresponding substitution occurs if $S_1>2$.
\begin{align}\label{newequation2}
\cT_\lambda(\tilde\rho_{K+1}) & =  \sum_{i_1} \hdots \sum_{i_K} \chi_{\vec{i}} \otimes \int d\mu (X) X \left(  \bigotimes_{k=1}^K \ket{i_k}\bra{i_k}  \right)X^\dagger   \nonumber \\
& = \sum_{i_1} \hdots \sum_{i_K} \chi_{\vec{i}} \otimes \left( \int d\mu (X)  \bigotimes_{k=1}^K X_k \ket{i_k}\bra{i_k}  X_k^\dagger \right) \nonumber  \\
& = \sum_{i_1} \hdots \sum_{i_K} \chi_{\vec{i}} \otimes \left( \bigotimes_{k=1}^K \mathbb{N}_{i_k}^k \right) 
\end{align}
where $X_k$ are the representations of $U(S_k)$ (except for $X_1$, which is a representation of $U(S_1-1)$) on the corresponding state-spaces. That is, $X = X_1 \otimes \hdots \otimes X_K$. In the last equality we used the fact that $d\mu(X) =d\mu(X_1)\cdots d\mu(X_K)$, and that $\int d\mu(X_k) X_k \ket{i_k}\bra{i_k} X_k^\dagger$ is an intertwiner of $U(S_k)$ on $\cF_{i_k}(\cA_k)$ for $k >1$, and an intertwiner of $U(S_1-1)$ on $\cF_{i_1}(a_2,\hdots,a_{S_1})$ for $k=1$. These intertwiners are maximally mixed states, since the integration preserves the trace of the operator it acts on.
\end{proof}
\begin{corollary}\label{cor K+1_twirlstate_p}
Let the BS of the WFH configuration be characterized by the ordered partition $\lambda(S) = \sum_{i=1}^K S_i$, where $S_1=1$. The state \(\tilde\rho_{1,K} = \sum_{i_2} \hdots \sum_{i_K} \chi_{\vec{i}} \otimes \left( \bigotimes_{k=2}^K \ket{i_k}\bra{i_k} \right) \), where $\ket{i_k}$ is a Fock state in the mode $a_{S_1+\hdots+S_{k-1}+1}$, is indistinguishable from \(\rho\) by the WFH configuration.
\end{corollary}
\begin{proof}
The proof is identical to that of Cor.~\ref{cor K+1_twirlstate}, with the difference being that the integration in Eq.~\ref{newequation2} is with respect to the measure $d\mu(X) =d\mu(X_2)\cdots d\mu(X_K)$ and $X = X_2\otimes \hdots \otimes X_K$.
\end{proof}

Notice that $\tilde \rho_{K+1}$ effectively occupies $K+1$ modes, while $\tilde \rho_{1,K}$ occupies $K$ modes. We see that if \(\tilde\rho_{K+1}\) (or $\tilde \rho_{1,K}$) is determined by the measurement statistics, then the corresponding \(\rho_t\) is determined as well since the unknown parameters are captured by the set of matrices $\chi_{\vec{i}} $. Since \(\tilde\rho_{K+1}\) ($\tilde \rho_{1,K}$) only occupies $K+1$ ($K$) modes depending on the value of $S_1$, we will work with this state to simplify future calculations. Finally, it is worth mentioning that, since our task is state characterization, it is reasonable to assume that $S_1>1$ as even in an experiment designed to have only a single mode that is acted on by the $B_1$ component of the BS, part of the state could still occupy some of these modes due to leakage or mode mismatch or background light. However, considerations of computational or statistical limitations might lead one to use $S_1=1$ in their model. Thus, we do study the $S_1=1$ case in Chap. \ref{chap invertibility}.

%% file: chapter_invertibility.tex
\chapter{Measurement Statistics Determine the Twirled State}\label{chap invertibility}

In this chapter we use the mathematical machinery of generating functions introduced in Sec. \ref{qo genfun} of Chap. \ref{chap quantum optics} to ascertain the conditions under which the twirled state is determined, or not determined, by a particular WFH configuration. The twirled state is determined by the measurement statistics of a WFH configuration if the latter uniquely specify it. We introduce the generating functions that are useful for analyzing the different kinds of measurement configurations described in Sec. \ref{sec genfunc for WFH} of Chap. \ref{chap sym and twirl}. We start by analyzing the WFH configurations where the BS acts identically on all modes in Sec. \ref{sec BS with trivial part} (we refer to such a BS as a ``BS characterized by a trivial partition"). We derive a set of results about when the twirled state can be determined by the WFH configuration - in particular, we show that for a twirled state with a finite maximum photon number, a finite set of different LO amplitudes are sufficient to determine the state using the statistics of only one of the counters. For the case where the statistics of both counters are used, a single LO magnitude suffices to determine the twirled state, provided the measurement statistics are known for all phases of the LO. We consider WFH configurations with BSs that are characterized by arbitrary partitions in Sec. \ref{sec BS with nontrivial part}. We treat the WFH configurations with one and two photon counters separately. Similarly, we make a special effort to consider the configurations with a BS characterized by a partition where $S_1=1$. We find that when only one counter is present, the measurement statistics of a WFH configuration can determine the twirled state if and only if the partition size is $K=1$ ($K\leq 2$) when $S_1>1$ ($S_1=1$). If the statistics of both counters are used, then the twirled state can be determined if and only if the size of the partition is $K\leq 2$ ($K\leq 3$) when $S_1>1$ ($S_1=1$). In the case where the size of the partition is $K=2$, for a state with finite maximum photon number, the corresponding twirled state can be determined from a finite set of different LO amplitudes.

\section{Generating Functions for WFH Configurations}\label{sec genfunc for WFH}

We first discuss the generating functions associated with WFH configurations with a BS that is characterized by a trivial partition, $B =  I_{S} \otimes \begin{pmatrix} \eta & \zeta \\ \zeta  & -\eta \end{pmatrix}$. See Sec.~\ref{qo genfun} of Chap. \ref{chap quantum optics} for a refresher on generating functions, whether operator valued or complex valued, and on the isomorphism between them that we exploit to ease our calculations. We design a generating function at the outputs of the BS that generates the observables associated with the photon counters as the coefficients of its power series. In particular, we want to capture the expectations of the products of all normally ordered powers of $\hat{n'}_1 = \sum_{i=1}^S (a'_i)^\dagger a'_i $ and $\hat{n'}_2 = \sum_{i=1}^S (b'_i)^\dagger b'_i$. This is accomplished by the generating function
\begin{align}\label{eq genfun_operval_BSout}
H(u,v) = \sum_{k=0}^\infty \sum_{l=0}^\infty \frac{1}{k!l!} \normord{\hat{n'}_1^k} \normord{\hat{n'}_2^l} u^k v^l ,
\end{align}
 where $u$ and $v$ are the formal generating function arguments representing the two photon counters. We can associate a complex valued generating function with $H(u,v)$ due to the isomorphism between the polynomials of mode operators and the polynomials of complex variables described in Sec.~\ref{qo order} of Chap. \ref{chap quantum optics}. In particular, let $\vec{\alpha}$
and $\vec{\gamma}$ be the formal variables for the two families of mode operators. Then,
\begin{align}\label{eq genfun_BSout}
  H(u,v,\vec{\alpha},\vec{\alpha}^*,
  \vec{\gamma},\vec{\gamma}^*) & = \bra{\vec{\alpha}} \bra{\vec{\gamma}} H(u,v) \ket{\vec{\alpha}} \ket{\vec{\gamma}} \nonumber \\
  & = \sum_{k=0}^\infty \sum_{l=0}^\infty \frac{1}{k!l!} \abs{\vec{\alpha}}^{2k} \abs{\vec{\gamma}}^{2k}  u^k v^l \nonumber \\
    &= e^{|\vec{\alpha}|^{2}u + |\vec{\gamma}|^{2}v}.
\end{align}

The action of the BS transforms the variables corresponding to the mode operators as
\begin{align}\label{eq: mode_op_trans_BS}
  \vec{\alpha}\oplus \vec{\gamma}
  &\mapsto (\eta\vec{\alpha}+\zeta \vec{\gamma})
     \oplus (\zeta \vec{\alpha}-\eta \vec{\gamma}).
\end{align}
Performing the corresponding substitution
yields the generating function for the modes at the input paths of the BS:
\begin{align}\label{eq genfun_trivpart}
  H_{\textrm{in}}(u,v, \vec{\alpha},\vec{\alpha}^*,
  \vec{\gamma},\vec{\gamma}^*)
    &=\exp(|\eta\vec{\alpha}+\zeta\vec{\gamma}|^{2}u
                   + |\zeta \vec{\alpha}-\eta\vec{\gamma}|^{2}v).
\end{align}
It is here that the usefulness of using generating functions, and, in particular, using the expectations of generating functions with normally ordered coefficients over coherent states, can be seen. The simple transformation performed above represents an infinite list of transformations of the operators $\normord{\hat{n'}_1^k} \normord{\hat{n'}_2^l}$ by the BS to obtain the corresponding operators on the input paths. Instead of keeping track of all these transformations, the expectation of the generating function in Eq.~\ref{eq genfun_operval_BSout} over coherent states allows us to represent the set of operators as a simple exponential function, with each operator in the set obtainable by taking the corresponding derivative evaluated at the origin and substituting the mode operators for the corresponding complex variables in a normal order. Then, any transformation of the operators which is linear in the mode operators and respects the normal ordering, such as that enacted by a BS, corresponds to a simple variable transformation in the exponent as in Eq.~\ref{eq genfun_trivpart}. The transformed operators can now be accessed by evaluating the partial derivatives of $H_\textrm{in}$ at the origin and making the appropriate substitutions. 

Performing a partial trace of the operator-valued generating function corresponding to $ H_{\textrm{in}}$ over the probe modes gives the generating function on the input state. Since the probe is in a coherent state $\vec{\gamma}$, the partial trace results in the same $ H_{\textrm{in}}(u,v, \vec{\alpha},\vec{\alpha}^*,
  \vec{\gamma},\vec{\gamma}^*)$, but where $\vec{\gamma}$ is now strictly a vector of complex variables, while $\vec{\alpha}$ and its powers remain a substitute for the corresponding polynomials of mode operators on the input state. The probe state $\vec{\gamma}$ is prepared in $\gamma\vec{e}_{1}$, where $\vec{e}_1 = (1,0,\hdots,0)^T$, and thus the generating function only depends on $\gamma$ and $\gamma^*$. In particular, for a specific probe amplitude $\gamma$, the operators associated with the coefficients of $ H_{\textrm{in}}(u,v, \vec{\alpha},\vec{\alpha}^*, \vec{\gamma},\vec{\gamma}^*)$ are the set of observable operators for that value of $\gamma$.

The family of operators associated with this generating function
does not change if the formal variables $u$ and $v$ are scaled by
non-zero constants.  Furthermore, we are also free to rescale the prepared mode amplitude by a nonzero
constant. The effect of this rescaling is that the physical amplitudes
that must be prepared are related to the amplitudes in the generating
function by a constant. This scaling freedom makes it possible to write
equivalent generating functions. Two such functions are helpful to us.
In both cases, we substitute $u\leftarrow u/\eta^{2}$ and $v\leftarrow v/\zeta^{2}$ in $H_{\textrm{in}}$.
Define $\xi = \zeta/\eta$.
The first function is
\begin{align}\label{eq genfun_trivpart_1}
  H_{1, \textrm{in}} 
  &=\exp(|\vec{\alpha}+\xi\vec{\gamma}|^{2}u
                   + |\vec{\alpha}-\xi^{-1}\vec{\gamma}|^{2}v).
\end{align}
Thus, our analysis requires only the parameter $\xi$ of the BS. For the second function, we also substitute $\gamma \leftarrow \gamma/\xi$ in $H_{\textrm{in}}$, which
gives
\begin{align}\label{eq genfun_trivpart_2}
  H_{2, \textrm{in}} 
  &=\exp(|\vec{\alpha}+\vec{\gamma}|^{2}u
                   + |\vec{\alpha}-\xi^{-2}\vec{\gamma}|^{2}v).
\end{align}
We use this form when only information from one photon counter is to be used,
which corresponds to setting $v=0$ in the generating function.

For WFH configurations that have BSs characterized by non-trivial partitions we again start with the generating function $H(u,v,\vec{\alpha},\vec{\alpha}^*, \vec{\gamma},\vec{\gamma}^*)$ at the outputs of the BS. For a BS characterized by the ordered partition $\sum_{i=1}^K S_i$ we write $\vec{\alpha}$ and $\vec{\gamma}$ as $\vec{\alpha} = \oplus_{i=1}^K \vec{\alpha_i}$ and $\vec{\gamma} = \oplus_{i=1}^K \vec{\gamma_i}$, where $\vec{\alpha_i}$ and $\vec{\gamma_i}$ have dimension $S_i$. The action of the BS with $B = \bigoplus_{i=1}^K I_{S_i} \otimes \begin{pmatrix} \eta_i & \zeta_i \\ \zeta_i  & -\eta_i \end{pmatrix}$, in the Heisenberg picture, transforms the formal variables as
\begin{align}
 \bigoplus_{i=1}^K (\vec{\alpha_i} \oplus  \vec{\gamma_i}) \mapsto  \bigoplus_{i=1}^K \left( (\eta_i\vec{\alpha_i}+\zeta_i \vec{\gamma_i})
     \oplus (\zeta_i \vec{\alpha_i} - \eta_i \vec{\gamma_i}) \right).
\end{align}
The generating function resulting from this transformation at the inputs of the BS is
\begin{align}\label{eq genfun_nontrivpart_a}
G_{\textrm{in}}(u,v,\vec{\alpha},\vec{\gamma},\vec{\alpha}^*,\vec{\gamma}^*) &= \exp(\abs{ \bigoplus_{i=1}^K \eta_i \vec{\alpha_i} + \zeta_i \vec{\gamma_i}}^2 u +\abs{ \bigoplus_{i=1}^K \zeta_i \vec{\alpha_i} -\eta_i \vec{\gamma_i}}^2 v ) \nonumber \\
 &= \exp( \sum_{i=1}^K \abs{\eta_i \vec{\alpha_i} + \zeta_i \vec{\gamma_i}}^2 u + \sum_{i=1}^K \abs{\zeta_i \vec{\alpha_i} -\eta_i \vec{\gamma_i}}^2 v ).
\end{align}
Performing a partial trace over the probe modes results in the same form of $G_{\textrm{in}}$, but where $\vec{\gamma}$ is strictly a vector of complex variables. $G_{\textrm{in}}$ can be simplified using the fact that $\vec{\gamma} = \vec{\gamma}_1 \oplus \vec{0}= \gamma \vec{e_1}$. Then,
\begin{align}\label{eq genfun_nontrivpart}
G_{\textrm{in}}(u,v,\vec{\alpha},\vec{\gamma},\vec{\alpha}^*,\vec{\gamma}^*) &= \exp(  \abs{\eta_1 \vec{\alpha_1} + \zeta_1  \vec{\gamma_1}}^2 u + \abs{\zeta_1 \vec{\alpha_1} - \eta_1 \vec{\gamma_1}}^2 v  +  \sum_{i=2}^K (\eta_i^2  u +  \zeta_i^2 v) \abs{ \vec{\alpha_i}}^2   ).
\end{align}
 When only one of the counters is used, that corresponds to setting $v=0$ in $G_\textrm{in}$. We denote the resulting generating function by $G_{\textrm{in},2}$:
\begin{align}\label{eq genfun_nontrivpart_onedet}
G_{\textrm{in},2}(u,\vec{\alpha},\vec{\gamma},\vec{\alpha}^*,\vec{\gamma}^*) &= \exp(  \left( \abs{\eta_1 \vec{\alpha_1} + \zeta_1  \vec{\gamma_1}}^2  +  \sum_{i=2}^K \eta_i^2  \abs{ \vec{\alpha_i}}^2 \right) u  ).
\end{align}
We treat $\gamma$ and $\gamma^*$ as variables of the generating functions here as well.

As explained in Sec. \ref{qo pc} the family of expectations of the operators corresponding to the variable expressions $\abs{\vec{\alpha}}^{2k}$ and $\abs{\vec{\gamma}}^{2l}$ for non-negative integers $k$ and $l$ at the outputs of the BS determine and are determined by the total photon number distribution of the photon counters. Since $H(u,v,\vec{\alpha},\vec{\alpha}^*,\vec{\gamma},\vec{\gamma}^*)$ generates these operators, for a given WFH configuration the generating function on the input state, obtained after transforming $H$ by the BS and tracing out the probe modes, generates a set of operators on the input state the expectations of which capture all the information available from the measurement statistics for the given probe state.

\section{WFH Configurations With a BS Characterized by a Trivial Partition}\label{sec BS with trivial part}
In this section we treat WFH configurations that utilize a BS characterized by a trivial partition. We first consider the WFH measurement configuration with one photon
counter and selectable LO amplitude $\gamma$. The subsection that comes after assumes both photon counters are available.
\subsection{Results With One Photon Counter}
 The unknown input state associated with a BS of this form is represented as a $2$-mode twirled state $\tilde \rho_2$. Let
$F(u,\gamma,\gamma^*,\vec{\alpha},\vec{\alpha}^*)=\exp(|\vec{\alpha}+\gamma\vec{e}_{1}|^{2}u)$
be the generating function $H_{2,\textrm{in}}$ with $v=0$. We denote
$\vec{\alpha} = (\alpha_1)\oplus\vec{\beta}$ so that $F$ can be expressed as
\begin{align}
 F(u,\gamma,\gamma^*,\vec{\alpha},\vec{\alpha}^*) & =
\exp( ( |\alpha_1+\gamma |^{2}+\abs{\vec{\beta}}^2 ) u) \nonumber \\
  & = \exp((|\vec{\alpha}|^{2}+ \alpha_1 \gamma^* + \alpha_1^*\gamma + |\gamma|^{2})u) 
  \end{align}
    The coefficients of $u^{d}$ of this generating function are polynomials of
$\vec{\alpha}, \vec{\alpha}^*$, $\gamma$ and
$\gamma^*$.  The polynomials of $\vec{\alpha},
\vec{\alpha}^*$, obtained from these coefficients by evaluation at any given
$\gamma$, represent polynomials of mode operators whose expectations can be
inferred from the observed counter outcome distributions at this
$\gamma$. Our first claim is that if these are available for all $\gamma$ in a neighborhood of $0$, then
the expectation of operators corresponding to any polynomial in
$|\vec{\beta}|^{2}$, $\alpha_1$ and $\alpha_1^*$ can be
determined. We later show that these expectations determine $\tilde\rho_2$. For this scenario, we can treat both $u$ and $\gamma$ as
generating function variables. 

\begin{theorem}\label{thm:1CntrNbhd}
  Consider a WFH configuration characterized by a BS with a trivial partition and with one photon counter. If the counter
  outcome distribution is known exactly for all $\gamma$ in a neighborhood
  of $0$, then the expectations of the operators corresponding to
  any polynomial in $|\vec{\beta}|^{2}$, $\alpha_1$ and $\alpha_1^*$
  can be determined.
\end{theorem}
\begin{proof}
  Let $F_1(u,\gamma)$ be the generating function
  \begin{align}
    F_{1}(u,\gamma) &= \exp(-|\gamma|^{2}u)F(u,\gamma) 
      = \exp((|\vec{\alpha}|^{2}+ \alpha_1\gamma^*+  
          \alpha_1^*\gamma)u) .
  \end{align}
  Then $F_1$ is observable at each $\gamma$ since the polynomial
  coefficients of the powers of $u$ change by a non-zero scalar.
  Let $m(a,b,c) = |\vec{\alpha}|^{2a}\alpha_1^{b}(\alpha_1^*)^{c}$,
  where $a,b,c$ are non-negative integers. We can extract $m(a,b,c)$ from
  $F_{1}$ by differentiation and evaluation as follows:
  \begin{align}\label{eq F_1_deriv}     
         \left[\partial_{u}^{a}\frac{1}{u^{a+b}}\left[\partial_{\gamma}^{c}\partial_{\gamma^*}^{b} F_{1}\right]_{\gamma=0}\right]_{u=0}
&=         \left[\partial_{u}^{a}\frac{1}{u^{a+b}}\left[ (u \alpha_1)^b (u \alpha_1^*)^c  F_{1}\right]_{\gamma=0}\right]_{u=0} \nonumber \\     
 &=          \left[\partial_{u}^{a}  \exp(|\vec{\alpha}|^{2}u) \right]_{u=0} \alpha_1^b (\alpha_1^*)^c \nonumber \\ 
     &=          |\vec{\alpha}|^{2a}\alpha_1^b (\alpha_1^*)^c = m(a,b,c),
  \end{align}
  This shows that the operator corresponding to $m(a,b,c)$ is observable given that the counter
  outcome distributions are known for all $\gamma$ in a neighborhood of
  $0$. To complete the proof, consider an arbitrary $m'(a,b,c)=|\vec{\beta}|^{2a}\alpha_1^{b}(\alpha_1^*)^{c}$.
    We can write
   \begin{align}
      m'(a,b,c) &= (|\vec{\alpha}|^{2}-|\alpha_1|^{2})^{a}
         \alpha_1^{b}(\alpha_1^*)^{c} \nonumber\\      
      &=
          \sum_{j=0}^{a}\binom{a}{j}|\vec{\alpha}|^{2j}|\alpha_1|^{2(a-j)}
          \alpha_1^{b}(\alpha_1^*)^{c} \nonumber\\
      &= \sum_{j=0}^{a}\binom{a}{j} m(j,b+a-j,c+a-j).\label{eq:m'_from_m}
   \end{align}
   It follows that the $m'(a,b,c)$ can be expressed as linear combinations
   of the $m(a,b,c)$. 
\end{proof}
We remark here that in the proof above we treated $\gamma$ and $\gamma^*$ as actual variables and thus the evaluations of the derivatives in Eq.~\ref{eq F_1_deriv} are actual transformations of the complex-valued function $F_1$. The observability of the operators corresponding to the $m(a,b,c)$ is deduced from the fact that these operators are determined by their expectations with the set of all coherent states (see Sec.~\ref{qo genfun} for further discussion). We do not make this remark anymore when we treat the probe magnitude as a generating function variable that can be used for differentiation.

Thm.~\ref{thm:1CntrNbhd} shows that the operators corresponding to the $m'(a,b,c)=|\vec{\beta}|^{2a}\alpha_1^{b}(\alpha_1^*)^{c}$ are observable when the measurement statistics are known for all $\gamma$ in a neighborhood of $0$. Recall that at the beginning of Chap.~\ref{chap quantum optics} we made a brief note that the family of states we consider in this thesis are the states with rapidly decaying Wigner functions. This restriction on the set of states was necessary to ensure that the expectations of all polynomials of mode operators are defined. We also mentioned that Ref.~\cite{hernandez2021rapidly} shows that all rapidly decaying Wigner functions are Schwartz functions, so this family of states is quite large and includes the important class of Gaussian states \cite{serafini2017quantum}. For the discussion of this section, this restriction means that the expectations of all operators corresponding to the $m'(a,b,c)$ are finite for $\tilde \rho_2$. The question is if these expectations determine $\tilde \rho_2$? At least for the set of two-mode twirled states with Husimi functions that have determinate complex moment sequences, the answer is affirmative. See Sec.~\ref{qo husimi} for a refresher on the relevant concepts.

The first thing to note is that since $\tilde \rho_2$ is in vacuum for the modes $\geq 3$, the operator corresponding to $m'(k, j,i )$ can be replaced by $ (a_1^\dagger)^i a_1^j (a_2^\dagger)^k  a_2^k$ as the expectations of all the other monomials after expanding the sum $\sum_{l=2}^S a_l^\dagger a_l$, and normally ordering the summands, vanish. Further, notice that we can use the commutation relations to express any anti-normally ordered monomial of the form $   a_1^i (a_1^\dagger)^j   a_2^k (a_2^\dagger)^k$ as a (finite) linear combination of the normally-ordered monomials $(a_1^\dagger)^i a_1^j (a_2^\dagger)^k  a_2^k$. This is because the application of the commutation relation for a given mode does not change the difference in the powers of the creation and annihilation operators for that mode. Therefore, the expectations of the $   a_1^i (a_1^\dagger)^j   a_2^k (a_2^\dagger)^k$ are also determined by the measurement statistics under the assumptions of Thm.~\ref{thm:1CntrNbhd}. We note that, for the rest of this section, by a ``linear combination" we mean a finite linear combination.

\begin{corollary}\label{cor:1CntrNbhd}
Consider a WFH configuration characterized by a BS with a trivial partition and with one photon counter. Assume the measurement statistics are known for all $\gamma$ in a neighborhood of $0$. Denote the Husimi function of $\tilde \rho_2$ by $Q(\vec{\delta},\vec{\delta}^*)$, where $\vec{\delta} = (\delta_1,\delta_2)$. If the complex moment sequence of $Q(\vec{\delta},\vec{\delta}^*)$ is determinate, then $\tilde \rho_2$ is determined by the measurement statistics.
\end{corollary}

\begin{proof}
As discussed in the paragraphs before the corollary statement, the assumptions of the statement about the available measurement statistics imply, according to Thm.~\ref{thm:1CntrNbhd} and due to the fact that $\tilde \rho_2$ occupies the first two modes, that the expectations of all operators of the form $   a_1^i (a_1^\dagger)^j   a_2^k (a_2^\dagger)^k$ are determined. Further, the expectations of the anti-normally operators not of this form are zero. To see this, notice that every operator in this set is of the form $   a_1^i (a_1^\dagger)^j   a_2^k (a_2^\dagger)^l$, where $k \neq l$. Since $\tilde \rho_2 $ is of the form $\tilde \rho_2 = \sum_i \chi_i \otimes \ket{i}\bra{i}$, first taking the partial trace of its product with $   a_1^i (a_1^\dagger)^j   a_2^k (a_2^\dagger)^l$ over mode $1$ leaves us with a (possibly infinite) sum of expectations of $a_2^k (a_2^\dagger)^l$, each with a Fock state in mode $2$. The latter all vanish since $a_2^k (a_2^\dagger)^l$ takes a Fock state to a different Fock state when $k \neq l$.

We now use the correspondence between the expectations of anti-normally ordered monomials of mode operators and the moments of the Husimi function. For $\tilde \rho_2$, this correspondence is given by
\begin{align}
\tr (\tilde \rho_2 a_1^i (a_1^\dagger)^j   a_2^k (a_2^\dagger)^l) =  \int d\vec{\delta} d\vec{\delta}^* Q(\vec{\delta},\vec{\delta}^*)  \delta_1^i (\delta_1^*)^j   \delta_2^k (\delta_2^*)^l
\end{align}
for all non-negative integers $i,j,k,l$. The set of integrals on the right-hand side of the identity comprise the set of complex moments of $Q(\vec{\delta},\vec{\delta}^*) $. By the preceding discussion, these are all determined by the form of $\tilde \rho_2 $ and by the measurement statistics. And, since, by the assumption of the corollary statement, the complex moment sequence of $Q(\vec{\delta},\vec{\delta}^*) $ is determinate, $Q(\vec{\delta},\vec{\delta}^*) $ is determined by the measurement statistics. Therefore, $\tilde \rho_2$ is determined as well.
\end{proof}

Analytic arguments can be used to show that the expectations of the
operators associated with polynomials in $|\vec{\beta}|^{2}$,
$\alpha_1$ and $\alpha_1^*$ can be determined given the counter
probability distributions for $\gamma$ in any neighborhood, not just
neighborhoods of $0$. But practical concerns imply that these
expectations be determined from a finite set of $\gamma$ only.  In
this case, it is likely not possible to determine the desired
expectations unconditionally. Instead, we consider expectations for
states with less than $N$ photons in the modes. With this bound on
photon number, because of normal ordering, operators associated with
monomials in $\vec{\alpha},\vec{\alpha}^*$, where either the total degree of the creation operators or the total degree of the annihilation operators in the monomial is greater than $N$, are guaranteed to have zero expectation. It
therefore suffices to construct the $m'(a,b,c)=|\vec{\beta}|^{2a}\alpha_1^{b}(\alpha_1^*)^{c}$ in the set $\cM'=\{m'(a,b,c): a+\max(b,c) \le N\}$. A stronger result
is the next theorem, for which we need the following
definitions. Let $\cP_{N}$ be the vector space spanned by the monomials $\gamma^k (\gamma^*)^l$ for $\gamma\in\cmplx$ with $k \leq N$, $l \leq N$. Let
$\cM=\{m(a,b,c): a+\max(b,c) \le N\}$, where $m(a,b,c) = |\vec{\alpha}|^{2a}\alpha_1^{b}(\alpha_1^*)^{c}$. Let us further introduce $\cM'_{>} = \{m'(a,b,c): b \leq c, a+c \leq N \} \subset \cM'$ and $\cM_{>} = \{m(a,b,c): b \leq c, a+c \leq N \} \subset \cM'$ for future results. Note the implicit dependence of $\cM$, $\cM'$, $\cM_>$ and $\cM'_>$ on $N$.

\begin{theorem}\label{thm:1CntrNbnd}
  Let a finite set $\Gamma\subseteq\cmplx$ satisfy the following
  condition: For all $p\in\cP_{N}$ such that $p(\Gamma) = \{0\}$, we
  have $p=0$. Let the input state have maximum $N$ photons. Then, given the counter outcome distributions for all
  elements $\gamma \in \Gamma$, the members of $\cM$ and $\cM'$ are
  observable.
\end{theorem}

\begin{proof}
 Eq. ~\ref{eq:m'_from_m} shows that $\cM'$ is in the span
  of $\cM$, so it suffices to prove the statement for $\cM$. To see the former, note that for non-negative integers $a,b,c$ the corresponding $m'(a,b,c) \in \cM'$ and $m(a,b,c) \in \cM$ iff $a+\max(b,c) \leq N$, and then notice that the $m(j,b+a-j,c+a-j)$ appearing in the expansion of $m'(a,b,c)$ in Eq. ~\ref{eq:m'_from_m} satisfy this property - that is, $j+\max(b+a-j,c+a-j) = j+\max(b,c) +a -j = a+\max(b,c)$. The
  assumption on $\Gamma$ in the theorem statement implies that the map $p\in\cP_{N}\mapsto
  (p(\gamma,\gamma^*))_{\gamma\in \Gamma}$ is linear and
 injective. It follows that for any given monomial $p=\gamma^{l}(\gamma^*)^{k}\in\cP_{N}$, there is a sequence $(c_{\gamma})_{\gamma\in\Gamma}$ such that for all
  monomials $q\in\cP_{N}$ different from $p$,
  \begin{align}
      \sum_{\gamma\in\Gamma}c_{\gamma}q(\gamma,\gamma^*)&=0,\nonumber\\
      \sum_{\gamma\in\Gamma}c_{\gamma}p(\gamma,\gamma^*)&= 1.
  \end{align}
  We can sum the generating function $F_{1} = \exp((|\vec{\alpha}|^{2}+ \alpha_1\gamma^*+  
          \alpha_1^*\gamma)u)$ for values of $\gamma\in\Gamma$
  accordingly, and define $F_{1,p}=\sum_{\gamma\in\Gamma}c_{\gamma}F_{1}(u,\gamma) $.
  For $n\leq 2N$, the coefficient of $u^{n}$ of $F_{1,p}$ is given by
  \begin{align}
    (n!) \mathrm{coeff}_{u^{n}}F_{1,p} &= \left[ \partial_u^n  \sum_{\gamma\in\Gamma}c_{\gamma}F_{1}(u,\gamma) \right]_{u=0} \nonumber \\
      &= \sum_{\gamma\in\Gamma}c_{\gamma} \left[ \partial_u^n  F_{1}(u,\gamma)\right]_{u=0} \nonumber \\
      & = \sum_{\gamma\in\Gamma}c_{\gamma} (|\vec{\alpha}|^{2}+ \alpha_1\gamma^*+  
          \alpha_1^*\gamma)^n\nonumber \\
          & = \sum_{i+j+m=n} \binom{n}{m,i,j} \alpha_1^{i}(\alpha_1^*)^{j}|\vec{\alpha}|^{2m} ( \sum_{\gamma\in\Gamma}c_{\gamma}  (\gamma^*)^i \gamma^j).
  \end{align}
If $n<k+l$, $n\leq N$, the whole expression vanishes because $\sum_{\gamma\in\Gamma}c_{\gamma}  (\gamma^*)^i \gamma^{j}$ vanishes for all values of $i$ and $j$ in the sum. For $n\geq k+l$ the expectations of the operators associated with the summands on the right-hand side vanish for either $i+m >N$ or $j+m>N$ (see the discussion preceding the theorem statement). Therefore, the value of $ \sum_{\gamma\in\Gamma}c_{\gamma}  (\gamma^*)^i \gamma^j $ when either $i>N$ or $j>N$ does not matter. And $ \sum_{\gamma\in\Gamma}c_{\gamma}  (\gamma^*)^i \gamma^j $ vanishes for all $(i,j) \neq (k,l)$, $i\leq N, j \leq N$. Therefore,
\begin{align}
(n!) \mathrm{coeff}_{u^{n}}F_{1,p}          & = \binom{n}{n-k-l,k,l} \alpha_1^{k}(\alpha_1^*)^{l}|\vec{\alpha}|^{2(n-(k+l))} =\nonumber \\
          & = \binom{n}{n-k-l,k,l} m(n-(k+l),k,l)
\end{align} 
 when $n\geq k+l$, $n-k \leq N$ and $n-l \leq N$, and $(n!) \mathrm{coeff}_{u^{n}}F_{1,p}=0$ otherwise. This implies the observability of $\cM'$, since each member of $\cM'$ can be written as $m(n-(k+l),k,l)$ for some $(n,k,l)$ that satisfies $2N \geq n \geq k+l$, $k \geq n-N$ and $l \geq n-N$.
  \end{proof}
Thm. \ref{thm:1CntrNbnd} and its proof indicate that one can find a minimum number of distinct LO amplitudes that suffice to determine $\cM'$. 
\begin{prop}\label{cor:1CntrNbnd}
There exist $\Gamma$ of size $(N+1)^2$ satisfying the conditions of Thm. \ref{thm:1CntrNbnd}.
\end{prop}
\begin{proof}
$\cP_N$ is a vector space of two-variable polynomials with dimension $(N+1)^2$. The condition on $\Gamma$ - namely, that the linear map $p \in \cP_N \mapsto (p(\gamma,\gamma^*))_{\gamma \in \Gamma}$ is injective, is the same condition as the one required for the existence of a unique polynomial $p \in \cP_N$ for interpolating the data points $(\gamma,\gamma^*,p(\gamma,\gamma^*))_{\gamma \in \Gamma}$. Picking the ordered polynomial basis 
\begin{align}
\{1,\gamma,\hdots, \gamma^N, \gamma^*,\gamma^*\gamma,\hdots, \gamma^* \gamma^N,\hdots,(\gamma^*)^N,(\gamma^*)^N\gamma,\hdots,(\gamma^*)^N\gamma^N \},
\end{align}
 the map $p \in \cP_N \mapsto (p(\gamma,\gamma^*))_{\gamma \in \Gamma}$ is represented by the matrix
\begin{align}\label{eq: Gamma_matrix}
& J(N,\Gamma) =\nonumber \\
& \left( \begin{array}{cccc|cccc|c|cccc}
1  & \gamma_1 & \hdots & \gamma_1^N & \gamma_1^* & \gamma_1\gamma_1^* & \hdots & \gamma_1^N \gamma_1^* & \hdots &  (\gamma_1^*)^N & \gamma_1 (\gamma_1^*)^N & \hdots & \gamma_1^N (\gamma_1^*)^N \\
\vdots & \vdots & \hdots & \vdots & \vdots & \vdots & \hdots & \vdots & \hdots & \vdots & \vdots & \hdots & \vdots \\
1  & \gamma_{N_p} & \hdots & \gamma_{N_p}^N & \gamma_{N_p}^* & \gamma_{N_p}\gamma_{N_p}^* & \hdots & \gamma_{N_p}^N \gamma_{N_p}^* & \hdots &  (\gamma_{N_p}^*)^N & \gamma_{N_p} (\gamma_{N_p}^*)^N & \hdots & \gamma_{N_p}^N (\gamma_{N_p}^*)^N 
\end{array} \right),
\end{align}
 where $N_p$ is the size of $\Gamma$. Here, $J(N,\Gamma) $ is acting on a vector of size $(N+1)^2 $ that is composed of the coefficients of $p$. The condition is then that this matrix be left-invertible so that the coefficients can be obtained from $(p(\gamma,\gamma^*))_{\gamma \in \Gamma}$. We now show that $M = (N+1)^2$ mutually distinct points can be found such that the corresponding $J(N,\Gamma)$ is invertible (in this case $J(N,\Gamma)$ is a square matrix, so left-invertibility is equivalent to invertibility).
 
We can relate the complex matrix $J(N,\Gamma) $ to the real matrix 
\begin{align}\label{eq: Gamma_matrix_xy}
R(N,\Gamma) = \left( \begin{array}{cccc|cccc|c|cccc}
1  & x_1 & \hdots & x_1^N & y_1 & x_1y_1 & \hdots & x_1^N y_1 & \hdots &  y_1^N & x_1 y_1^N & \hdots & x_1^N y_1^N \\
\vdots & \vdots & \hdots & \vdots & \vdots & \vdots & \hdots & \vdots & \hdots & \vdots & \vdots & \hdots & \vdots \\
1  & x_{N_p} & \hdots & x_{N_p}^N & y_{N_p} & x_{N_p}y_{N_p} & \hdots & x_{N_p}^N y_{N_p} & \hdots &  y_{N_p}^N & x_{N_p} y_{N_p}^N & \hdots & x_{N_p}^N y_{N_p}^N 
\end{array} \right),
\end{align} 
where $x_i = \textrm{Real}(\gamma_i)$ and $y_i = \textrm{Im}(\gamma_i)$, by a right-multiplying $R(N,\Gamma) $ with an invertible matrix. This is because the monomials in $\{ \gamma^n (\gamma^*)^m \}_{0 \leq n \leq N, 0 \leq m \leq N}$ can be expressed as linear combinations of the monomials in $\{ x^n y^m \}_{0 \leq n \leq N, 0 \leq m \leq N}$, and vice versa. The former is accomplished by writing any $\gamma^n (\gamma^*)n$ as $(x+iy)^n (x-iy)^m$ and expanding, while the latter is accomplished by writing any $x^n y^m$ as $(\frac{\gamma+\gamma^*}{2})^n (\frac{\gamma-\gamma^*}{2i})^m $ and expanding. Thus, the invertibility of $R(N,\Gamma)$ implies the invertibility of $J(N,\Gamma) $. The former is equivalent to the existence of a unique solution to a particular two-variable interpolation problem. More specifically, it is an instance of Lagrange interpolation, a discussion of which can be found in textbooks such as \cite{lorentz1992multi}. Cor.~5.6.3 in \cite{lorentz1992multi} states that Lagrange interpolation is almost regular, which in our case means that a unique solution exists for almost all sets of $(N+1)^2$ points in the $2$-dimensional plane spanned by $x$ and $y$. 
\end{proof}
Notice that Prop.~\ref{cor:1CntrNbnd} does not show that $(N+1)^2$ distinct probe amplitudes are necessary to determine $\cM'$. It only shows that the particular strategy of extracting $\cM'$ from $F_1$ described in the proof of Thm. \ref{thm:1CntrNbnd} requires at least $(N+1)^2$ distinct probe amplitudes. We suspect that there exists a strategy to extract $\cM'$ from the measurement statistics using a fewer number of distinct probe amplitudes.

We have shown that given an input state with at most $N$ photons, a finite number of probe amplitudes are sufficient to determine the expectations of the operators associated with the member of $\cM'$. And the operators corresponding to the $m'(a,b,c)$ not in $\cM'$ have zero expectations. We want to show that $\cM'$ determines (and is determined by) $\tilde\rho_2$. Here is a stronger result.
\begin{lemma}\label{lem:twirldet}
Assume $\tilde \rho_2$ has at most $N$ photons. Then, the set of expectations of the operators associated with the $m'(a,b,c) \in \cM'_>$ determines and is determined by $\tilde \rho_2$. Further, the expectations of the operators associated with the expressions in any strict subset of $ \cM'_>$ are not sufficient to determine $\tilde \rho_2$, unless the subset chosen is $\cM'_> \setminus m'(0,0,0)$.
\end{lemma}
\begin{proof}
That the expectations of the operators associated with the  $m'(a,b,c) \in \cM'_>$ are determined by $\tilde \rho_2$ is self-evident. To show the converse, we start with the same observations used in Cor.~\ref{cor:1CntrNbhd}. Namely, since $\tilde \rho_2$ occupies modes $1$ and $2$, the expectation of the operator corresponding to $m'(i,j,k)$ is equal to the expectation of $(a_1^\dagger)^k a_1^j  (a_2^\dagger)^i a_2^i$. Further, the anti-normally ordered monomials $ a_1^k (a_1^\dagger)^j   a_2^i(a_2^\dagger)^l$ have null expectations for $i \neq l$, and for $i=l$ can be expressed as linear combinations of the normally-ordered monomials $(a_1^\dagger)^k a_1^j  (a_2^\dagger)^i a_2^i$. The latter fact, in our case, can be restated as follows. The expectation of any anti-normally ordered monomial of the form $ a_1^k (a_1^\dagger)^j   a_2^i(a_2^\dagger)^i$ can be written as a linear combination of the expectations of operators associated with the expressions in $\cM'$. 

Using the correspondence of $\tilde\rho_2$ with its Husimi function, the set of expectations of the monomials $ a_1^k (a_1^\dagger)^j   a_2^i(a_2^\dagger)^l$ for all tuples $(k,j,i,l)$ correspond to the set of all moments of the Husimi function. As described in Sec.~\ref{qo husimi}, any state with bounded photon number has a determinate complex moment sequence. Thus, the set of expectations $ a_1^k (a_1^\dagger)^j   a_2^i(a_2^\dagger)^l$ for all tuples $(k,j,i,l)$ determine $\tilde \rho_2$. According to the above observation, this means that the expectations of all operators associated with the $m'(a,b,c) \in \cM'$ determine $\tilde \rho_2$. Next, we show that the operators associated with the $m'(a,b,c) \in \cM'_> \subset \cM'$ suffice. 

Consider any $m'(i,j,k) \in \cM' \setminus \cM'_>$. Then, $j>k $. The expectation of the corresponding operator is equal to
\begin{align}\label{eq: m(i,j,k)<=>m(j,i,k)}
\langle (a_1^\dagger)^k a_1^j  (a_2^\dagger)^i a_2^i \rangle & = \tr (\tilde \rho_2 (a_1^\dagger)^k a_1^j  (a_2^\dagger)^i a_2^i ) \nonumber \\
& = \left( \tr ( \left(\tilde \rho_2 (a_1^\dagger)^k a_1^j  (a_2^\dagger)^i a_2^i \right)^\dagger)\right)^* \nonumber \\
& = \left( \tr (  (a_1^\dagger)^j a_1^k  (a_2^\dagger)^i a_2^i \tilde \rho_2 )\right)^* \nonumber \\
& = \langle (a_1^\dagger)^j a_1^k  (a_2^\dagger)^i a_2^i \rangle^*.
\end{align}
Thus, for any $m'(i,j,k) \in \cM' \setminus \cM'_>$, to obtain the expectation of the corresponding operator, one needs to take the complex conjugate of the expectation of the operator associated with $m'(i,k,j) \in \cM'_>$. 

Now we prove the final statement. The number of non-zero parameters in $\tilde \rho_2$ is equal to $(N+1)(N+2)(2N+3)/6$. The trace condition is given by the expectation of the operator corresponding to $m'(0,0,0)$, and thus the latter does not contribute additional information about the state. Including the trace condition, the number of unknown parameters of $\tilde \rho_2$ is $(N+1)(N+2)(2N+3)/6-1$. The expectation of any operator associated with some $m'(i,j,k) \in \cM'_>$ is a linear combination of the non-zero parameters of $\tilde \rho_2$. Therefore, at least $(N+1)(N+2)(2N+3)/6-1$ different $m'(i,j,k) \in \cM'_> \setminus m'(0,0,0)$ are necessary to determine $\tilde \rho_2$. Counting the number of members of $\cM'_> \setminus m'(0,0,0)$ one obtains $(N+1)(N+2)(2N+3)/6-1$. Therefore, every $m'(i,j,k) \in \cM'_> \setminus m'(0,0,0)$ is necessary to determine $\tilde \rho_2$.
\end{proof}

\begin{corollary}\label{cor:twirldet}
  Let $\Gamma$ satisfy the assumption in the statement of
  Thm.~\ref{thm:1CntrNbnd}. Then $\tilde\rho_2$ with at most
  $N$ photons is determined by the measurement statistics of the counter at probe amplitudes in $\Gamma$.
\end{corollary}

\begin{proof}
According to Thm.~\ref{thm:1CntrNbnd}, the measurement statistics at all $\gamma \in \Gamma$ determine the expectations of the operators associated with the $m'(a,b,c) \in \cM'$. And the latter determine $\tilde \rho_2$ according to Lem.~\ref{lem:twirldet}.
\end{proof}

Notice that the condition on the invertibility of the relationship between the measurement statistics at $\Gamma$ and $\tilde \rho_2$ assumes, at first glance, that the photon-number probabilities $p_{k}$ are known for all $k$ at each probe amplitude in $\Gamma$. This is because the proof of Cor.~\ref{cor:twirldet} assumes the operators $ a_1^k (a_1^\dagger)^l  \normord{\hat{n}_2^c} $, where $c+\max (k,l) \leq N$, are observable, and that requires knowing the expectations of some of the powers of the total number operator at the output of the BS according to the transformation rule of the mode operators by the BS. And the powers of the total number operator are given as infinite sums of the projectors $\hat{D}_k$. However, since $\tilde \rho_2$ has maximum photon number $N$, one can always find a finite set of photon number probabilities of the counter at various probe amplitudes in $\Gamma$ that are sufficient to determine $\tilde \rho_2$. To see this, observe that the counter outcome probabilities at a particular $\gamma \in \Gamma$ are given by the equation $p_{k}(\gamma) = \ tr_\textsl{s}(\Pi_{k}(\gamma) \tilde \rho_2)$ and are thus linear combinations of the non-zero elements of $\tilde \rho_2$. There is a finite number of the latter, and, since we know $\{\{p_{k}(\gamma)\}_{k=0}^\infty\}_{\gamma \in \Gamma}$ determines $\tilde \rho_2$ by Cor.~\ref{cor:twirldet}, one is guaranteed to find a finite subset of $\{\{p_{k}(\gamma)\}_{k=0}^\infty\}_{\gamma \in \Gamma}$ that can be inverted to obtain the elements of $\tilde \rho_2$. Since the number of non-zero parameters in $\tilde \rho_2$ is equal to $(N+1)(N+2)(2N+3)/6$, the minimum size of this subset must be equal to $(N+1)(N+2)(2N+3)/6-1$, that is to the number of unknown parameters when the trace condition on $\tilde \rho_2$ is taken into account. This observation deserves its own statement.
\begin{corollary}\label{cor:min_p_k_invertibility}
  Let $\Gamma$ satisfy the assumption in the statement of
  Thm.~\ref{thm:1CntrNbnd}. Let $p_k(\gamma)$ denote the probability of observing $k$ photons at probe amplitude $\gamma$. Let $\tilde\rho_2$ have maximum 
  $N$ photons. Then, there exist $(N+1)(N+2)(2N+3)/6-1$ different probabilities in $\{\{p_{k}(\gamma)\}_{k=0}^\infty\}_{\gamma \in \Gamma}$ that determine $\tilde \rho_2$.
\end{corollary}
\begin{proof}
See the discussion preceding the statement of the corollary.
\end{proof}
We leave it as an open problem to find specific sets of photon-number probabilities which determine $\tilde \rho_2$. A related open problem is to find an upper bound on $k$ in the corollary statement for a given $\Gamma$.

The results derived so far assumed that both the magnitude and the phase of the probe can be adjusted. Now we consider what can be determined from the measurement statistics for a fixed probe magnitude. 
\begin{theorem}\label{thm: magfixed_onedet_trivpart}
Assume the measurement statistics can be obtained for probe states with fixed magnitude \(R\) and with any phase \(\theta\). Then, $\tilde \rho_2$ can be determined from the counter outcome distribution iff the maximum photon number of $\tilde \rho_2$ is \(N \leq 1\).
\end{theorem}
\begin{proof}
As explained at the end of Sec. \ref{sec genfunc for WFH}, for a particular value of $\gamma$, $F(u,\gamma,\gamma^*,\vec{\alpha},\vec{\alpha}^*)$, and by extension $F_1(u,\gamma,\gamma^*,\vec{\alpha},\vec{\alpha}^*)$, generate a set of operators the expectations of which capture all of the information available by the measurement statistics. Thus, it suffices to show that the expectations of the set of observable coefficients of $F_1$ associated with the powers of $u$ for all values of \(\theta\) do not determine $\tilde \rho_2$ when $N>1$, and determine it when $N=1$ (the $N=0$ case is trivial).
 
The coefficient of $u^r$ is $(\abs{\vec{\alpha}}^2+\alpha_1 \gamma^* + \alpha_1^* \gamma)^r/n!$. We substitute $\gamma = R e^{i\theta}$ into the coefficients, expand the parenthesis, and collect the resulting terms by phase: 
\begin{align}\label{eq coeffF1}
(n!)\textrm{coeff}_{u^r} F_1 & =  \sum_{k+l = 0}^r \binom{r}{k,l,r-k-l} \abs{\vec{\alpha}}^{2(r-k-l)} \alpha_1^k (\alpha_1^*)^l (\gamma^*)^k \gamma^l \nonumber \\
& =    \sum_{k+l = 0}^r \binom{r}{k,l,r-k-l} \abs{\vec{\alpha}}^{2(r-k-l)}  \alpha_1^k (\alpha_1^*)^l  R^{k+l} e^{-i(k-l)\theta} \nonumber \\
& = \sum_{t=-r}^{r}e^{-i\theta t} \sum_{l=\max(0,-t) }^{\min(\lfloor \frac{r-t}{2} \rfloor , -t+ \lfloor \frac{r+t}{2} \rfloor )}R^{2l+t} \binom{r}{l,l+t,r-t-2l} (\alpha_1^*)^{l}\alpha_1^{l+t} \abs{\vec{\alpha}}^{2(r-t-2l)} \nonumber \\
&=\sum_{t=-r}^{r}e^{-i\theta t}\sum_{l=\max(0,-t) }^{\min(\lfloor \frac{r-t}{2} \rfloor ,-t+  \lfloor \frac{r+t}{2} \rfloor )}R^{2l+t} \binom{r}{l,l+t,r-t-2l} m(r-t-2l,l+t,l) \nonumber \\
&= \sum_{t=-r}^{r}e^{-i\theta t} Q_{r,t}
\end{align}
where $Q_{r,t}$ is the coefficient of $e^{-i \theta t }$. Given
    the coefficient of $u^r$ of $F_1$ at $2r+1$ different phases $\theta$, the $Q_{r,t}$
    can be extracted by interpolation, so the expectations of the operators
    corresponding to the $Q_{r,t}$ can be determined from the available
    statistics. Conversely, the measurement statistics for any $\theta$ can be determined from these expectations. 

According to Lem.~\ref{lem:twirldet} it is necessary for every operator corresponding to every expression in $\cM'_{>} $ to be observable by the measurement statistics to determine $\tilde\rho_2$ completely (the identity observable corresponding to the trace condition and associated with $m'(0,0,0)$ is always observable). Note that a derivation similar to Eq. ~\ref{eq:m'_from_m}, but instead expressing $m(a,b,c)$ as $(\abs{\vec{\beta}}^2+\abs{\alpha_1}^2)^a \alpha_1^b (\alpha_1^*)^c $ and expanding, shows that $\cM$ is in the span of $\cM'$ (and that $\cM_{>}$ is in the span of $\cM'_{>}$). We show below that $\cM_{>}$, and by extension $\cM'_{>}$, cannot be determined from the set of $Q_{r,t}$ when $N \geq 3$. In particular, notice that the parameter $t$ is the difference in the powers of $\alpha_1$ and $\alpha_1^*$ in $m(r-t-2l,l+t,l)$, and it is non-positive for all members of $ \cM_> $. Therefore, the different values of $t = 0,-1,\hdots,-N$ partition $\cM_{>}$ into mutually non-intersecting subsets. Notice also that $r$ is the total degree of $m(r-t-2l,l+t,l)$, and thus can be used to partition each subset into smaller subsets. So the different tuples $(r,t)$ partition $\cM_>$ into mutually non-intersecting subsets. Let us denote these subsets by $\cM_{>,r,t}=\{m(a,b,c) \mid b = c+t, a+b+c =r  \} $. Then, $Q_{r,t}$ is a linear combination of the members of the subset $\cM_{>,r,t}$. Therefore, it suffices to consider the tuple $(r,t)=(N,0)$ and show that some of the expressions in $\cM_{>,N,0}$ cannot be determined from $Q_{N,0}$. Let us write $Q_{N,0}$ explicitly:
\begin{align}
Q_{N,0} & = \sum_{l=0 }^{\lfloor N/2  \rfloor}R^{2l}\binom{N}{l,l,N-2l} m(N-2l,l,l).
\end{align}
For $N \geq 2$, $Q_{N,0}$ is a linear combination of more than one member of $\cM_{>,N,0}$, and thus the operators associated with the expressions in $\cM_{>,N,0}$ cannot be individually observable.

For $N=1$, the set of $Q_{r,t}$'s with non-positive $t$, and excluding those $Q_{r,t}$ corresponding to operators with zero expectations, is $\{Q_{0,0},Q_{1,-1}, Q_{1,0}, Q_{2,0}   \}$. Each of these besides $Q_{2,0}$ is proportional to one of the members of $\cM_{>} \setminus m(0,1,1)= \{m(0,0,0),m(0,0,1), m(1,0,0) \}$, while $Q_{2,0}$ is a linear combination of $m(0,1,1)$ and $m(2,0,0)$. The latter corresponds to an operator with a zero expectation, and thus the expectation of the operator corresponding to $m(0,1,1)$ can also be obtained. 
\end{proof}

\subsection{Results With Two Photon Counters}

We now consider the WFH configuration with two photon counters. Thms.~\ref{thm:1CntrNbhd},~\ref{thm:1CntrNbnd}, Prop.~\ref{cor:1CntrNbnd}, as well as Cors.~\ref{cor:1CntrNbhd},~\ref{cor:twirldet} and~\ref{cor:min_p_k_invertibility} naturally apply to this configuration as well since the configuration with one photon counter is equivalent to forgetting the measurement outcomes of one of the photon counters in the two counter configuration. It might be possible to show that a lower bound on the size of $\Gamma$ exists when both counters are utilized by a similar derivation that we did in the last section for the configuration with one counter. But we do not consider that problem. Using an additional counter increases the amount of information in the measurement statistics for a given probe amplitude. But for the sets of probe amplitudes assumed in Thm.~\ref{thm:1CntrNbhd} or Thm.~\ref{thm:1CntrNbnd} this information is already available in the measurement statistics of the single counter as $\tilde \rho_2$ is determined by the latter. We find that a situation where the second
photon counter helps is when the magnitude of the probe is fixed, in which case $\tilde \rho_2$ can be determined with a two counter measurement configuration.

The generating function associated with the WFH configuration with both photon counters is $H_{1,\textrm{in}}(u,v,\vec{\alpha},\vec{\alpha}^*,\gamma,\gamma^*)$ defined in Eq.~\ref{eq genfun_trivpart_1}. The observable operators associated with this configuration are given by the coefficients of the $u^n v^m$. Any linear combination of these operators is also observable. Performing a transformation of the variables $u$, $v$ of $H_{1,\textrm{in}}$ results in coefficients of the powers of the new variables which are linear combinations of the coefficients of the old variables. We use this fact repeatedly in the proof of the next theorem.

\begin{theorem}\label{thm: twodet_R_fixed}
Assume the Husimi function corresponding to $\tilde \rho_2$ has a determinate complex moment sequence. Assume the measurement statistics can be obtained for probe states with fixed magnitude \(R\) and with any phase \(\theta\). Then, $\tilde \rho_2$ is determined from the measurement statistics at all values of $\theta$.
\end{theorem}
\begin{prop}\label{prop: twodet_R_fixed}
Denote the reduced state in the mode matching the probe, obtained from $\tilde \rho_2$, by $\rho_r$. Assume the Husimi function corresponding to $\rho_r$ has a determinate complex moment sequence. The expectations of the operators associated with the expressions $(\abs{\vec{\alpha}}^2- \abs{\vec{\gamma}}^2)^r $ at the outputs of the BS for non-negative integers $r$ suffice to determine $\rho_r$
\end{prop}
\begin{proof}[Proof of Thm~\ref{thm: twodet_R_fixed} and Prop.~\ref{prop: twodet_R_fixed}]
We expand the exponent in the definition of $H_{1, \textrm{in}}$, taking advantage of $\vec{\gamma} = \gamma \vec{e}_1$:
\begin{align}
  H_{1, \textrm{in}} 
  &=\exp(|\vec{\alpha}+\xi\vec{\gamma}|^{2}u + |\vec{\alpha}-\xi^{-1}\vec{\gamma}|^{2}v) \nonumber \\
& = \exp( ( \abs{\vec{\alpha}}^2 + \xi^2 \abs{\gamma}^2 + \xi \alpha_1 \gamma^* +\xi \alpha_1^* \gamma )u + (\abs{\vec{\alpha}}^2 + \xi^{-2} \abs{\gamma}^2 -  \xi^{-1} \alpha_1 \gamma^*  -  \xi^{-1}  \alpha_1^* \gamma) v) \nonumber \\
& =  \exp(  \abs{\vec{\alpha}}^2 (u+v) +\abs{\gamma}^2 ( \xi^2 u +\xi^{-2} v) + \alpha_1 \gamma^*(\xi  u - \xi^{-1} v) +  \alpha_1^* \gamma (\xi  u - \xi^{-1} v ) )
\end{align}
Define the generating function $F_2(u,v,\vec{\alpha},\vec{\alpha}^*,\gamma,\gamma^*) = H_{1, \textrm{in}} \exp(-\abs{\gamma}^2 ( \xi^2 u +\xi^{-2} v) ) $. Then,
\begin{equation}\label{equation 5.4}
F_2 =  \exp(  \abs{\vec{\alpha}}^2 (u+v)  + \alpha_1 \gamma^*(\xi  u - \xi^{-1} v) +  \alpha_1^* \gamma (\xi  u - \xi^{-1} v ) ).
\end{equation}
$F_2$ is observable since it is a product of $H_{1,\textrm{in}}$ with a scalar-valued (that is, not depending on $\vec{\alpha},\vec{\alpha}^*$) generating function.

We treat $u$ and $v$ as real variables. Next, we define the independent variables \(t = u+v \) and \(s = \xi  u - \xi^{-1} v\), and express \(\gamma\) in polar form as \(R e^{i\theta}\). Then, $F_2$ is expressed as
\begin{equation}\label{equation 5.5}
F_2 =  \exp(  \abs{\vec{\alpha}}^2 t  + \alpha_1 R e^{- i\theta}s +  \alpha_1^*R e^{i\theta} s )
\end{equation}
Now, since \(R\) is fixed we can incorporate it into \(s\) and define the complex variable \(z = s e^{i\theta}\) with its complex conjugate \(z^* = s e^{-i\theta}\) to express $F_2$ in the new variables as
\begin{equation}\label{equation 5.6}
F_2( t,z,z^*,\vec{\alpha},\vec{\alpha}^*) = \exp(  \abs{\vec{\alpha}}^2 t  + \alpha_1 z^* +  \alpha_1^* z)
\end{equation}
The coefficient of $F_2$ corresponding to the monomial $t^a (z^*)^b z^c$ can be seen to be proportional to $m(a,b,c)$. And thus the set of all $m(a,b,c)$ are observable. It was shown in the proof of Cor.~\ref{cor:1CntrNbhd} that $\tilde \rho_2$ is determined by the $m(a,b,c)$ if its Husimi function has a determinate complex moment sequence.

To prove the proposition statement, we first observe that it is sufficient to know the expectations of operators corresponding to the expressions $( \alpha_1^*)^k \alpha_1^l$ to determine the reduced state \(\rho_r\). This is because the set of these expectations corresponds to the set of all moments of the Husimi function of \(\rho_r\) as described in Sec.~\ref{qo husimi}, and we have assumed that the Husimi function has a determinate moment sequence. Next, notice that Eq.~\ref{equation 5.6} implies that these expectations can be obtained from $F_2( t,z,z^*,\vec{\alpha},\vec{\alpha}^*)$ at \(t=0\) - that is, from $F_2( 0,z,z^*,\vec{\alpha},\vec{\alpha}^*) = \exp(  \alpha_1 z^* +  \alpha_1^* z)$. Since $t = u+v$, the line $t=0$ corresponds to $u= -v$. Thus, at the outputs of the BS we only require the part of the generating function $H = \exp (\abs{\vec{\alpha}}^2 u + \abs{\vec{\gamma}}^2 v) $ (Eq.~\ref{eq genfun_BSout}) where $v = -u$. The coefficients of the powers of $u$ in $ \exp ( (\abs{\vec{\alpha}}^2 - \abs{\vec{\gamma}}^2 )u)$ are precisely the ones in the statement of the proposition.
\end{proof}

So far we have assumed that our detectors are photon number resolving, but in practice such detectors might not be available. Click detectors are generally significantly cheaper and more widely available. These are designed to give a binary answer upon measurement - either there were no photons registered (the detector does not click), or there were one or more photons registered (the detector clicks). This motivates us to investigate the conditions when the WFH configuration determines, or does not determine, $\tilde \rho_2$ when the photon counters are substituted with click detectors that do not distinguish the modes. In the following a BS is ``balanced" when the magnitudes of transmission and reflection coefficients are $1/\sqrt{2}$, and ``unbalanced" otherwise.

\begin{prop}\label{prop 5}
Assume the input state has at most \(N\) photons, and the measurement statistics are available for any probe amplitude. Then, \(\tilde\rho_2\) can be determined from the measurement statistics of the click detectors iff \(N \leq 2\) ($N \leq 1$) if the BS is unbalanced (balanced).
\end{prop}
\begin{proof}
The proof proceeds by considering explicitly the linear equations resulting from Born's rule (Eq.~\ref{eq Born_D_kl_2}). Since $\tilde \rho_2$ occupies only modes $1$ and $2$, it suffices to restrict the projectors of the click detectors to the two-mode subspace. Let us denote the ``no click" outcome with '0' and the ``click" outcome with '1'. Then, with 2 detectors there are 4 possible outcomes: \((0,0), (1,0), (0,1)\) and \((1,1)\), where the first and second elements of each tuple denote the outcomes of counter 1 and counter 2 in Fig.~\ref{fig:bsconfig}, treated as click detectors, respectively. The projectors corresponding to these outcomes, listed in the same order, are given by $\hat{D}_{00}$, $\hat{D}_{I0} = I_1 \otimes \ket{0,0}\bra{0,0} - \hat{D}_{00}$, $\hat{D}_{0I} = \ket{0,0}\bra{0,0} \otimes I_2 - \hat{D}_{00}$, $ \hat{D}_{II} = I_1 \otimes I_2 -  \hat{D}_{00}$, where $I_1$ and $I_2$ are the identity operators on the output paths going to counter 1 and counter 2, respectively. Here we have ordered the state-spaces corresponding to the $4$ modes in the tensor product according to the ordered mode basis $(a'_1,a'_2,b'_1,b'_2)$. Notice that $\hat{D}_{00}$ and $\hat{D}_{II}$ sum to the identity, and thus the information provided by the probability of outcome $(1,1)$ is redundant. Further, to simplify our calculations, we can choose to work with the set of projectors $\hat{D}_{00}$, $\hat{D'}_{I0} = \hat{D}_{I0} + \hat{D}_{00} = I_1 \otimes \ket{0,0}\bra{0,0}$ and $\hat{D'}_{0I} = \hat{D}_{0I} + \hat{D}_{00} = \ket{0,0}\bra{0,0} \otimes I_2$ instead of the mutually orthogonal set corresponding to the measurement outcomes, since these span the same space of operators.

For a given probe amplitude $\gamma$ and BS unitary $U$, the operators on the input state corresponding to the projectors can be computed from the relation in Eq.~\ref{eq: Pi_kl_def}. In particular, the operator corresponding to $\hat{D}_{00} = \ket{0,0}\bra{0,0} \otimes \ket{0,0}\bra{0,0}$ is computed to be
\begin{align}
\Pi_{00}(\gamma) &= \tr_\textsl{p}((I_1 \otimes \ket{\gamma,0}\bra{\gamma,0} ) U^{\dagger}\hat{D}_{00}U) \nonumber \\
& =  \tr_\textsl{p}((I_1 \otimes \ket{\gamma,0}\bra{\gamma,0} ) \ket{0,0}\bra{0,0} \otimes \ket{0,0}\bra{0,0}) \nonumber \\
& = \abs{\bra{0}\ket{\gamma}}^2  \ket{0,0}\bra{0,0} \nonumber \\
& = e^{-\abs{\gamma}^2} \ket{0,0}\bra{0,0}.
\end{align}
The calculation of the operators on the input modes corresponding to $\hat{D'}_{0I} $ and $\hat{D'}_{I0} $ is slightly more involved. The BS unitary in mode space is $B = \begin{pmatrix} \eta & \zeta \\ \zeta  & -\eta \end{pmatrix} \otimes  \begin{pmatrix} 1 & 0 \\ 0  & 1 \end{pmatrix}  $ with our choice of ordering of the modes. We adopt the technique used in Eq.~\ref{lem Uirreps eq} in the calculation below. $\ket{\vec{0}}$ stands for the vacuum on the full 4-mode space inside the partial trace, and for the 2-mode input space when the trace operation has been completed. Writing $\hat{D'}_{I0} =  \sum_{k_1+k_2 = 0}^\infty \ket{k_1,k_2}\bra{k_1,k_2} \otimes \ket{0,0}\bra{0,0}$, and using the linearity of the partial trace, the corresponding operator on input modes is computed as
\begin{align}
& \Pi_{I0}(\gamma)  =  \sum_{k_1+k_2 = 0}^\infty  \tr_\textsl{p}((I_1 \otimes \ket{\gamma,0}\bra{\gamma,0} ) U^{\dagger}  \ket{k_1,k_2}\bra{k_1,k_2} \otimes \ket{0,0}\bra{0,0} U) \nonumber \\
&=  \sum_{k_1+k_2 = 0}^\infty   \tr_\textsl{p} ((I_1 \otimes \ket{\gamma,0}\bra{\gamma,0} )  \frac{1}{k_1!k_2!} U^\dagger ( b_1^{\dagger})^{k_1}( b_2^{\dagger})^{k_2} \ket{\vec{0}}\bra{\vec{0}} ( b_1)^{k_1}( b_2)^{k_2}  U)  \nonumber \\
& =  \sum_{k_1+k_2 = 0}^\infty \frac{1}{k_1!k_2!}  \tr_\textsl{p} ((I_1 \otimes \ket{\gamma,0}\bra{\gamma,0} )  ( \zeta a_1^{\dagger}-\eta b_1^\dagger)^{k_1}(  \zeta a_2^{\dagger}-\eta b_2^\dagger)^{k_2} \ket{\vec{0}}\bra{\vec{0}}  (  \zeta a_1- \eta b_1)^{k_1}(  \zeta a_2- \eta b_2)^{k_2} ) \nonumber \\
& = e^{-\abs{\gamma}^2} \sum_{k_1+k_2 = 0}^\infty \frac{1}{k_1!k_2!}    ( \zeta a_1^{\dagger}-\eta \gamma^*)^{k_1}(  \zeta a_2^{\dagger} )^{k_2} \ket{\vec{0}}\bra{\vec{0}}  (  \zeta a_1- \eta \gamma)^{k_1}(  \zeta a_2)^{k_2} \nonumber \\
& =e^{-\abs{\gamma}^2} \sum_{k_1+k_2 = 0}^\infty \frac{\zeta^{2k_2}}{k_1!}    ( \zeta a_1^{\dagger}-\eta \gamma^*)^{k_1} \ket{0}\bra{0}  (  \zeta a_1- \eta \gamma)^{k_1} \otimes \ket{k_2}\bra{k_2}. 
\end{align}
A similar calculation for the operator $\Pi_{0I}(\gamma)$ corresponding to the projector $\hat{D'}_{0I}$ yields
\begin{align}
\Pi_{0I}(\gamma) & = e^{-\abs{\gamma}^2}\sum_{k_1+k_2 = 0}^\infty \frac{\eta^{2k_2}}{k_1!}    ( \eta a_1^{\dagger}+\zeta \gamma^*)^{k_1} \ket{0}\bra{0}  (  \eta a_1+ \zeta \gamma)^{k_1} \otimes \ket{k_2}\bra{k_2}.
\end{align}

We use the representation of $\tilde \rho_2$ in Cor.~\ref{cor K+1_twirlstate}, which for $K=1$ has the form $ \tilde\rho_{2} = \sum_{n=0}^N  \chi_{n} \otimes  \ket{n}\bra{n}$. The expectations of the operators are then linear combinations of the parameters of the matrices $\chi_{n} $. In particular, 
\begin{align}\label{eq: exp_pi_00}
\langle  \Pi_{00}(\gamma) \rangle & = e^{-\abs{\gamma}^2} \bra{0,0}  \tilde\rho_{2}   \ket{0,0} \nonumber \\
& = e^{-\abs{\gamma}^2} \chi_{0,00},
\end{align}
so $ \Pi_{00}(\gamma) $ determines only the element $\chi_{0,00}$ for any $\gamma$. We expand the expectation of $\Pi_{I0}(\gamma) $ to obtain
\begin{align}\label{eq: exp_pi_10}
 & \tr(  \Pi_{I0}(\gamma)  \tilde\rho_{2}  ) =e^{-\abs{\gamma}^2} \sum_{n} \tr( \chi_{n} \otimes  \ket{n}\bra{n}  \Pi_{I0}(\gamma)  ) \nonumber \\
& = e^{-\abs{\gamma}^2} \sum_{n}  \sum_{k_1+k_2 = 0}^\infty \frac{\zeta^{2k_2}}{k_1!} \left[ \bra{0}  (  \zeta a_1- \eta \gamma)^{k_1} \chi_{n}   ( \zeta a_1^{\dagger}-\eta \gamma^*)^{k_1} \ket{0} \right] \abs{\bra{n}\ket{k_2}}^2 \nonumber \\
& = e^{-\abs{\gamma}^2} \sum_{n} \sum_{k_1}\frac{\zeta^{2n}}{k_1!}  \sum_{i =0}^{k_1} \sum_{j=0}^{k_1} \binom{k_1}{i}\binom{k_1}{j} \zeta^{i+j} (-\eta\gamma^*)^{k_1-i} (-\eta  \gamma)^{k_1-j}  \sqrt{i! j!} \bra{j}\chi_n\ket{i} \nonumber \\
& = \sum_{i=0}^N \sum_{j=0}^N \zeta^{i+j} \sqrt{i! j!}  \left[e^{-\abs{\gamma}^2} \sum_{k_1 = max(i,j)}^\infty \frac{(-\eta)^{2k_1-i-j}}{k_1!}\binom{k_1}{i}\binom{k_1}{j}  ( \gamma^*)^{k_1-i} \gamma^{k_1-j}  \right] (\sum_{n=0}^{N-\max(i,j)} \zeta^{2n}\chi_{n,ji}).
\end{align}
The expression is a linear combination of the terms $\sum_{n=0}^{N-\max(i,j)} \zeta^{2n}\chi_{n,ji}$, so at best it is possible to determine these terms from the expectations at different $\gamma$. A similar calculation for $\langle \Pi_{0I}(\gamma) \rangle$ yields
\begin{align}\label{eq: exp_pi_01}
\sum_{i=0}^N \sum_{j=0}^N \eta^{i+j} \sqrt{i! j!}  \left[ e^{-\abs{\gamma}^2}\sum_{k_1 = max(i,j)}^\infty \frac{\zeta^{2k_1-i-j}}{k_1!}\binom{k_1}{i}\binom{k_1}{j}  ( \gamma^*)^{k_1-i} \gamma^{k_1-j}  \right] (\sum_{n=0}^{N-\max(i,j)} \eta^{2n}\chi_{n,ji})
\end{align}
This expression is similarly a linear combination of the terms $\sum_{n=0}^{N-\max(i,j)} \eta^{2n}\chi_{n,ji}$, so at best it is possible to determine these terms. If the BS is balanced, $\langle \Pi_{0I}(\gamma) \rangle$ and $\langle \Pi_{I0}(\gamma) \rangle$ are identical. 

To see that for a BS balanced $\tilde \rho_2$ is not determined when $N>1$, it is sufficient to give an example of a $\tilde \rho_2$ with $N=2$. In particular, assume that $\tilde \rho_2$ is diagonal. Then there are $6$ non-zero parameters, but the trace condition, $\langle  \Pi_{00}(\gamma) \rangle $, as well as the terms $\sum_{n=0}^{N-\max(i,j)} \eta^{2n}\chi_{n,ji}$ for the indices $(i,j) = (0,0),(1,1),(2,2)$, comprise $5$ equations in these parameters. When the BS is unbalanced, we show that $\tilde \rho_2$ cannot be determined when $N>2$ by considering a diagonal $\tilde \rho$ with at most $N=3$ photons. The number of non-zero parameters is $10$. The trace condition and $\langle  \Pi_{00}(\gamma) \rangle $ supply one equation each. The terms $\sum_{n=0}^{N-\max(i,j)} \eta^{2n}\chi_{n,ji}$ and $\sum_{n=0}^{N-\max(i,j)} \zeta^{2n}\chi_{n,ji}$ supply $4$ equations each for the indices $(i,j) = (0,0),(1,1),(2,2),(3,3)$, but notice that for the index $(i,j) = (3,3)$ the equations are the same (both give $ \chi_{0,33}$). Together, these comprise $9$ equations for $10$ unknowns. Using this style of investigation, it is straightforward to conclude that when $N=2$ and the BS is unbalanced the unknown parameters of $\tilde\rho_2$ can be obtained from the sums $\sum_{n=0}^{N-\max(i,j)} \eta^{2n}\chi_{n,ji}$ and $\sum_{n=0}^{N-\max(i,j)} \zeta^{2n}\chi_{n,ji}$ together with the trace condition and $\langle  \Pi_{00}(\gamma) \rangle $. If the BS is balanced, the unknown parameters of $\tilde\rho_2$ can be obtained when $N=1$. The requirement is that Eqs.~\ref{eq: exp_pi_10} and~\ref{eq: exp_pi_01} be invertible with respect to the terms $\sum_{n=0}^{N-\max(i,j)} \eta^{2n}\chi_{n,ji}$ and $\sum_{n=0}^{N-\max(i,j)} \zeta^{2n}\chi_{n,ji}$. This is what we show next.

The argument we bring concerns Eq.~\ref{eq: exp_pi_10}, and an identical argument can be made for Eq~\ref{eq: exp_pi_01}. We observe that the terms in the square brackets in the expansions~\ref{eq: exp_pi_10} are convergent power series for each tuple $(i,j)$. They converge for every value of $\gamma$ since their linear combinations are equal to probability values, and thus their derivatives of any order in a neighborhood of $0$ are well-defined. Notice that for the tuple $(i,j)$ the monomials $\gamma^k (\gamma^*)^l$ in the corresponding power series all have a difference of $i-j$ between the powers of $\gamma$ and $\gamma^*$. For each $\delta = i-j\geq 0$, apply the operators $\partial^{i-j+k}_{\gamma}\partial^k_{\gamma^*}$ for $k=0,\hdots, N-(i-j)$, and evaluate at $\gamma=0$, and for each $\delta = i-j <  0$ apply the operators $\partial^{k}_{\gamma}\partial^{k+j-i}_{\gamma^*}$ for $k=0,\hdots, N-(j-i)$, and evaluate at $\gamma=0$. For each value of $\delta$ one obtains $N-\abs{\delta}+1$ equations involving the terms $\sum_{n=0}^{N-\max(i,j)} \zeta^{2n}\chi_{n,ji}$ with $i-j=\delta$. There are $N-\abs{\delta}+1$ such terms for each $\delta$, and the set of equations determines them. This is not hard to check by hand for $N\leq 2$.
\end{proof}
A natural extension of this measurement configuration with click detectors is to modify this configuration by putting a cascading array of BSs at each output of the original BS where the probe and the input state are interfered. Then, one would put click detectors at the output paths of every BS. If there are $k$ click-detectors for each output of the original BS, this allows for distinguishing up to $k$ photons in each output. Studying the symmetries and observables of such a configuration is beyond the scope of this work, however.

\section{WFH Configurations With Arbitrary BS}\label{sec BS with nontrivial part}

In this section we identify a set of conditions under which, for a given WFH configuration, the available measurement statistics determine or do not determine $\tilde \rho_{K+1}$ ($\tilde \rho_{1,K}$ when $S_1=1$). These conditions depend on $\lambda(S)$, on the number of counters utilized, and on the set of probe amplitudes at which the measurement statistics are available. The generating function on the input state is given by $G_{\textrm{in}}(u,v,\vec{\alpha},\vec{\alpha}^*, \vec{\gamma},\vec{\gamma}^*)$ in Eq.~\ref{eq genfun_nontrivpart} when both counters are used, and by $G_{\textrm{in},2}(u,\vec{\alpha},\vec{\alpha}^*, \vec{\gamma},\vec{\gamma}^*)$ in Eq.~\ref{eq genfun_nontrivpart_onedet} when one of the counters is used. As discussed in the previous section, the expectations of the family of operators associated with $G_\textrm{in}$ or $G_{\textrm{in},2}$ (depending on the number of counters utilized) for each value of $\gamma$ contain all the information in the measurement statistics about the input state for that $\gamma$. Therefore, if $\tilde \rho_{K+1}$ ($\tilde \rho_{1,K}$ when $S_1=1$) is not determined by these expectations, then it cannot be determined by the measurement statistics. 

\subsection{Results With Two Photon Counters}

\subsubsection{BS Characterized by a Partition With $S_1>1$}
Let us first determine the set of operators associated with $G_\textrm{in}$ when $\gamma$ is treated as a generating function variable - that is, it is assumed that the measurement statistics are available for all $\gamma$ in a neighborhood of $0$. Recall that we introduced the notation $\vec{\alpha} = \oplus_{i=1}^K \vec{\alpha_i}$, where $\vec{\alpha_i}$ is a vector of size $S_i$, and where the mode operators associated with the variables in $\vec{\alpha_i}$ transform according to $B_i$. We currently assume $S_1 >1$. Then we can write $\vec{\alpha}_1 = \alpha_1 \oplus \vec{\beta}$, where $\alpha_1$ is the variable associated with $a_1$. We expand the first two terms in the exponent of $G_\textrm{in}$ according to Eq.~\ref{eq genfun_nontrivpart}:
\begin{align}\label{eq G_in second}
G_\textrm{in} &= \exp ( \abs{\eta_1 \alpha_1+ \zeta_1 \gamma}^2 u +  \abs{\zeta_1 \alpha_1- \eta_1 \gamma}^2 v +( \eta_1^2 u + \zeta_1^2 v )\abs{\vec{\beta}}^2 + \sum_{i=2}^K (\eta_i^2  u +  \zeta_i^2 v) \abs{ \vec{\alpha_i}}^2) \nonumber \\
& =  \exp ( \eta_1\zeta_1(\alpha_1 \gamma^* +\alpha_1^* \gamma )(u -v) +  ( \zeta_1^2 u + \eta_1^2 v )\abs{\gamma}^2  + ( \eta_1^2 u + \zeta_1^2 v )\abs{\vec{\alpha_1}}^2 + \sum_{i=2}^K (\eta_i^2  u +  \zeta_i^2 v) \abs{ \vec{\alpha_i}}^2).
\end{align}
We introduce the generating function $G_1 = G_\textrm{in} \exp( -( \zeta_1^2 u + \eta_1^2 v )\abs{\gamma}^2)$. The coefficients of the momomials in the variables $u$, $v$, $\gamma$ and $\gamma^*$ of $G_1$ are observable. The coefficient of the momomial $\gamma^k (\gamma^*)^l$ of $G_1$ is a generating function in the variables $u$, $v$ and, after multiplication with $k!l!$, is given by\begin{align}\label{eq:G_1kl}
G_{1,(k,l)} & = \partial_\gamma^k \partial_{\gamma^*}^l G_1 \Bigg|_{\gamma=0} \nonumber \\
& = \partial_\gamma^k \partial_{\gamma^*}^l \exp ( \eta_1\zeta_1(\alpha_1 \gamma^* +\alpha_1^* \gamma )(u -v)   + ( \eta_1^2 u + \zeta_1^2 v )\abs{\vec{\alpha_1}}^2 + \sum_{i=2}^K (\eta_i^2  u +  \zeta_i^2 v) \abs{ \vec{\alpha_i}}^2) \Bigg|_{\gamma=0} \nonumber \\
& = (\eta_1\eta_2 (u-v) )^{k+l} (\alpha_1^*)^k \alpha_1^l  \exp \Bigg( \Bigg. \eta_1\zeta_1(\alpha_1 \gamma^* +\alpha_1^* \gamma )(u -v)   + ( \eta_1^2 u + \zeta_1^2 v )\abs{\vec{\alpha_1}}^2  \nonumber \\
& + \sum_{i=2}^K (\eta_i^2  u +  \zeta_i^2 v) \abs{ \vec{\alpha_i}}^2 \Bigg. \Bigg) \Bigg|_{\gamma=0} \nonumber \\
& = (\eta_1\zeta_1 (u-v) )^{k+l} (\alpha_1^*)^k \alpha_1^l  \exp ( ( \eta_1^2 u + \zeta_1^2 v )\abs{\vec{\alpha_1}}^2 + \sum_{i=2}^K (\eta_i^2  u +  \zeta_i^2 v) \abs{ \vec{\alpha_i}}^2).
\end{align}

For each $k$ and $l$ the coefficient of the monomial $\gamma^k (\gamma^*)^l u^a v^b$ of $G_1$ corresponds to the coefficient of $u^a v^b$ of the corresponding $G_{1,(k,l)}/(k!l!)$. However, the expressions for these coefficients are somewhat complicated, and we can simplify our task by considering the generating function $G'_{1,(k,l)} = G_{1,(k,l)}/(\eta_1\zeta_1 (u-v) )^{k+l}$. Here, the division should be treated formally, so that 
\begin{align}\label{eq:G'_1kl}
G'_{1,(k,l)} = (\alpha_1^*)^k \alpha_1^l \exp ( ( \eta_1^2 u + \zeta_1^2 v )\abs{\vec{\alpha_1}}^2 + \sum_{i=2}^K (\eta_i^2  u +  \zeta_i^2 v) \abs{ \vec{\alpha_i}}^2)
\end{align}
is defined everywhere. $G'_{1,(k,l)} $ is observable. To see this, consider the variable transformation of $u$ and $v$ into $u-v$ and another expression independent of $u-v$ (call it $t(u,v)$). The coefficients of $G_{1,(k,l)}$ associated with these new variables are observable. Then, notice that the coefficient of $G'_{1,(k,l)}$ associated with the monomial $(u-v)^n t^m$ is the same as the coefficient of $G_{1,(k,l)}$ associated with the monomial $(u-v)^{n+k+l} t^m$. Let us denote the coefficients of $G'_{1,(k,l)} $ in the variables $u$ and $v$ by $g_K(k,l,a,b)/(a!b!)$:
\begin{align}\label{eq:g_K}
& g_K(k,l,a,b)  =\left[ \partial_u^a \partial_v^b G'_{1,(k,l)} \right]_{u=0,v=0} \nonumber \\
& = \left[  (\alpha_1^*)^k \alpha_1^l  (\sum_{i=1}^K \eta_i^2 \abs{\vec{\alpha}_i}^2  )^a (\sum_{i=1}^K \zeta_i^2 \abs{\vec{\alpha}_i}^2  )^b \exp ( \sum_{i=1}^K (\eta_i^2  u +  \zeta_i^2 v) \abs{ \vec{\alpha_i}}^2) \right]_{u=0,v=0} \nonumber \\
& =( \alpha_1^*)^k \alpha_1^l  (\sum_{i=1}^K \eta_i^2 \abs{\vec{\alpha}_i}^2  )^a (\sum_{i=1}^K \zeta_i^2 \abs{\vec{\alpha}_i}^2  )^b \nonumber \\
& = ( \alpha_1^*)^k \alpha_1^l   \left[ \sum_{c_1+\hdots+c_K = a} \binom{a}{c_1,\hdots,c_K}  \prod_{i=1}^K  \eta_i^{2c_i} \abs{\vec{\alpha}_i}^{2c_i} \right] \left[ \sum_{c_1+\hdots+c_K = b} \binom{a}{c_1,\hdots,c_K}  \prod_{i=1}^K  \zeta_i^{2c_i} \abs{\vec{\alpha}_i}^{2c_i} \right] \nonumber \\
& = ( \alpha_1^*)^k \alpha_1^l \sum_{c_1+\hdots + c_K = a+b} C(c_1,\hdots,c_K) \prod_{i=1}^K \abs{\vec{\alpha}_i}^{2c_i},
\end{align}
where $C(c_1,\hdots,c_K)$ is an expression in terms of the powers of $\eta_i^2$ and the $\zeta_i^2$ that is determined by the product of the two multinomial expansions in the second to last line of Eq.~\ref{eq:g_K}. 

The discussion above implies that the measurement statistics at all $\gamma$ in a neighborhood of $0$ determine and are determined by the expectations of the operators corresponding to $g_K(k,l,a,b)$. We find that when $K$ is large enough, one cannot determine $\tilde \rho_{K+1}$ from these expectations. For this purpose we introduce the following sets of expressions in the variables $\vec{\alpha},\vec{\alpha}^*$: $\cM'_K = \{ m'_K(k,l,\vec{c}) = (\alpha_1^*)^k \alpha_1^l \abs{\vec{\beta}}^{2 c_1} \prod_{i=2}^K \abs{\vec{\alpha}_i }^{2c_i} \mid k,l,c_i \in \mathbb{Z}_{\geq 0}  \}$ and $\cM_K = \{ m_K(k,l,\vec{c}) = (\alpha_1^*)^k \alpha_1^l  \prod_{i=1}^K \abs{\vec{\alpha}_i }^{2c_i} \mid k,l,c_i \in \mathbb{Z}_{\geq 0}   \}$. The members of $\cM_K$ are linear combinations of the members of $\cM'_K$ and vice versa. In particular, one can write any $m_K(k,l,\vec{c})\in \cM_K$ as
\begin{align}\label{eq:m_K intermsof m'_K}
(\alpha_1^*)^k \alpha_1^l  \prod_{i=1}^K \abs{\vec{\alpha}_i }^{2c_i} & = (\alpha_1^*)^k \alpha_1^l \abs{\alpha_1 \oplus \vec{\beta} }^{2c_1}  \prod_{i=2}^K \abs{\vec{\alpha}_i }^{2c_i} \nonumber \\
& = (\alpha_1^*)^{k} \alpha_1^{l}( \abs{\alpha_1}^2+\abs{ \vec{\beta} }^{2})^{c_1}  \prod_{i=2}^K \abs{\vec{\alpha}_i }^{2c_i} \nonumber \\
&= \sum_{i=0}^{c_1} \binom{c_1}{i} (\alpha_1^*)^{k+i} \alpha_1^{l+i} \abs{ \vec{\beta} }^{2(c_1-i)} \prod_{i=2}^K \abs{\vec{\alpha}_i }^{2c_i}.
\end{align}
And conversely,
\begin{align}\label{eq:m'_K intermsof m_K}
(\alpha_1^*)^k \alpha_1^l \abs{ \vec{\beta} }^{2c_1} \prod_{i=2}^K \abs{\vec{\alpha}_i }^{2c_i} & = (\alpha_1^*)^k \alpha_1^l (\abs{\vec{\alpha_1}}^2 -  \abs{\alpha_1}^2)^{c_1}  \prod_{2=1}^K \abs{\vec{\alpha}_i }^{2c_i} \nonumber \\
& =  \sum_{i=0}^{c_1} \binom{c_1}{i} (-1)^i (\alpha_1^*)^{k+i} \alpha_1^{l+i} \abs{ \vec{\alpha}_1 }^{2(c_1-i)} \prod_{i=2}^K \abs{\vec{\alpha}_i }^{2c_i}.
\end{align}
Let us further introduce the following sets of monomials: $\cM_{K,\geq} = \{ m_K(k,l,\vec{c}) \mid k \geq l  \}  \subset \cM_K$, and $\cM'_{K,\geq} = \{ m'_K(k,l,\vec{c}) \mid k \geq l  \} \subset \cM'_K$. It can be seen from the Eqs.~\ref{eq:m'_K intermsof m_K} and~\ref{eq:m_K intermsof m'_K} that monomials in $\cM_{K,\geq}$ are linear combinations of monomials in $\cM'_{K,\geq}$ and vice versa. We also introduce the subsets $\cM'_{K}(N) \subset \cM'_K$ composed of the $m'_K(k,l,\vec{c})$ with $\max(k,l)+c_1+\hdots+c_K \leq N$, and the subsets $\cM'_{K,\geq}(N) \subset \cM'_{K,\geq}$ composed of the $m'_K(k,l,\vec{c})$ with $k+c_1+\hdots+c_K \leq N$ for each $N$.

\begin{lemma}\label{lem: M_K<=>rho_S1>1}
Assume the Husimi function of $\tilde \rho_{K+1}$ has a determinate complex moment sequence. Then, the set of expectations of all operators associated with the expressions in $\cM'_{K,\geq}$ (or, equivalently, in $\cM_{K,\geq}$) determine and are determined by $\tilde \rho_{K+1}$. Further, the expectations of the operators associated with the expressions in any strict subset of $\cM'_{K,\geq}$ are not sufficient to determine $\tilde \rho_{K+1}$, unless the subset is $\cM'_{K,\geq} \setminus m_K'(0,0,0)$.
\end{lemma}
\begin{proof}
That $\tilde \rho_{K+1}$ determines the expectation of every operator associated with $\cM'_{K,\geq}$ is evident. Further, if every $m'_K(k,l,\vec{c}) \in  \cM'_{K,\geq}$ is observable then so is every $m'_K(k,l,\vec{c}) \in  \cM'_{K}$. This is due to the fact that the expectation of the operator corresponding to any $m'_K(k,l,\vec{c}) \in \cM'_{K} \setminus  \cM'_{K,\geq}$ is the complex conjugate of the expectation of the operator corresponding to $m'_K(l,k,\vec{c}) \in  \cM'_{K,\geq}$ (natural extension of Eq.~\ref{eq: m(i,j,k)<=>m(j,i,k)} to arbitrary $K$). Since $\tilde \rho_{K+1}$ occupies the modes associated with the variables $\alpha_1$, $\beta_1$, $\alpha_{2,1},\hdots, \alpha_{K,1}$, the expectation of the operator corresponding to $ (\alpha_1^*)^k \alpha_1^l \abs{\vec{\beta}}^{2 c_1} \prod_{i=2}^K \abs{\vec{\alpha}_i }^{2c_i} $ is equal to the expectation of the operator corresponding to the monomial $ (\alpha_1^*)^{k} \alpha_1^{l} \abs{\beta_1}^{2c_1}\prod_{i=2}^K \abs{\alpha_{i,1}}^{2c_i}$. Further, the monomials in the same variables as the $ (\alpha_1^*)^{k} \alpha_1^{l} \abs{\beta_1}^{2c_1}\prod_{i=2}^K \abs{\alpha_{i,1}}^{2c_i}$, but not of the same form, are associated with operators the expectations of which vanish. This is because in every such monomial the power difference between at least one of the variables in $\{\beta,\alpha_{2,1},\hdots,\alpha_{K,1}\}$ and its complex conjugate is not zero. The form of $\tilde \rho_{K+1}$ then insures the vanishing of the expectations. Thus, the expectation of every normally-ordered monomial of mode operators in the modes occupied by $\tilde \rho_{K+1}$ is assumed to be available. Therefore, the same applies to every anti-normally ordered monomial of mode operators in the modes occupied by $\tilde \rho_{K+1}$ since these can be expressed as linear combinations of normally-ordered monomials of mode operators by repeated use of commutation relations. By the identification of the set of moments of the Husimi function and the set of expectations of anti-normally ordered monomials of mode operators, the complex moment sequence of the Husimi function is also available. According to the assumptions of the lemma statement, the latter is determinate, and, hence, the Husimi function is determined. Therefore, $\tilde \rho_{K+1}$ is determined.

To prove the second part of the statement, assume some $m'_K(k',l',\vec{c'}) \in  \cM'_{K,\geq} \setminus m_K'(0,0,0)$ is not observable. Then, $m'_K(l',k',\vec{c'})$ is also not observable. Consider a $\tilde \rho_{K+1}$ with at most $N = k'+c'_1+\hdots+c'_K$ photons. Then, the expectations of the operators associated with the $m'_K(k,l,\vec{c}) \in \cM'_{K} \setminus  \cM'_{K} (N)$ are zero. So, if the expectations associated with the observable operators in $\cM'_{K} (N) $ do not determine $\tilde \rho_{K+1}$, then the latter cannot be determined. Since the expectation of every operator associated with $ (\cM'_{K}(N) \setminus  \cM'_{K,\geq} (N)) \setminus m'_K(l',k',\vec{c'})$ is determined by the set of expectations of the operators associated with $ \cM'_{K,\geq }(N) \setminus m'_K(k',l',\vec{c'})$, if the latter do not determine $\tilde \rho_{K+1}$ then $\tilde \rho_{K+1}$ cannot be determined. Observe that the number of members of $\cM'_{K,\geq} (N) $ is equal to the number of elements of $\tilde \rho_{K+1}$ in its upper triangular part. Further, the expectation of every operator associated with $\cM'_{K,\geq} (N) $ is a linear combination of these elements. Since one of these expectations is not available, and that expectation is not the trace condition, $\tilde \rho_{K+1}$ cannot be determined. 
\end{proof}

\begin{theorem}\label{thm 2det_nontrivpart}
Assume the BS is characterized by a partition with size $K$ and $S_1>1$. Assume the measurement statistics are available for all $\gamma$ in a neighborhood of $0$. Assume the Husimi function of $\tilde \rho_{1,K}$ has a determinate complex moment sequence.  Then, $\tilde \rho_{K+1}$ can be determined from the measurement statistics of the counters iff $K \leq 2$.
\end{theorem}
\begin{proof}
As explained above, the expectations of the operators associated with the $g(k,l,a,b)$ contain all the information in the measurement statistics. We first prove that $\tilde \rho_{K+1}$ cannot be determined when $K >2 $ by showing that not every monomial in $\cM_{K,\geq}$ can be obtained from the $g_K(k,l,a,b)$. At the risk of being redundant, we note that if $\tilde \rho_{K+1}$ was characterized, then the expectation of every operator would be determined, and, therefore, if there exist operators associated with some $m_K(k,l,\vec{c}) $ the expectations of which cannot be ascertained, then $\tilde \rho_{K+1}$ cannot be fully characterized. 

 It suffices to prove that not every operator corresponding to the expressions in $\cM_{K,\geq}$ is observable for $K =3$, since $\cM_{3,\geq} \subset \cM_{K,\geq}$ for larger $K$. Eq.~\ref{eq:g_K} can be written as:
\begin{align}
g_3(k,l,a,b) = \sum_{c_1+c_2+c_3 = a+b} C(c_1,c_2,c_3) m_3(k,l,c_1,c_2,c_3).
\end{align}
For a given pair $(k,l)$, and for a given non-negative integer $c$, the set of $m_3(k,l,c_1,c_2,c_3)$ with $c_1+c_2+c_3=c$ appear only in the equations associated with the set of $g_3(k,l,a,b)$ where $a+b=c$. Even for $c=1$ there are $2$ equations but $3$ unknowns. 

To prove that for $K \leq 2$ the operators associated with the expressions in $\cM_{K,\geq}$ are observable, we consider $G'_{1,(k,l)}  $ from Eq.~\ref{eq:G'_1kl}, which, for $K=2$, has the form
\begin{align}
G'_{1,(k,l)} & =  (\alpha_1^*)^k \alpha_1^l  \exp ( ( \eta_1^2 u + \zeta_1^2 v )\abs{\vec{\alpha_1}}^2 + (\eta_2^2  u +  \zeta_2^2 v) \abs{ \vec{\alpha_2}}^2).
\end{align}
We perform the variable transformations $u' = \eta_1^2 u + \zeta_1^2 v$ and $v' = \eta_2^2 u + \zeta_2^2 v$. Since $\eta_1 \neq \eta_2$, $u'$ and $v'$ are independent variables. Then, notice that the coefficient of the monomial $(u')^{c_1} (v')^{c_2}$ is equal to $m_2(k,l,c_1,c_2)/(c_1! c_2!)$. Thus, each member of $\cM_{2,\geq}$ is observable. Since $\cM_{1,\geq} \subset  \cM_{2,\geq}$, the same is true for $K=1$ as well.  
\end{proof}

The measurement configurations discussed in Thm.~\ref{thm 2det_nontrivpart} are particularly relevant for this thesis, as the experiment we describe in Chap.~\ref{chap experiment} is modeled by a BS characterized by a partition with $K=2$ and $S_1 \neq 1$. Therefore, it is important to consider this set of configurations in more detail and derive results similar to Thm.~\ref{thm:1CntrNbnd} and Prop.~\ref{cor:1CntrNbnd}. In particular, we assume that there are at most $N$ photons in the input state so that the expectation of the operator associated with any $m_2(k,l,c_1,c_2)$ where $\max(k,l)+c_1+c_2 > N$ vanishes. Let $\cP_N$ again denote the vector space spanned by the monomials $\gamma^k (\gamma^*)^l$ with $k \leq N$, $l \leq N$ for $\gamma \in \mathbb{C}$. Further, let $\cM_{2}(N)$ denote the subset of $\cM_2$ composed of the $m_2(k,l,c_1,c_2)$ with $\max(k,l)+c_1+c_2 \leq N$, and we have already defined $\cM'_{2}(N)$ to be the subset of $\cM'_2$ composed of the $m'_2(k,l,c_1,c_2)$ with $\max(k,l)+c_1+c_2 \leq N$. According to Eqs.~\ref{eq:m_K intermsof m'_K} and~\ref{eq:m'_K intermsof m_K} $\cM_{2}(N)$ and $\cM'_{2}(N)$ can be expressed in terms of each other. Since the expectations of the operators associated with the $m_2(k,l,c_1,c_2)$ where $\max(k,l)+c_1+c_2 > N$ vanish, according to Lem.~\ref{lem: M_K<=>rho_S1>1}, $\tilde \rho_3$ is determined by the expectations of the operators associated with the expressions in $\cM_{2}(N)$ (equivalently, in $\cM'_{2}(N)$). We show that a finite set of probe amplitudes is sufficient to determine the latter.

\begin{theorem}\label{thm:2CntrNbnd_K=2}
Assume the BS is characterized by a partition with size $2$ and $S_1>1$. Assume $\tilde \rho_{3}$ has at most $N$ photons. Let a finite set $\Gamma\subseteq\cmplx$ satisfy the following
  condition: For all $p\in\cP_{N}$ such that $p(\Gamma) = \{0\}$, we
  have $p=0$. Then, given the outcome distributions of both counters at all $\gamma \in \Gamma$, the expectations of the operators associated with the members of $\cM_2(N)$ and $\cM'_2(N)$ can be determined. Therefore, $\tilde \rho_3$ is determined by the measurement statistics.
\end{theorem}
\begin{proof}
It suffices to prove the statement for $\cM_2(N)$. The generating function $G_1$ is given by
\begin{align}
G_1(u,v,\gamma,\gamma^*) =\exp ( \eta_1\zeta_1(\alpha_1 \gamma^* +\alpha_1^* \gamma )(u -v)   + ( \eta_1^2 u + \zeta_1^2 v )\abs{\vec{\alpha_1}}^2 +  (\eta_2^2  u +  \zeta_2^2 v) \abs{ \vec{\alpha_2}}^2).
\end{align}  
The assumption on $\Gamma$ in the theorem statement implies that for any given monomial $p=\gamma^{l}(\gamma^*)^{k}\in\cP_{N}$ there exist coefficients $(c_{\gamma})_{\gamma\in\Gamma}$ such that for all
  monomials $q\in\cP_{N}$ different from $p$,
  \begin{align}
      \sum_{\gamma\in\Gamma}c_{\gamma}q(\gamma,\gamma^*)&=0,\nonumber\\
      \sum_{\gamma\in\Gamma}c_{\gamma}p(\gamma,\gamma^*)&= 1.
  \end{align}
We consider the sum $G_{1,p}=\sum_{\gamma\in\Gamma}c_{\gamma}G_1(u,v,\gamma,\gamma^*) $. We expand $G_1$ in terms of the $G_{1,(i,j)}$ in Eq.~\ref{eq:G_1kl} to obtain
  \begin{align}\label{eq: G_1,p}
   G_{1,p} & =
      \sum_{\gamma\in\Gamma}c_{\gamma} \sum_{i=0}^\infty \sum_{j=0}^\infty  \frac{1}{i! j!} G_{1,(i,j)} (\gamma^*)^i \gamma^j \nonumber \\
& = \sum_{i=0}^\infty \sum_{j=0}^\infty  \frac{1}{i! j!} G_{1,(i,j)}\sum_{\gamma\in\Gamma}c_{\gamma} (\gamma^*)^i \gamma^j.
  \end{align}
 For $i \leq N$, $j \leq N$ and $(i,j) \neq (k,l)$ the sum $\sum_{\gamma\in\Gamma}c_{\gamma} (\gamma^*)^i \gamma^j$ vanishes. Furthermore, since $G_{1,(i,j)}$ is proportional to a product of a power series with the monomial $(\alpha_1^*)^i \alpha_1^j $, when $i>N$ or $j>N$ every coefficient of $G_{1,(i,j)}$ is associated with an operator with zero expectation. Therefore, taking the expectations of the operator-valued generating functions associated with both sides of Eq.~\ref{eq: G_1,p}, the expectation of the operator-valued generating function associated with $G_{1,(k,l)}$ can be determined. Since, according to the proof of Thm.~\ref{thm 2det_nontrivpart}, the $m_2(k,l,c_1,c_2)$ are in the span of the coefficients of $G_{1,(k,l)}$, and $G_{1,(k,l)}$ is observable for every $k \leq N$, $l \leq N$, every member of $\cM_2(N)$ is observable. 
\end{proof}
Note that since the polynomial space is $(N+1)^2$-dimensional, there exist $\Gamma$ of size $(N+1)^2$ that satisfy the conditions of Thm.~\ref{thm:2CntrNbnd_K=2}. The proof is given in Prop.~\ref{cor:1CntrNbnd}. Thus, $(N+1)^2$ different probe amplitudes are sufficient to determine $\tilde\rho_3$ for a BS characterized by a partition with size $K=2$ and $S_1>1$. Further, a result similar to Cor.~\ref{cor:min_p_k_invertibility} can be derived for $\tilde\rho_3$ with at most $N$ photons. The number of non-zero parameters of $\tilde\rho_3$ is equal to $\frac{ (N+1)(N+2)^2(N+3)  }{12}$.
%\begin{align}\label{eq num_of_param_rho_3}
%& (N+1)^2 + 2 (N)^2 + 3 (N-1)^2 + \hdots + N (2)^2 + N+1  = \sum_{k=1}^{N+1} \sum_{i=1}^k i^2  \nonumber \\
%& = \sum_{k=1}^{N+1} \frac{k(k+1)(2k+1)}{6} \nonumber \\
%& = \sum_{k=1}^{N+1} \frac{k^3}{3}+\sum_{k=1}^{N+1} \frac{k^2}{2}+\sum_{k=1}^{N+1} \frac{k}{6} \nonumber \\
%& = \frac{(N+1)^2(N+2)^2}{12} + \frac{(N+1)(N+2)(2N+3)}{12} + \frac{(N+1)(N+2) }{12} \nonumber \\
%& = \frac{ (N+1)(N+2)^2(N+3)  }{12}.
%\end{align}

\begin{corollary}\label{cor:min_p_kl_invertibility_S1>1}
  Let $\Gamma$ satisfy the assumption in the statement of
  Thm.~\ref{thm:2CntrNbnd_K=2}. Let $p_{kl}(\gamma)$ denote the probability of observing $k$ and $l$ photons by the counter 1 and counter 2, respectively, at the probe amplitude $\gamma$. Let $\tilde\rho_3$ have at most $N$ photons. Then, there exist $ \frac{ (N+1)(N+2)^2(N+3)  }{12}-1$ different probabilities in $\{\{p_{kl}(\gamma)\}_{k,l=0}^\infty\}_{\gamma \in \Gamma}$ that determine $\tilde \rho_3$.
\end{corollary}
\begin{proof}
The number of unknown parameters in $\tilde \rho_3$ is equal to $ u(N) = \frac{ (N+1)(N+2)^2(N+3)  }{12}-1$, when the trace condition is taken into account. According to Thm~\ref{thm:2CntrNbnd_K=2} the measurement statistics at all $\gamma \in \Gamma$ determine $\tilde \rho_3$. The measurement statistics are encoded in the set of probabilities $p_{kl}(\gamma)$, which are linear combinations of the non-zero parameters of $\tilde \rho_3$ according to Eq.~\ref{eq Born_D_kl_2}. Since the number of unknown parameters is $u(N)$, one can find a finite subset of the $p_{kl}(\gamma)$ of size $u(N) $ that determines $\tilde \rho_3$.
\end{proof}
We leave it as an open problem to find specific finite sets of probabilities of the outcomes of the counters and specific sets $\Gamma$ for a given $N$ that determine $\tilde \rho_3$. Several similar problems can also be posed. For example, given a $\Gamma$ that satisfies the conditions of Thm.~\ref{thm:2CntrNbnd_K=2}, find an upper bound on the outcomes $k$ and $l$ such that there exists a set of probabilities, with outcomes less than the corresponding upper bounds, that determines $\tilde \rho_3$.

\subsubsection{BS Characterized by a Partition With $S_1=1$}

So far we have considered measurement configurations that have a BS characterized by a partition where $S_1>1$. Let us now consider the BSs characterized by partitions where $S_1=1$. In this case $\tilde \rho_{1,K}$ occupies $K$ modes - in particular, $\tilde \rho_{1,K} = \sum_{i_2} \hdots \sum_{i_K} \chi_{\vec{i}} \otimes \left( \bigotimes_{k=2}^K \ket{i_k}\bra{i_k} \right) $, where for a particular $k=2,\hdots,S$ the $\ket{i_k}$ are Fock states in the mode $a_{S_1+\hdots + S_{K-1}+1}$. We are guided by the same overall strategy we utilized for $S_1>1$. Let us introduce the following sets of monomials in the variables $\vec{\alpha}$ and $\vec{\alpha}^*$: $\cM_{K,1} =\{m_{K,1}(k,l,c_2,\hdots,c_K) =  (\alpha_1^*)^k \alpha_1^l  \prod_{i=2}^K \abs{\vec{\alpha}_i }^{2c_i} \mid k,l,c_i \in \mathbb{Z}_{\geq 0}   \}$ and $\cM_{K,1,\geq} = \{m_{K,1}(k,l,c_2,\hdots,c_K) \mid k \geq l  \} \subset \cM_{K,1}$. We further define the subsets $\cM'_{K,1}(N) \subset \cM'_{K,1}$ that are composed of the $m'_{K,1}(k,l,c_2,\hdots,c_K)$ with $\max(k,l)+c_2+\hdots+c_K \leq N$, and the subsets $\cM'_{K,1,\geq}(N) \subset \cM'_{K,1,\geq}$ composed of the $m'_{K,1}(k,l,c_2,\hdots,c_K)$ with $k+c_2+\hdots+c_K \leq N$ for each $N$.
\begin{lemma}\label{lem: M_K<=>rho_S1=1}
Assume the Husimi function of $\tilde \rho_{1,K}$ has a determinate complex moment sequence. Then, the set of expectations of all operators associated with the expressions in $\cM_{K,1,\geq}$ determine and are determined by $\tilde \rho_{1,K}$. Further, the expectations of the operators associated with the expressions in any strict subset of $\cM'_{K,1,\geq}$ are not sufficient to determine $\tilde \rho_{1,K}$, unless the subset is $\cM'_{K,1,\geq} \setminus m_{K,1}'(0,0,0,\hdots,0)$.
\end{lemma}
\begin{proof}
The lemma can be proved using the same strategy as in the proof of Lem.~\ref{lem: M_K<=>rho_S1>1}. We outline the steps without mentioning all the details which can be inferred from the proof of Lem.~\ref{lem: M_K<=>rho_S1>1}. The proof of the first statement is built around showing that, if every member of $\cM'_{K,1,\geq}$ is observable, the expectation of every anti-normally ordered monomial of mode operators in the modes occupied by $\tilde \rho_{1,K}$ is available. Then, since these expectations correspond to the moments of the Husimi function, and the latter has a determinate moment sequence, it is determined by this set of expectations. Since any anti-normally ordered monomial of mode operators is a linear combination of normally-ordered monomials of mode operators, this set of expectations is determined by the set of expectations of all normally-ordered monomials of mode operators in the modes occupied by $\tilde \rho_{1,K}$. The expressions associated with the latter have the form $(\alpha_1^*)^k \alpha_1^l  \prod_{i=2}^K \alpha_{i,1}^{c_i} (\alpha^*_{i,1})^{c'_i}$. For the expressions where $c_i \neq c_i'$ for at least one $i$, the expectations of the corresponding operators vanish due to the form of $\tilde \rho_{1,K}$, while for any expression where $c_i = c_i'$ for all $i$, the expectation of the corresponding operator is equal to the expectation associated with $m_{K,1}(k,l,c_2,\hdots,c_K) $. And the expectation associated with any $m_{K,1}(k,l,c_2,\hdots,c_K) \notin  \cM'_{K,1,\geq}$ is determined by the expectation associated with $m_{K,1}(l,k,c_2,\hdots,c_K) \in  \cM'_{K,1,\geq}$. 

The proof of the second statement is built around showing that, if some \[m_{K,1}(k',l',c'_2,\hdots,c'_K) \in \cM'_{K,1,\geq} \setminus m_{K,1}'(0,0,0,\hdots,0) \] is not observable, then an arbitrary $\tilde \rho_{1,K}$ with at most $N = k'+c_2'+\hdots+c_K'$ photons cannot be determined. The reason is that the expectations of the operators associated with $\cM'_{K,1,\geq}(N)$ are linear combinations of the elements in the upper triangular part of $\tilde \rho_{1,K}$, and the number of expectations equals the number of elements. Thus the upper triangular part of $\tilde \rho_{1,K}$ cannot be determined from the subset of the expectations associated with $\cM'_{K,1,\geq}(N) \setminus m_{K,1}(k',l',c'_2,\hdots,c'_K)$. And the expectations of all operators associated with $\cM'_{K,1,\geq} \setminus \cM'_{K,1,\geq}(N) $ are zero, and therefore, do not add any information about $\tilde \rho_{1,K}$.
\end{proof}

Notice that for the case $S_1=1$ the second line of Eq.~\ref{eq G_in second} is still valid (with $\abs{\vec{\alpha}}^2 = \abs{\alpha_1}^2$). Similarly, Eqs.~\ref{eq:G_1kl},~\ref{eq:g_K} and~\ref{eq:m'_K intermsof m_K} are also valid as their derivations do not assume a restriction on the dimension of $\vec{\alpha}_1$. Remember that the $g_K(k,l,a,b)$ in Eq.~\ref{eq:g_K} are observable when the measurement statistics are available for all $\gamma$ in a neighborhood of $0$. Moreover, the expectations of the operators corresponding to the $g_K(k,l,a,b)$ contain all the information in the measurement statistics. In this case $g_K(k,l,a,b)$ becomes
\begin{align}\label{eq:g_K, S_1=1}
g_K(k,l,a,b) & =   \sum_{c_1+\hdots + c_K = a+b} C(c_1,c_2,\hdots,c_K) ( \alpha_1^*)^{k+c_1} \alpha_1^{l+c_1}  \prod_{i=2}^K \abs{\vec{\alpha}_i}^{2c_i} \nonumber \\
& =  \sum_{c_1+\hdots + c_K = a+b}  C(c_1,c_2,\hdots,c_K) m_{K,1}(k+c_1,l+c_1,c_2,\hdots,c_K).
\end{align}

\begin{theorem}\label{thm 2det_nontrivpart, S_1=1}
Assume the BS is characterized by a partition with size $K$ and $S_1=1$. Assume the measurement statistics are available for all $\gamma$ in a neighborhood of $0$. Assume the Husimi function of $\tilde \rho_{1,K}$ has a determinate complex moment sequence. Then, $\tilde \rho_{1,K}$ can be determined from the measurement statistics of the counters iff $K \leq 3$.
\end{theorem}
\begin{proof}
The proof proceeds in the same manner as the proof of Thm.~\ref{thm 2det_nontrivpart}. We first prove that $\tilde \rho_{1,K}$ cannot be determined when $K >3 $ by showing that there exist members of $\cM_{K,1,\geq}$ which cannot be written as linear combinations of the $g_K(k,l,a,b)$. It suffices to show this for $K =4$, in which case 
\begin{align}\label{g_4(k,l,a,b)}
g_{4}(k,l,a,b) 
& =  \sum_{c_1+c_2+c_3 + c_4 = a+b}  C(c_1,c_2,c_3,c_4) m_{4,1}(k+c_1,l+c_1,c_2,c_3,c_4).
\end{align}
Consider the subset $\cM_{4,1,\geq}^0 = \{ m_{4,1}(k,k,c_2,c_3,c_4 )\mid k \geq 0, c_i \geq 0 \textrm{ for all }i\} \subset \cM_{4,1,\geq}$. $\cM_{4,1,\geq}^0$ only affects the $g_{4}(k,l,a,b) $ where $k=l$. Conversely, $g_{4}(k,k,a,b) $ only depends on the members of $\cM_{4,1,\geq}^0$. Consider the set of $g_{4}(k,k,a,b) $ with $k+a+b =1$. There are $3$ members in this set. The $g_{4}(k,k,a,b) $ in this set only depend on the $m_{4,1}(k,k,c_2,c_3,c_4 ) \in \cM_{4,1,\geq}^0$ with $k+c_2+c_3+c_4 =1$, and there are $4$ such $ m_{4,1}(k,k,c_2,c_3,c_4 )$. Conversely, the set of $m_{4,1}(k,k,c_2,c_3,c_4 )$ with $k+c_2+c_3+c_4 =1$ only affects the $g_{4}(k,k,a,b) $ with $k+a+b =1$. Thus, the former cannot be expressed in terms of the latter.

To show that for $K \leq 3$ every expression in $\cM_{K,1,\geq}$ can be expressed as a linear combination of the $g_K(k,l,a,b)$, it suffices to show this for $K =3$. Then, starting with $G'_{1,(k,l)} $ in Eq.~\ref{eq:G'_1kl}, 
\begin{align}
G'_{1,(k,l)} & = (\alpha_1^*)^k \alpha_1^l  \exp ( ( \eta_1^2 u + \zeta_1^2 v )\abs{\alpha_1}^2 +  (\eta_2^2  u +  \zeta_2^2 v) \abs{ \vec{\alpha_2}}^2+  (\eta_3^2  u +  \zeta_3^2 v) \abs{ \vec{\alpha_3}}^2).
\end{align}
Since $\eta_2 \neq \eta_3$, $\eta_2^2 u + \zeta_2^2 v$ and $\eta_3^2 u + \zeta_3^2 v$ are linearly independent. So we can introduce the variable transformation $u' = \eta_2^2 u + \zeta_2^2 v$ and $v' = \eta_3^2 u + \zeta_3^2 v$. Then, the expression $ \eta_1^2 u + \zeta_1^2 v $ in the exponent is a linear function of $u'$ and $v'$: $ \eta_1^2 u + \zeta_1^2 v = h_1 u' + h_2 v'$ for some $h_1$ and $h_2$. The coefficient of the monomial $(u')^a(v')^b$, after multiplying by $a! b!$, is given by
\begin{align}\label{der_u'_v'_G'_kl}
\left[ \partial_{u'}^a \partial_{v'}^b G'_{1,(k,l)} \right]_{u'=0, v'=0} & = (\alpha_1^*)^k \alpha_1^l  \left[ \partial_{u'}^a \partial_{v'}^b\exp ( (h_1 u' + h_2 v' ) \abs{\alpha_1}^2 +  u' \abs{ \vec{\alpha_2}}^2+  v' \abs{ \vec{\alpha_3}}^2) \right]_{u'=0, v'=0} \nonumber \\
 & = (\alpha_1^*)^k \alpha_1^l  \left[ \partial_{u'}^a \partial_{v'}^b\exp (    u'(h_1 \abs{\alpha_1}^2 + \abs{ \vec{\alpha_2}}^2 ) +v'(h_2 \abs{\alpha_1}^2 + \abs{ \vec{\alpha_3}}^2)      ) \right]_{u'=0, v'=0} \nonumber \\
  & = (\alpha_1^*)^k \alpha_1^l (h_1 \abs{\alpha_1}^2 + \abs{ \vec{\alpha_2}}^2 )^a (h_2 \abs{\alpha_1}^2 + \abs{ \vec{\alpha_3}}^2)^b.
\end{align}
Let us denote these terms by $g'_{3}(k,l,a,b)$. The observability of every member of $\cM_{3,1,\geq}$ can be shown by recursion. Because the recursive sequence is somewhat complicated, we build it gradually for the reader by considering consecutively larger subsets of $\cM_{3,1,\geq}$. For each $d \geq 0$ let us introduce the set of sets $\cM_{3,1,\geq}^{d} = \{\cM_{3,1,\geq}^{d,a,b  } = \{ m_{3,1}(k+d,k,a,b )  \mid k \geq 0 \} \}_{a,b=0}^\infty$. The $\cM_{3,1,\geq}^{d} $ do not intersect and their union is $\cM_{3,1,\geq}$. The same is true for the $\cM_{3,1,\geq}^{d,a,b  }$. We show that, for a given $d$, the corresponding $\cM_{3,1,\geq}^{d}$ is determined by $\{g'_{3}(k+d,k,a,b)\}_{a,b,k=0^\infty}$. In the following, by ``one can obtain the expression" we mean the expression can be written as a linear combination of observable expressions. We start with $\cM_{3,1,\geq}^{d,0,0  }$. It can be seen that $m_{3,1}(k+d,k,0,0 )$ is equal to $g'_{3}(k+d,d,0,0)$ and thus $\cM_{3,1,\geq}^{d,0,0  }$ is observable. Next, consider the $\cM_{3,1,\geq}^{d,a,0  }$ for $a\geq 1$. Any $m_{3,1}(k'+d,k',a',0 )$ can be written as a linear combination of $g'_{3}(k'+d,d,a',0)$ and the $m_{3,1}(k+d,k,a,0 )$ with $(a,k) =(a'-1,k'+1),(a'-2,k'+2) \hdots,(0,k'+a')$ by rearranging the terms across the equality sign in the expansion of $g'_{3}(k'+d,d,a',0)$. Thus, starting with $a'=1$, one can obtain any $m_{3,1}(k+d,k,a,0 )$ by recursion. The same argument shows that every member of $\cM_{3,1,\geq}^{d,0,b  }$ for $b\geq 1$ is a linear combination of the $g'_{3}(k+d,d,0,b)$. 

We generalize this recursive approach of obtaining the members of $\cM_{3,1,\geq}^{d}$ to any tuple $(a,b)$ as follows. We have already shown that $\cM_{3,1,\geq}^{d,a,0  }$ and $\cM_{3,1,\geq}^{d,0,b  }$ are observable sets for any $a$ and $b$, respectively. We start by considering the $\cM_{3,1,\geq}^{d,a,1  }$ for all $a$. Without going too much into the details, any $m_{3,1}(k'+d,k',a',1 )$ can be written as a linear combination of $g'_{3}(k'+d,d,a',1)$, some of the members belonging to $\{\cM_{3,1,\geq}^{d,a,0  }\}_a$, and the $m_{3,1}(k+d,k,a,1 )$ with $(a,k) =(a'-1,k'+1),(a'-2,k'+2) \hdots,(0,k'+a')$ by rearranging the terms across the equality sign in the expansion of $g'_{3}(k'+d,d,a',1)$. Thus, starting with $a'=0$, every $m_{3,1}(k'+d,k',a',1 )$ can be obtained by recursion. Finally, for an arbitrary $b$, $m_{3,1}(k'+d,k',a',b' )$ can be written as a linear combination of $g'_{3}(k'+d,d,a',b')$, some of the members belonging to $\{\{\cM_{3,1,\geq}^{d,a,b  }\}_{a=0}^\infty\}_{b=0}^{b'-1}$, and the $m_{3,1}(k+d,k,a,b' )$ with $(a,k) =(a'-1,k'+1),(a'-2,k'+2) \hdots,(0,k'+a')$ by rearranging the terms across the equality sign in the expansion of $g'_{3}(k'+d,d,a',b')$. Thus, starting with $b'= 1$, the $m_{3,1}(k'+d,k',a',b' )$ can be obtained by recursion. 
\end{proof}

\subsection{Results With One Photon Counter}

When one counter is used, the generating function corresponding to the WFH configuration is $G_{\textrm{in},2}$ in Eq.~\ref{eq genfun_nontrivpart_onedet}. We first consider BSs characterized by partitions where $S_1>1$. We assume the measurement statistics are available for all $\gamma$ in a neighborhood of $0$ throughout this section. Since $G_{\textrm{in},2} = G_{\textrm{in}} |_{v=0}$ we can use the tools we developed in the last section for the measurement configurations with both photon counters. In particular, the generating functions $G'_{1,(k,l)}$ in Eq.~\ref{eq:G'_1kl} are observable at $v=0$:
\begin{align}\label{eq:G_1kl_v=0}
G'_{1,(k,l)} |_{v=0}
& = (\alpha_1^*)^k \alpha_1^l \exp (  \sum_{i=1}^K \eta_i^2   \abs{ \vec{\alpha_i}}^2 u).
\end{align}
Let us denote the coefficients of the powers of $u$ of $G'_{1,(k,l)} |_{v=0}$ by $f_K(k,l,a)/(a!)$, so that
\begin{align}\label{f_K(k,l,a)}
f_K(k,l,a) & = \left[  \partial_u^a  G'_{1,(k,l)} |_{v=0}   \right]_{a=0} \nonumber \\
& = (\alpha_1^*)^k \alpha_1^l  ( \sum_{i=1}^K \eta_i^2   \abs{ \vec{\alpha_i}}^2)^a \nonumber \\
& =  (\alpha_1^*)^k \alpha_1^l \sum_{c_1+\hdots +c_K=a} \binom{a}{c_1,\hdots,c_K} \prod_{i=1}^K   \eta_i^{2c_i}   \abs{ \vec{\alpha_i}}^{2c_i}  \nonumber \\
& = \sum_{c_1+\hdots +c_K=a} \binom{a}{c_1,\hdots,c_K} \prod_{i=1}^K   \eta_i^{2c_i}   m_K(k,l,c_1,\hdots,c_K).
\end{align}
The expectations of the operators associated with the $f_K(k,l,a)$ contain all the information in the measurement statistics.

\begin{theorem}\label{thm 1det_nontrivpart}
Assume the BS is characterized by a partition with size $K$ and $S_1>1$. Assume the measurement statistics are available for all $\gamma$ in a neighborhood of $0$. Assume the Husimi function of $\tilde \rho_{K+1}$ has a determinate complex moment sequence. Then, $\tilde \rho_{K+1}$ can be determined from the measurement statistics of the counter iff $K =1$.
\end{theorem}
\begin{proof}
We have already proven that for $K=1$ the measurement statistics determine $\tilde \rho_{K+1}$ in Cor.~\ref{cor:1CntrNbhd}. We need to be shown that for $K>1$ there exist members of $\cM_{K,\geq}$ that cannot be obtained from $G_{\textrm{in},2}$. 
It suffices to show this for $K=2$ since $\cM_{2,\geq} \subset \cM_{K,\geq}$ for larger $K$. Since the expectations of the operators associated with the $f_2(k,l,a)$ contain all the information in the measurement statistics, it suffices to show that some members of $\cM_{2,\geq}$ cannot be written as linear combinations of the $f_2(k,l,a)$. In this case
\begin{align}
f_2(k,l,a) & = \sum_{c_1=0}^a \binom{a}{c_1}    \eta_1^{2c_1}   \eta_2^{2(a-c_1)}   m_2(k,l,c_1,a-c_1).
\end{align}
Notice that for a given $k$ and $l$ the $m_2(k,l,c_1,c_2)$ with $c_1+c_2=a$ only appear in the expansion of $f_2(k,l,a)$. So there is a single equation for $a+1$ unknowns.
\end{proof}

We now consider BSs characterized by partitions where $S_1=1$. In this case the $f_K(k,l,a)$ are given by
\begin{align}\label{f_K(k,l,a)_S_1=1}
f_K(k,l,a) & = (\alpha_1^*)^k \alpha_1^l  ( \eta_1^2   \abs{ \alpha_1}^2+ \sum_{i=2}^K \eta_i^2   \abs{ \vec{\alpha_i}}^2)^a \nonumber \\
& =  (\alpha_1^*)^k \alpha_1^l \sum_{c_1+\hdots +c_K=a} \binom{a}{c_1,\hdots,c_K} \eta_1^{2c_1}   \abs{ \alpha_1}^{2c_1} \prod_{i=2}^K   \eta_i^{2c_i}   \abs{ \vec{\alpha_i}}^{2c_i}  \nonumber \\
& = \sum_{c_1+\hdots +c_K=a} \binom{a}{c_1,\hdots,c_K} \prod_{i=1}^K   \eta_i^{2c_i}   m_{K,1}(k+c_1,l+c_1,c_2,\hdots,c_K).
\end{align}

\begin{theorem}\label{thm 1det_nontrivpart_S_1=1}
Assume the BS is characterized by a partition with size $K$ and $S_1=1$. Assume the measurement statistics are available for all $\gamma$ in a neighborhood of $0$. Assume the Husimi function of $\tilde \rho_{1,K}$ has a determinate complex moment sequence. Then, $\tilde \rho_{1,K}$ can be determined from the measurement statistics of the counter iff $K \leq 2$.
\end{theorem}
\begin{proof}
We first consider $K = 2$ (the $K=1$ case corresponds to a single mode state, which is covered by Cor.~\ref{cor:1CntrNbhd}). We show that every member of $\cM_{2,1,\geq}$ can be written as a linear combination of the $f_2(k,l,a)$ in Eq.~\ref{f_K(k,l,a)_S_1=1}:
\begin{align}
f_2(k,l,a) 
& = (\alpha_1^*)^k \alpha_1^l  ( \eta_1^2   \abs{ \alpha_1}^2+ \eta_2^2   \abs{ \vec{\alpha_2}}^2)^a.
\end{align}
We define $h_1 = \eta_1^2/\eta_2^2  $. Then,
\begin{align}
f_2(k,l,a)/ \eta_2^2 
& = (\alpha_1^*)^k \alpha_1^l  ( h_1   \abs{ \alpha_1}^2+  \abs{ \vec{\alpha_2}}^2)^a.
\end{align}
So, for each $k$, $l$ and $a$, $f_2(k,l,a)/ \eta_2^2 $ is of the same form as the right hand side of Eq.~\ref{der_u'_v'_G'_kl} when $b=0$. We have already shown in the second part of the proof of Thm~\ref{thm 2det_nontrivpart, S_1=1} that the $\cM_{3,1,\geq}^{d,a,0}$ for all $d$ and $a$ can be obtained from the set of expressions of the form $ (\alpha_1^*)^{k+d} \alpha_1^k  ( h_1   \abs{ \alpha_1}^2+  \abs{ \vec{\alpha_2}}^2)^a$. It is left to notice that $m_{2,1}(k+d,k,a) = m_{3,1}(k+d,k,a,0)$, and thus the $m_{2,1}(k+d,k,a)$ can similarly be obtained from the $f_2(k+d,k,a)/ \eta_2^2 $ for all $k$, $d$ and $a$. 

To prove the other direction of the theorem statement, it suffices to consider $K =3$. $f_3(k,l,a)$ has the form
\begin{align}
f_3(k,l,a) 
& = \sum_{c_1+c_2 +c_3=a} \binom{a}{c_1,c_2,c_3}  \eta_1^{2c_1} \eta_2^{2c_2} \eta_3^{2c_3}   m_{3,1}(k+c_1,l+c_1,c_2,c_3).
\end{align}
For a given pair of integers $d \geq 0 $ and $r$ the expressions $m_{3,1}(d+k+c_1,d+c_1,c_2,c_3)$, where $k+ c_1+c_2+c_3=r$, only appear in the expansions of the $f_3(d+k,k,a)$ where $k+a=r$. Even for $r=1$ there are more unknowns ($m_{3,1}(d,0,1,0)$, $m_{3,1}(d,0,0,1)$ and $m_{3,1}(d+1,1,0,0)$ to be exact) than equations (the expansions of $f_3(d+1,1,0) $ and $f_3(d,0,1)$ to be exact).
\end{proof}

%% file: chapter_simulations.tex
\chapter{Numerical Simulations With Finite Data}\label{chap simulations}

In the previous chapter we derived a set of analytical results about the relationship between the photon number distribution of the counter(s) and the unknown input state for different kinds of WFH measurement configurations. We found that for WFH configurations with a BS characterized by a trivial partition, the two-mode twirled state with maximum photon number $N$ is completely determined by the measurement statistics of one of the detectors at $(N+1)^2$ different probe amplitudes (Thm.~\ref{thm:1CntrNbnd} and Cor.~\ref{cor:twirldet}). We also found that for WFH configurations with a BS characterized by a partition size $K=2$ and $S_1>1$, the measurement statistics of both detectors at $(N+1)^2$ different probe amplitudes determine the three mode twirled state (Thm.~\ref{thm:2CntrNbnd_K=2} together with Cor.~\ref{cor:twirldet}). The sufficient condition on the $ (N+1)^2$ mutually distinct probe amplitudes $\Gamma = \{\gamma_i\}_{i=1}^{(N+1)^2}$ for the above results to hold is that the corresponding matrix in Eq.~\ref{eq: Gamma_matrix} be full rank. However, remember that we only proved a sufficiency condition, and we indeed find numerically that in general a smaller number of probe amplitudes suffice to determine the corresponding twirled state. Further, it suffices to use only a part of the measurement statistics for each amplitude, by assuming the photon counters cannot distinguish the photon number beyond some value. In practice, we check this by constructing a positive-operator-valued measure (POVM) for each probe amplitude for a given WFH configuration and checking the informational completeness (IC-ness) of this set of POVMs with respect to the corresponding twirled state. We briefly describe our definition of a POVM, as well as the notion of the IC-ness of a POVM or of a set of POVMs in Sec.~\ref{sec POVM and IC}.

In this chapter we first aim to describe the numerical models we constructed for computing the probabilities of the outcomes of the counter(s) for a given twirled state and WFH measurement configuration as well as for obtaining an estimate of the twirled state from a set of simulated or experimentally obtained statistical data. We constructed two classes of models corresponding to WFH configurations with a BS characterized by a trivial partition, and to WFH configurations with a BS characterized by partition size $K=2$ and $S_1>1$. Second, we report some simulations we did with the first class of models using a balanced BS to test the accuracy of the models and the performance of the reconstruction algorithm. These tests should be seen in the context of the aims of the next chapter, where an instant of the second class of models is used to describe an experiment and to reconstruct $\tilde \rho_3$ from the experimental data. 

We describe how we model the measurement configurations in Sec.~\ref{sec algorithm sketch}. We start by assuming the WFH configurations are ideal, in the sense that the probabilities of measuring any number of photons by the counters are given by Eq.~\ref{eq Born_D_kl_2}. The operator corresponding to detecting $k$ and $l$ photons by counter 1 and counter 2, respectively, is given by $\Pi_{kl} = \tr_\textsl{p}( (I \otimes \sigma)U^\dagger \hat{D}_{kl} U ) $ (Eq.~\ref{eq: Pi_kl_def}), where $\hat{D}_{kl}$ is given by Eq.~\ref{eq D_kl}, $\sigma = \ket{\gamma}\bra{\gamma}\otimes \ket{\vec{0}}\bra{\vec{0}} $ is a coherent state in mode 1 of the probe, and $U$ is the unitary corresponding to the BS. When only counter 1 is used, the operator corresponding to detecting $k$ photons is given by $\Pi_{k} = \tr_\textsl{p}( (I \otimes \sigma)U^\dagger \sum_{l=0}^\infty \hat{D}_{kl} U )  = \sum_{l=0}^\infty \Pi_{kl}$ (Eq.~\ref{eq: Pi_k_def}). Notice the implicit dependence of $\Pi_k$ and $\Pi_{kl}$ on $U$ and $\gamma$. When the context calls for it, we explicitly show the dependence on $\gamma$ by writing $\Pi_k(\gamma)$ and $\Pi_{kl}(\gamma)$ instead. We further assume that there is a maximum photon number $N_c$ such that when a photon number greater than $N_c$ is measured by a photon counter, it can only tell us that the measured number of photons was greater than $N_c$. The operators associated with such measurements are described in Subsec.~\ref{subsec POVMs_ideal}. We refer to a POVM associated with such a measurement configuration for a given probe amplitude as an ideal POVM. We later consider detection in the presence of losses, where the POVM elements corresponding to photon number measurements differ from the ideal POVM elements. We refer to POVMs composed of such elements as non-ideal POVMs.

According to Cor.~\ref{cor:min_p_k_invertibility} for a WFH configuration with a BS characterized by a trivial partition and one counter there exists a set of $(N+1)^2$ different probe amplitude $\Gamma$ such that a finite set of operators $\Pi_k (\gamma)$, where $\gamma \in \Gamma$, suffice to determine $\tilde \rho_2$. The reasoning in Sec.~\ref{sec BS with trivial part} that led to Cor. ~\ref{cor:min_p_k_invertibility} can be extended straightforwardly to the WFH configuration with a BS characterized by a trivial partition, and with both counters present. Thus, there exists a finite subset of $\{ \{ \Pi_{kl}(\gamma) \}_{k,l =0}^\infty \}_{\gamma \in \Gamma}$ that determines $\tilde \rho_2$ when it has maximum photon number $\leq N$. Similarly, according to Cor.~\ref{cor:min_p_kl_invertibility_S1>1} for a WFH configuration with a BS characterized by partition size $K=2$ with $S_1>1$ one can also find a finite subset of $\{ \{ \Pi_{kl}(\gamma) \}_{k,l =0}^\infty \}_{\gamma \in \Gamma}$ that determines $\tilde \rho_3$ when it has maximum photon number $\leq N$. Note that the BS unitary implicit in the $\Pi_{kl}(\gamma)$ or the $\Pi_{k}(\gamma)$ corresponds to that of the matching WFH configuration. We can compose sets of POVMs from these finite subsets of the $\Pi_{kl}(\gamma)$ or of the $\Pi_{k}(\gamma)$ as described in Subsec.~\ref{subsec POVMs_ideal}.

 The results of the analysis with simulated data for a WFH configuration with a BS characterized by a trivial partition are presented in Sec.~\ref{sim BS_trivpart}. In each simulation the true input state with maximum photon number $N$ is decided first. Then, a set of POVMs that is IC is constructed, with the IC condition checked numerically. This procedure is described for a set of ideal POVMs in Subsec.~\ref{subsec POVMs_ideal}, and we use the same procedure for non-ideal POVMs described in Subsec.~\ref{subsec POVMs_nonideal}. For the chosen true input state, we first calculate the corresponding twirled state analytically and compute the probabilities of the different measurement outcomes according to the chosen set of POVMs. We then utilize a pseudo-random generator to sample uniformly from the measurement configuration and perform a maximum likelihood estimation (MLE) to reconstruct the twirled state. The description of the MLE procedure is given in Subsec.~\ref{subsec max_like}. The quantum state fidelity is generally used as a figure of merit -  for two arbitrary states $\rho_1$ and $\rho_2$ the fidelity is given by $\cF (\rho_1,\rho_2) = (\tr (\sqrt{  \sqrt{\rho_1} \rho_2 \sqrt{\rho_1}    }))^2 = (\tr (\sqrt{  \sqrt{\rho_2} \rho_1 \sqrt{\rho_2}    }))^2$. We used python as the main programming language.

\section{POVMs and Informational Completeness}\label{sec POVM and IC}

The concept of a POVM is central to the theory of quantum measurement and quantum measurement devices. An in-depth discussion of the concept can be found in Ref.~\cite{busch1995oper}. For a given measurement device, if the number of possible different outcomes of a measurement is finite, and the measured state is finite-dimensional, then the following definition of a POVM is usually used.

\begin{definition}\label{POVM def_simple}
A POVM $E$ on a finite-dimensional Hilbert space $\cH$ is a finite collection of positive semi-definite matrices $\{ E_i \}$ such that $\sum_i E_i $ equals the identity on $\cH$. Each element of $E$ is associated with a distinct measurement outcome, such that the probability of the measurement outcome corresponding to $E_i$ for a given state $\rho$ is equal to $\tr(\rho E_i)$. \cite{nielsen:qc2001a}
\end{definition}

 Since we assume that the unknown state has a maximum photon number bounded by $N$ we can associate the expectation of any operator on the state-space with the expectation of a finite dimensional matrix. Further, we assume that the counters cannot distinguish the outcomes corresponding to photon numbers greater than some value $N_c$. Therefore, for example, for a WFH configuration with an ideal single counter, all measurement outcomes of a photon number $\geq N_c$ for a given probe amplitude are assigned to the same POVM element. Thus, the number of POVM elements in our models is always finite, and, therefore, this definition of a POVM is appropriate for our purposes. 

We say a POVM $E = \{E_i\}_{i=1}^n$ as defined in Def.~\ref{POVM def_simple} is informationally complete (IC) with respect to the state $\rho \in \cH$ if the probabilities $p_i = \tr(\rho E_i)$ determine $\rho$. More specifically, $E$ gives rise to the linear map $\cH \mapsto \cP(n)$, where $\cP(n)$ is the $n-1$ dimensional probability simplex, by the trace operation. Then, $E$ is IC with respect to $\rho$ (or with respect to $\cH$) if this linear map is left-invertible. It can be seen that the IC condition is necessary for the state to be determined by the measurement outcome probabilities. 

It is sometimes the case that a POVM is not IC, but a set of several different POVMs that can be constructed together determine the state. In this work a POVM is associated with a particular measurement setting - defined as a particular WFH configuration with a particular probe state. In general, we find that a single POVM might not be IC, but using a set of probe amplitudes for a given WFH configuration (that is, using several measurement settings) can result in measurement statistics that determines the corresponding twirled state. We refer to a set of POVMs as a ``measurement context". If the set of linear maps associated with the probabilities for the different POVMs can be inverted to express the twirled state in terms of the set of probabilities, then we say that the measurement context is IC.

\section{Description of the Numerical Procedure}\label{sec algorithm sketch}

In this chapter and in the next, we assume the input states have a maximum photon number $N$. More specifically, $\tilde \rho_2$ has the form:
\begin{align}\label{tilde rho_2 max_N}
\tilde \rho_2 =  \sum_{i=0}^N \chi_{i} \otimes \ket{i}\bra{i},
\end{align}
where the $\chi_{i}$ are matrices in mode $1$ of sizes $(N-i+1)\times (N-i+1)$. Let us refer to the convex space of states of this form by $\cR_{2,N}$. Further, we denote the vector space spanned by these states by $\cR^s_{2,N}$. Similarly, $\tilde \rho_3$ has the form:
\begin{align}\label{tilde rho_3 max_N}
\tilde \rho_3 = \sum_{i_2=0}^N \left( \sum_{i_1=0}^{N-i_2} \chi_{i_1,i_2} \otimes \ket{i_1}\bra{i_1} \right) \otimes \ket{i_2}\bra{i_2},
\end{align}
where the $ \chi_{i_1,i_2}$ are matrices in mode $1$ of sizes $(N-i_1-i_2+1)\times (N-i_1-i_2+1)$. We express $\tilde \rho_3$ as in Eq.~\ref{tilde rho_3 max_N} so that it is evident that one can think of $\tilde \rho_3$ as $N+1$ non-normalized density matrices of the form of $\tilde \rho_2$ in modes $1$ and $2$, each multiplied with a Fock state in mode $3$. We denote the convex space of states of this form by $\cR_{3,N}$, and the vector space spanned by them by $\cR^s_{3,N}$.

\subsection{Ideal POVMs and Informational Completeness}\label{subsec POVMs_ideal}

 It is shown below that the individual operators $\Pi_{kl}(\gamma)$ or $\Pi_k(\gamma)$ have finite-dimensional supports, but there is no finite-dimensional subspace of the underlying Hilbert space that contains the supports of the total collection of these operators. Nevertheless, we can model all of these operators as finite dimensional matrices of the same size depending on $N$ and on the partition of the BS. This is because we assume the state-spaces are constrained to $\cR_{2,N}$ or $\cR_{3,N}$ when the BS is characterized by partition sizes $K=1$ or $K=2$, respectively, and thus the supports of these operators are likewise constrained to these state-spaces. This argument extends to the operators in the non-ideal POVMs we construct later.

We treat the POVMs associated with a BS characterized by a trivial partition first. The action of the BS in mode space is given by Eq.~\ref{eq: BS_general} where $K=1$ and $S_1=2$. According to our discussion in Sec.~\ref{sec genfunc for WFH} of the last chapter, we can use the fact that the photon counters cannot detect the overall phase and incorporate the phase that the BS imparts on the probe modes into $\gamma$ to express $B$ as $B = \begin{pmatrix} 1 & 0 \\ 0 & 1 \end{pmatrix} \otimes \begin{pmatrix} \eta & \zeta \\ \zeta  & -\eta \end{pmatrix}$, where $\eta$ and $\zeta$ are real parameters. Let us now expand $\Pi_{kl}$:
\begin{align}\label{Pi_kl expanded_start}
 \Pi_{kl} & = \tr_\textsl{p}( (I \otimes \ket{\gamma}\bra{\gamma}\otimes \ket{0}\bra{0} ) U^\dagger \hat{D}_{kl} U ) \nonumber \\
& = \tr_\textsl{p} \left( (I \otimes \ket{\gamma}\bra{\gamma} \otimes \ket{0}\bra{0} ) U^\dagger \sum_{k_1+k_2=k}\sum_{l_1+l_2=l} \ket{k_1,k_2}\bra{k_1,k_2} \otimes \ket{l_1,l_2}\bra{l_1,l_2} U \right) \nonumber \\
& =\sum_{k_1+k_2=k}\sum_{l_1+l_2=l} \bra{\gamma} \bra{0} U^\dagger \frac{1}{\sqrt{k_1! k_2! l_1! l_2!}} (a_1^\dagger)^{k_1}(a_2^\dagger)^{k_2}  (b_1^\dagger)^{l_1}  \nonumber \\
& \hdots (b_2^\dagger)^{l_2}  \ket{\vec{0}}\bra{\vec{0}}  \frac{1}{\sqrt{k_1! k_2! l_1! l_2!}} a_1^{k_1} a_2^{k_2} b_1^{l_1} b_2^{l_2}    U \ket{\gamma} \ket{0} .
\end{align}
We can now write $  \ket{\vec{0}}\bra{\vec{0}}$ in the last line as $ U \ket{\vec{0}}\bra{\vec{0}}U^\dagger $ and use Eq.~\ref{lem Uirreps eq} to obtain
\begin{align}\label{Pi_kl expanded}
\Pi_{kl} & = \sum_{k_1+k_2=k}\sum_{l_1+l_2=l} \frac{1}{k_1! k_2! l_1! l_2!} \bra{\gamma} \bra{0} (\eta a_1^\dagger + \zeta b_1^\dagger )^{k_1}  (\zeta a_1^\dagger - \eta b_1^\dagger)^{l_1} (\eta a_2^\dagger + \zeta b_2^\dagger )^{k_2} \nonumber \\
& \hdots (\zeta a_2^\dagger - \eta b_2^\dagger)^{l_2}   \ket{\vec{0}}\bra{\vec{0}} (\eta a_1+ \zeta b_1 )^{k_1}  (\zeta a_1 - \eta b_1)^{l_1} (\eta a_2 + \zeta b_2 )^{k_2}  (\zeta a_2 - \eta b_2)^{l_2} \ket{\gamma} \ket{0}  \nonumber \\
& =  \sum_{k_1+k_2=k}\sum_{l_1+l_2=l} \frac{1}{k_1! k_2! l_1! l_2!} (\eta a_1^\dagger + \gamma^* )^{k_1}  (\zeta a_1^\dagger - \eta \gamma^*)^{l_1} (\eta a_2^\dagger )^{k_2} \nonumber \\
& \hdots (\zeta a_2^\dagger )^{l_2}   \ket{\vec{0}}\bra{\vec{0}} (\eta a_1+ \zeta \gamma )^{k_1}  (\zeta a_1 - \eta \gamma)^{l_1} (\eta a_2)^{k_2}  (\zeta a_2 )^{l_2} \nonumber \\
& =  \sum_{k_1+k_2=k}\sum_{l_1+l_2=l} \frac{\eta^{2k_2} \zeta^{2l_2}}{k_1! k_2! l_1! l_2!} \left( (\eta a_1^\dagger + \gamma^* )^{k_1}  (\zeta a_1^\dagger - \eta \gamma^*)^{l_1}  \ket{0}\bra{0} (\eta a_1+ \zeta \gamma )^{k_1}  (\zeta a_1 - \eta \gamma)^{l_1} \right) \nonumber \\
& \hdots (a_2^\dagger )^{k_2+l_2}  \ket{0}\bra{0} a_2^{k_2+l_2}.  
\end{align}
Let us look closely at the last line of \ref{Pi_kl expanded}. The terms in the large parenthesis are matrices in mode $1$ with maximum photon number $k_1+l_1$. Each such matrix is multiplied with a Fock state projector onto $k_2+l_2$ photons in mode $2$. Thus, the maximum photon number in each term in the linear combination equals $k_1+l_1+k_2+l_2=k+l$. We can combine the terms with the same number of photons in mode $2$ to write the last line of \ref{Pi_kl expanded} as a linear combination of matrices in mode $1$ each multiplied by a different Fock state projector in mode $2$. Let us denote these matrices by $\Pi_{kl,n}$, so that $\Pi_{kl} = \sum_{n=0}^{k+l} \Pi_{kl,n} \otimes \ket{n}\bra{n}$. It can be seen that $\Pi_{kl}  \in \cR^s_{2,k+l}$. 

Now, since $\tilde \rho_2$ has maximum photon number $N$, the summands of $\Pi_{kl}$ associated with Fock states with total photon number that's greater than $N$, have zero expectations with respect to $\tilde \rho_2$. More specifically, when $k+l \geq N$,
\begin{align}\label{eq:Born_rule_p_kl}
p_{kl} &= \tr (\tilde \rho_2 \Pi_{kl} ) \nonumber \\
& = \tr ( ( \sum_{i=0}^N \chi_{i} \otimes \ket{i}\bra{i} )  \sum_{n=0}^{k+l} \Pi_{kl,n} \otimes \ket{n}\bra{n})  \nonumber \\
& = \sum_{i=0}^N  \sum_{n=0}^{k+l} \tr (\chi_i \Pi_{kl,n} )  \abs{\bra{i} \ket{n}}^2  \nonumber  \\
& = \sum_{i=0}^N  \tr (\chi_i \Pi_{kl,i} ).
\end{align}
Thus, the summands $\Pi_{kl,n}$ with $n>N$ dot have a contribution to the trace product. Further, notice that $\chi_i$ has maximum photon number $N-i$, while $\Pi_{kl,i} $ has maximum photon number $k+l-i$. Then, if we expand $\Pi_{kl,i}$ as $\Pi_{kl,i} = \sum_{x=0}^{k+l-i}\sum_{y=0}^{k+l-i} \Pi_{kl,i} (x,y) \ket{x} \bra{y} $, this implies that the elements $\Pi_{kl,i} (x,y)$ where either $x>N-i$ or $y>N-i$ do not contribute to the trace product with $\chi_i$. Thus, we can restrict $\Pi_{kl}$ to $\cR^s_{2,N}$ without it affecting our computations. Similarly, if $k+l <N$, we can again model $\Pi_{kl}$ as a matrix in $\cR^s_{2,N}$ by adding zeros where necessary.

To make our simulations more relevant to experiment, the POVMs we construct are associated with counters that cannot resolve photon numbers greater than some number $N_c$. See Sec.~\ref{qo pc} of Chap.~\ref{chap quantum optics} for how the measurement process of such a counter is related to the measurement process by the usual ideal counter. There, we characterize such a photon counter by the projectors $\hat{D}_k$ when $k \leq N_c$ photons are detected, together with the identity resolving projector $\hat{D}_> = I - \sum_{k=0}^{N_c} \hat{D}_k$ when a photon number $>N_c$ is detected. We still consider such counters as ideal counters since they do not experience losses or other types of noise. We now describe the operators on the input state associated with such photon counters when they are used in a WFH configuration with a given probe $\sigma$. Let us first discuss the WFH configurations where both photon counters are utilized. This means substituting the projectors $\hat{D}_{kl}$ in Eq.~\ref{eq: Pi_kl_def} by the projectors associated with joint detection by two photon counters, each capable of resolving up to $N_c$ photons. When $k\leq N_c$ and $l \leq N_c$ photons are detected by the two photon counters, the resulting operator on the input modes is given by $\Pi_{kl}$. When $k\leq N_c$ and $l > N_c$ photons are measured, the projector associated with the second counter is $\hat{D}_>$, and thus the operator on the input modes can be expressed in terms of the $\Pi_{kl}$ as
\begin{align}\label{eq Pi_k,>}
\Pi_{k,>} &= \tr_\textsl{p} ( (I\otimes \sigma)U^\dagger ( \hat{D}_k \otimes \hat{D}_>) U ) \nonumber \\
& = \tr_\textsl{p} ( (I\otimes \sigma)U^\dagger ( \hat{D}_k \otimes   \sum_{l=N_c+1}^{\infty} \hat{D}_l   ) U ) \nonumber \\
& = \sum_{l=N_c+1}^\infty \Pi_{kl}.
\end{align}
Similarly, the operator associated with the measurement of $k > N_c$ and $l \leq N_c$ photons is given by $\Pi_{>,l} =  \sum_{k=N_c+1}^\infty \Pi_{kl}$. When both photon counters register greater than $N_c$ photons, the resulting operator on input modes is 
\begin{align}\label{eq Pi_>,>}
\Pi_{>,>} &= \tr_\textsl{p} ( (I\otimes \sigma)U^\dagger ( \hat{D}_> \otimes \hat{D}_>) U ) \nonumber \\
& = \tr_\textsl{p} ( (I\otimes \sigma)U^\dagger (   \sum_{k=N_c+1}^{\infty} \hat{D}_k   \otimes   \sum_{l=N_c+1}^{\infty} \hat{D}_l   ) U ) \nonumber \\
& =  \sum_{k=N_c+1}^\infty \sum_{l=N_c+1}^\infty \Pi_{kl}.
\end{align}
When a single probe amplitude $\gamma$ is used, the corresponding POVM is composed of all the operators on the input modes corresponding to all possible measurement outcomes of the two counters. For the types of ideal counters we assume in this section, the POVM is thus given by
\begin{align}\label{POVM(gamma_i)_ideal}
\textrm{POVM}(\gamma) =  \{  \Pi_{kl}(\gamma) \}_{k=0,l=0}^{N_c,N_c}   \cup  \{ \Pi_{k,>}  (\gamma) \}_{k=0}^{N_c} \cup  \{ \Pi_{>,l}  (\gamma) \}_{l=0}^{N_c} \cup \Pi_{>,>}.
\end{align}

Now, for a given WFH configuration, each POVM corresponds to a different ``measurement setting" - that is, to the use of a particular probe amplitude for making measurements. When $N_p$ different probe amplitudes $\Gamma = \{\gamma_i\}_{i=1}^{N_p}$ are used in the same WFH configuration, we construct the POVM for each measurement setting, and refer to the whole set of POVMs as the measurement context $\cC(\Gamma)$:
\begin{align}\label{POVM(Gamma)}
\cC(\Gamma) = \bigcup_{i=1}^{N_p}  \textrm{POVM}(\gamma_i) 
\end{align}
It can be the case that the number of measurements (or shots) conducted in each measurement setting is different. In our simulations we always use the same number of samples for all probe amplitudes for reconstruction, while the experimental data described in the next chapter results in some variation across the different probe amplitudes. It could be the case that using different sample sizes for the different $\textrm{POVM}(\gamma_i)$, either intentionally or because of experimental circumstances, would result in a better reconstruction, but we do not investigate this problem.

For the WFH configurations where one photon counter is utilized, the operator on input modes corresponding to the projector $\hat{D}_k$ of the counter when $k \leq N_c$ photons are registered is given by $\Pi_k = \sum_{l=0}^\infty \Pi_{kl} $. When $k>N_c$ photons are registered by the counter, the corresponding operator is given as
\begin{align}
\Pi_> &=\tr_\textsl{p} ( (I\otimes \sigma)U^\dagger (  \hat{D}_> \otimes \sum_{l=0}^\infty  \hat{D}_l) U ) \nonumber \\
& = \sum_{l=0}^\infty \Pi_{>,l} \nonumber \\
& = \sum_{k=N_c+1}^\infty \sum_{l=0}^\infty \Pi_{kl}.
\end{align}
The POVM for a single probe amplitude is constructed by the concatenation of these operators.

Now, when it comes to encoding these operators into an algorithm, there is an issue that arises for the WFH configurations with a single counter that does not exist when both counters are used. The elements of the matrices $\Pi_{kl}$ can be encoded with accuracy that can be in principle arbitrarily high by computing the matrices in Eq.~\ref{Pi_kl expanded}. And notice that the matrices $\Pi_{k,>}$, $\Pi_{>,l}$ as well as $\Pi_{>,>}$ can be expressed as finite combinations of the $\Pi_{kl}$ and the identity, and thus the accuracy of their encoding is also constrained by the rounding error of the numerical implementation. In the case of the $\Pi_k$, however, each of the elements of these matrices is formally given as an infinite sum since $\Pi_k = \sum_{l=0}^\infty \Pi_{kl} $. One therefore has to approximate this series by summing up to a certain value of the index $l$. We only present simulations performed with a WFH configuration with both counters, mainly because the simulations are meant as a preparation for the data analysis of the experiment, and the latter, being modeled by a configuration with a BS characterized by a partition size $K=2$ and $S_1>1$, requires the use of both counters to determine the twirled state. Another reason is simply the time constraint.
 
The measurement context in \ref{POVM(Gamma)} is IC if the linear relationship between the probabilities $p_{kl} (\gamma_i)$ and $\tilde \rho_2$ is invertible (see Sec.~\ref{sec POVM and IC} for more details). This is equivalent to the condition that the elements of $\cC(\Gamma)$ span $\cR^s_{2,N}$. One way to check the IC-ness of a POVM computationally is by putting each POVM element in vector form by first stacking horizontally the rows of each of the $N+1$ submatrices in mode 1 (e.g. the $\Pi_{kl,n}$ for the element $\Pi_{kl}$) and then concatenating the rows. We then vertically stack the vectorized POVM elements and check the rank of the resulting matrix. The measurement context is IC if the rank is equal to the number of columns - that is, to $(N+1)(N+2)(2N+3)/6$.

For the WFH configurations with a BS with partition size $K=2$, the discussion above generalizes straightforwardly. A derivation similar to Eq.~\ref{Pi_kl expanded} shows that $\Pi_{kl}$ can be represented as a matrix in $\cR^s_{3,k+l}$, and investigating the dependence of $p_{kl}$ on $\Pi_{kl}$ as in Eq.~\ref{eq:Born_rule_p_kl} and the surrounding discussion reveals that one can restrict $\Pi_{kl}$ to $\cR^s_{3,N}$.  Namely, $\Pi_{kl}$ can be expressed as $\Pi_{kl} = \sum_{i_2=0}^N \left( \sum_{i_1=0}^{N-i_2} \Pi_{kl,i_1,i_2} \otimes \ket{i_1}\bra{i_1} \right) \otimes \ket{i_2}\bra{i_2}$, where each $\Pi_{kl,i_1,i_2}$ is a matrix in mode 1 with maximum photon number $N-i_1-i_2$. Using photon counters that cannot resolve photon number greater than $N_c$ results in a similar set of operators as for the WFH configuration with a BS characterized by a trivial partition. Namely, the $\Pi_{kl}$ for $k \leq N_c$ and $l \leq N_c$ are available, while the operators corresponding to the measurement outcomes $(k \leq N_c, l>N_c)$, $( k > N_c, l \leq N_c)$ and $ (k >N_c, l >N_c) $ are again given by the expressions for the $\Pi_{k,>}$, the $\Pi_{>,l}$ and $\Pi_{>,>}$, respectively. Naturally, the $\Pi_{kl}$ appearing in these expressions (e.g. the last line of Eq.~\ref{eq Pi_k,>} for $\Pi_{k,>}$) are now the operators produced by the WFH configuration with a BS with partition size $K=2$. Since, for a given probe amplitude $\gamma$ and $N$, the operators for the two different classes of WFH configurations are in one-to-one correspondence, the POVMs are constructed in an identical fashion as for the WFH configuration with a BS characterized by a trivial partition. Similarly, the IC-ness of the sets of POVMs is checked numerically by the same procedure, with the difference being that now each POVM element belongs to $\cR^s_{3,N}$ and therefore its vectorization is more involved. Also, now the rank of the matrix composed of the row-wise concatenated vectorizations of the elements of the POVMs needs to have rank $(N+1)(N+2)^2(N+3)/12 $, which is the number of non-zero elements in $\tilde \rho_3$.

\subsection{Treatment With Non-Ideal POVMs}\label{subsec POVMs_nonideal}
So far we have discussed constructing POVMs with elements that are derived from the projectors onto total-photon number states at the output(s) of the BS of the WFH configuration. In practice, there are always losses accompanying the detection of photon-number. A lossy photon counter, with the property that all modes experience the same loss, is well described by the composition of the ideal photon counter with a BS in front that has a transmission coefficient of $1-\nu$ \cite[Chap.~4]{leonhardt:qc1997a}. Here $\nu$ is called the efficiency of the counter. For a single lossy counter the projector $\hat{D'}_k$ corresponding to the detection of $k$ photons is given by:
\begin{align}
\hat{D'}_k = \sum_{m=k}^\infty \binom{m}{k} \nu^{m} (1-\nu)^{m-k} \hat{D}_m.
\end{align}
Then, when two such photon counters are used in a WFH configuration, the corresponding operator on the input modes is calculated similarly as for the ideal counters. Further, we can relate these operators to the $\Pi_{kl}$ of the corresponding WFH configuration with ideal counters. More specifically, the operator corresponding to the detection of $k$ and $l$ photons by counters $1$ and $2$, respectively, is given by:
\begin{align}\label{eq Pi_kl lossy}
\Pi'_{kl} & = \tr_\textsl{p} ((I \otimes \sigma) U^\dagger( \hat{D'}_{1,k} \otimes \hat{D'}_{2,l}) U ) \nonumber  \\
& = \tr_\textsl{p} ((I \otimes \sigma) U^\dagger(  \sum_{m=k}^\infty  \binom{m}{k} \nu_1^{m} (1-\nu_1)^{m-k} \hat{D}_m  \otimes \sum_{n=l}^\infty  \binom{n}{l} \nu_2^{n} (1-\nu_2)^{n-l} \hat{D}_n )U ) \nonumber  \\
& = \sum_{m=k}^\infty  \sum_{n=l}^\infty  \binom{m}{k}  \binom{n}{l}\nu_1^{m} (1-\nu_1)^{m-k} \nu_2^{n} (1-\nu_2)^{n-l}\tr_\textsl{p} ((I \otimes \sigma) U^\dagger(\hat{D}_m\otimes \hat{D}_n )U ) \nonumber  \\
& = \sum_{m=k}^\infty  \sum_{n=l}^\infty \binom{m}{k}  \binom{n}{l} \nu_1^{m} (1-\nu_1)^{m-k} \nu_2^{n} (1-\nu_2)^{n-l} \Pi_{mn}.
\end{align}
Here we used the first subscript in $\hat{D'}_{1,k}$ and in $\hat{D'}_{2,l}$ to indicate the photon counter, and we denoted the efficiencies of counters $1$ and $2$ by $\nu_1$ and $\nu_2$, respectively. When the WFH configuration utilizes only one counter (We can assume counter 1 is used without loss of generality), the operator on input modes corresponding to measuring $k$ photons can be similarly related to $\Pi_k$ as follows:
\begin{align}\label{eq Pi_k lossy}
\Pi'_k &=  \tr_\textsl{p} ((I \otimes \sigma) U^\dagger( \hat{D'}_k \otimes I) U ) \nonumber  \\
& = \tr_\textsl{p} ((I \otimes \sigma) U^\dagger(  \sum_{m=k}^\infty  \binom{m}{k}  \nu_1^{m} (1-\nu_1)^{m-k} \hat{D}_m \otimes I  )U ) \nonumber  \\
& = \sum_{m=k}^\infty  \binom{m}{k}  \nu_1^{m} (1-\nu_1)^{m-k} \tr_\textsl{p} ((I \otimes \sigma) U^\dagger(\hat{D}_m \otimes I )U ) \nonumber  \\
& = \sum_{m=k}^\infty  \binom{m}{k}  \nu_1^{m} (1-\nu_1)^{m-k} \Pi_m.
\end{align}
We denote the probabilities corresponding to these operators by $p'_{kl}$ and $p'_{k}$. For example, for a WFH configuration with a BS characterized by a trivial partition, these are computed as $p'_{kl} = \tr (\tilde \rho_2 \Pi'_{kl} )$ and $p'_{k} \tr (\tilde \rho_2 \Pi'_{k} )$.

Now, notice that $\Pi'_{kl}$ and $\Pi'_{k}$ are expressed as infinite sums of the $\Pi_{kl}$ and $\Pi_{k}$, respectively. Therefore, they can also be expressed as matrices in $\cR^s_{2,N}$ or $\cR^s_{3,N}$, depending on the partition of the BS. In practice, we cannot compute infinite sums and need to approximate Eqs.~\ref{eq Pi_kl lossy} and~\ref{eq Pi_k lossy} by summing only a finite number of terms in them. In our simulations we use probe amplitudes that have an average photon number at most equal to $3$. Further, we also restrict to $N \leq 10$ maximum number of photons in the input state. This implies that the probabilities $p_{kl}$ ($p_k$) corresponding to the operators $\Pi_{kl}$ ($\Pi_{k}$) should be negligible when, for example, $k+l>25$ ($k>25$). This in turn means that approximating Eqs.~\ref{eq Pi_kl lossy} and~\ref{eq Pi_k lossy} by summing the finite sums
\begin{align}
\Pi'_{kl} & =\sum_{m=k}^{25}  \sum_{n=l}^{25-m} \binom{m}{k}  \binom{n}{l} \nu_1^{m} (1-\nu_1)^{m-k} \nu_2^{n} (1-\nu_2)^{n-l} \Pi_{mn}
\end{align}
and
\begin{align}
\Pi'_k &= \sum_{m=k}^{25}  \binom{m}{k}  \nu_1^{m} (1-\nu_1)^{m-k} \Pi_m,
\end{align}
respectively, will have negligible impact on the probabilities $p'_{kl}$ and $p'_k$. As discussed in the next subsection, the similarity of the approximated probabilities to the true probabilities should not have a significant impact on the estimated state. Notice that in the above approximations we implicitly assumed that $\Pi'_{kl} = 0$ ($\Pi'_{k} =0$) when $k+l>25$ ($k>25$).

Finally, we again assume that the counters cannot distinguish the number of photons greater than some number $N_c$. Therefore, for a given measurement setting with probe amplitude $\gamma$, the corresponding POVM is constructed similarly as for the ideal photon counters. For example, when both photon counters are used,
\begin{align}\label{POVM(gamma_i)}
\textrm{POVM}(\gamma) =  \{  \Pi'_{kl}(\gamma) \}_{k=0,l=0}^{N_c,N_c}   \cup  \{ \Pi'_{k,>}  (\gamma) \}_{k=0}^{N_c} \cup  \{ \Pi'_{>,l}  (\gamma) \}_{l=0}^{N_c} \cup \Pi'_{>,>},
\end{align}
where the operators $\Pi'_{>,l}$, $\Pi'_{k,>}$ and $\Pi'_{>,>}$ are given by the same expressions as the $\Pi'_{>,l}$, $\Pi'_{k,>}$ and $\Pi'_{>,>}$ in the last subsection, respectively, but with the $\Pi'_{kl}$ substituted in the places of the $\Pi_{kl}$. We again use multiple measurement settings to define a measurement context, which is composed of the set of POVMs of each measurement setting. We use the same notation as in Eq.~\ref{POVM(Gamma)} to refer to such a set of POVMs for the set of probe amplitudes $\Gamma$. The IC-ness of $\cC(\Gamma) $ is numerically checked by the same procedure as for the ideal POVMs.

\subsection{Maximum Likelihood Estimation}\label{subsec max_like}
We take a general approach in the discussion of this subsection. We assume there is a true twirled state $\rho_{true}$ which can be prepared and measured by the WFH configuration any number of times. The measurement configuration is either equipped with a BS characterized by a trivial partition or with a BS characterized by a partition size $K=2$ and $S_1>1$, The state space is either $\cR_{2,N}$ or $\cR_{3,N}$ depending on the partition of the BS of the measurement configuration and it is implicit that the elements of the POVMs in the measurement context belong to $\cR^s_{2,N}$ or to $\cR^s_{3,N}$, respectively. We assume that the measurement configuration has been prepared in $N_p$ different measurement settings corresponding to the probe amplitudes in $\Gamma = \{\gamma_i\}_{i=1}^{N_p}$. We assume we have a sequence of $M = \sum_{i=1}^{N_p} M_i$ measurement outcomes, where $M_i$ is the number of measurement outcomes using the measurement setting with probe amplitude $\gamma_i$. We assume that all $M_i \geq 1$. For brevity of description, we denote the POVM elements in the measurement context by $E_i$ in some arbitrary order, so that $\cC(\Gamma) = \{E_i\}_i$, and the probabilities of the corresponding measurement outcomes by $p(i)$, respectively. The POVMs are constructed by using either ideal counters or lossy counters in the WFH configuration. Then, we count the number of measurement outcomes in the sequence corresponding to $E_i$, and denote it by $m(i)$. We denote the corresponding frequencies by $f(i)= m(i)/M$.

To obtain an estimate of the true twirled state from the sequence of samples, we use maximum likelihood estimation (MLE). An introduction to MLE can be found in textbooks such as \cite[Chap.~7]{casella2002stat}. MLE is a procedure for maximizing the likelihood function. The latter as a function depends on both the parameter space (which in our case is the set of twirled states) and the set of possible samples for a given sample size. By a sample we mean a sequence of outcomes. More specifically, the likelihood function for a particular value of the parameter is a probability distribution over all possible sequences of outcomes of the given sample size. In MLE, one does not consider the likelihood function a probability distribution, as one takes the sample as fixed. Thus, the likelihood function, given a particular sample, is a function on parameter space such that its value for a particular parameter is equal to the probability of that parameter producing this sample. Let us illustrate this with an example relevant to our problem. For a given measurement setting associated with the probe amplitude $\gamma_i$ consider $\textrm{POVM}(\gamma_i) $ as a map from the parameter space (that is, the space of twirled states) to probability distributions. The map for each element of $\textrm{POVM}(\gamma_i) $ is given by the trace with the twirled state. Then, given the sequence of $M_i$ outcomes that are sampled from the probability distribution resulting from $\rho_{true}$, the likelihood function is
\begin{align}\label{eq like_single_POVM}
\prod_{j=1}^{M_i} \tr (\rho E_{o_j})
\end{align}
where $o_j$ is the $j$'th outcome in the sequence, and $\rho$ is a twirled state. Thus, the likelihood can be seen to be a function of $\rho$ and has the interpretation as the probability of observing the sequence of outcomes for each twirled state $\rho$. The task of MLE is to find the $rho$ (or the set of $\rho$s) where the likelihood is maximal. This is equivalent to finding the distribution that maximizes \ref{eq like_single_POVM}. Since we found that a single POVM is not IC in general, the likelihood function will attain its maximum for a range of density matrices.

In our problem there are $N_p$ measurement settings and thus $N_p$ different likelihoods. Therefore, our task is to maximize them all simultaneously. This is done by maximizing the function that is the product of these likelihoods, 
\begin{align}\label{eq like_mult_POVM}
\prod_{i=1}^{N_p} \prod_{j=1}^{M_i} \tr (\rho E_{o_{ij}}),
\end{align}
where $o_{ij}$ is the $j$'th measurement outcome in the sequence obtained in the measurement setting with $\gamma_i$. To see that this is the natural extension of the MLE described in the previous paragraph, observe that we assume the sample size for each measurement setting is fixed. In that case \ref{eq like_mult_POVM} gives the probability of observing the sequence of outcomes $o_{ij}$ for each $\rho$. This is because \ref{eq like_mult_POVM} for a given $\rho$ is a product of $N_p$ probability distributions, where the $i$'th distribution is over the sequences of measurement outcomes of length $M_i$ for the measurement setting with $\gamma_i$, and thus it is itself a probability distribution. Thus, the twirled state that maximizes \ref{eq like_mult_POVM} is the most ``likely" one to have given rise to the observed sequence for this set of measurement settings and sample sizes for each setting.

In practice it is much easier to work with the natural logarithm of \ref{eq like_mult_POVM}, which we call the ``log-likelihood". The state that maximizes \ref{eq like_mult_POVM} also maximizes its natural logarithm, since the latter is a monotonic function. Using the notation introduced in the beginning of this section the log-likelihood of our problem can be represented as
\begin{align}\label{eq log_like}
\cL(\rho) &= \log \left( \prod_{i=1}^{N_p} \prod_{j=1}^{M_i} \tr (\rho E_{o_{ij}}) \right) \nonumber \\
&= \log\left( \prod_{i} \tr(\rho E_{i})^{m(i)} \right) \nonumber \\
& =   \sum_i m(i) \log(\tr (E_i \rho )).
\end{align}
There are several things to note about $\cL$. First, it is a function on a compact convex space of density matrices. Second, if $\rho_{true}$ lies in the interior of $\cR_{2,N}$ or $\cR_{3,N}$, $\cL(\rho)$ is infinitely differentiable as a function of $\rho$ in a neighborhood of $\rho_{true}$ since all probabilities $\tr (E_i \rho )$ that appear in $cL$ for a given sequence of outcomes cannot be zero. Third, since all $M_i \geq 1$, $\cL(\rho)$ as a distribution over the set of samples is different for different $\rho$. The latter is due to the fact that $\cC(\Gamma)$ is IC. To see this, consider the marginal of $\cL(\rho)$ for a given $\rho$ over the sequence of outcomes composed of the first outcome of each measurement setting. This marginal is the Cartesian product of the $N_p$ probability distributions resulting from the set of $\textrm{POVM}(\gamma_i) $. Since these distributions determine $\rho$, their Cartesian product has to be different for different $\rho$. And if any marginal of $\cL(\rho)$ is unique for each $\rho$, that implies that $\cL(\rho)$ is also unique for each $\rho$. Finally, for each measurement setting the elements of the corresponding subsequence of outcomes are independent and identically distributed (i.i.d.). Alternatively, one could say that the elements of the whole sequence are independent but non-identically distributed (i.n.i.d.), and specify how exactly they are non-identical in this instance. These properties of $\cL(\rho)$ guarantee that the corresponding MLE procedure is a ``consistent" estimator \cite[Chap.~18]{stuart1999advanced}. Consistency in our case means that as the $M_i$ tend to infinity (that is, in the ``asymptotic data limit") the state that maximizes $\cL(\rho)$ converges to $\rho_{true}$.

Now, for a finite sample size there exists a (not-necessarily unique) state $\rho_{max}$ which maximizes the corresponding $\cL$. In the asymptotic data limit, $\rho_{max}$ converges to $\rho_{true}$. One generally assumes that the $M_i$ are large enough such that $\rho_{max}$ is unique with high probability, and is also very close to $\rho_{true}$. The task of a MLE algorithm is to obtain a state $\hat \rho$ which is very close to $\rho_{max}$ and therefore is a good estimate of $\rho_{true}$. The algorithm we utilize uses an iterative scheme to approach $\rho_{max}$. See \cite{hradil:qc2004a} and \cite{rehacek2007diluted} for derivations and proofs of the claims that follow. The discussion in the rest of the paragraph is valid for any finite-dimensional quantum state reconstruction using MLE, but we present the results in \cite{hradil:qc2004a} and \cite{rehacek2007diluted} as they apply to our problem. It can be shown that $\rho_{max}$ satisfies the equality $\hat{R}(\rho_{max}) \rho_{max} \hat{R}(\rho_{max}) =\rho_{max}$, where 
\begin{align}\label{eq R_expression}
\hat{R}(\rho) = \frac{1}{M} \sum_i \frac{m(i)}{\tr (\rho E_i)} E_i
\end{align}
is a state dependent positive semi-definite matrix. This result can be obtained from the application of Jensen's inequality to $\cL(\rho)$ and considering the properties of the state where the inequality becomes an equality \cite[Sec.~3.3]{hradil:qc2004a}. The form of Eq.~\ref{eq R_expression} inspires the following iterative scheme. Start with a random initial state $\hat{\rho}_0$ and iterate according to:
\begin{align}
\hat{\rho}_{k+1} = \hat{R}(\hat{\rho}_{k}) \hat{\rho}_{k} \hat{R}(\hat{\rho}_{k}),
\end{align}
where $\hat{\rho}_k$ is the estimate obtained at the $k$'th step of the iteration. This iterative procedure is referred to as the ``$R \rho R$ algorithm". Unfortunately, the $R \rho R$ algorithm does not guarantee the convergence to $\rho_{max}$, or even that the log-likelihood will not decrease during any iteration step. However, we find in our simulations that in general $R \rho R$ brings the estimate quite close to $\rho_{max}$ before we observe either a decrease of $\cL$ in an iteration or a stalling of its value. A more robust algorithm is achieved by instead iterating as:
\begin{align}\label{eq diluted_RrhoR}
\hat{\rho}_{k+1} = \frac{I + \epsilon \hat{R}(\hat{\rho}_{k})}{1+\epsilon} \hat{\rho}_{k}  \frac{I + \epsilon \hat{R}(\hat{\rho}_{k})}{1+\epsilon},
\end{align}
where $I$ is the identity and $\epsilon$ is a small constant. It can be shown (see \cite{rehacek2007diluted}) that for a small enough $\epsilon$ the iteration is guaranteed not to decrease $\cL$ during a given step. This algorithm is referred to as the ``diluted $R \rho R$" algorithm. Notice that the $R \rho R$ algorithm can be thought of as the limit of the diluted $R \rho R$ algorithm when $\epsilon \rightarrow \infty$.

An important part of any iterative MLE procedure is specifying the termination conditions. Two considerations are relevant here. One reason to terminate the iterations is that the iterations are stuck at the same value of $\cL$ for too long and thus do not improve the estimate or the improvements are marginally small. This consideration can be encoded into the procedure by choosing a small value $\Delta \cL$ beforehand and terminating the iterations when the increase in $\cL$ becomes less than $\Delta \cL$.  The second reason is that we have a good quantitative measure which allows us to conclude that the estimate is very close to $\rho_{max}$, and thus continuing the iterations does not make sense from the point of view of cost-benefit analysis. It turns out that a good quantity for judging the closeness of the current estimate to $\rho_{max}$ can be calculated using the results of \cite{glancy2012gradient}. At the iteration step $k$ it can be shown that $\cL(\rho_{max})- \cL(\hat{\rho}_k)  $ is upper bounded by $\texttt{r}_k = \max_{\rho } \tr (\rho \hat{R}(\hat{\rho}_k) )-1$. The latter is the derivative of $\cL$ at $\hat{\rho}_k$ along the straight path connecting $\hat{\rho}_k$ with $\rho_{max}$ with the domain of the path parameter being $[0,1]$. More specifically, if we parametrize $\rho_{\lambda} = \hat{\rho}_k (1-\lambda) + \rho_{max} \lambda$, then $\texttt{r}_k =\at{ \frac{d \cL}{d\lambda}}{\lambda=0}$. $\texttt{r}_k$ can be computed as $\texttt{r}_k = \max(\textrm{eig}( \hat{R}(\hat{\rho}_k) )-1$. One can thus calculate $\texttt{r}_k$ at each iteration step and terminate the procedure when $\texttt{r}_k$ is smaller than some predefined value $\texttt{r}$. We refer to $\texttt{r}$ and to the condition $\texttt{r}_k < \texttt{r}$ as the ``stopping criterion".

We use combinations of the two algorithms described above to construct our MLE procedures for different simulations. We start by choosing the stopping criterion and the value of $\Delta \cL$. The initial estimate $\hat{\rho}_0$ is always set to the maximally mixed state. If at any point of the procedure the stopping criterion is satisfied, the procedure terminates and the values of the estimate at the step of termination is taken as our estimate of $\rho_{true}$. The procedures always start with applying the $R\rho R$ algorithm until the increase in $\cL$ during one iteration is less than $\Delta \cL$. If this happens, we switch to a diluted $R\rho R$ algorithm with a high value of $\epsilon$ (usually $\epsilon = 10^{30}$ is used). The iterations with this value of $\epsilon$ continue again until the increase in $\cL$ is less than $\Delta \cL$ during a single iteration, at which point we choose a smaller value of $\epsilon$. Assuming the stopping criterion is not met, this process goes on until the value of epsilon becomes negligibly small, at which point we conclude that the procedure has stalled and we terminate the algorithm.

\section{Analysis for a Balanced BS Characterized by a Trivial Partition}\label{sim BS_trivpart}

In this section we present the results of some reconstructions we performed using simulated data with a WFH configuration with a BS characterized by a trivial partition and which has equal transmission and reflection coefficients. For each reconstruction procedure the parameters $N$, $N_c$ and the true twirled state $\rho_{true}$ are chosen first. Then, the probe amplitudes $\Gamma = \{\gamma_i\}_{i=1}^{N_p}$ are chosen and the corresponding $\cC(\Gamma) $ is constructed. Only IC measurement contexts are used. Then, the probabilities $p(i)$ for all $E_i \in \cC(\Gamma) $ are computed, and a pseudo-random generator is used to generate $M_i$ samples from the probability distribution corresponding to $\textrm{POVM}(\gamma_i)$, for each $i$. In our simulations we always set the $M_i$ equal to each other - that is, each $M_i$ is equal to $M/N_p$, so $M$ is chosen to be divisible by $N_p$.

We start by describing the kinds of $\rho_{true}$ that we used in our simulations in Subsec.~\ref{subsec input states}. Coherent states are of particular interest, as they are in a sense the simplest states to use to check the performance of our reconstruction procedure. This is because the states at the outputs of the BS are also coherent states with amplitudes which are linearly related to the amplitudes of the coherent states in the input paths, and thus the photon number statistics of both counters are Poissonian and depend on only three parameters. So, if the reconstruction is poor for a coherent input state, the same procedure is likely going to be poor for more sophisticated kinds of input states that have a higher-dimensional parameter space with a greater variety of distributions. The simulations with coherent states truncated in maximum photon number and using ideal counters are described in Subsec.~\ref{subsec coh state analysis}, while the last subsection describes simulations with other kinds of input states, as well as when the counters are assumed to have losses.

 The simulations are not exhaustive even for the specific type of measurement configuration considered here, and were in part meant to verify the correctness of the programming. Another purpose of the simulations was to set reasonable expectations for how the reconstruction process from the experimental data (described in the next chapter) might perform with the given values of $M_i$, $N_p$, $N_c$, the loss coefficients of the counters, as well as for the chosen regulating parameters $\texttt{r}$ and $\Delta \cL$.

\subsection{Upper Bounds on the Photon Number of Input States}\label{subsec input states}

Here, we describe some of the states we use in our simulations. First, however, we should consider the fact that the states that are of interest in practice often do not have a bound on their maximum photon number. Idealized examples are coherent states produced by a laser source or squeezed states produced using non-linear media. When simulating with these states, we have to truncate them at some maximum photon number $N$. More specifically, the parameters of the density matrices of these states in Fock basis corresponding to greater than $N$ photons are set to zero and the rest is normalized to have unit trace. We want the truncated states to be good approximations of the original states, and for this reason we choose $N$ and the parameters of these states such that the fidelity between the truncated and original states is high.

 Let us first look at coherent states in the mode matching the LO. Since these states occupy mode $1$ only, the twirling map leaves them invariant. There is an implicit tensor product with the vacuum in mode $2$ in the following expressions. For the coherent state $\ket{\alpha}$, its truncated version is given by
\begin{align}
\ket{\alpha(N)} =\sqrt{ \frac{1}{ \sum_{i=0}^N \abs{\alpha}^{2i}/i!  }}  \sum_{n=0}^N\frac{ \alpha^{n}}{\sqrt{n!}} \ket{n}.
\end{align}
The fidelity between $\ket{\alpha}$ and $\ket{\alpha_N}$ is then
\begin{align}\label{eq fid_coh_truncoh}
\cF(\ket{\alpha},\ket{\alpha(N)}) &=  \abs{ \bra{\alpha}\ket{\alpha(N)}  }^2 \nonumber \\
&=  \abs{ e^{-\abs{\alpha}^2/2}\sum_{m=0}^\infty \frac{(\alpha^*)^m}{\sqrt{m!}} \bra{m} \sqrt{ \frac{1}{ \sum_{i=0}^N \abs{\alpha}^{2i}/i!  }}  \sum_{n=0}^N\frac{ \alpha^{n}}{\sqrt{n!}} \ket{n} }^2 \nonumber \\
& = e^{-\abs{\alpha}^2} \abs{ \sqrt{ \frac{1}{ \sum_{i=0}^N \abs{\alpha}^{2i}/i!  }}  \sum_{n=0}^N \frac{\abs{\alpha}^{2n}} {n!} }^2 \nonumber \\
& = e^{-\abs{\alpha}^2} \sum_{n=0}^N \frac{\abs{\alpha}^{2n}}{n!},
\end{align}
and is seen to depend on only the magnitude of $\ket{\alpha}$ and $N$. In our simulations we consider $N \leq 5$ since we find that the computation time for reconstruction becomes impractical for our purposes for larger $N$. When $\abs{\alpha} = 0.9$ the fidelity $\cF(\ket{\alpha},\ket{\alpha(N)}) $ is approximately equal to $0.9998$, so we find that $\ket{\alpha(5)}$ is a good approximation for $\ket{\alpha}$ when $\abs{\alpha} \leq 0.9$. This claim can be wrong if the POVM is heavily biased to sample from the part of $\ket{\alpha}$ that is orthogonal to $\ket{\alpha(5)}$, but this should not be the case in our simulations as we use probe states with low magnitude ($\abs{\gamma} \leq 2$) in conjunction with counters that can distinguish up to $N_c =9$ photons. This reasoning can be applied to the states described below as well. 

%The two-mode twirled state corresponding to the two-mode coherent state $\ket{\alpha_1}\otimes \ket{\alpha_2}$ is given by
%\begin{align}
%\sum_{n=0}^\infty e^{-\abs{\alpha_2}^2} \frac{\abs{\alpha_2}^{2n}}{n!} \ket{\alpha_1}\bra{\alpha_1} \otimes  \ket{n}\bra{n}.
%\end{align}
%Therefore, the truncated version corresponds to
%\begin{align}
%\cN \sum_{n=0}^N  \frac{\abs{\alpha_2}^{2n}}{n!} \left( \sum_{i=0}^{N-n} \sum_{j=0}^{N-n}  \frac{\alpha_1^i (\alpha_1^*)^j}{  \sqrt{i!j!}}  \ket{i}\bra{j} \right) \otimes  \ket{n}\bra{n},
%\end{align}
%where the normalization $\cN$ is equal to $ 1/\left( \sum_{i=0}^N \frac{ ( \abs{\alpha_1}^2+\abs{\alpha_2}^2    )^{i}}{i!} \right)$. Note that truncating the original state and then twirling it results in the same state as twirling the original state and then truncating it. Expressing the truncated and original states in block diagonal form The fidelity between them is computed to be
%\begin{align}
%
%\end{align}

Another class of states we consider are two-mode squeezed vacuum (TMSV) states. These states have the form
\begin{align}
\ket{\textrm{TMSV}} = \frac{1}{\cosh (r)} \sum_{n=0}^\infty (-e^{i\phi} \tanh(r))^{n} \ket{nn}.
\end{align}
The truncated versions are given by
\begin{align}
\ket{\textrm{TMSV}(N)}  = \sqrt{\frac{1}{\sum_{i=0}^N  \tanh^{2i}(r)  }} \sum_{n=0}^N (-e^{i\phi} \tanh(r))^{n} \ket{nn}.
\end{align}
The state fidelity between $\ket{\textrm{TMSV}(N)} $ and $\ket{\textrm{TMSV}} $ can be calculated as for the coherent states in Eq.~\ref{eq fid_coh_truncoh} and equals
\begin{align}
\cF( \ket{\textrm{TMSV}(N)} , \ket{\textrm{TMSV}} ) = \frac{\sum_{i=0}^N  \tanh^{2i}(r)}{\cosh^2 (r)}.
\end{align}
It can be seen that the fidelity only depends on $r$, which is usually called the squeezing parameter. When $r \leq 0.5$ the fidelity is $\geq 0.9999$ for the maximum photon number $N=10$. The corresponding truncated twirled state is calculated to be
\begin{align}
\frac{1}{\sum_{i=0}^N  \tanh^{2i}(r)  } \sum_{n=0}^{ \lfloor N/2 \rfloor } \tanh^{2n}(r) \ket{n}\bra{n} \otimes \ket{n}\bra{n}.
\end{align}

We also consider ``cat" states of the form $\ket{\textrm{cat}} =\sqrt{ \frac{1}{2(1-e^{-4 \abs{\alpha}^2 })} }(\ket{\alpha}\ket{-\alpha} - \ket{-\alpha}\ket{\alpha} ) $, which has the corresponding twirled state
\begin{align}\label{eq twirled_cat}
\frac{e^{-\abs{\alpha}^2} }{2(1-e^{-4 \abs{\alpha}^2 })} \sum_{n=0}^\infty \left(   \ket{\alpha}\bra{\alpha} + \ket{-\alpha} \bra{-\alpha} + (-1)^{n+1} \ket{-\alpha}\bra{\alpha} + (-1)^{n+1} \ket{\alpha}\bra{-\alpha}        \right) \otimes \ket{n}\bra{n}.
\end{align}
The fidelity between the truncated state $\ket{\textrm{cat}(N)}$ and $\ket{\textrm{cat}}$ is again related to the normalization constant after truncation. Namely, $\cF = \abs{  \bra{ \textrm{cat}}\ket{\textrm{cat}(N)}  }^2 = \abs{ \frac{1}{\cN} \bra{  \textrm{cat}(N)}\ket{\textrm{cat}(N)}  }^2 = \frac{1}{\cN^2}$, where the normalization constant is computed to be
\begin{align}
\cN^2 = \frac{  2(e^{2\abs{\alpha}^2} - e^{-2\abs{\alpha}^2} )} {\sum_{n=0}^N\sum_{m=0}^{N-n} \frac{\abs{\alpha}^{n+m}}{n!m!} ( (-1)^n-(-1)^m)},
\end{align}
and can be seen to depend on only the absolute value of $\alpha$.
When truncated at $N=5$, the state fidelity between $\ket{\textrm{cat}}$ and $\ket{\textrm{cat}(N)}$ is $\geq 0.9998$ provided $\abs{\alpha} \leq 0.7$.

\subsection{ Analysis With Single Mode Coherent States  }\label{subsec coh state analysis}

We did the bulk of our simulations with truncated ($N=5$) single-mode coherent states in mode $1$, to test the performance and the accuracy of the reconstruction procedure under various conditions. The BS is balanced, and both counters are utilized. Some of the parameters that are varied include the set of probe amplitudes $\Gamma$, the sample size $M$, and the various conditions that are used to direct the iterations of the reconstruction algorithm. The latter are described as follows. As discussed earlier, the parameter $\texttt{r}_k = \max(\textrm{eig}( \hat{R}(\hat{\rho}_k) )-1$ at the $k$'th iteration serves as an upper bound on how far $\cL$ is from $\cL(\rho_{max})$ at the current step of the iterations. Therefore, it makes sense to choose, at the beginning of the algorithm, a small enough value $\texttt{r}$ such that the algorithm terminates when $\texttt{r}_k \leq \texttt{r}$. However, it is possible that the iterations stagnate such that even after an arbitrarily long time, $\texttt{r}_k$ does not fall below $\texttt{r}$. To remedy this, we choose a small value $\Delta \cL$ and continue the iterations at the current value of $\epsilon$ after $k$'th step when $\cL(\hat{\rho}_{k}) -\cL(\hat{\rho}_{k-1}) \geq \Delta \cL$. So, if the log-likelihood does not increase enough, we conclude that the iterations have stagnated at the current value of $\epsilon$, and we proceed by halving the value of $\epsilon$. We terminate the algorithm when either $\epsilon \leq 10^{-30}$, at which point the iterations essentially do not affect the state according to Eq.~\ref{eq diluted_RrhoR}, or when $\texttt{r}_k \leq \texttt{r}$. We consider different values of $\Delta \cL$ and $\texttt{r}$ for different input states and sample sizes and observe the performance of the algorithm.

We start by using the truncated coherent state with magnitude $\abs{\alpha} = 0.2$ and phase $\phi = \pi/4$, which was chosen for no particular reason, and investigate how sample size affects the reconstruction accuracy. We use the following randomly chosen set of probe amplitudes that produce an IC set of POVMs: $\Gamma = \{ 0.9, 1.1e^{i\pi/10},1.3 e^{i \pi/5},1.5e^{i3\pi/10},1.7e^{i2\pi/5}   \}$. We first use $\Delta \cL=10^{-8}$ and $\texttt{r} = 1/M$. This implies that the iterations terminate when 
\begin{align}\label{eq r_singl-mode-coh}
  \max(\textrm{eig}(\hat{R}( \hat{\rho}_k )   ))-1 = \max(\textrm{eig}(  \frac{1}{M} \sum_i \frac{m(i)}{\tr (\hat{\rho}_k E_i)} E_i))-1 \leq 1/M, 
\end{align}
which can be rewritten as $ \max(\textrm{eig}(\sum_i \frac{m(i)}{\tr (\hat{\rho}_k E_i)} E_i)) \leq M+1$. The left-hand side reaches its minimum value of $M$ when $m(i) = \tr (\hat{\rho}_k E_i)$, but this value might not be reached even when $\hat{\rho}_k$ is optimal (that is, $\hat{\rho}_k=\rho_{max}$). Indeed, even the stopping criterion might be unreachable with $\rho_{max}$. Therefore, there is a degree of arbitrariness in choosing the stopping criterion, in part because we cannot know $\rho_{max}$ beforehand. Nevertheless, we find that Eq.~\ref{eq r_singl-mode-coh} is satisfied for a large range of values of $M$. The accuracy of the output of the algorithm is measured by comparing it with $\rho_{true}$ using the state fidelity measure. The results are expressed in Table~\ref{table 1}. The number of trials is small for larger values of $M$ due to the fact that the computation of $\cL$, $\hat{R}$, and other various calculations take significantly longer in each iteration step. This is primarily due to the increase in the number of non-zero $m(i)$ when the sample size is increased. The fewer number of trials for larger $M$ should not be problematic, however, since we expect the variance of $\hat{\rho}$ to decrease proportionally to the increase in the size of $M$. The primary observation is that $\cF(\hat{\rho},\rho_{true}  )  $ increases as $M$ increases and then seems to plateau out after $M=10^{6}$.

\begin{table}[h!]
\centering
\renewcommand{\arraystretch}{1.5}
\begin{tabular}{||c | c | c||} 
 \hline
Average fidelity ($\cF(\hat{\rho},\rho_{true}  )  $)   &	  Sample size ($M$)	  &   Number of trials averaged over  \\ [0.5ex] 
 \hline\hline
0.959	& $10^3$	  & 50  \\
\hline
0.9872  & 	$10^4$	& 50 \\
\hline
0.996  &	$10^5$	& 13 \\
\hline
0.9977	  & $10^6$   &	 7 \\
\hline
0.9977  & 	$10^7$  &	2 \\ [1ex] 
 \hline
\end{tabular}
\caption{The fidelity between the estimate and the true state for different sample sizes. The true state is $\rho_{true} = \ket{0.2 e^{i\pi/4}} \bra{0.2 e^{i\pi/4} } \otimes \ket{0}\bra{0}$.  $\Delta \cL = 10^{-8}$ and $\texttt{r} = 1/M$ are used. }
\label{table 1}
\end{table}

Another observation, not listed in the table, is that the iterations terminate differently for different values of $M$. When $M$ is large, the algorithm terminates for a value of $\texttt{r}_k$ that is well above $\texttt{r}$ - that is, the iterations pass over all values of $\epsilon$ until $\epsilon  = 10^{-30}$. We conclude that $\Delta \cL = 10^{-8}$ is too loose of a requirement to terminate the iterations for a give value of $\epsilon$. So, while the fidelity value of $0.9977$ is not low for $M=10^7$ samples, it is plausible that higher values of the fidelity are achievable if the iterations are allowed to continue even when the incremental increase in $\cL$ is less than $10^{-8}$. We found that by decreasing $\Delta \cL$ by several orders of magnitude, the iterations do reach a point where $\texttt{r}_k \leq 1/M$ even when $M=10^{7}$, but this is accompanied by a substantial increase in computation time. For example, for $M=10^7$ we had to decrease $\Delta \cL$ down to $10^{-13}$ to see $\texttt{r}_k$ fall below $10^{-7}$. When $M$ is small, on the other hand, $\texttt{r}_k$ does fall below $\texttt{r}$ during the iterations at some value of $\epsilon$. In fact, for $M \leq 10^4$ this happens for $\epsilon = \infty$ and at a step where $\cL(\hat{\rho}_k) -\cL(\hat{\rho}_{k-1}) > 10^{-5} >> \Delta \cL$. But this does not mean that a higher fidelity cannot be reached, since $\texttt{r} = 1/M$ might be a loose requirement for termination for smaller sample sizes. 

To investigate the impact of the condition for the continuation of iterations at a given value of $\epsilon$ and of the condition for termination on the fidelity of the output with the true state for different sample sizes, we performed four different kinds of simulations. These are distinguished by different choices of $\Delta \cL$ and $\texttt{r}$ as described below:
\begin{enumerate}
\item $\Delta \cL = 10^{-8}$ and $\texttt{r} = 1/M$.
\item $\Delta \cL = 0$ and $\texttt{r} = 1/M$.
\item $\Delta \cL = 10^{-8}$ and $\texttt{r} = 0$.
\item $\Delta \cL = 10^{-12}$ and $\texttt{r} = 0$.
\end{enumerate}
The simulations where $\texttt{r}=0$ (cases $3$ and $4$) terminate when $\epsilon$ reaches $10^{-30}$, while case $2$ terminates when $\texttt{r}_k$ falls below $1/M$. Case $4$ was meant to investigate how high of a fidelity can be reached by this algorithm when it is allowed to run as long as possible, but we found that the algorithm does not terminate even after several days when $\Delta \cL = 10^{-13}$ is chosen. So, we instead chose $\Delta \cL = 10^{-12}$ to compare the performance with case $2$ for larger values of $M$. The results are shown in Table~\ref{table 2} for the input state with magnitude $0.2$ and phase $\pi/10$, which was again chosen for no particular reason.

\begin{table}[!h]
\centering
\renewcommand{\arraystretch}{1.5}
\begin{tabular}{|| c | p{0.15\linewidth} | p{0.3\linewidth} | p{0.3\linewidth}  ||} 
 \hline
$M$	& Number of trials averaged over & 	Fidelity with $\Delta \cL=10^{-8}$ and $\texttt{r} = 1/M$ (case $1$)	 &  Fidelity with $\Delta \cL =0$ and $\texttt{r} = 1/M$
(case $2$)	 \\ [0.5ex] 
 \hline\hline
$10^3$	& 50 & 	0.960	& 0.965	\\
\hline
$10^4$	& 25 & 	0.987	& 0.989\\
\hline
$10^5$	& 13	 & 0.996	& 0.997	\\
\hline
$10^6$ & 	7	& 0.9977	& 0.9990	\\
\hline
$10^7$	& 2 &	0.9977 &	0.9997  \\ [1ex] 
 \hline
\end{tabular}
\caption{The average fidelity between the estimate and the true state for different values of $M$, $\Delta \cL$ and $\texttt{r}$. The true state is $\rho_{true} = \ket{0.2 e^{i\pi/10}} \bra{0.2 e^{i\pi/10} } \otimes \ket{0}\bra{0}$.}\label{table 2}
\end{table}
\addtocounter{table}{-1}
\begin{table}[!h]
\centering
\renewcommand{\arraystretch}{1.5}
\captionsetup{list=no}
\begin{tabular}{|| c | p{0.15\linewidth} | p{0.2\linewidth} | p{0.2\linewidth} | p{0.25\linewidth}  ||} 
 \hline
$M$	& Number of trials averaged over & Fidelity with $\Delta \cL=10^{-8}$ and $\texttt{r} = 0$
(case $3$)	& Fidelity with $\Delta \cL=10^{-12}$ and $\texttt{r} = 0$
(case $4$)	& $\cL(\hat{\rho}_k)  –\cL(\hat{\rho}_{k-1})$ at the end of the iterations for case $2$ \\ [0.5ex] 
 \hline\hline
$10^3$	& 50 & 	0.980	& 0.979	& $5.0 \cdot 10^{-5}$\\
\hline
$10^4$	& 25 & 	 0.993	& 0.994	& $1.8 \cdot 10^{-6}$\\
\hline
$10^5$	& 13	 	& 0.997 & 	0.997 &	$3.2 \cdot 10^{-8}$\\
\hline
$10^6$ & 	7	& 0.9975	& 0.9988	& $8.5\cdot 10^{-10}$\\
\hline
$10^7$	& 2  &	0.9975 &	0.9995	& $5.3 \cdot 10^{-13}$
 \\ [1ex] 
 \hline
\end{tabular}
\caption{The average fidelity between the estimate and the true state for different values of $M$, $\Delta \cL$ and $\texttt{r}$. The true state is $\rho_{true} = \ket{0.2 e^{i\pi/10}} \bra{0.2 e^{i\pi/10} } \otimes \ket{0}\bra{0}$. (continued)}
\end{table}

From our discussion above, we expect that for smaller $M$ cases $1$ and $2$ should produce the same average fidelity since they both terminate when $\texttt{r}_k$ falls below $1/M$. We do observe that the average fidelity is slightly greater in the case $2$, especially for $M=10^3$. This could be due to the large variance in the estimator for such a small number of samples. We also expect to observe the converse for large $M$, and indeed the average fidelity increases for $M=10^6$ and $M=10^7$ as $\cL$ is lowered to zero. The last column shows the observed change in the log-likelihood in the last iteration step before termination for case $2$. It can be seen that for $M=10^3$ the change is significantly larger than $10^{-8}$ and, therefore, it is not surprising that the fidelity increases to $0.980$ when $\texttt{r} $ is set to zero, in which case the iterations terminate when the change in log-likelihood is less than $10^{-8}$ for all values of $\epsilon$. This measurable increase in fidelity when $\texttt{r} $ is set to zero is also observable for $M=10^4$. When $M=10^5$ we observe that there is no change in the average fidelity when $\texttt{r} $ is set to zero, and this is explained by the fact that the change in log-likelihood at the point where $\texttt{r}  $ falls below $1/M$ is of the same order as $\Delta \cL = 10^{-8}$. For these smaller values of $M$ we notice that decreasing $\Delta \cL$ to $10^{-12}$, which is the lowest we can go given the time constraint, does not affect the average fidelity as compared to case $3$. This makes us conclude that continuing the iterations after the change in the log-likelihood reaches below $10^{-8}$ does not increase the performance of the algorithm for $M \leq 10^5$. These results make sense statistically, since an increase in the number of samples is accompanied by a decrease in statistical noise and a decrease in the variance of the maximum likelihood estimate.

When $M\geq 10^6$ we find that on average the best reconstruction occurs in case $2$, but the reconstruction accuracy is only slightly worse in case $4$. This is likely due to the number of trials being small, as the case $4$ is designed to iterate for a long time, and is thus expected to achieve the best reconstruction accuracy on average. We conclude that, for $M\geq 10^6$, using the regulating conditions of case $2$ results in reconstruction accuracy that is indistinguishable from the accuracy that results from letting the iterations run for a long time. The results of the reconstructions in cases $1$ and $3$ are also of similar accuracy, which we ascribe to the fact that the iterations always terminate without satisfying the stopping condition $\texttt{r}_k \leq 1/M$. Another observation is that when allowing the simulations to run until $\texttt{r}_k$ falls below $1/M$ (case $2$), the change in the log-likelihood calculated at the step of termination decreases significantly when $M$ is increased by orders of magnitude. In particular, for $M=10^7$, when the largest average fidelity is obtained, the iterations increase the likelihood only by $5.3*10^{-13}$, which is only half of the $\Delta \cL = 10^{-12}$ used in case $4$. The two simulations in case $2$ for this value of $M$ took more than an hour, and since the reconstruction accuracy is not significantly different than in case $4$, we conclude that for near-optimal performance, given the time constraint, we can use $\Delta \cL = 10^{-12}$ together with $\texttt{r}=1/M$ in our further simulations with $M\geq 10^6$.

\begin{figure}[!htb]
    \centering
\includegraphics[width=110mm]{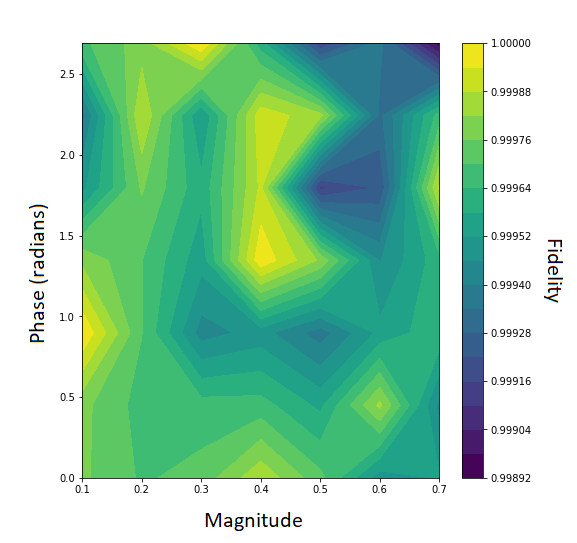}
    \caption{Fidelity for different single-mode coherent input states. The horizontal (vertical) axis denotes the magnitude (phase) of the input state. The set of probe amplitudes composing the measurement context is $ \Gamma = \{ 0.3, 0.6e^{i \pi/6}, 0.9 e^{i 2\pi/6}, 1.2 e^{3\pi/6}, $ $1.5 e^{i 4\pi/6}, 1.8 e^{ i 5\pi/6}   \} $. $M=10^7$ samples were drawn for each estimate.    }
    \label{fig: fig_coh_cG_1}
\end{figure}

\begin{figure}[!htb]
    \centering
\includegraphics[width=110mm]{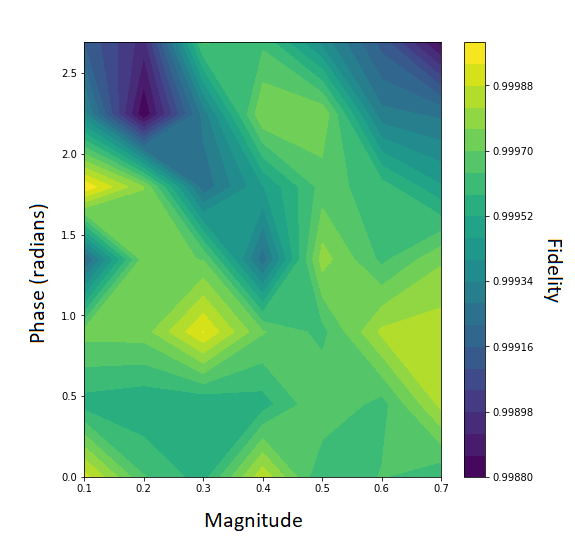}
    \caption{Fidelity for different single-mode coherent input states. The horizontal (vertical) axis denotes the magnitude (phase) of the input state. The set of probe amplitudes composing the measurement context is $ \Gamma = \{   3 e^{i 0.79\pi}, 1.9 e^{i 0.5\pi}, 1.4 e^{i 0.6\pi},$ $1.7 e^{i 0.55\pi},1.5 e^{i 0.31\pi},2.9 e^{i 0.63\pi}       \}  \). $M=10^7$ samples were drawn for each estimate.    }
    \label{fig: fig_coh_cG_2}
\end{figure}

\begin{figure}[!htb]
    \centering
\includegraphics[width=110mm]{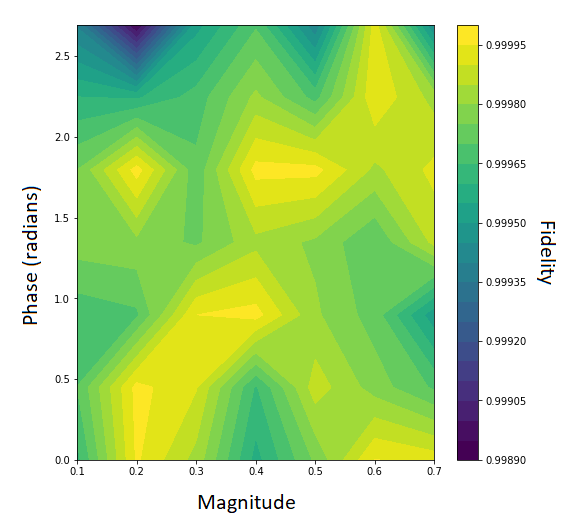}
    \caption{Fidelity for different single-mode coherent input states. The horizontal (vertical) axis denotes the magnitude (phase) of the input state. There are $N_p = 6$ probe states composing the measurement context, each having magnitude $0.5$ and a phase randomly drawn from $[0,\pi]$. $M=10^7$ samples were drawn for each estimate.   }
    \label{fig: fig_coh_cG_3}
\end{figure}

So far we have fixed the input state to observe the effect of the regulating parameters of the algorithm on the quality of the reconstructed state. We have only done such studies with two low-magnitude coherent states due to the slow speed of the reconstruction. But we do not have a reason to expect that the conclusions we have drawn above do not extend to arbitrary low-magnitude coherent states since our observations are in agreement with statistical intuition. Thus, here we only use $\Delta \cL = 10^{-12}$ and $\texttt{r}=1/M$, and aim to observe the reconstruction accuracy for different coherent states with the same set of probe amplitudes. We draw $M=10^7$ samples once for each coherent state, perform the reconstruction and calculate $\cF(\hat{\rho},\rho_{true})$. We show the results for three different $\Gamma$, each having $N_p=6$ probe amplitudes. Fig.~\ref{fig: fig_coh_cG_1} shows the results of the simulations with $\Gamma = \{ 0.3, 0.6e^{i \pi/6}, 0.9 e^{i 2\pi/6}, 1.2 e^{3\pi/6}, 1.5 e^{i 4\pi/6}, 1.8 e^{ i 5\pi/6} \}$, while in Fig.~\ref{fig: fig_coh_cG_2} we used probe amplitudes randomly drawn from the $2D$ grid spanned by magnitudes from the range $[0.3,3]$ and phases from the range $[0, \pi]$. It can be seen that in both cases the fidelity is greater than $0.9988$ for input states with magnitudes ranging from $0.1$ to $0.7$ and phases spanning $[0,\pi]$. It turns out that for $N=5$ there exist IC sets of POVMs where the magnitudes of all probes are the same. We used such an IC set of POVMs composed of $N_p=6$ probes with magnitudes $0.5$ and randomly chosen phases from $[0,\pi]$. The results are shown in Fig.~\ref{fig: fig_coh_cG_3}, and it can be seen that fixing the magnitude of the probe does not seem to observably affect the accuracy of the reconstruction, and the lowest fidelity we observe is $\approx 0.9989$. We performed these simulations for several other randomly chosen sets of probe amplitudes and observed similar values for the fidelities in the same range of input state magnitudes and phases.

\subsection{Analysis With Other Input States and Non-Ideal Counters}\label{subsec other_inputs_and_imp_cntrs}

The results of the simulations with single mode coherent states give us confidence that we have correctly implemented the numerical model and the reconstruction algorithm. Further, the high values of state fidelity achieved for different coherent states when the sample size is large ($\geq 10^6$), and with different sets of probe amplitudes, indicate that the measurement configuration generally allows for constructing good statistical models. By ``good" we mean that the relationship between the parameter space and the statistical distribution of the data is such that one is very likely to end up with a likelihood function that has high Fisher information and is maximized near the true state for a given dataset. Similarly, the fact that these high values of state fidelity are achieved under regulating conditions that allow for large number of iterations during the likelihood maximization process implies that the specific algorithm we chose to use performs well in finding the maximum of the likelihood with our parametrization of the state space and of the POVMs. One can argue that these simulations, for a given set of probe amplitudes and sample size, were done with a small subset of the corresponding statistical model that is characterized by only two separate parameters (the magnitude and the phase of the coherent state), and therefore do not say much about how ``good" the statistical model is or how well the reconstruction algorithm will perform for a dataset chosen at random from the statistical model. But notice that we did not encode any assumptions about the parameter space during the reconstruction process - the parameter space over which the likelihood maximization was performed was the whole of $\cR_{2,N}$. Nevertheless, to increase our confidence in our approach, we performed simulations with different kinds of input states described in Subsec.~\ref{subsec input states}.

We considered the truncated TMSV state with $r=0.5$ and $N=8$, such that the fidelity with the corresponding TMSV state is $\approx 0.9996$. In this case there exist IC sets of POVMs with $N_P=10$ probe amplitudes. We obtained one such set of POVMs by randomly drawing from the magnitudes in the range $[0.3,3]$ and from the phases in the range $[0,\pi]$. More specifically,
\begin{align}
 \Gamma &= \{  2.4e^{i 0.31\pi}, 1.5e^{i 0.57 \pi}      , 1.9e^{i 0.03 \pi}      , 1.1  e^{i 0.07 \pi}      ,  2.8 e^{i 0.59\pi}      , 2.6e^{i  0.98 \pi}      , \nonumber  \\  & 0.9e^{i 0.02\pi}      , 2.3e^{i 0.78 \pi}     , 0.9   e^{i 0.86\pi}      ,  1.3       e^{i 0.35 \pi} \}.\nonumber
\end{align}
 We drew $M=10^7$ samples and used the regulating parameters $\Delta \cL = 10^{-12}$ and $\texttt{r}=1/M$, and obtained a fidelity of $\approx 0.996$ between the true and the estimated state. The reconstruction took more than $4$ hours and terminated when $\texttt{r}_k$ reached below $\texttt{r}$. The slow reconstruction speed and the lower value of fidelity achieved than in the case of single-mode coherent states is most likely due to the fact that the vast majority of the entries of the true twirled state are zero. Another reason is the larger value of $N$ which significantly increases the parameter space, decreases the Fisher information for the same $M$, and significantly increases the computation time for a single iteration. 

We performed another set of simulations with the truncated (at $N=5$) twirled cat state introduced in Eq.~\ref{eq twirled_cat}. We used the minimum number of different probe states that seem to be necessary for an IC set of POVMs, $N_p =6$. We again used $\texttt{r}=1/M$ and $M=10^7$. For $\alpha = 0.5 e^{i \pi/4}$, chosen for no particular reason, and for $\Gamma = \{ 2.8e^{i 0.71\pi }, 2.3e^{i 0.48 \pi }, 1.3e^{i 0.25 \pi }, 2 e^{i 0.06 \pi },2.3 e^{i 0.23 \pi }, 1.6e^{i 0.33 \pi }  \}$ we achieved a fidelity of $\approx 0.997$ when $\Delta \cL = 0$. For the same $\alpha$, using \[\Gamma = \{  2.5e^{i 0.89 \pi }, 0.6e^{i 0.23 \pi }, 1.8e^{i 0.2 \pi }, 0.7e^{i 0.21 \pi }, 0.5e^{i 0.19 \pi }, e^{i 0.85 \pi}  \} \] and $\Delta \cL = 10^{-12}$ we achieved a fidelity of $\approx 0.9990 $. Both simulations terminated with $\texttt{r}_k$ falling below $1/M$. The difference in the achieved fidelities for the two different $\Gamma$ implies that there is an observable sensitivity to the choice of $\Gamma$. To see how the size of $\Gamma$ affects the reconstruction accuracy, we simulated for the same $\alpha$ but used $N_p=10$ and $N_p=15$ probe states, each drawn randomly from the magnitudes in the range $[0.3,3]$ and from the phases in the range $[0,\pi]$. The fidelities obtained were $\approx 0.9996 $ for $N_p=10$ and $\approx 0.9997$ for $N_p=15$. In both cases the iterations terminated when the $\epsilon$ reached $10^{-30}$. The implication seems to be that for the same $\texttt{r}$ the neighborhood of the likelihood function around its maximum is closer to (the twirled version of) $\rho_{true}$ when more probe amplitudes are used. We performed more simulations with several other values of $\alpha$ using $N_p=6$ randomly selected probe states and observed similar reconstruction accuracy, with the fidelities averaging around $0.998$.

Finally, we briefly considered simulations with lossy counters described by the POVM elements in Eq.~\ref{eq Pi_kl lossy}. The BS is again balanced, and both counters are utilized. To estimate the effect of the efficiency $\nu$ (which is the same for both counters) on the reconstruction accuracy, we performed simulations for different values of $\nu$ in the range $[0.5,1]$ for the same truncated ($N=5$) single-mode coherent state as an input state, and for the same $\Gamma$ using $M=10^6$ samples. The regulating parameters $\Delta \cL = 10^{-12}$ and $\texttt{r}=1/M$ were used. The fidelities observed are shown in Table~\ref{table 3} for $4$ different single-mode coherent states and $\Gamma$ composed of $N_p=6$ probe states. The members of the latter were randomly drawn from the magnitudes in the range $[0.3,3]$ and from the phases in the range $[0,\pi]$. The simulations where the observed fidelities are much closer to $1$ compared to the rest (in particular, for $\nu=0.9$ for the input state $\alpha=0.5 e^{i\pi/4}$, where the fidelity has $5$ nines after the decimal point) terminated when $\epsilon$ reached $10^{-30}$. But these are not the only cases where $\texttt{r}$ was not reached (though in general the fidelities observed are higher when the algorithm runs until $\epsilon$ reaches $10^{-30}$), so we cannot explain the significantly higher fidelities for these few cases by the fact that the number of iterations before termination was several times higher. The most plausible explanation is that the likelihood function for the particular sample peaks at a $\rho_{max}$ that is much closer to $\rho_{true}$. Otherwise, it seems the efficiency of the counters does not impact the accuracy of the reconstruction significantly.

\begin{table}[h!]
\centering
\renewcommand{\arraystretch}{1.5}
\begin{tabular}{|| c |  p{0.4\linewidth} | p{0.4\linewidth} ||} 
 \hline
Efficiency &    $\alpha = 0.5 e^{i \pi/4}$ and $\Gamma =$ \begin{align} \{  2.8e^{i 0.31 \pi },2.8e^{i 0.06 \pi }, 0.4e^{i 0.76 \pi }, \nonumber \\ 1.9e^{i 0.02 \pi }, 2.1e^{i 0.77 \pi },0.3e^{i 0.48 \pi }  \} \nonumber \end{align} &  $\alpha = 0.9 e^{i \pi/10}$ and  $\Gamma =$ \begin{align} \{  2.6e^{i 0.76 \pi },2.5e^{i 0.83 \pi },2.1e^{i 0.33 \pi },  \nonumber \\  0.5e^{i 0.34 \pi },  2.4e^{i 0.3 \pi },1.1e^{i 0.89 \pi } \} \nonumber \end{align}  \\ [0.5ex] 
 \hline\hline
1	&   0.998890   & 	 0.998511 	\\
\hline
0.9	& 0.999997 & 	0.999826	 \\
\hline
0.8	& 	 0.999294   & 0.999000 	 \\
\hline
0.7 & 	0.999621	& 0.998930	\\
\hline
0.6	& 0.999304 &	0.998369   \\
\hline
0.5	& 0.999417 &	0.998334  \\  [1ex] 
 \hline
\end{tabular}
\caption{The fidelity between the estimate and the true state for different values of the efficiency $\nu$, and for different single-mode coherent states and probe amplitudes. $M=10^7$, $\Delta \cL=10^{-12}$ and $\texttt{r}=1/M$. The fidelities reported are from a single estimate.}
\label{table 3}
\end{table}
\addtocounter{table}{-1}
\begin{table}[!h]
\centering
\renewcommand{\arraystretch}{1.5}
\captionsetup{list=no}
\begin{tabular}{|| c |  p{0.4\linewidth} | p{0.4\linewidth} ||} 
 \hline
Efficiency &           $\alpha = - 0.7 $ and  $\Gamma =$ \begin{align} \{  1.9e^{i 0.97 \pi },2.6e^{i 0.55 \pi }, 2.2e^{i 0.87 \pi },  \nonumber \\ 2.4e^{i 0.31 \pi }, 1.7e^{i 0.77 \pi },2.8e^{i 0.53 \pi }  \} \nonumber \end{align}  &  $\alpha = 0.5 e^{i 5\pi/6}$ and  $\Gamma =$ \begin{align} \{  2.6e^{i 0.87 \pi },1.3e^{i 0.32 \pi },  1.0e^{i 0.45 \pi }, \nonumber \\ 1.8e^{i 0.51 \pi }, 1.6e^{i 0.02 \pi },2.1e^{i 0.8 \pi } \} \nonumber \end{align}  \\ [0.5ex] 
 \hline\hline
1		& 0.998296	&  0.999362 	\\
\hline
0.9		& 0.997922	&  0.999749 \\
\hline
0.8	 	& 0.999325	&  0.999957 \\
\hline
0.7 & 0.998457	&  0.998770	\\
\hline
0.6	& 	0.998462 &	 0.999625  \\
\hline
0.5	&	0.998282 &  0.999751  \\  [1ex] 
 \hline
\end{tabular}
\caption{The fidelity between the estimate and the true state for different values of the efficiency $\nu$, and for different single-mode coherent states and probe amplitudes. $M=10^7$, $\Delta \cL=10^{-12}$ and $\texttt{r}=1/M$. The fidelities reported are from a single estimate. (continued)}
\end{table}

\begin{figure}[!htb]
    \centering
\includegraphics[width=110mm]{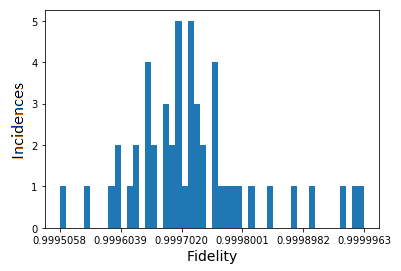}
    \caption{Histogram of the fidelity for the single-mode coherent state $\alpha=0.5 e^{i\pi/4}$ and for $\Gamma = \{  2.8e^{i 0.31 \pi },2.8e^{i 0.06 \pi },0.4e^{i 0.76 \pi },1.9e^{i 0.02 \pi },2.1e^{i 0.77 \pi },0.3e^{i 0.48 \pi }  \}$. Efficiency of the counters is $\nu=0.9$. $50$ simulated reconstructions are shown, where $M=10^7$ for each simulation. }
    \label{fig: hist_fid_sim_1}
\end{figure}

To test the last hypothesis, we performed $50$ simulations for the single-mode coherent state with $\alpha=0.5 e^{i\pi/4}$ and with $\Gamma = \{  2.8e^{i 0.31 \pi },2.8e^{i 0.06 \pi },0.4e^{i 0.76 \pi },1.9e^{i 0.02 \pi },2.1e^{i 0.77 \pi },0.3e^{i 0.48 \pi }  \}$ at the efficiency $\nu=0.9$. We want to observe the variations in the fidelity depending on the drawn sample, in order to determine whether the very high fidelity observed above is due to the value of $\nu$ or due to the particular sample that was drawn in that trial. We obtained an average fidelity of $0.999718$ with a standard deviation of $0.000100$. The histogram is shown in Fig.~\ref{fig: hist_fid_sim_1}. It can be seen that the estimates are clustered around the fidelity of $0.9997$, which is near the same order of accuracy as the results for the other input states in Table~\ref{table 3}.

%% file: chapter_experiment.tex
\chapter{Analysis of the Experiment}\label{chap experiment}

In this chapter we describe the experiment that was performed by David S. Phillips, Thomas Gerrits and Michael Mazurek at NIST, Boulder, and our subsequent data analysis. The data analysis was abandoned midway after some systemic discrepancies were discovered in the data that we could not trace back to the conditions of the experiment. Nevertheless, these systemic discrepancies are not large enough to prevent what one might consider a good agreement between the reconstructed state and the (assumed) true state. 

The experiment was initially meant to model the WFH configuration where the BS is characterized by a trivial partition and thus the input state could be described as a two-mode twirled state in $\cR_{2,N}$. A pulsed laser was used to simultaneously produce the LO and input states. Only coherent input states were generated. The mode of the LO was set to a fixed polarization. The polarization of the input state was controlled with respect to the polarization of the LO. The mode of the input state matching the LO (mode $1$) is defined by having the same polarization and temporal structure as the LO. All occupied modes in the orthogonal polarization, as well as the modes in the same polarization as the LO but with zero temporal overlap with it, are orthogonal to mode $1$. According to our results at the end of Sec.~\ref{sec twirl} we could assume only one of these modes is occupied and call it mode $2$. However, after the experiment was conducted, it was found that the BS acts differently depending on the polarization. The WFH configuration that models the experiment belongs to the family with a BS characterized by a partition size $K=2$, and, therefore, we need to model the input state as living in $\cR_{3,N}$. Mode $1$ is still the same, but the set of modes in the same polarization as the LO but with orthogonal temporal shapes are distinguishable from the set of modes in the orthogonal polarization to the LO, as the BS acts differently on these two sets of modes. According to our theoretical results, we can assume only one mode in the former set of modes is occupied, and similarly for the latter set of modes. These two modes are denoted as mode $2$ and mode $3$, respectively.

We start by giving a summary of the experiment in Sec.~\ref{sec desc_exp} (a more detailed description can be found in David S. Phillips' thesis \cite[Chap.~7]{phillips2020advanced}). The following two sections are devoted to data analysis. Sec.~\ref{sec desc_gen_data} describes how the data for reconstruction is generated - namely, how the relative frequencies corresponding to the outcomes of a chosen measurement context are obtained. It also discusses how the ``best guess" of the input state is obtained. Sec.~\ref{sec desc_recon} describes the results from performing maximum likelihood reconstruction with this data.

\section{Description of the Experiment}\label{sec desc_exp}

\begin{figure}[!htb]
    \centering
\includegraphics[width=150mm]{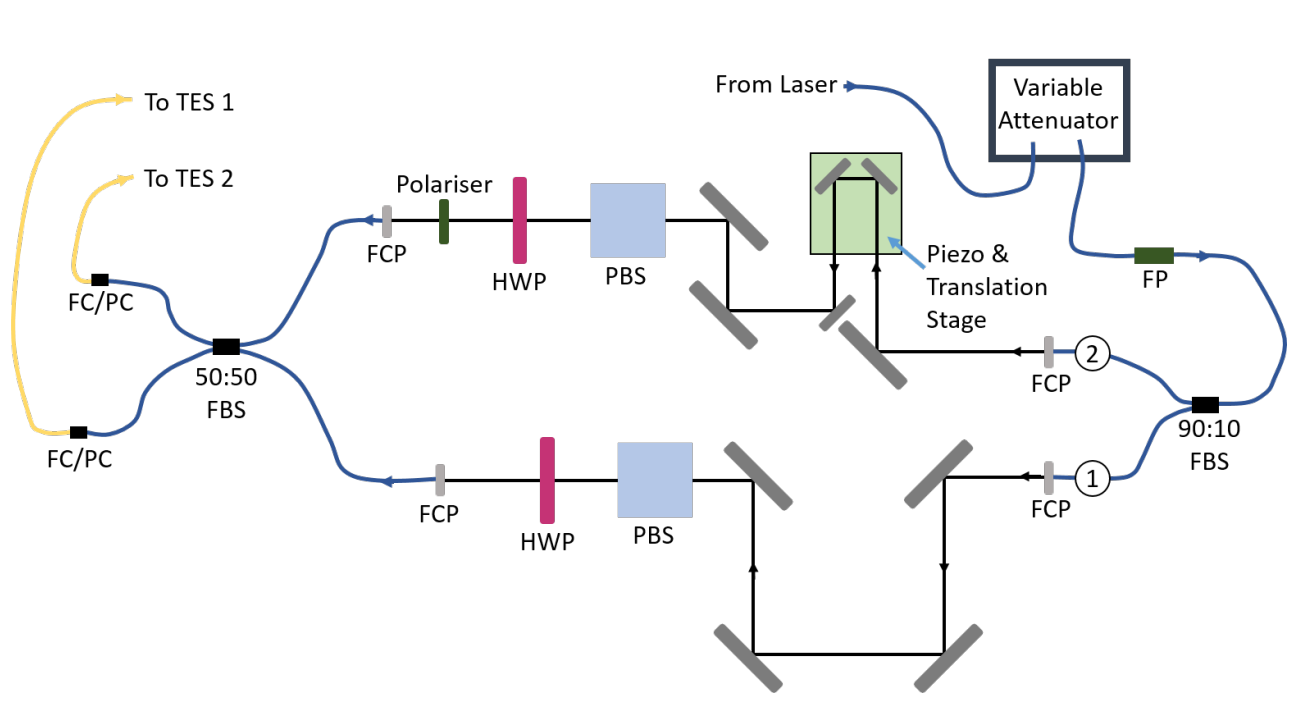}
\caption{Schematic of the experimental setup (taken from David S. Phillips' thesis \cite[Fig.~7.2]{phillips2020advanced}). Some of the abbreviations of the components, starting from the left, are as follows: TESs stand for transition-edge sensors, FBS stands for a fiber-optic BS, HWP stands for a half-wave plate, PBS stands for a polarizing BS, and FP stands for a fiber-optic polarizer. The description is in the main text, and further information can be found in Ref.~\cite{phillips2020advanced}. }
    \label{fig: exp_setup}
\end{figure}

We divide our description of the experimental setup into four parts: (1) generation of the input and the LO states, (2) manipulation of the phase and the magnitude of the LO, (3) implementation of the BS action, and (4) measurement by the counters. The schematic of the optical setup is shown in Fig.~\ref{fig: exp_setup}. A pulsed laser with a wavelength of $1550 \textrm{ nm}$ and pulse length of $100 \textrm{ ps}$ was used to generate both the LO and the input state. A function generator was used to trigger the laser pulses at $100 \textrm{ kHz}$, resulting in $10^5$ measurement shots per second. This number is chosen to give enough time for the photon counters to reset after each shot. The laser light was first passed through a polarization-maintaining (PM) and single-mode (SM) fiber-optic cable to shape it into a well-defined mode. A variable attenuator was used to decrease the magnitude of the coherent pulse to having an average of at most several photons. This caused some de-polarization, and thus a fiber polarizer (FP) was used before a $90:10$ fiber beam splitter (FBS). The $90\%$ arm of the FBS was used to generate the LO state (the upper arm in Fig.~\ref{fig: exp_setup}), while the $10\%$ arm was used for constructing the input state (the lower arm in Fig.~\ref{fig: exp_setup}).

To generate the input state, the corresponding output of the FBS was first passed through a formation of four mirrors to control the path length and the beam alignment. Afterwards the pulse was passed through a polarizing beam splitter (PBS), which separates the pulse into orthogonal polarization components. We refer to the polarization that is passed through to the next stage of the setup by ``vertical", and this is aligned with the polarization of the FP. The orthogonal ``horizontal" component is discarded. The vertically polarized input is then passed through a half-wave plate (HWP) to set its polarization with respect to the vertical (different polarizations correspond to different input states). To construct the LO state, the $90\%$ arm of the FBS is passed through an arrangement of mirrors, two of which are attached to a stage controlled by a piezoelectric device, which is in turn attached to a more coarsely controllable translation stage. This arrangement is used to alter the path length of the LO with a precision that is of the order of a $\textrm{nm}$, and thus allows for an adjustment of the relative phase between the input and the LO states. We describe the intricacies of this arrangement in more detail in the next section. After phase adjustment, the LO state is passed through a sequence composed of a PBS, a HWP and a polarizer to adjust the magnitude of the pulse. The polarizer is set to vertical polarization so that the horizontal polarization component of the LO is blocked from passing through. The magnitude of the pulse in the vertical polarization, and thus the magnitude of the pulse after the polarizer, can be controlled by tuning the angle of the HWP. The input state as well as the LO state in the model are assumed to be the states in the input fibers of the FBS right before interaction.

The next part of the setup is the BS, which is realized by a PM FBS that was designed to act as a $50:50$ BS on the vertically polarized modes. Measurements by power meters at the outputs of the FBS without using the variable attenuator and directing the light in a vertical polarization to one of the input ports showed that the splitting ratio of the BS on the vertical modes was $49.46:50.54$. The same measurement with horizontally polarized light gave a splitting ratio of $73.58:26.42$. We used these numbers in our model, but the error in them is estimated to be up to $1\%$. The pulses at the outputs of the FBS are then sent to two transition-edge sensors (TESs), which we describe below. The fibers leading to the TESs were not PM, and the experimenters used mating couplers to connect them with the outputs of the FBS. The resulting losses from the couplings were measured and incorporated into the POVM elements corresponding to the outcomes of the TESs (described in Subsec.~\ref{subsec POVM_const}). 

The TESs operate by keeping a film of superconducting material at the critical temperature and bias current - that is, the temperature and current at which the transition occurs from the superconducting to the conducting regime. This transition is accompanied by a rapid increase of the resistance from a value of zero to a finite value, which then changes very little upon further increase of the temperature. Due to the shape of the dependence of the resistance on the temperature around the critical point, very small increases in the temperature can be detected by monitoring the resistance across the superconducting film. This is precisely what happens when one or several photons are absorbed by the film. The critical temperature of a TES is typically around $100 \textrm{ mK}$, and an adiabatic demagnetisation refrigerator is used to cool it to such a low temperature and provide good temperature control. The change in the resistance across the superconducting film is detected by measuring the change in the magnetic field generated by an inductor that is in series with the TES and through which a constant current is maintained. The measurement of the magnetic field is performed by an array of superconducting quantum interference devices (SQUIDs), which are held at a temperature of about $10\textrm{ K}$. The design of a TES is its own field of study, and we will not go into further detail here. More information about the design of the circuits incorporating the TES, the signal readout and its analysis can be found in \cite{gerrits2016super}. For the purposes of this thesis, it suffices to mention that the experimenters provided us with a measurement of a quantity proportional to the energy absorbed by the TES from the light pulse. We describe our analysis with these values in the next section.

The experiment lasted five days, with a different input coherent state $\ket{\alpha}$ prepared and measured for each day. An effort was made to keep the magnitudes of the input states the same at $\abs{\alpha} \approx \sqrt{0.7}$, and the states prepared on different days differed by the angle of the polarization with respect to the vertical polarization as set by the HWP. The HWP angles used were $54.8^\circ$, $64.80^\circ$, $76.74^\circ$, $84.59^\circ$ and $90.66^\circ$. The relationship between the HWP angle ($\theta_{HWP}$) and the polarization angle with respect to the vertical ($\theta_p$) was measured to be $\theta_{p} = 2 \theta_{HWP}-109.6^\circ$ by using power meters instead of TESs and bypassing the attenuator. The chosen HWP angles were meant to test the whole polarization range, with $\theta_{HWP} = 54.8^\circ$ corresponding to vertical polarization and $\theta_{HWP} =90.66^\circ$ corresponding to a polarization angle of $71.72^\circ$ with respect to the LO polarization. At the beginning of each day, the input state was prepared and measured by the experimental setup by blocking the LO arm, thus providing us with TES data to obtain a more accurate estimate of the input state to compare the reconstructed state with. We refer to these datasets as ``input state only" datasets.

Three different LO magnitudes were prepared for each input state, and individual measurements of these (by blocking the input state from entering the FBS) were made. We thus obtained TES data to estimate the magnitudes of the LO states. We refer to these datasets as ``LO state only" datasets, while the measurements performed corresponding to the WFH configuration are referred to as ``interference experiments". In the interference experiment, the piezoelectric stage was used to sweep the position of the holding mirrors (shown in Fig.~\ref{fig: exp_setup}) repeatedly. More precisely, the position of the piezoelectric stage is linearly related to the voltage supplied to the piezoelectric device, and a saw-tooth voltage function was fed to the device. The period of the function was set to $\approx 0.8$ seconds, and the height of the saw-tooth with respect to the beginning point of the sweep was chosen such that the resulting voltage difference roughly corresponded to $8.4 $ radians of phase change of the LO pulse (thus giving us data from the full range of the relative phase). TES data was collected in $0.8$ second intervals, which corresponded to $83856$ shots, with each interval followed by $0.2$ seconds of no measurements, until around $3 \cdot 10^8$ measurement instances were collected for each LO magnitude. We refer to a dataset taken within a single interval as a single ``piezo ramp" dataset, even though that is not strictly accurate, as the period of the saw-tooth function is slightly different from the acquisition window. The fast speed of the ramp of the piezoelectric stage was chosen to mitigate the effect of the drift of the stage. In particular, the experimenters found that the stage can drift significantly (in terms of its impact on the relative phase) within dozens of seconds at the same voltage, and this made it impossible to rely on the voltage input to estimate the relative phase. Therefore, the data was divided into consecutive pairs of piezo ramps, with the $83856$ data points from the first ramp used to obtain the relationship between the voltage and phase, and the second set of $83856$ data points used for reconstruction. Therefore, the phase stability required was of the order of $2$ seconds, which is achievable by the measurement setup.

\section{Generating Data for Reconstruction}\label{sec desc_gen_data} 

In this section we describe how the ``true" input and the ``true" LO states for the different interference experiments were estimated from the ``input state only" and ``LO state only" datasets, respectively. We also describe how the different POVMs for the different TES measurement outcomes at different LO amplitudes were constructed, as well as how the relative frequencies corresponding to the measurements by different POVM elements were obtained from the interference experiments. The relative frequencies were then used for state reconstruction analysis (described in the next section).

\subsection{Construction of the POVMs}\label{subsec POVM_const}

\begin{figure}[!htb]
    \centering
\includegraphics[width=150mm]{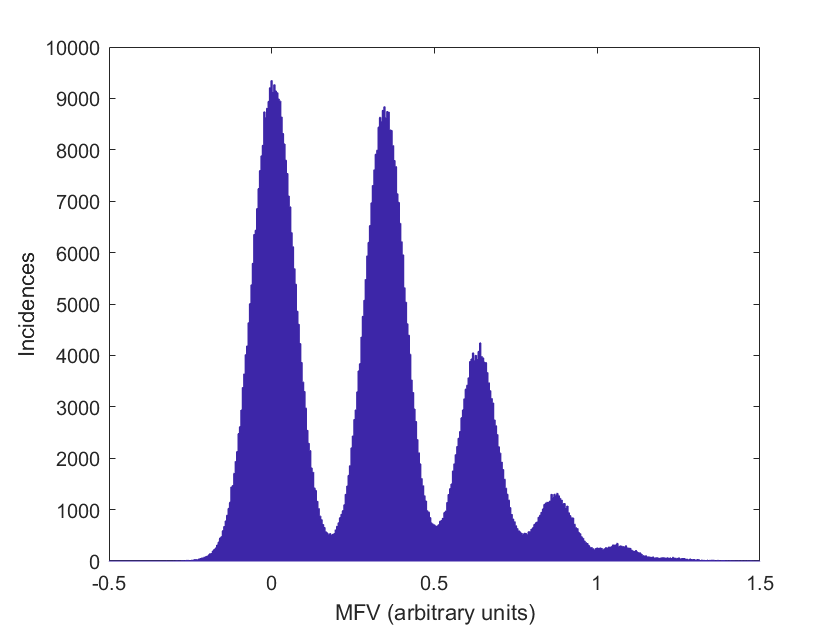}
    \caption{Example of a histogram of MFVs for many shots. The dataset shown is obtained from one of the TESs during one of the ``LO state only" measurements for the input state with $\theta_{HWP} = 54.8^\circ$. }
    \label{fig: hist_LO3_54deg}
\end{figure}

The construction of the POVMs corresponding to the measurement outcomes of the TESs as part of the measurement configuration of the experiment is a multistep process. The first stage was performing tomography of the TES detectors. This was done by the experimenters according to the method described in Ref.~\cite{brida2012quantum} on a different day before the start of the experiment. Their method uses a set of known coherent states as inputs to the TES. As mentioned in the last section, the analysis of the TES output for a given shot results in a single number, which we refer to as the matched filter value (MFV), and which contains information about the number of photons absorbed by the TES from the light pulse. As an example, a histogram of MFVs of one of the TESs for one of the ``LO state only" datasets is shown in Fig.~\ref{fig: hist_LO3_54deg}. It can be clearly seen that the MFVs are naturally grouped according to photon number, and that there are overlaps between the groupings. Broadly speaking, there are two kinds of noise that affect the estimation of the photon number measurement of the state from the MFV outcome. The first kind of noise has to do with all sorts of processes occurring in the complicated circuitry that links the act of absorption of the light pulse to the final readout. Using the example above, if one were to assume that the state absorbed is a Fock state with a well-defined frequency, the corresponding distribution of the MFVs would be one of the Gaussian-looking profiles in Fig.~\ref{fig: hist_LO3_54deg}. Because of the overlap between the different profiles, this source of noise would produce an uncertainty in the photon number of the Fock state. The second kind of noise has to do with losses and background light present in the absorption process. This includes losses in the coupling with the fiber leading to the TES. In this case, even if the Gaussian profiles had near-zero overlap, so that the number of photons absorbed could be deduced with high certainty, there would still be uncertainty in the photon number measurement on the input state. Thus, the tomography of the TESs needs to take these two kinds of sources of noise into account and link the MFVs to the projectors onto photon numbers on the input state.

The method of detector tomography that was used is based on assigning photon numbers to MFVs. In particular, the method produces a set of MFV intervals corresponding to each photon number. Using Fig.~\ref{fig: hist_LO3_54deg} for illustration again, the edges of the intervals (we refer to these as ``bounds") would be near the troughs in the histogram. While more advanced data analysis can be imagined given the knowledge of the TES design, this grouping strategy requires minimal processing and is based on the statistically sound assumption that the absorption of a given number of photons produces an MFV with Gaussian error. The TES detector tomography also produces a $10 \times 26$ matrix which encodes the conditional probabilities $P(n|m)$ of detecting $n = 0,\hdots,8$ or $n>8$ photons when $m=0,\hdots,24$ or $m>24$ photons are present in the fiber leading to the TES. We refer to these matrices as $T_1$ and $T_2 $ for the two TESs. There are additional losses in the couplings between the outputs of the FBS and the fibers leading to the TESs, which were estimated by the experimenters. The losses are modeled as an additional Markov process as described in Subsec.~\ref{subsec POVMs_nonideal}, to obtain the conditional probabilities for each TES of detecting a particular photon number when a certain number of photons is present at the corresponding output port of the FBS. More specifically, let us denote the updated conditional probabilities of detecting $n = 0,\hdots,8$ or $n>8$ photons when $m=0,\hdots,24$ or $m>24$ photons are present at the outputs of the FBS by $T'_1$ and $T'_2$ for the two TESs. These were approximated as
\begin{align}\label{eq T_i'}
T'_{i,nm} = \sum_{j=0}^{m} T_{i,nj} \binom{m}{j} \nu_i^j (1-\nu_i)^{m-j},
\end{align}
where $i=1$ or $i=2$, $\nu_i$ is the transmission coefficient for the $i$'th TES, and the approximation assumes that there are no more than $25$ photons in the outputs of the FBS. This approximation is justified given the low magnitudes of the LO and input states. The transmission coefficients were measured by the experimenters to be $\nu_1 =0.9586  $ and $\nu_2 =0.9413 $.

Then, for the $i$'th TES, the positive operator on the state at the matching output port of the FBS corresponding to the measurement of $k$ photons is given by
\begin{align}
\hat{D''}_{i,k} = \sum_{m=0}^{25}  T'_{i,km} \hat{D}_{i,m}. 
\end{align}
The final POVM elements corresponding to $k$ and $l$ photon number measurements by the two TESs for a given LO amplitude are calculated similarly to Eq.~\ref{eq Pi_kl lossy}. Let us denote these by $\Pi''_{kl}$. Skipping some of the steps in the derivation that are already familiar from Subsec.~\ref{subsec POVMs_nonideal}, we obtain
\begin{align}
\Pi''_{kl} & = \tr_\textsl{p} \left(\left(I \otimes \ket{\gamma}\bra{\gamma}\otimes \ket{\vec{0}}\bra{\vec{0}}  \right) U^\dagger( \hat{D''}_{1,k} \otimes \hat{D''}_{2,l}) U \right) \nonumber  \\
& = \sum_{m=0}^{25}  \sum_{n=0}^{25}  T'_{1,km}   T'_{2,ln}   \Pi_{mn},
\end{align}
where $ \Pi_{mn} = \tr_\textsl{p}( (I \otimes \ket{\gamma}\bra{\gamma}\otimes \ket{\vec{0}}\bra{\vec{0}} ) U^\dagger \hat{D}_{kl} U ) $. As a reminder, since our BS is characterized by the partition size $K=2$, the unitary $U$ is acting on $3$ modes. We encode its action in mode space as
\begin{align}\label{eq: B exp}
 B= \begin{pmatrix}  \sqrt{0.4946}   &     \sqrt{0.5054}  \\ -\sqrt{0.5054}   &   \sqrt{0.4946}   \end{pmatrix} \oplus \begin{pmatrix}  \sqrt{0.4946}   &     \sqrt{0.5054}  \\ -\sqrt{0.5054}   &   \sqrt{0.4946}   \end{pmatrix} \oplus \begin{pmatrix}  \sqrt{0.2642}   &     \sqrt{0.7358}  \\ -\sqrt{0.7358}   &   \sqrt{0.2642}   \end{pmatrix}  
 \end{align}
 using the experimentally measured splitting ratios for the ``vertical" and ``horizontal" polarizations. For a given LO amplitude $\gamma$, these $\Pi_{kl}$ are calculated similarly to Eq.~\ref{Pi_kl expanded} and are encoded as elements of $\cR_{3,N}^s$. We do not present this calculation here.

The circuit by which the resistance across the TES is measured utilizes a SQUID array, which is cooled anew each day. This reset of the cooling chamber causes a slight shift in the MFVs by an overall scaling factor. More precisely, the output of the SQUID indirectly measures the change in the resistance of the TES, and we assume that the effect of the reset on this measurement is small enough that it can be assumed to be linear to first order. This necessitates a correction in the bounds for the MFVs by an overall scaling factor. We find that a correction in the bounds throughout the day is also desirable. In particular, the interference experiment with a particular LO magnitude is performed right after the corresponding ``LO state only" measurement. The latter can be used to estimate the correct scaling factor, as described in the next paragraph, which is then used to assign photon numbers to the TES data for the corresponding interference experiment. The ``input state only" dataset is also used to estimate the correct scaling factor at the time of its measurement, which is only used to gain an estimate of the true state to compare the reconstruction of the twirled state with.

 We perform model selection to obtain the estimates of the scaling factors for the bounds of both TESs. More specifically, since we know that both the LO and the input state are prepared in coherent states, the photon number distributions at the outputs of the FBS are Poissonian with a fixed Poisson parameter (given by the square of the coherent state magnitude) for each output for each of the four datasets (that is, the three ``LO state only" and the ``input state only" datasets). The probability of measuring $m$ photons in output port $i$ is determined by $ \hat{D}_{i,m}$. We have access to the $\hat{D''}_{i,k}$, which we use to relate the relative frequencies for the output of each TES to the corresponding Poisson parameter. We refer to this statistical model as the ``Poissonian model". The optimization is based on calculating likelihood ratios for different scaling factors. In particular, a given scaling factor gives different photon number frequencies because the scaling factor influences the MFV intervals corresponding to each photon number. We use the Poissonian model to estimate the most likely Poisson parameter for these frequencies and calculate the value of the likelihood function for this parameter. We do this for each of the four datasets. We then consider the ratio of this value of the likelihood function over the value of the likelihood function of the unconstrained model (that is, when the frequencies are taken as the probabilities). We maximize this likelihood ratio over a range of scaling factors, and choose the scaling factor that gives the largest likelihood ratio for each dataset. 
 
To estimate the accuracy of the Poissonian model when the optimal scaling factors are used, we performed parametric bootstrapping on the ``LO state only" datasets corresponding to the input state with $\theta_{HWP} = 76.74^\circ$. More specifically, for each of the three datasets, we computed the coherent state magnitude that maximizes the likelihood function and calculated the corresponding probabilities for each TES. We then simulated random samples from these probabilities the same number of times as the original number of experimental data points (which was of the order of $10^6$). We did this $1000$ times for each LO state. After generating the likelihood ratios for the bootstrapped distributions (that is, the ratio of the maximized likelihood function according to the Poissionian model over the unconstrained likelihood function) we looked at the deviation of the original likelihood ratio from the bootstrapped likelihood ratio distribution. We found that overall the model mismatch is not too bad, with the largest deviation being about $7\sigma$ for the LO state with the largest magnitude. While this deviation might seem to be very big, the likelihood ratio test in this case is very sensitive given the number of data points. It is possible that this deviation can be partly explained by the residual memory effects of the TESs. The “LO state only” data is also taken at $100\textrm{ kHz}$, and we found that $10 \textrm{ }\mu\textrm{s}$ is not enough time to completely eliminate the effects from the preceding absorption event.

The magnitude-squares of the LO states that were incorporated into the POVM elements were calculated from the values that maximized the likelihoods of the Poissonian models using the corresponding ``LO state only" dataset for each TES. More specifically, the magnitude-square of the LO state is assumed to be equal to the sum of the magnitude-squares at the output ports of the FBS, and the latter are estimated from the corresponding TES data by the procedure described above. In the next subsection we describe how the relative phases between the LO and input states were estimated from the ``piezo ramp" datasets. 

\subsection{Estimation of the Relative Phase and Generation of Frequencies}

\begin{figure}[!htb]
    \centering
    \subfigure[]
    {
\includegraphics[width=160mm]{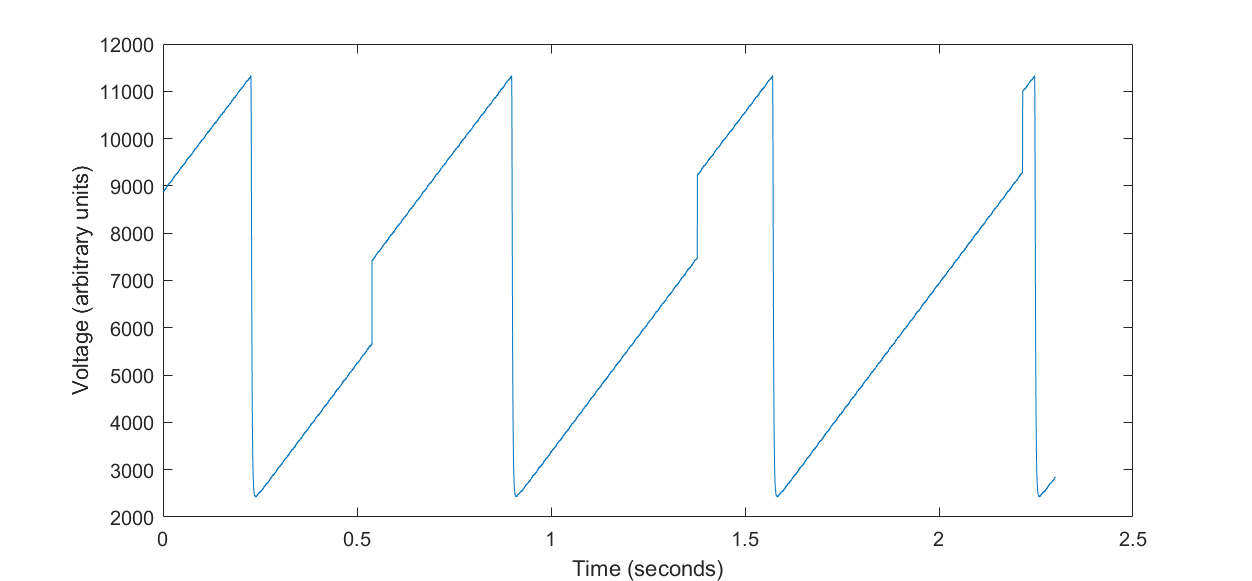}
    }
    \\
    \subfigure[]
    {
\includegraphics[width=160mm]{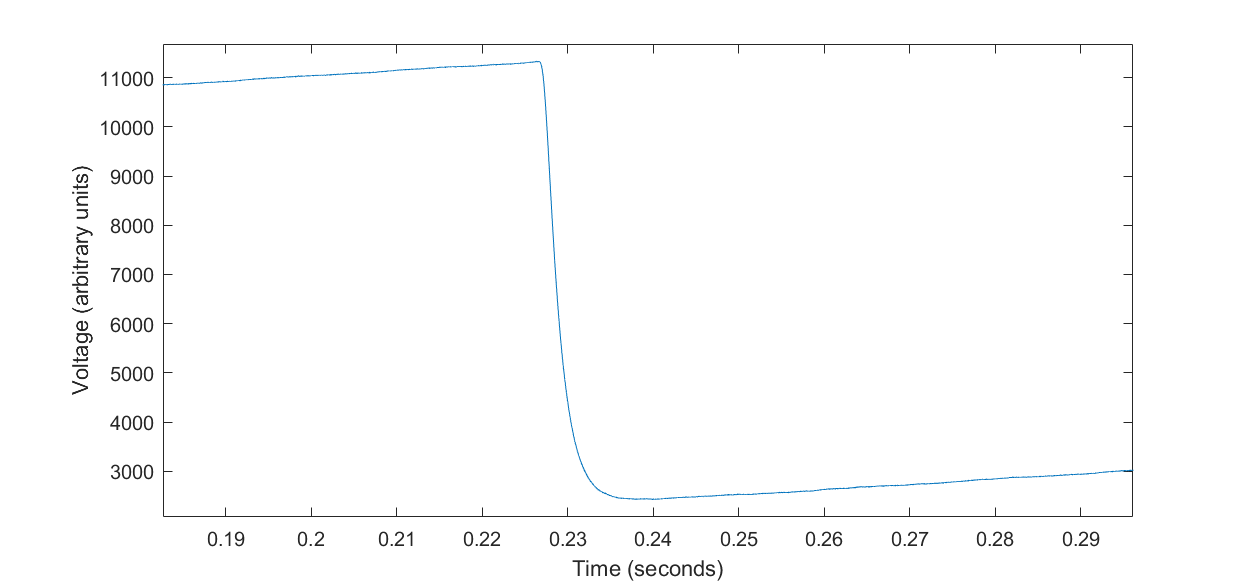}
    }
    \caption{(a) The trace of the voltage measured across the piezoelectric device over several seconds. The second figure (b) shows a zoomed-in segment of (a) for a better view of the transient response of the device to an instantaneous shift in the voltage input.}
  \label{fig: expiezoramp}.
\end{figure}

A segment of the voltage function fed to the piezoelectric device is shown in Fig.~\ref{fig: expiezoramp}. The vertical axis is the voltage measured across the device, while the horizontal axis is time (more specifically, the ``experimental time" taken by the clock that stops ticking when no measurements are made). One can observe the existence of fast transients (a zoomed-in segment is shown in part (b) of the figure), which occur when the voltage fed by the function generator starts a new saw-tooth cycle by making a sudden jump from the largest value to the lowest. The piezo cannot change its position instantaneously, and this results in the recorded transient response. One can also notice that there are instantaneous jumps in the voltage, occurring at roughly the same intervals as the transients. These are due to the $0.2$ seconds of no data taking within each second during which the piezoelectric device continuous its cycle. More specifically, data of the voltage readout was taken continuously, but the part of this data when no measurements were made was edited out. The data taken between two such jumps is what we denote as a single ``piezo ramp" dataset. We use the first ``piezo ramp" dataset to estimate the relationship between the voltage and the phase of the LO, which is assumed to be linear during the non-transient part of the time trace. We use this estimate to assign phases to the voltage values in the consecutive ``piezo ramp" dataset, and we repeat this for all the data we have.

\begin{figure}[!htb]
    \centering
\includegraphics[width=170mm]{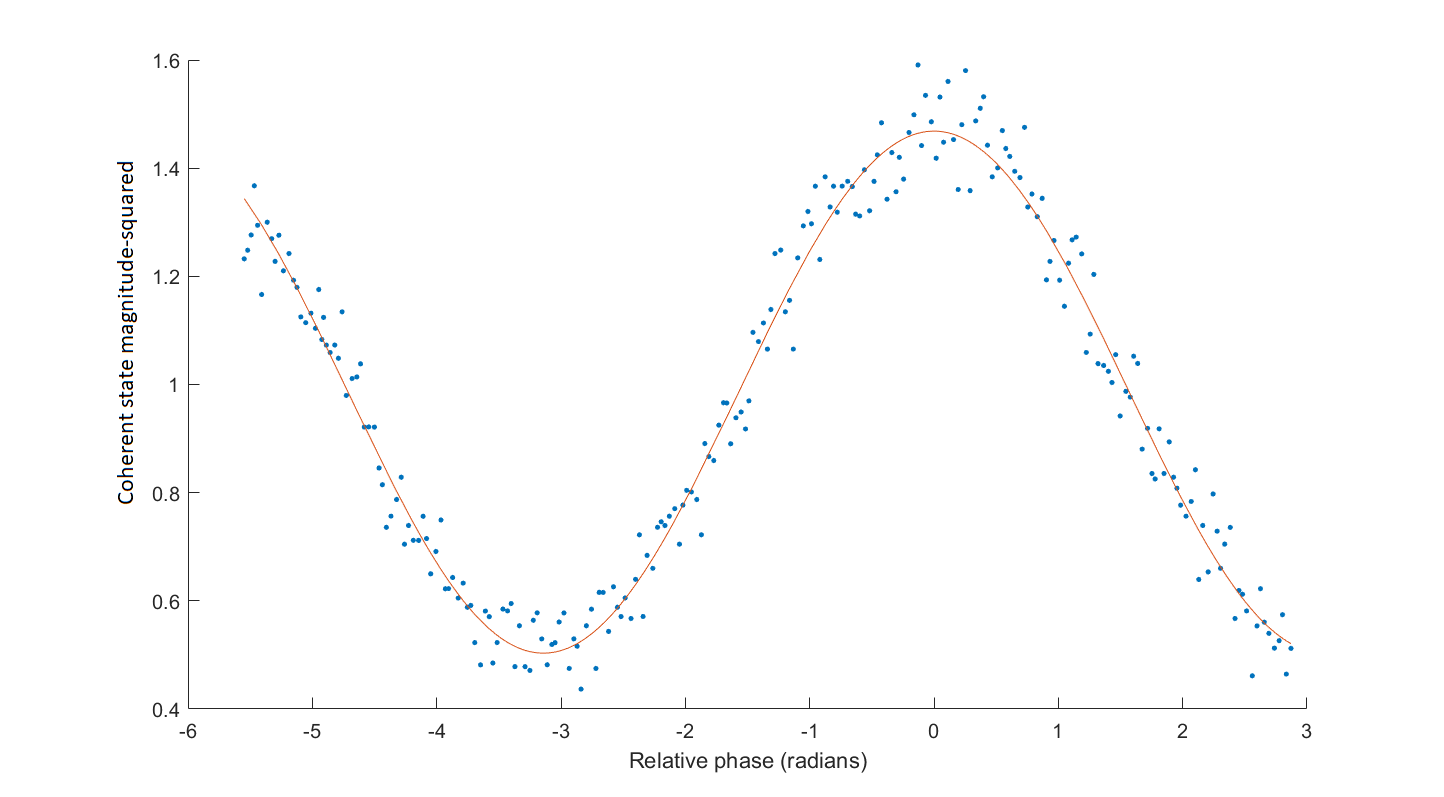}
    \caption{Fitting a sinusoid to the estimated coherent state magnitudes at one of the outputs of the FBS at a range of relative phase values. The HWP angle of the input state is $\theta_{HWP} = 76.74^\circ$. }
    \label{fig: ex_piezorampfit}
\end{figure}

The relationship between the relative phase $\phi$ and the voltage readout $V$ is assumed to be linear and modeled by the equation $\phi = a\cdot V+b$. The estimation of the constants $a$ and $b$ for a given ``piezo ramp" dataset is done as follows. Since both inputs to the FBS are coherent states, we expect the states at the outputs of the FBS to be coherent states with magnitudes that have sinusoidal dependence on the relative phase (more explicit description is provided in the next subsection). The algorithm first removes the part of the data corresponding to the transient voltage response described above, and assigns photon numbers to the MFVs of both TESs according to the estimated bounds for each voltage value. Once that is done, it orders the data of the TESs according to the corresponding voltage values, and divides the remaining data into $241$ consecutive segments. The number of data points in each segment averages to about $348$, which corresponds to almost $2^\circ$ of relative phase change. The number of data points in each segment is not always the same, since the procedure that removes the transient part of the piezo trace does not leave the same number of data points each time. We then make the assumption that the magnitude of the coherent state at the outputs of the FBS can be considered constant throughout the aforementioned $2^\circ$ window, so that we can use the Poissonian model described in the previous subsection to estimate the magnitude of the coherent state at each output using maximum likelihood estimation with each of the $348$ data points. We also calculate the Fisher Information for each estimate, which is used to estimate the corresponding variance by assuming the variance is equal to the inverse of the Fisher Information. We then use the estimated coherent state magnitudes for each segment, and perform nonlinear weighed least squares regression, with the weights given by the inverses of the variances, to a general sinusoid $c \cdot \sin (a \cdot V+b) +d$ for each TES. This allows us to obtain two estimates for each of the parameters $a$ and $b$, and we take the averages of the two for the final estimates of these parameters. An example of a single fit for one of the TESs for the input state with $\theta_{HWP} = 76.74^\circ$ is shown in Fig.~\ref{fig: ex_piezorampfit}.

There are several potential issues with this procedure. First, the nonlinear least-squares fitting of a sinusoid is not a convex optimization problem and the algorithm can return a bad estimate. Second, the coherent state magnitudes are not constant over a $2^\circ$ phase change period, which affects the accuracy of the Poissonian model. Third, the inverse Fisher Information may not be a very accurate estimate of the variance when the number of data points is only about $348$, which affects the accuracy of the weights assigned to the points in the least-squares procedure. We performed some simulations to investigate the accuracy of the last assumption. In particular, we simulated the maximum likelihood estimation procedure for $348$ datapoints $1000$ times for a set of different magnitude-squares using the Poissonian model to generate the datapoints. We found that the difference between the sample variance (based on the $1000$ estimates) and the inverse Fisher information is generally within $2 \%$.

We keep track of the sum of the squares of the weighed residuals, which should approximately follow a chi-squared distribution. More specifically, this sum is a chi-squared distribution when the distributions of each of the $241$ estimates of the magnitudes are Gaussians with variances equal to the corresponding weights. If the model were exact, we expect the average of the sum of the squares of the weighed residuals to be equal to the number of degrees of freedom of the model. The latter is calculated to be $241-4=237$, where we subtracted the number $4$ to account for the $4$ parameters that define the sinusoid. We find that, very rarely, the optimization algorithm converges to a very bad fit. We therefore use a cutoff value of $310$ for the sum of the squared weighed residuals to reject the fits associated with a higher value of this quantity. The corresponding p-value is $\approx 0.1\%$. For such fits, the next ``piezo ramp" dataset is also discarded. This seems like an unnecessary waste of data, but we find that the fits that are rejected are very few. We found that the observed average chi-squared value had a p-value of around $25\%$ instead of the $50\%$ one would expect from a perfect model. We originally assumed that this was consistent with our expectations, since there is a clear model mismatch as argued above. But, as described in the next section, there is a more profound model mismatch that cannot be explained with these arguments, and which eventually convinced us to abandon the further analysis of the experiment.

At the end of the procedure described above, for a given input and LO state, we obtain a collection of photon-number outcomes associated with phase values for each TES. We divide the phase range into $2^\circ$ consecutive intervals starting from $[0^\circ,2^\circ)$, and group the data of the photon-number outcomes into these intervals according to where their associated phase values fall. We then count the number of outcomes in each phase interval for each TES, and divide these numbers by the total number of outcomes to obtain the corresponding relative frequencies. For the interference experiment with the weakest LO magnitude, we first divide the data into two parts (described in the next subsection) and then generate the relative frequencies for each part. This is necessary to have a separate set of data of photon-number outcomes for estimating the true state (again, described in the next subsection). We make the assumption that the relative LO phase at which the frequencies in a particular phase interval were generated is equal to the value of the phase in the middle of that interval. We do this for several reasons. First, it is expensive to generate the POVM elements, to store them and to use them to construct the functions in the $R\rho R$ algorithm, and it would be practically impossible to use a separate POVM element for each phase value as the number of samples is of the order of $10^7$. Second, this should not have a large effect on the reconstruction, as we expect any two POVM elements with the same photon-number outcomes but differing up to $2^\circ$ in phase to not be too different from each other. This is confirmed by parametric bootstrapping studies described in the next section. Perhaps the best argument for choosing this phase interval is that we want to make use of the order of several dozen LO amplitudes out of which to construct our measurement context, and we want to have a reasonably large number of samples for our POVMs. In this case, we can obtain around $2.5 \cdot 10^7$ samples for each of the measurement contexts we decide to construct.

\subsection{Determining the ``True" Input State}\label{subsec det_true_state}
Here we describe how we use part of the data of photon-number outcomes from the interference experiment with the weakest LO state, as well as the ``input state only" dataset to obtain a ``best guess" of the true input state. For this purpose, we assume that the input state is in a coherent state, and we denote it by $\ket{\psi_\textrm{in}}$. The corresponding truncated (at $N=4$ maximum photon number), normalized and then twirled three-mode state is denoted as $\rho_{true}$. The latter is needed for comparing the best guess with the state obtained by the maximum likelihood reconstruction procedure (described in the next section) using state fidelity as a measure.

Before we describe our estimation procedure, a set of observations and definitions are in order that also apply to the maximum likelihood reconstruction procedure. The first observation is that the LO and the input state may undergo individual phase changes before their interaction in the middle of the FBS. We thus define the input state as the state within its corresponding input branch of the FBS right before the action of the BS. The absolute value of the phase of the input state is not observable, and for convenience we set it to $0$. That is, we choose our reference frame such that the absolute phase is $0$. We similarly define the LO state as the state in its corresponding input branch of the FBS right before the interaction with the input state. With these definitions and choice of the reference frame, and assuming the unitary in Eq.~\ref{eq: B exp} describes the BS action, the relative phase $\phi$ between the LO and the input state is equal to the absolute phase of the LO.

The magnitude of the input state is estimated using the same procedure as for the LO states. Namely, the Poissonian model is used to estimate the magnitudes of the pulses at the outputs of the FBS from the ``input state only" dataset, and the magnitude-square of the input state is computed as the sum of the estimated magnitude-squares at the outputs. It is left to describe how the polarization angle and the temporal overlap with the LO were estimated. Remember that we denoted the polarization axis of the LO by ``vertical". The input state with magnitude $\abs{\alpha}$ can be written as
\begin{align}\label{eq: inputstate_exp}
\ket{\psi_{\textrm{in}}} = \ket{\abs{\alpha} \cos(\theta_p) \cos(\theta_t)}_1 \otimes  \ket{\abs{\alpha} \cos(\theta_p) \sin(\theta_t)}_2 \otimes \ket{\abs{\alpha} \sin(\theta_p) }_3,
\end{align}
where the subscripts of the kets denote the corresponding mode number, $\theta_p$ is the polarization angle with respect to the LO polarization, and $\theta_t$ signifies the temporal overlap between the modes of the LO and the modes of the input state in the vertical polarization. With this parametrization, the LO state is written as $\ket{\gamma}_1 \otimes \ket{0}_2 \otimes \ket{0}_3$. We omit the tensor products in our descriptions in the rest of this section for brevity.

After the action of the FBS, the states at the output ports of the FBS are given by
\begin{align}
\ket{\psi_{\textrm{out},1}} = \ket{  B_{1,11} \abs{\alpha} \cos(\theta_p) \cos(\theta_t) + B_{1,12} \gamma }_1  \ket{ B_{1,11}  \abs{\alpha} \cos(\theta_p) \sin(\theta_t)  }_2  \ket{ B_{2,11} \abs{\alpha} \sin(\theta_p)   }_3,
\end{align}
and 
\begin{align}
\ket{\psi_{\textrm{out},2}} = \ket{  B_{1,21} \abs{\alpha} \cos(\theta_p) \cos(\theta_t) + B_{1,22} \gamma }_1  \ket{ B_{1,21}  \abs{\alpha} \cos(\theta_p) \sin(\theta_t)  }_2  \ket{ B_{2,21} \abs{\alpha} \sin(\theta_p)   }_3,
\end{align}
where $B_1 =  \begin{pmatrix}  \sqrt{0.4946}   &     \sqrt{0.5054}  \\ -\sqrt{0.5054}   &   \sqrt{0.4946}   \end{pmatrix}   $ and $B_2 =  \begin{pmatrix}  \sqrt{0.2642}   &     \sqrt{0.7358}  \\ -\sqrt{0.7358}   &   \sqrt{0.2642}   \end{pmatrix} $ according to Eq.~\ref{eq: B exp}. For the ``input state only" datasets, the LO amplitude is zero, and the magnitudes of $ \ket{\psi_{\textrm{out},1}}$ and $ \ket{\psi_{\textrm{out},2}}$ are constant. We estimate $\theta_p$ from the estimates of these magnitudes (which are obtained using the Poissonian model). In particular, we obtain a single equation with single unknown ($\theta_p$) from the data of each TES, which we solve and take their average as the estimate of our phase. These equations are
\begin{align}
\textrm{Mag}^2_1 &= \abs{  B_{1,11} \abs{\alpha} \cos(\theta_p)}^2 + \abs{  B_{2,11} \abs{\alpha} \sin(\theta_p) }^2 \nonumber \\
& = \abs{\alpha}^2 (   B_{1,11}	^2 \cos^2(\theta_p)+ B_{2,11}^2 \sin^2(\theta_p) )
\end{align}
and
\begin{align}
\textrm{Mag}^2_2 &= \abs{  B_{1,21} \abs{\alpha} \cos(\theta_p)}^2 + \abs{  B_{2,21} \abs{\alpha} \sin(\theta_p)}^2 \nonumber \\
& = \abs{\alpha}^2 (   B_{1,21}^2 \cos^2(\theta_p)+ B_{2,21}^2 \sin^2(\theta_p) ),
\end{align}
where $\textrm{Mag}^2_1$ and $\textrm{Mag}^2_2$ stand for the squares of the magnitudes for counters 1 and 2, respectively. Note that we also have an estimate of $\theta_p$ from the measured HWP angle, but we find that these do not match too closely, which was another factor in the decision to terminate the data analysis. We detail this at the end of the next section. 

To estimate the temporal overlap $\abs{ \cos (\theta_t)}^2$ between the LO and the part of the input state in the vertical polarization we use the first $1/12$'th of the photon-number data generated from the interference experiment with the weakest LO state to estimate $\theta_t$. The rest of the data generated from all three interference experiments are used for reconstructing the three-mode twirled state using the maximum likelihood estimation procedure. Now, the squares of the magnitudes at the outputs of the FBS are functions of $\theta_t$. In particular,
\begin{align}
\textrm{Mag}^2_1 &= \abs{  B_{1,11} \abs{\alpha} \cos(\theta_p) \cos(\theta_t) + B_{1,12} \gamma }^2 + \abs{  B_{1,11} \abs{\alpha} \cos(\theta_p)\sin(\theta_t) }^2 + \abs{  B_{2,11} \abs{\alpha} \sin(\theta_p)}^2,
\end{align}
and
\begin{align}
\textrm{Mag}^2_2   &= \abs{  B_{1,21} \abs{\alpha} \cos(\theta_p) \cos(\theta_t) + B_{1,22} \gamma }^2 + \abs{  B_{1,21} \abs{\alpha} \cos(\theta_p)\sin(\theta_t) }^2 + \abs{  B_{2,21} \abs{\alpha} \sin(\theta_p)}^2.
\end{align}
Notice that the dependence on the phase of the LO is due to the appearance of $\gamma$ in the first summands. We remind the reader that we use roughly half of the data from the intereference experiment for each LO state to generate a set of samples of photon-number for both TESs associated with $180$ equally spaced different values of the relative phase. We calculate the relative frequencies from the photon-number data obtained with this procedure from the first $1/12$'th duration of the experiment. We use the Poissonian model on each set of frequencies to obtain an estimate of the magnitude of the coherent state at both outputs of the FBS for all phase points. We then perform nonlinear least squares optimization to fit the equations above to the squares of the magnitudes, and obtain two estimates of $\theta_t$. We take their average as our final estimate.

We should note that the procedure for estimating $\rho_{true}$ is fairly involved, uses multiple datasets, relies on the accuracy of the assumed BS coefficients and on detector tomography, and on the procedure for the estimation of the relative phase corresponding to the photon-number outcomes. Thus, our ``best guess" of the input state cannot be assumed to be a good guess unless we are sure that the steps leading to it have high accuracy. As we show later, we have good evidence to believe that this is not so. Therefore, the analysis comparing the twirled state estimate from the maximum likelihood reconstruction with $\rho_{true}$, shown in the next section, should not be treated as tests of the performance of the reconstruction process. 

Our analysis was mostly restricted to the input state associated with the HWP setting $\theta_{HWP} = 76.74^\circ$. The corresponding $\ket{\psi_{\textrm{in}}}$ estimate has magnitude $\abs{\alpha} =  0.8324$, polarization angle $\cos(\theta_p) =0.7388 $ (or $\theta_p = 42.37^\circ$) and overlap angle $\cos(\theta_t) = 0.7138$. For comparison, the polarization angle $\theta_p$ extracted from $\theta_{HWP}$ is $43.88^\circ$, which differs from the estimate that we trust more by almost $1.5^\circ$. Of course, there is significant uncertainty in these numbers, which is mostly due to the error propagation from the uncertainties in the BS coefficients, in the operators in Eq.~\ref{eq T_i'} that are used to estimate the magnitude in each output, in the LO magnitude and phase, which are all used to obtain $\ket{\psi_{\textrm{in}}}$. We truncate $\ket{\psi_{\textrm{in}}}$ at $N=4$ maximal photon-number, followed by normalization. The fidelity between the truncated state $\rho_{true}$ and $\ket{\psi_{\textrm{in}}}$ is given by Eq.~\ref{eq fid_coh_truncoh}, which in our case equals $0.9992$. We describe the ``true state" estimates of the other $4$ input states at the end of the next section in the context of the discussion of the discrepancies that caused us to terminate the analysis.

\section{State Reconstruction Analysis}\label{sec desc_recon}

%Notes to self: 
%
%- The TES trace analysis was done in the first half of 2020 roughly
%
%- concurrently, I was doing analysis on the ``old" data and that is pretty much all I have.
%
%- I wrote ``script of procedure" in early 2021, but I performed the steps manually in the first half of 2020 to obtain the results on the ``old" data. During this the old POVM bounds and matrices were used. The latter are 10 x 26.
%
%- Thomas uploaded a small chunk of ``new" data together with updated POVMs (6 x 26 in size matrices ) and the new bounds around Nov 2020.
%
%- I did perform analysis on the ``new" data, but it seems I did that in early 2021 (probably waiting for my computer and being busy with Comps 3). But it seems the only analysis I did was to check whether the discrepancy of the interference plot was fixed since I had access to only a small chunk of the data, and we concluded it wasn't, and thus there was no need for Thomas to upload the rest of the ``new" data. So we stopped working on the analysis after Manny's email in march.
%
%- But Manny's email followed my email on the inconsistencies between the BS coefficients and HWP angles and the temporal overlaps, not really the phase dependence in the total magnitude during the interference experiments - though clearly both played a factor in this decision. 

In this section we present the results of reconstruction of the three-mode twirled state $\tilde \rho_3$ using maximum likelihood estimation. We first describe how we constructed IC measurement contexts, and then describe and compare the results of the reconstruction with the relative frequencies associated with the different measurement contexts. We spend significant time investigating the extent and causes of model mismatch, and argue towards the end why we decided to abandon the further analysis of the experiment.

\subsection{Construction of Measurement Contexts}

According to our strategy of binning the experimental outcomes according to the corresponding phase of the LO, there are three LO magnitudes and $180$ phases for each magnitude to construct POVM elements from. For a WFH configuration with a BS that is characterized by partition size $K=2$ and $S_1>1$, according to Thm.~\ref{thm:2CntrNbnd_K=2} and Prop.~\ref{cor:1CntrNbnd}, $(N+1)^2 =25$ different LO amplitudes are sufficient to construct an IC measurement context. We numerically find (using the procedure described in Sec.~\ref{subsec POVMs_ideal}) that even a smaller number is sufficient, but when only $3$ LO magnitudes are available, the number of phases associated with each LO magnitude must be at least $10$. This is, of course, conditioned on our choice to use a maximum photon number of four. Let us denote the three magnitudes by $R_1$, $R_2$ and $R_3$ by the increasing order of their values. For the input state with $\theta_{HWP} = 76.74^\circ$ these were estimated to be $R_1 =1.082$, $R_2 = 1.286$ and $R_3 = 1.531$. To have an equal representation across the magnitudes, we decided to use $10$ different phase values for each magnitude to construct $18$ different measurement contexts that have mutually distinct POVM elements. We perform reconstruction with each measurement context to observe the sensitivity of the estimation procedure to different datasets.

The $18$ different IC measurement contexts were constructed as follows. Let us refer to these measurement contexts by a number ranging from $1$ to $18$. The measurement context $\#1$ is constructed from the following LO amplitudes: The phase values $ \{(1+36i)^\circ \mid i=0,1,\hdots,9\} $ are used for the magnitude $R_1$, while for the magnitudes $R_2$ and $R_3$ we use the phase values  $\{(13+36i)^\circ \mid i=0,1,\hdots,9\}$ and $ \{(25+36i)^\circ \mid i=0,1,\hdots,9\}$, respectively. So the phase values for the different magnitudes are mutually different, and the set of phase values for each magnitude are equidistantly distributed along the whole phase range. The other $17$ measurement contexts are constructed similarly: the phase values for $R_1$ are chosen to be $ \{(n+36i)^\circ \mid i=0,1,\hdots,9\} $, for $R_2$ they are $\{(n+12+36i)^\circ \mid i=0,1,\hdots,9\}$, and for $R_3$ they are $ \{(n+24+36i)^\circ \mid i=0,1,\hdots,9\}$ where $n \in \{3,5,\hdots,35\}$ and corresponds to the the measurement context $\# (n+1)/2 $. The reader can verify that these $18$ measurement contexts are built with mutually different LO amplitudes. Further, this choice of measurement contexts insures that all samples that were generated from the interference experiments are used in our analysis.

\subsection{Reconstruction With Different Measurement Contexts}
For the reconstruction we use the diluted $R\rho R$ algorithm described in the last chapter. As a reminder, this algorithm starts iterating with the $R\rho R$ algorithm until either a stopping condition is met or the increase in the value of the logarithmic likelihood during the iteration is less than some predetermined value $\Delta \cL$. If the latter, the algorithm transitions to a diluted $R\rho R$ regime, with the value of the initial dilution parameter $\epsilon$ being $10^{30}$. If the stopping condition is not met and the change in logarithmic likelihood during an iteration becomes less than $\Delta \cL$, the value of $\epsilon$ is halved. This continues until either the stopping condition is met at some point of the iterations, or the value of $\epsilon$ reaches $10^{-30}$ at which point we terminate the iterations. The stopping condition is based on calculating the maximum eigenvalue of $\hat{R}$ (Eq.~\ref{eq R_expression}) and terminating the algorithm if it is less than $1+\texttt{r}$ where $\texttt{r}$ is decided beforehand, as described in the next paragraph. The number of samples for each measurement context averages to around $26.5 \cdot 10^6$ which are almost equally divided among the $30$ different LO amplitudes. Our simulations with coherent states in the previous chapter suggest that if the experiment perfectly matched the model, we should expect such a sample size to result in a value of fidelity $\geq 0.999$ between the estimate and $\rho_{true}$ if the estimate is close enough to the state that maximizes the likelihood. Note that $\rho_{true}$ is truncated, but the state that generated the experimental data, $\ket{\psi_{\textrm{in}}}$, was not truncated, which is different from the simulations where we used the truncated state to simulate data. However, since the fidelity between $\ket{\psi_{\textrm{in}}}$ and $\rho_{true}$ is $\geq 0.9992$, the fidelity between the estimate and $\rho_{true}$ should still be near $ 0.999$ if there is no model mismatch.

In our analysis we decided to set $\Delta \cL$ to zero, and used a value of $\texttt{r}=200/M$, where $M$ is the number of samples, for the stopping condition. Our simulations in the last chapter suggest that this should result in as good of a performance as we can expect from our algorithm. The value of $\texttt{r}$ was chosen to terminate the reconstruction in a reasonable time (around one hour). Some experimentation with smaller values of $\texttt{r}$ showed that the termination time of the algorithm would increase substantially with little gain in the performance. In particular, we found that the value of $\texttt{r}=1/M$ which we commonly employed with our simulations in the last chapter did not result in a termination of the algorithm even after several days. We did not observe such an issue with our simulations. One likely reason is that all our simulations were with a WFH configuration with a BS characterized by a trivial partition, which required us to encode our variables and perform computations on matrices in $\cR_{2,N}^s$. In contrast, we encode our state and perform computation on matrices in $\cR_{3,N}^s$ which have an order $N$ times more parameters - increasing the computation time and accumulated error in each iteration step. 

\begin{figure}[!htb]
    \centering
\includegraphics[width=160mm]{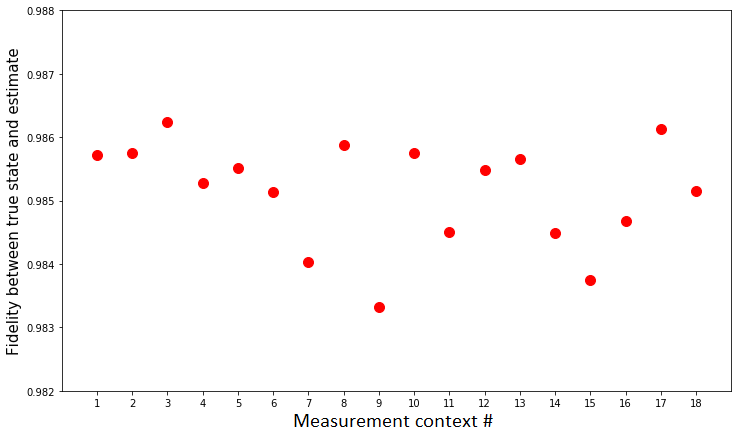}
    \caption{The fidelity between the estimate and $\rho_{true}$ for each of the $18$ measurement contexts. The HWP angle of the input state is $\theta_{HWP} = 76.74^\circ$. }
    \label{fig: fid_true_vs_est_7476Deg}
\end{figure}

\begin{figure}[!htb]
    \centering
\includegraphics[width=140mm]{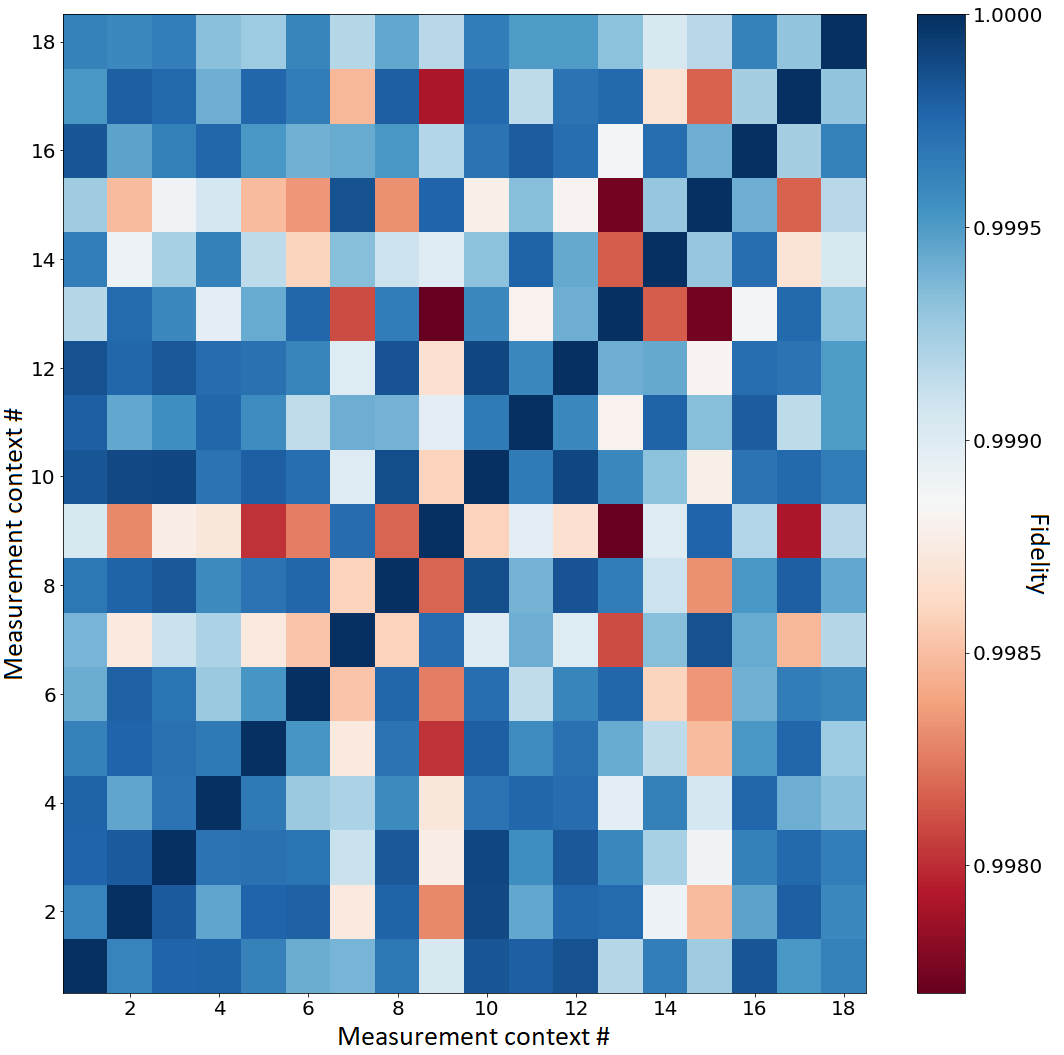}
    \caption{The fidelities between the $18$ estimates. The HWP angle of the input state is $\theta_{HWP} = 76.74^\circ$. }
    \label{fig: cross_fid_7674Deg}
\end{figure}

The fidelity between each of the $18$ estimates and $\rho_{true}$ is shown in Fig~\ref{fig: fid_true_vs_est_7476Deg}. It can be seen that the fidelities are consistently within $0.985 \pm 0.002 $. While these are not bad results, they fall short of the values (consistently around $0.999$) we observed during our simulations with coherent states in the last chapter. This is expected because, as we argued towards the end of Subsec.~\ref{subsec det_true_state}, $\rho_{true}$ cannot be taken to be an accurate guess of the true state. The fact that the fidelities observed for all $18$ estimates are similar, even though the data they were reconstructed from are mutually distinct, suggests that the estimates are close to each other. Indeed, comparing the estimates with each other, we find that the fidelity between any two estimates is above $0.9975$ and there are pairs which have a fidelity well above $0.9995$ (Fig.~\ref{fig: cross_fid_7674Deg}). These facts suggest that we are not handicapped by statistical noise or convergence issues in the reconstruction algorithm. Rather, we are observing a model mismatch, such that the frequencies generated from the data for all phase values and magnitudes of the LO are not represented very well by the total photon statistics at the outputs of the FBS resulting from the interference of $\rho_{true}$ with the LO by the FBS with the coefficients we assume in our model. In the next subsection we aim to investigate the degree of the model mismatch and where the model errs significantly.

\subsection{Studies With Parametric Bootstrapping}

One common method to inspect the validity of a statistical model, is to perform parametric bootstrapping. Parametric bootstrapping uses the estimate obtained from a particular measurement context and associated experimental frequencies to calculate the probabilities given a model and employs a random number generator to generate samples from these probabilities. The number of generated samples is equal to the number of samples generated by the experiment, and the bootstrapped estimate is obtained from these samples by the same procedure that was used to obtain the original estimate from the experimental samples. Ideally, this is done many times to obtain a number of bootstrapped estimates and thus to get an idea about the distribution of the estimator when the true state is the original estimate. Bootstrapping can be used to generate confidence intervals for the estimate, but if there is significant model mismatch, the confidence intervals may have coverage probabilities that are different from their nominal confidence levels. 

\begin{figure}[!htb]
    \centering
\includegraphics[width=160mm]{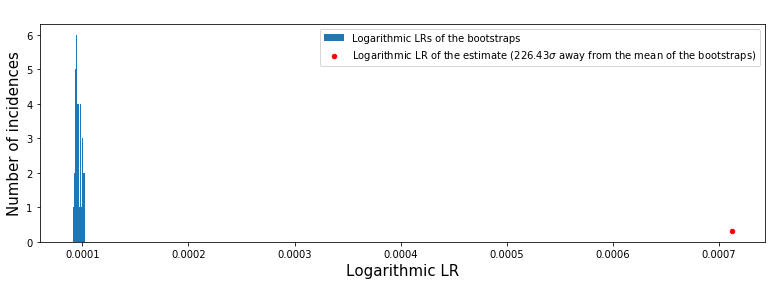}
    \caption{The histogram of the logarithmic likelihood ratio (LR) of the bootstraps with the logarithmic LR of the estimate shown as a dot for illustration. $60$ bootstraps were generated. Further details are in the text.}
    \label{fig: hist like_ratio_boots}
\end{figure}

We use parametric bootstrapping to calculate and compare likelihood ratios (LRs). In particular, for the original estimate and for each bootstrap we have the final value of $\cL$ as well as the logarithmic likelihood of the unconstrained model (that is, where the probabilities are taken to be equal to the frequencies).  We compare the LRs of the bootstraps with the LR of the estimate. If the model is accurate, the sample size is large enough, and the estimate is close to the state that maximizes $\cL$, we expect the difference between the LRs of the bootstraps and the LR of the estimate to be near zero. A histogram based on $60$ bootstraps generated for the estimate obtained using the measurement context $\#1$ is shown in Fig.~\ref{fig: hist like_ratio_boots}, where the logarithmic LR is defined as $-2(\cL-\cL_u)$ with $\cL_u $ being the logarithmic likelihood of the unconstrained model. The logarithmic LR of the estimate is $226$ times the sample standard deviation of the bootstrap values away from the mean of the bootstraps.

We performed further bootstrapping studies to be more confident that we are observing model mismatch. We simulated the experimental model and the bootstrapping procedure described above, with the ``true state" being the estimate obtained using the measurement context $\#1$ and the corresponding frequencies. Namely, we used this ``true state" to generate samples by a random number generator according to the measurement context $\#1$ and then ran the reconstruction algorithm on these frequencies to generate an estimate. We then used this estimate to generate $12$ bootstraps and calculated the logarithmic LR for each of them. The logarithmic LR value for the estimate lies within the range of the logarithmic LR values of the bootstraps - in sharp contrast with the $226.43 \sigma$ deviation observed in the analysis of the experimental data. 

We tried to investigate how much the various approximations and assumptions used in the data generation procedure might contribute to the model mismatch. In particular, one approximation we make is binning the outcomes of the TESs into $2^\circ$ phase bins, and assigning the LO amplitude with the mean value of the bin to the outcomes. To get an idea about the impact of this binning procedure when the model matches the process of data generation, we again started with the estimate obtained from the experimental data using the measurement context $\#1$. In contrast to the simulations described in the previous paragraph, we used a measurement context consisting of $10$ times more LO phases to generate the samples. In particular, we chose $10$ equally spaced phase values falling within $1^\circ$ in each direction of each LO phase appearing in the measurement context $\#1$. We generated samples using this measurement context followed by binning the samples into $2^\circ$ phase bins as in the generation of experimental data, and used the measurement context $\#1$ for reconstruction. We then produced $24$ bootstraps from the estimate using the measurement context $\#1$ for generation of frequencies and looked at the logarithmic LR of the estimate and the bootstraps. We again found that the logarithmic LR of the estimate falls in the range of those of the bootstraps, indicating that the effect of phase binning on the estimator is of the order of the statistical noise from an order of $10^7$ samples. 

We report yet another investigation we did with phase binning. In the reconstruction procedure from the experimental data, instead of using the measurement context $\#1$, we used slightly different POVM elements for each LO amplitude. In particular, the POVM element corresponding to the observation of $k$ and $l$ photons from a particular phase bin was constructed as an equally weighed average of operators corresponding to the same photon number observations but associated with different phases in the phase bin. To clarify, the new POVM elements are the numerical integrals of the precise-phase POVM elements in the corresponding phase intervals. We used $10$ equally spaced phase values for each phase interval for the numerical integration. We used this modified measurement context to generate the estimate from the experimental data, followed by parametric bootstrapping with the same model using $55$ bootstraps. We again observed $\approx 240\sigma$ deviation in the logarithmic LR of the estimate from the mean of the bootstraps. 

We conclude that the model mismatch we observe when comparing the LRs of the bootstraps with the LR of the estimate obtained from the experimental data cannot be due to assigning slightly different phase values to the samples and using a measurement context associated with the corresponding LO phases. The sensitivity of the statistical distribution of logarithmic LRs to such deviations in the POVMs is negligible when the sample size is $\approx 2.7\cdot10^7$.

\subsection{Investigation of the Discrepancies}

This is a good place to recount which of the parameters that we use in our data analysis are given to us from measurements preceding the experiment (that we take as being their true values), and which of the parameters that we use in the reconstruction are estimated from the experimental data. The parameters we assume are true are the reflection and transmission coefficients of the BS for each polarization, the matrices $T_1$ and $T_2$, the transmission coefficients in the couplings of the FBS outputs to the fibers leading to the TESs which we incorporate into $T_1$ and $T_2$ producing $T'_1$ and $T'_2$ as in Eq.~\ref{eq T_i'}, and the bounds for the TESs (used for assigning photon numbers to the outcomes) up to a scaling factor. The rest of the parameters that are used in the data analysis - namely, the scaling factors for the bounds, and the magnitudes and phases of the LO, are estimated from the ``LO state only" datasets and from a part of the datasets of the interference experiments using the parameters above. 

While it is not clear how/if the bounds for the TESs or the matrices $T'_1$ and $T'_2$ can be double-checked using the available data, the BS coefficients corresponding to the polarization of the LO are straightforward to verify (assuming the bounds and the matrices $T'_1$ and $T'_2$ are exact) from the ``LO state only" datasets using the Poissonian model for both TESs. There are $15$ such datasets - one for each of the $3$ LO magnitudes for each of the $5$ different input states. The estimates for the splitting ratio from these datasets are in a tight range from $48.38:51.62 $ to $48.61:51.39 $, while the corresponding splitting ratio used in the model is $49.46:50.54$ - a $1.9\%$ difference from the average of the estimates. 

A stronger disagreement between the value of a quantity measured before the experiment and our estimate of that quantity based on the experimental data pertains to the temporal overlap $\theta_t$ in the input state. In particular, before each interference experiment in each day, the temporal overlap between the input state and the LO was maximized by measuring the visibility. A temporal overlap of $\geq 95\%$, corresponding to $\cos(\theta_t) \geq 0.974$, for each input state was reported. In contrast, we found by the procedure described in Subsec.~\ref{subsec det_true_state} that the values of $\cos(\theta_t)$ were well below that number for all input states. These numbers, along with the other parameters that describe the estimate of the true state, are detailed in Table~\ref{table: true_states} for completeness. The value of the temporal overlap reported is obtained according to the procedure described in Subsec.~\ref{subsec det_true_state}. Another thing to note is that the values of the temporal overlap according to the two counters were found to be somewhat different from each other, and this difference became more pronounced for larger $\theta_p$ values. In particular, for the input state with the HWP setting $\theta_{HWP} = 90.66^\circ$, which has the highest value of $\theta_p$, the temporal overlaps were $33.09\%$ and $38.63\%$ according to the two different TESs.

\begin{table}[!htb]
\centering
\renewcommand{\arraystretch}{1.5}
\begin{tabular}{|| p{3.5cm} | l | l | p{3cm} | l |l ||} 
 \hline
Input state (HWP setting)   &	  $\abs{\alpha}$	  &  $\theta_p$ & $\theta_p$ according to $\theta_{HWP}$ & $\cos(\theta_t)$ & Temporal overlap \\ [0.5ex] 
 \hline\hline
$54.8^\circ$ &  0.8145 & $0^\circ$ & $0^\circ$ & 0.8237 & $67.85 \%$ \\
\hline
$64.8^\circ$ & 0.8080 & $15.97^\circ$ & $20^\circ$ & 0.7897 & $62.36\%$ \\
\hline
$76.74^\circ$ & 0.8324 & $42.37^\circ$ & $43.88^\circ$ & 0.7138 & $50.95\%$ \\
\hline
$84.59^\circ$ & 0.8337 & $58.66^\circ$ & $59.58^\circ$ & 0.6853 & $46.97\%$ \\
\hline
$90.66^\circ$ & 0.8289 & $71.05^\circ$ & $71.72^\circ$ & 0.5986 & $35.84\%$ \\ [1ex] 
\hline
\end{tabular}
\caption{The estimates of the input states using the method described in Subsec.~\ref{subsec det_true_state}. $\abs{\alpha}$, $\theta_p $ and $\theta_t$ form the input state according to Eq.~\ref{eq: inputstate_exp}. The temporal overlap is defined as $\cos^2(\theta_t)\cdot 100\%$.} 
\label{table: true_states}
\end{table}

One useful test we performed to investigate the nature of the model mismatch was to look at the sum of the magnitude-squares estimated at the outputs of the FBS for different values of the relative phase during the interference experiment. In particular, if our model of the experiment is correct, we expect the interference fringes to combine to form a flat horizontal line. Remember that we used the data of the TESs during one acquisition window to estimate the relationship between the voltage across the piezo and the relative phase between the LO and the input state. We then used this estimate to assign a relative phase to the outcomes of the TESs during the consequent acquisition window and binned the outcomes into $\approx 2^\circ$ phase bins according to their corresponding phase value. We performed this for the whole dataset of each interference experiment for the input state with $\theta_{HWP} = 76.74^\circ$. Thus, the magnitude-squares at the outputs of the FBS for different values of the relative phase can be estimated using the Poissonian model on the samples for each phase bin. We can do this for the samples extracted from the whole dataset as well as for the samples extracted from a single acquisition window. We show the scatter plots of the estimated magnitude-squares in each phase bin for both TESs, as well as their sum at each phase value, for a sample size coming from $8$ acquisition windows of the interference experiment with the weakest LO magnitude in Fig.~\ref{fig: intens_scatters}a. The next subfigure (Fig.~\ref{fig: intens_scatters}b) is based on the whole set of samples extracted from the same interference experiment. It can be seen that the combination of the magnitude=squares exhibits a clear sinusoidal pattern across the phase values. The relative height of this sinusoid is $\approx 2\%$ of the mean of the sum of the magnitude-squares. Similar results are observable for the other two LO magnitudes.

\begin{figure}[ht!]
    \centering
\includegraphics[width=170mm]{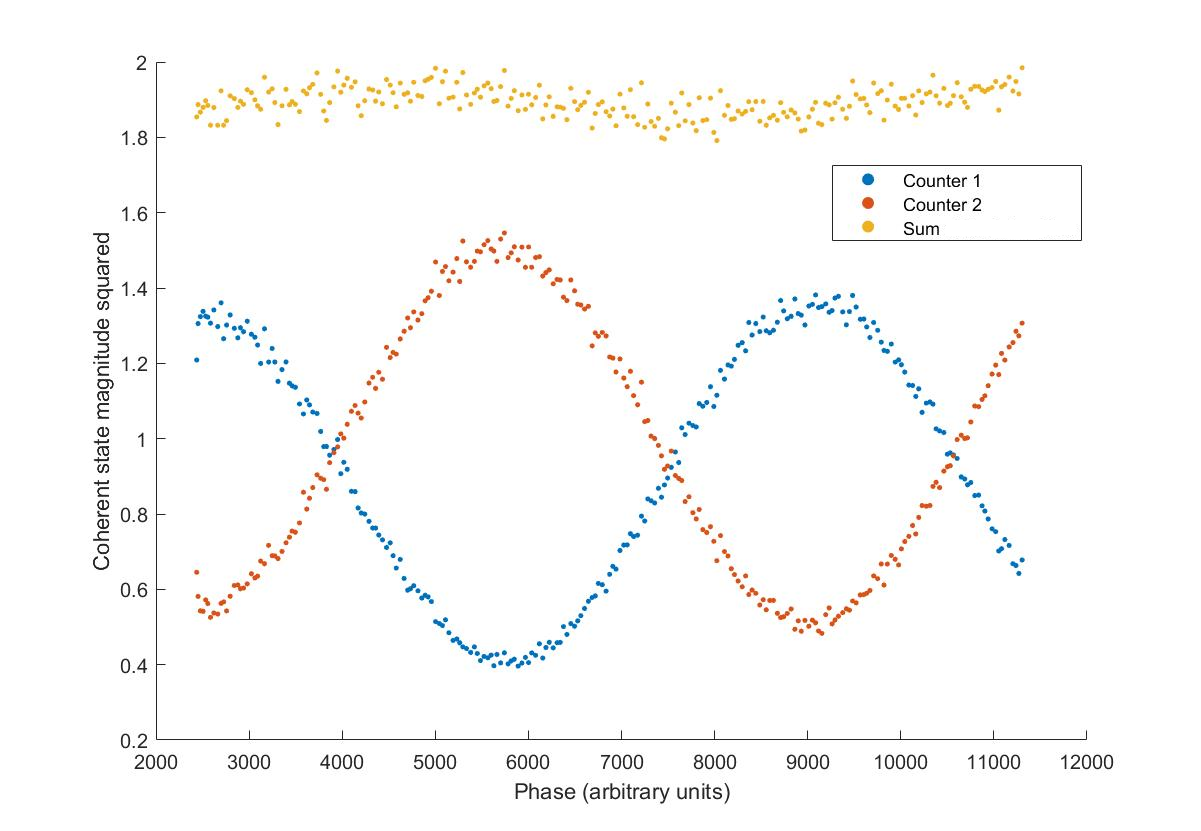}
    \caption{(a) The estimates of the magnitude-squares at the outputs of the FBS and their sum using samples from $8$ acquisition windows of an interference experiment. The input state is $\theta_{HWP} = 76.74^\circ$ and the LO has magnitude $1.082$.}
  \label{fig: intens_scatters}.
\end{figure}
\begin{figure}[ht!]\ContinuedFloat
\captionsetup{list=no}
    \centering
\includegraphics[width=175mm]{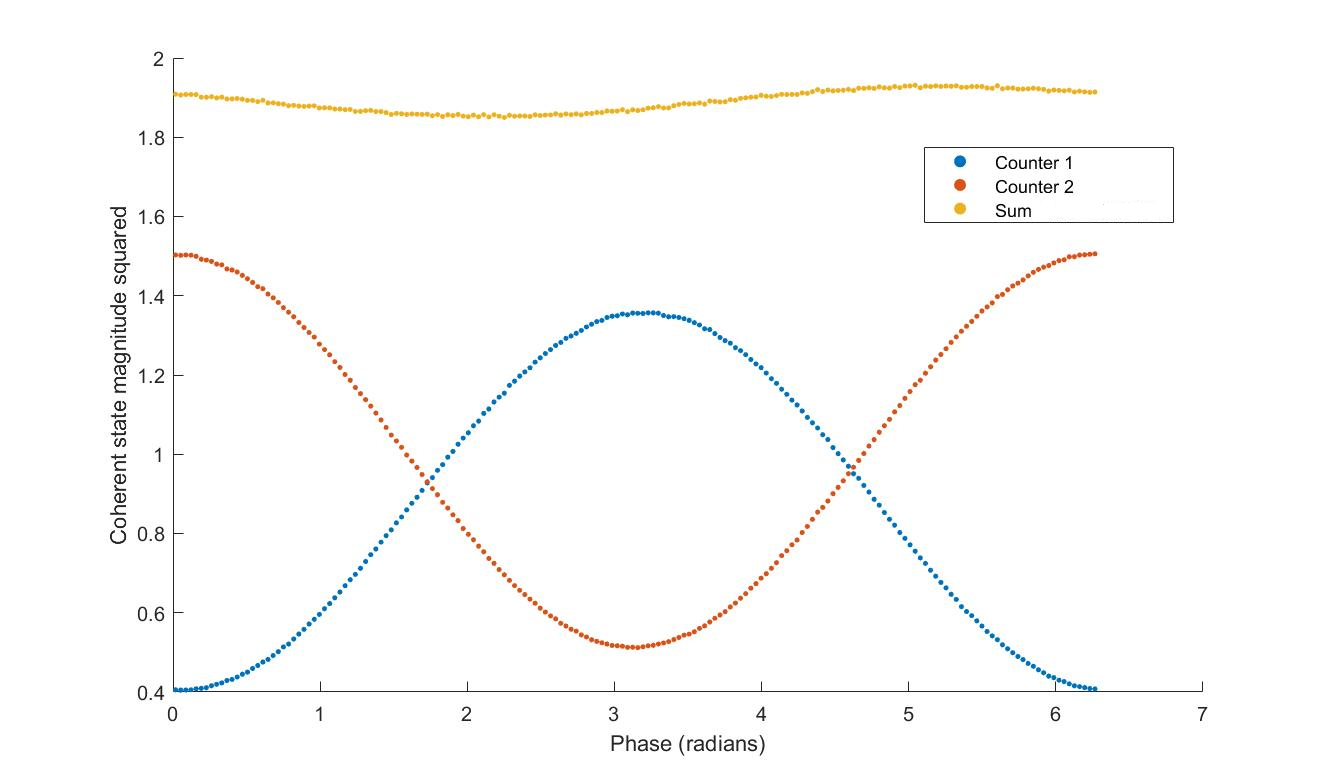}
    \caption{(b) The estimates of the magnitude-squares at the outputs of the FBS and their sum using all the samples extracted from the data of an interference experiment. The input state is $\theta_{HWP} = 76.74^\circ$ and the LO has magnitude $1.082$. (cont.)}
\end{figure}

One culprit could be the memory effect in the TESs. In particular, it was established that the TES does not fully reset during the $10 \textrm{ }\mu\textrm{s}$ time interval between the shots. This implies that the initial conditions of the TES signal at the beginning of the absorption of the following pulse are slightly higher than the fully relaxed conditions, which causes the MFV estimate to be slightly higher than it would be otherwise. This in turn can cause our estimates of the coherent state amplitudes using the Poissonian model to be higher than they are. If the memory effect is stronger for the larger magnitudes, it is plausible that this could cause an apparent phase shift in the interference fringes. We worked on methods to filter out this bias in the final MFV estimate, and the experimental group redid the MFV extraction analysis from the signals of the TESs using this improved method for a part of the experimental data for each input state. In fact, the estimated parameters for the ``true state" reported in Table~\ref{table: true_states} for all input states besides the one with $\theta_{HWP} = 74.74^\circ$ were extracted using the new datasets. However, this did not appear to affect the dependence of the sum of the magnitude-squares on relative phase, and a similar degree of discrepancy was still observable.

The shift in the phase between the sinusoids corresponding to the two TESs seems to be around $4^\circ$ shy of $180^\circ$. It is interesting to explore how much the model mismatch would express itself in the value of the LR if the phases assigned to POVMs are all shifted in the same direction from the correct values by $4^\circ$. In particular, we perform a bootstrapping simulation, starting with the estimate produced using the measurement context $\#1$ as our true state. We use the measurement context $\#1$ to generate random samples but use a modified version of the measurement context $\#1$ with all LO phases shifted by $4^\circ$ in the same direction to perform reconstruction. We generate $24$ bootstraps from the resulting estimate, but using the correct LO phases. We find that the deviation of the logarithmic LR of the estimate is $\approx 2.7 \sigma$ from the mean of the bootstraps. So it seems that the large model mismatch that we observe in the reconstruction from the experimental data cannot be explained by assuming that for whatever reason there is a systemic error in the assignment of the relative phase to the outcomes of the TESs.

These discrepancies, and our inability to explain them and, ideally, to incorporate them in our model, forced us to abandon the data analysis of the experiment. Currently, there is an ongoing effort to set up and conduct a similar experiment to demonstrate the theory developed in this thesis, and care is taken to utilize the lessons from our failures to account for many possible sources of uncertainty that affect the data analysis.

%% file: appendix_opordering.tex
\chapter{A Result About Anti-Normally Ordered Operators}\label{app antinormord}

Here we derive a result about the relationship of the anti-normally ordered powers of the total number operator and its usual powers. We could not find this result in the literature, and since we derive a similar result for the normally-ordered case, it seems like a good opportunity to include this result in this thesis as well.

We use the notation introduced in Sec.~\ref{qo order}. Let us introduce the anti-normal ordering transformation on expressions of mode variables as the map that rearranges the terms in each monomial by moving the variables representing creation operators to the right of the monomial. We use triple vertical dots to represent anti-normal ordering - namely, if $P$ is a polynomial of mode variables then $\anormord{P}$ is the corresponding anti-normally ordered operator. A calculation similar to \ref{eq:monexprcov} shows that anti-normal ordering commutes with PLTs. In particular:
\begin{align}\label{eq:polyanormPLT}
 U^{\dagger}\reallywidehat{\anormord{P(\vec{\boldsymbol{a}},\vec{\boldsymbol{a}^{\dagger}})}}U
   &=\reallywidehat{\anormord{P(U_M\vec{\boldsymbol{a}},U^*_M\vec{\boldsymbol{a}^{\dagger})}}}.
\end{align}

\begin{lemma}
  \begin{align}
    \reallywidehat{\anormord{\boldsymbol{n}^{k}}} &=(\hat n+(S-1) + k)_{k},
  \end{align}
\end{lemma}

\begin{proof}
  The map $P(\boldsymbol{a}, \boldsymbol{a}^{\dagger}) \mapsto
  \reallywidehat{\anormord{P(\boldsymbol{a}, \boldsymbol{a}^{\dagger})}}$ is an intertwiner of the group of PLTs, and
  $\boldsymbol{n}^{k}$ is invariant under the PLT action on polynomial
  expressions. It follows that $\reallywidehat{\anormord{\boldsymbol{n}^{k}}}$ is also invariant under the action of PLTs, which in turn means (according to Cor.~\ref{cor PLTinvpolynum}) that it
  is a polynomial of the total number operator. We use induction to obtain our proof. One can verify the
  formula in the lemma for $S=1$ by direct calculation:
  \begin{align}
a_{1}^{k} (a_{1}^{\dagger})^{k}&=   \sum_{m=0}^{\infty} (m+1)\cdots(m+k)\ket{m}\bra{m} \nonumber \\
 & = \left[ \sum_{m_1 =0}^{\infty} (m_1+1) \ket{m_1}\bra{m_1}\right]\left[  \sum_{m_2 =0}^{\infty} (m_2+2) \ket{m_2}\bra{m_2}  \right]\cdots \left[  \sum_{m_k =0}^{\infty} (m_k+k) \ket{m_k}\bra{m_k}  \right] \nonumber \\
 &=  (a_{1}^{\dagger}a_{1}+1)(a_{1}^{\dagger}a_{1}+2)\cdots (a_{1}^{\dagger}a_{1}+k) =  (a_{1}^{\dagger}a_{1} +k)_{k}.
\end{align}
  
   For general
  $S>1$, let $\boldsymbol{n'}$ be the expression for the total number
  operator on the first $S-1$ modes and $\boldsymbol{n}_{S}$ the expression for the
  mode $S$ number operator. Then, assuming the formula holds for the first $S-1$ modes, one obtains
  \begin{align}\label{lem_antiord_eq1}
     \reallywidehat{\anormord{\boldsymbol{n}^{k}}}
          &=    
          \reallywidehat{\anormord{\sum_{j=0}^{k}\binom{k}{j}          \boldsymbol{n'}^{j} \boldsymbol{n}_{S}^{k-j}}}\nonumber\\
          &= 
          \sum_{j=0}^{k}\binom{k}{j}\reallywidehat{(\anormord{\boldsymbol{n'}^{j}}\;
            \anormord{\boldsymbol{n}_{S}^{k-j}})}\nonumber\\
          &=
          \sum_{j=0}^{k}\binom{k}{j}\reallywidehat{\anormord{\boldsymbol{n'}^{j}}}\;
            \reallywidehat{\anormord{\boldsymbol{n}_{S}^{k-j}}}\nonumber\\
          &=
          \sum_{j=0}^{k}\binom{k}{j}(\hat n'+S-2+j)_{j}
            (\hat n_{S}+(k-j))_{k-j}.
   \end{align}
   We want to show that the last line in the above is equal to $(\hat{n}+S-1+k)_k$. Since $\reallywidehat{\anormord{\boldsymbol{n}^{k}}}$ is a polynomial of $\hat{n}$, its expectations with all Fock state with total photon number $m$ are equal. Thus, it suffices to calculate its expectation on the state $\ket{m,0}$ with $m$ photons in the first $S-1$ modes and $0$ in the last:
   \begin{align}
       \bra{m,0}\reallywidehat{\anormord{\boldsymbol{n}^{k}}}\ket{m,0} &=
          \sum_{j=0}^{k}\binom{k}{j}(m+S-2+j)_{j}(k-j)_{k-j} \nonumber \\
          &= k! \sum_{j=0}^{k}\binom{m+S-2+j}{j} \nonumber\\
          &= k! \binom{m+S-1+k}{k} = (m+S-1+k)_{k}.
   \end{align}
 \end{proof}

%% file: arik_thesis.bbl
\begin{thebibliography}{10}

\bibitem{Olivares4}
{\sc A.~Allevi, M.~Bina, S.~Olivares, and M.~Bondani}, {\em Homodyne-like
  detection scheme based on photon-number-resolving detectors}, International
  Journal of Quantum Information,  (2017), p.~1740016.

\bibitem{bachor2019guide}
{\sc H.~A. Bachor and T.~C. Ralph}, {\em A Guide to Experiments in Quantum
  Optics}, Wiley-VCH, third~ed., Oct 2019.

\bibitem{Beck:qc2016}
{\sc K.~M. Beck, M.~Hosseini, Y.~Duan, and V.~Vuleti{\'c}}, {\em Large
  conditional single-photon cross-phase modulation}, Proceedings of the
  National Academy of Sciences, 113 (2016), pp.~9740--9744.

\bibitem{Olivares3}
{\sc M.~Bina, A.~Allevi, M.~Bondani, and S.~Olivares}, {\em Phase-reference
  monitoring in coherent-state discrimination assisted by a photon-number
  resolving detector}, Scientific Reports, 6 (2016), p.~26025.

\bibitem{Olivares2}
{\sc M.~Bina, A.~Allevi, M.~Bondani, and S.~Olivares}, {\em Homodyne-like
  detection for coherent state-discrimination in the presence of phase noise},
  Opt. Express, 25 (2017), pp.~10685--10692.

\bibitem{brida2012quantum}
{\sc G.~Brida, L.~Ciavarella, I.~P. Degiovanni, M.~Genovese, L.~Lolli, M.~G.
  Mingolla, F.~Piacentini, M.~Rajteri, E.~Taralli, and M.~G.~A. Paris}, {\em
  Quantum characterization of superconducting photon counters}, New Journal of
  Physics, 14 (2012), p.~085001.

\bibitem{busch1995oper}
{\sc P.~Busch, M.~Grabowski, and P.~Lahti}, {\em Operational Quantum Physics},
  Lecture Notes in Physics Monographs, Springer Berlin, Heidelberg, 1995.

\bibitem{casella2002stat}
{\sc G.~Casella and R.~L. Berger}, {\em Statistical Inference}, Thomson
  Learning, second~ed., 2002.

\bibitem{MAURODARIANO2003205}
{\sc G.~M. D'Ariano, M.~G.~A. Paris, and M.~F. Sacchi}, {\em Quantum
  tomography}, 2003.

\bibitem{diestel2014joys}
{\sc J.~Diestel and A.~Spalsbury}, {\em The Joys of {H}aar Measure}, Graduate
  studies in mathematics, American Mathematical Society, 2014.

\bibitem{Walmsley5}
{\sc G.~Donati, T.~J. Bartley, X.~min Jin, M.~D. Vidrighin, A.~Datta,
  M.~Barbieri, and I.~A. Walmsley}, {\em Observing optical coherence across
  {F}ock layers with weak-field homodyne detectors}, Nature communications, 5
  (2014), p.~5584.

\bibitem{einstein1905uber}
{\sc A.~Einstein}, {\em Über die von der molekularkinetischen {T}heorie der
  {W}ärme geforderte {B}ewegung von in ruhenden {F}lüssigkeiten suspendierten
  {T}eilchen}, Annalen der Physik, 322 (1905), pp.~549--560.

\bibitem{fulton:qc1991a}
{\sc W.~Fulton and J.~Harris}, {\em Representation Theory: A First Course},
  no.~129 in Graduate Texts in Mathematics, Springer, New York, 1991.

\bibitem{gerrits2016super}
{\sc T.~Gerrits, A.~Lita, B.~Calkins, and S.~W. Nam}, {\em Superconducting
  Transition Edge Sensors for Quantum Optics}, Springer International
  Publishing, Cham, 2016, pp.~31--60.

\bibitem{glancy2012gradient}
{\sc S.~Glancy, E.~Knill, and M.~Girard}, {\em Gradient-based stopping rules
  for maximum-likelihood quantum-state tomography}, New Journal of Physics, 14
  (2012), p.~095017.

\bibitem{glauber1953coherent}
{\sc R.~J. Glauber}, {\em Coherent and incoherent states of the radiation
  field}, Phys. Rev., 131 (1963), pp.~2766--2788.

\bibitem{glauber1963quantum}
{\sc R.~J. Glauber}, {\em The quantum theory of optical coherence}, Phys. Rev.,
  130 (1963), pp.~2529--2539.

\bibitem{haag1992local}
{\sc R.~Haag}, {\em Local Quantum Physics}, Springer, 1992.

\bibitem{hernandez2021rapidly}
{\sc F.~Hernández and C.~J. Riedel}, {\em Rapidly decaying {W}igner functions
  are {S}chwartz functions}, Journal of Mathematical Physics, 63 (2022),
  p.~022104.

\bibitem{hertz1887ueber}
{\sc H.~Hertz}, {\em {U}eber einen {E}influss des ultravioletten {L}ichtes auf
  die electrische {E}ntladung}, Annalen der Physik, 267 (1887), pp.~983--1000.

\bibitem{hradil:qc2004a}
{\sc Z.~Hradil, J.~Rehacek, J.~Fiurasek, and M.~Jezek}, {\em Maximum-likelihood
  methods in quantum mechanics}, in Quantum State Estimation, Springer-Verlag,
  New York, 2004, pp.~163--172.

\bibitem{rehacek2007diluted}
{\sc J.~\ifmmode \check{R}\else \v{R}\fi{}eh\'a\ifmmode~\check{c}\else
  \v{c}\fi{}ek, Z.~Hradil, E.~Knill, and A.~I. Lvovsky}, {\em Diluted
  maximum-likelihood algorithm for quantum tomography}, Phys. Rev. A, 75
  (2007), p.~042108.

\bibitem{Koashi}
{\sc M.~Koashi, K.~Kono, M.~Matsuoka, and T.~Hirano}, {\em Probing the
  two-photon phase coherence of parametrically down-converted photons by a
  local oscillator}, Phys. Rev. A, 50 (1994), pp.~R3605--R3608.

\bibitem{Walmsley1}
{\sc A.~Kuzmich, I.~A. Walmsley, and L.~Mandel}, {\em Violation of {B}ell's
  inequality by a generalized {E}instein-{P}odolsky-{R}osen state using
  homodyne detection}, Phys. Rev. Lett., 85 (2000), pp.~1349--1353.

\bibitem{leonhardt:qc1997a}
{\sc U.~Leonhardt}, {\em Measuring the Quantum State of Light}, Cambridge
  University Press, Cambridge, UK, 1997.

\bibitem{LEONHARDT2}
{\sc U.~Leonhardt, M.~Munroe, T.~Kiss, T.~Richter, and M.~Raymer}, {\em
  Sampling of photon statistics and density matrix using homodyne detection},
  Optics Communications, 127 (1996), pp.~144 -- 160.

\bibitem{lorentz1992multi}
{\sc R.~A. Lorentz}, {\em Multivariate {B}irkhoff Interpolation}, Lecture Notes
  in Mathematics, Springer Berlin, Heidelberg, 1992.

\bibitem{lvovsky2009continuous}
{\sc A.~I. Lvovsky and M.~G. Raymer}, {\em Continuous-variable optical
  quantum-state tomography}, Rev. Mod. Phys., 81 (2009), pp.~299--332.

\bibitem{maxwell:qc2013a}
{\sc D.~Maxwell, D.~J. Szwer, D.~Paredes-Barato, H.~Busche, J.~D. Pritchard,
  A.~Gauguet, K.~J. Weatherill, M.~P.~A. Jones, and C.~S. Adams}, {\em Storage
  and control of optical photons using {R}ydberg polaritons}, Phys. Rev. Lett.,
  110 (2013), p.~103001.

\bibitem{migdall2013experimental}
{\sc A.~Migdall, S.~V. Polyakov, J.~Fan, and J.~C. Bienfang}, {\em
  Single-Photon Generation and Detection}, vol.~45 of Experimental Methods in
  the Physical Sciences, Academic Press, 2013.

\bibitem{Muller}
{\sc C.~R. M{ü}ller, M.~A. Usuga, C.~Wittmann, M.~Takeoka, C.~Marquardt, U.~L.
  Andersen, and G.~Leuchs}, {\em Quadrature phase shift keying coherent state
  discrimination via a hybrid receiver}, New Journal of Physics, 14 (2012),
  p.~083009.

\bibitem{nielsen:qc2001a}
{\sc M.~A. Nielsen and I.~L. Chuang}, {\em Quantum Computation and Quantum
  Information}, Cambridge University Press, Cambridge, UK, 2001.

\bibitem{Olivares1}
{\sc S.~Olivares, A.~Allevi, G.~Caiazzo, M.~G.~A. Paris, and M.~Bondani}, {\em
  Quantum tomography of light states by photon-number-resolving detectors}, New
  Journal of Physics, 21 (2019), p.~103045.

\bibitem{peyronel:qc2013a}
{\sc T.~Peyronel, O.~Firstenberg, Q.-Y. Liang, S.~Hofferberth, A.~Gorshkov,
  T.~Pohl, M.~Lukin, and V.~Vuletic}, {\em Quantum nonlinear optics with single
  photons enabled by strongly interacting atoms}, Nature, 488 (2012),
  pp.~57--60.

\bibitem{phillips2020advanced}
{\sc D.~S. Phillips}, {\em Advanced measurements for quantum photonics and
  quantum technologies}, PhD thesis, University of Oxford, 2020.

\bibitem{Walmsley3}
{\sc G.~Puentes, A.~Datta, A.~Feito, J.~Eisert, M.~B. Plenio, and I.~A.
  Walmsley}, {\em Entanglement quantification from incomplete measurements:
  applications using photon-number-resolving weak homodyne detectors}, New
  Journal of Physics, 12 (2010), p.~033042.

\bibitem{Walmsley2}
{\sc G.~Puentes, J.~S. Lundeen, M.~P.~A. Branderhorst, H.~B.
  Coldenstrodt-Ronge, B.~J. Smith, and I.~A. Walmsley}, {\em Bridging particle
  and wave sensitivity in a configurable detector of positive operator-valued
  measures}, Phys. Rev. Lett., 102 (2009), p.~080404.

\bibitem{Resch}
{\sc K.~J. Resch, J.~S. Lundeen, and A.~M. Steinberg}, {\em Quantum state
  preparation and conditional coherence}, Phys. Rev. Lett., 88 (2002),
  p.~113601.

\bibitem{schmudgen2017moment}
{\sc K.~Schm{\"u}dgen}, {\em The Moment Problem}, Graduate Texts in
  Mathematics, Springer International Publishing, 2017.

\bibitem{serafini2017quantum}
{\sc A.~Serafini}, {\em Quantum Continuous Variables: A Primer of Theoretical
  Methods}, CRC Press, 2017.

\bibitem{slussarenko2019photonic}
{\sc S.~Slussarenko and G.~J. Pryde}, {\em Photonic quantum information
  processing: A concise review}, Applied Physics Reviews, 6 (2019), p.~041303.

\bibitem{steck2007quantum}
{\sc D.~Steck}, {\em Quantum and atom optics}, 2007.

\bibitem{stuart1999advanced}
{\sc A.~Stuart, J.~K. Ord, and S.~Arnold}, {\em Advanced Theory of Statistics,
  Volume 2A: Classical Inference and the Linear Model}, Oxford University
  Press, sixth~ed., 1999.

\bibitem{sudarshan1963equivalence}
{\sc E.~C.~G. Sudarshan}, {\em Equivalence of semiclassical and quantum
  mechanical descriptions of statistical light beams}, Phys. Rev. Lett., 10
  (1963), pp.~277--279.

\bibitem{Thekkadath}
{\sc G.~S. Thekkadath, D.~S. Phillips, J.~F.~F. Bulmer, W.~R. Clements,
  A.~Eckstein, B.~A. Bell, J.~Lugani, T.~A.~W. Wolterink, A.~Lita, S.~W. Nam,
  T.~Gerrits, C.~G. Wade, and I.~A. Walmsley}, {\em Tuning between
  photon-number and quadrature measurements with weak-field homodyne
  detection}, Phys. Rev. A, 101 (2020), p.~031801.

\bibitem{Lukin}
{\sc J.~D. Thompson, T.~L. Nicholson, Q.-Y. Liang, S.~H. Cantu, A.~V.
  Venkatramani, S.~Choi, I.~A. Fedorov, D.~Viscor, T.~Pohl, M.~D. Lukin, and
  V.~Vuletic}, {\em Symmetry-protected collisions between strongly interacting
  photons}, Nature, 542 (2017), pp.~206--209.

\bibitem{Shadow}
{\sc J.~Tiedau, V.~S. Shchesnovich, D.~Mogilevtsev, V.~Ansari, G.~Harder, T.~J.
  Bartley, N.~Korolkova, and C.~Silberhorn}, {\em Quantum state and mode
  profile tomography by the overlap}, New Journal of Physics, 20 (2018),
  p.~033003.

\bibitem{wallentowitz1996unbalanced}
{\sc S.~Wallentowitz and W.~Vogel}, {\em Unbalanced homodyning for quantum
  state measurements}, Phys. Rev. A, 53 (1996), pp.~4528--4533.

\bibitem{weedbrook2012gaussian}
{\sc C.~Weedbrook, S.~Pirandola, R.~Garc\'{\i}a-Patr\'on, N.~J. Cerf, T.~C.
  Ralph, J.~H. Shapiro, and S.~Lloyd}, {\em Gaussian quantum information}, Rev.
  Mod. Phys., 84 (2012), pp.~621--669.

\bibitem{wilf2005generating}
{\sc H.~S. Wilf}, {\em Generatingfunctionology}, A K Peters/CRC Press,
  third~ed., 2005.

\bibitem{xu2020exp}
{\sc H.~Xu, F.~Xu, T.~Theurer, D.~Egloff, Z.-W. Liu, N.~Yu, M.~B. Plenio, and
  L.~Zhang}, {\em Experimental quantification of coherence of a tunable quantum
  detector}, Phys. Rev. Lett., 125 (2020), p.~060404.

\bibitem{Walmsley4}
{\sc L.~Zhang, H.~B.~Coldenstrodt-Ronge, A.~Datta, G.~Puentes, J.~Lundeen,
  X.-M. Jin, B.~Smith, M.~Plenio, and I.~Walmsley}, {\em Mapping coherence in
  measurement via full quantum tomography of a hybrid optical detector}, Nature
  Photonics, 6 (2012), p.~364.

\bibitem{zhang2012recursive}
{\sc L.~Zhang, A.~Datta, H.~B. Coldenstrodt-Ronge, X.-M. Jin, J.~Eisert, M.~B.
  Plenio, and I.~A. Walmsley}, {\em Recursive quantum detector tomography}, New
  Journal of Physics, 14 (2012), p.~115005.

\end{thebibliography}
